\documentclass[hyper]{JHEP3}

\input{epsf}
\usepackage{epsfig}
\usepackage{amssymb}
\usepackage{amsfonts}
\usepackage{amsbsy}
\usepackage[all]{xy}
\usepackage{amsmath}
\usepackage{enumerate}

\usepackage{amssymb,amscd}
\usepackage{mathrsfs}
\usepackage{amsmath,amsthm}
\usepackage{xspace}

\def\a{\alpha}
\def\b{\beta}
\def\c{\check }

\def\re{\mbox{Re }}
\def\im{\mbox{Im }}

\def\IC{\mathbb{C}}
\def\IH{\mathbb{H}}

\def\IZ{{\mathbb{Z}}}
\def\IR{{\mathbb{R}}}
\def\IP{\mathbb{P}}
\def\IQ{\mathbb{Q}}

\def\fg{\mathfrak{g}}

\def\CC {{\cal C}}
\def\CI {{\cal I}}
\def\CM {{\cal M}}
\def\CN{{\cal N}}
\def\CK {{\cal K}}
\def\CN {{\cal N}}
\def\CR {{\cal R}}
\def\CD {{\cal D}}
\def\CF {{\cal F}}

\def\CP {{\cal P }}
\def\CL {{\cal L}}
\def\CV {{\cal V}}

\def\CX {{\cal X}}

\def\CO {{\cal O}}

\def\CG {{\cal G}}
\def\CH {{\cal H}}
\def\CB {{\cal B}}
\def\CS {{\cal S}}
\def\CA{{\cal A}}
\def\CK{{\cal K}}

\def\CU{{\cal U}}

\def\CT{{\cal T}}

\def\half{\frac{1}{2}}
\def\qtr{\frac{1}{4}}

\renewcommand{\Im}{{\rm Im }}
\renewcommand{\Re}{{\rm Re }}

\def\one{{\hbox{ 1\kern-.8mm l}}}

\def\sgn{{\rm sgn\,}}

\def\p{\partial}

\def\be{\bar{e}}

\def\zb{\bar {z}}

\def\half{\frac{1}{2}}

\def\hk{hyperk\"ahler\xspace}

\def\kahler{K\"ahler\xspace}
\newcommand{\abs}[1]{\lvert#1\rvert}
\newcommand{\norm}[1]{\lVert#1\rVert}
\def\pz{{(\zeta)}}
\newcommand{\ti}[1]{\textit{#1}}

\def\p{\partial}
\def\pb{\bar\partial}

\def\re{\mbox{Re }}
\def\im{\mbox{Im }}

\def\IC{\mathbb{C}}

\def\IZ{{\mathbb{Z}}}
\def\IR{{\mathbb{R}}}
\def\IP{\mathbb{P}}
\def\IQ{\mathbb{Q}}

\def\CC {{\cal C}}
\def\CI {{\cal I}}
\def\CM {{\cal M}}
\def\CN{{\cal N}}
\def\CK {{\cal K}}
\def\CN {{\cal N}}
\def\CR {{\cal R}}
\def\CD {{\cal D}}
\def\CF {{\cal F}}

\def\CP {{\cal P }}
\def\CL {{\cal L}}
\def\CV {{\cal V}}

\def\CX {{\cal X}}

\def\CO {{\cal O}}

\def\CG {{\cal G}}
\def\CH {{\cal H}}
\def\CB {{\cal B}}
\def\CS {{\cal S}}
\def\CA{{\cal A}}
\def\CK{{\cal K}}

\def\CU{{\cal U}}

\def\CT{{\cal T}}

\def\half{\frac{1}{2}}

\def\be{\begin{equation}
}

\def\ee{\end{equation}}

\newcommand{\inprod}[1]{\langle#1\rangle}
\newcommand\N{{\cal N}}
\renewcommand\sf{{\mathrm{sf}}}

\newcommand\eps{\epsilon}

\newcommand\RH{{\mathrm{RH}}}
\newcommand\WKB{{\mathrm{WKB}}}
\newcommand\hyper{{\mathrm{hyper}}}
\newcommand\vect{{\mathrm{vector}}}
\newcommand\BPS{{\mathrm{BPS}}}

\newcommand\flav{{\mathrm{flavor}}}
\newcommand\juggle{juggle\xspace}
\newcommand\juggles{juggles\xspace}
\newcommand\juggleing{juggling\xspace}
\newcommand\nsing{{l}}

\newcommand\id{{\mathbf{1}}}

\newcommand{\ccwarrow}{\text{\Large$\curvearrowleft$}}

\newcommand{\hkq}{/\!/\!/}

\newcommand{\bS}{\mathbf S}

\DeclareMathOperator{\End}{{End}}
\DeclareMathOperator{\Pexp}{{Pexp}}
\DeclareMathOperator{\Tr}{{Tr}}
\DeclareMathOperator{\Hom}{{Hom}}

\newcommand{\insfig}[3]{\begin{figure}[htbp] \centering \includegraphics[width=#1]{figures/#2-crop.pdf} \caption{#3} \label{fig:#2} \end{figure}}

\newcommand{\instwofigs}[6]{
\begin{figure}
\begin{minipage}[b]{0.5\linewidth} 
\centering \includegraphics[width=#1]{figures/#2-crop.pdf} \caption{#3} \label{fig:#2}
\end{minipage}
\hspace{0.4cm} 
\begin{minipage}[b]{0.45\linewidth}
\centering \includegraphics[width=#4]{figures/#5-crop.pdf} \caption{#6} \label{fig:#5}
\end{minipage}
\end{figure}
}

\newcommand{\insthreefigs}[9]{
\begin{figure}
\begin{minipage}[b]{0.31\linewidth}
\centering \includegraphics[height=#1]{figures/#2-crop.pdf} \caption{#3} \label{fig:#2}
\end{minipage}
\hspace{0.25cm}
\begin{minipage}[b]{0.31\linewidth}
\centering \includegraphics[height=#4]{figures/#5-crop.pdf} \caption{#6} \label{fig:#5}
\end{minipage}
\hspace{0.25cm}
\begin{minipage}[b]{0.31\linewidth}
\centering \includegraphics[height=#7]{figures/#8-crop.pdf} \caption{#9} \label{fig:#8}
\end{minipage}
\end{figure}
}

\title{Wall-crossing, Hitchin Systems, and the WKB Approximation}

\author{Davide Gaiotto$^1$, Gregory W. Moore$^2$, Andrew Neitzke$^3$\\
$^1$ School of Natural Sciences, Institute for Advanced Study, \\
Princeton, NJ 08540, USA\\
 $^2$ NHETC and Department of Physics and Astronomy,
Rutgers University,\\
Piscataway, NJ 08855--0849, USA\\
$^3$ Department of Physics, Harvard University,\\
Cambridge, MA 02138, USA\\
\\
{\tt dgaiotto@ias.edu, gmoore@physics.rutgers.edu, neitzke@physics.harvard.edu} }

\abstract{ We consider BPS states in a large class of $d=4$, $\N=2$
field theories, obtained by reducing six-dimensional $(2,0)$
superconformal field theories on Riemann surfaces, with defect
operators inserted at points of the Riemann surface. Further
dimensional reduction on $S^1$ yields sigma models, whose target
spaces are moduli spaces of Higgs bundles on Riemann surfaces with
ramification.  In the case where the Higgs bundles have rank $2$, we
construct canonical Darboux coordinate systems on their moduli
spaces.  These coordinate systems are related to one another by
Poisson transformations associated to BPS states, and have
well-controlled asymptotic behavior, obtained from the WKB
approximation. The existence of these coordinates implies the
Kontsevich-Soibelman wall-crossing formula for the BPS spectrum.
This construction provides a concrete realization of a general
physical explanation of the wall-crossing formula which was proposed
in \cite{Gaiotto:2008cd}.  It also yields a new method for computing
the spectrum using the combinatorics of triangulations of the
Riemann surface.}

\begin{document}

\bibliographystyle{utphys}


\section{A narrative table of contents}
\label{sec:IntroductionAndSummary}

Supersymmetric gauge theories have been a plentiful source of
delightful surprises both in theoretical physics and in mathematics.
A particularly rich class of theories are those with $d=4, \CN=2$
supersymmetry (henceforth referred to as $\CN=2$ supersymmetry). In
this context an important and distinguished subspace of the Hilbert
space is the space of BPS states. These states are in ``small'' or
``rigid'' representations of the supersymmetry algebra and this
rigidity leads to an amenability to analysis which is in turn   the
foundation for the exact results known about these theories.
Although BPS representations of $\CN=2$ are rigid, the BPS subspace
nevertheless depends nontrivially both on the ultraviolet parameters
as well as   on the choice of quantum vacuum of the theory, a
phenomenon known as wall-crossing.

Despite the fundamental nature of this BPS subspace, there is no
algorithm for computing it given an arbitrary $ \CN=2$ theory with
its choice of vacuum. Indeed the BPS spectrum is only known
explicitly in a small set of examples where special \emph{ad hoc}
techniques can be applied. The main result of this paper is a new
algorithm for determining the BPS spectrum of a certain infinite set
of $ \CN=2$ theories. The theories to which our methods apply are
described in Section \ref{sec:physics} and the new algorithm is
described in Section \ref{sec:UniversalStokesMatrix} of this paper.
Our result is promising because there are indications that
generalizations of the algorithm will apply to a much wider set of $
\CN=2$ theories.

The class of theories to which our main result applies are linear
quiver gauge theories with $SU(2)$ factor gauge groups at the nodes.
These are part of a larger class of distinguished $\CN=2$ gauge
theories, described extensively in Section \ref{sec:physics}. This
larger class of theories, which we call $\CS$ (for ``six''),
originates from compactifications of M5-branes  on a punctured
Riemann surface, $C$.\footnote{To be more precise, we consider the
low energy $(2,0)$ superconformal field theory resulting from the
decoupling of gravity. We then compactify this theory on $C$ with a
partial twisting of the $d=6$ $(2,0)$ superalgebra so as to preserve
$\CN=2$ supersymmetry.}  The superconformal $(2,0)$ theories have an
A-D-E classification and so we can label elements of $\CS$ by a
simply laced ``gauge group,'' a Riemann surface $C$, and a
decoration of the punctures of $C$ by ``defect operators.'' Theories
in class $\CS$  have the important property that they enjoy a close
relationship with Hitchin systems. This relation, which is
absolutely central to this paper, is revealed when one further
compactifies to three dimensions on a circle.  At low energies the
three-dimensional effective theory is a $d=3, \CN=4$ sigma model
with target space $\CM$. This target space may be identified, as a
Riemannian manifold, with  the moduli space of solutions to a
Hitchin system. To justify this, the essential observation is that
instead of compactifying on $C$ and then on $S^1$, we can --- by a
QFT version of the ``Fubini theorem'' --- construct the same
effective theory in three dimensions by first compactifying on $S^1$
and then on $C$. The first compactification on $S^1$ leads to a
five-dimensional supersymmetric Yang-Mills theory. The subsequent
compactification of the (twisted) $d=5$ super-Yang-Mills theory on
$C$ then leads to BPS equations which are well known to be the
Hitchin equations. In particular, if we begin with $K$ M5-branes
(i.e. the superconformal $u(K)$ $(2,0)$ theory in six dimensions)
then the Hitchin equations are equations
\eqref{eq:hitchin-1}-\eqref{eq:hitchin-3} below:
\begin{align}
F + R^2 [\varphi , \bar\varphi] & = 0,   \\
  \p_{\bar z} \varphi  + [ A_{\bar z}  ,\varphi] & = 0,  \\
    \p_z \bar\varphi  + [ A_z,\bar\varphi]  & =
 0,
\end{align}
where $R$ is the radius of the circle, $F$ is the fieldstrength of a
$u(K)$ gauge field $A$ on $C$ and $\varphi$ is the  $(1,0)$ part of
a $1$-form valued in the adjoint. $z$ is a local holomorphic
coordinate on $C$.

The description of the Hitchin system is incomplete without
specifying boundary conditions on $(A,\varphi)$ at the punctures of
$C$. At these punctures  the fields $(A,\varphi)$   have
singularities. Physically these singularities encode the somewhat
mysterious ``defect operators'' of the six-dimensional
superconformal theory (and in practice the defect operators are
defined by the specified singularities of $(A,\varphi)$).  The
simplest operators to consider -- and the ones upon which we focus
-- arise from intersections, at the punctures of $C$, of the
multiple $u(K)$ M5-brane theory with ``transverse'' singly-wrapped
M5-branes. By transverse we mean the following. In general the curve
$C$ is embedded in some hyperkahler manifold $Q$ as a holomorphic
curve. The gravitational decoupling limit allows us to replace $Q$
by a neighborhood of the zero-section of $T^*C$. The transverse
fivebranes fill the four-dimensional spacetime $\IR^{1,3}$ of the
$\CN=2$ theory and run along fibers of the projection $T^*C \to C$.
In Section \ref{sec:physics} we show how to translate this physical
picture into conditions on the singularities of $(A,\varphi)$. The
singularities are described in detail in Section
\ref{subsubsec:MapToHitchin}; see (\ref{eq:SimpleAtZero}),
(\ref{eq:SimpleAtInfty}), and (\ref{eq:SimplePoleResidue}), for the
case of regular singularities, and Section
\ref{subsubsec:Brane-bending}, equation (\ref{eq:ISPatInfty}),
(\ref{eq:GaugeFldInfty}), as well as Section
\ref{subsec:HitchinForAD}, for the case of irregular singularities.

The Hitchin system plays a central role throughout the paper and
Section \ref{sec:hitchin} of the paper summarizes the basic facts we
need about Hitchin systems. The mathematically-oriented reader can
skip Section \ref{sec:physics}  and proceed with the brief summary
in Section \ref{sec:hitchin}, although the rules for finding BPS
states might then appear somewhat unmotivated.

A particularly important set of examples of theories in the class
$\CS$ are provided by Witten's geometric construction of $ \CN=2$
theories using arrays of NS5- and D4-branes \cite{Witten:1997sc}.
These are often summarized by figures such as Figure
\ref{fig:witten-3}.  Much of Section \ref{sec:physics}  is merely a
review of Witten's construction and may be safely skipped by readers
familiar with \cite{Klemm:1996bj,Witten:1997sc}. We would note
however that Sections \ref{subsubsec:MapToHitchin},
\ref{subsubsec:LinConfQuivMatt}, and  \ref{subsubsec:Brane-bending}
contain some new points concerning how to use the physical picture
to describe the boundary conditions on $(A,\varphi)$.  In addition,
Sections \ref{sec:surprise} and \ref{sec:flavor} contain some novel
remarks on isomorphisms between Hitchin moduli spaces and on flavor
symmetries, respectively.  In particular, the isomorphisms of
Section \ref{sec:surprise} should be of some mathematical interest.
A similar class of isomorphisms has been independently noted
recently by Philip Boalch \cite{irkm}.

The geometry of the Hitchin system on $C$  beautifully encodes the
key data of the low energy effective Seiberg-Witten  theory of the
\emph{four-dimensional} $\CN=2$ theory in $\CS$. First, the
Seiberg-Witten curve,  which is a branched cover $\Sigma$ of $C$, is
nicely presented as the spectral curve of the Hitchin system. Thus
$\Sigma \subset T^*C$ and moreover the sheets of the cover may be
labeled by  the eigenvalues of $\varphi$. Quite generally, a
cotangent bundle $T^*C$ is canonically endowed with a symplectic
form which is further canonically trivialized by a one-form.
Restricting this one-form to $\Sigma$ one obtains   the
Seiberg-Witten differential, denoted $\lambda$.

Since our main theme is the BPS spectrum it behooves us to
understand how to describe this spectrum in the context of Hitchin
systems. Given the origin of the theories from M5-branes wrapped on
$C$ one can systematically understand the BPS states in terms of
strings in the six-dimensional theory, which in turn originate from
 open M2-branes ending on M5-branes \cite{Strominger:1996ac}.
 The translation of this description
 of BPS states to the language of  Hitchin systems
is described in Sections \ref{subsubsec:20AndCoulomb} and
\ref{sec:bps-strings}.  In the case $K=2$ we recover a well-known
construction of Klemm et. al. \cite{Klemm:1996bj}:  BPS states are
associated with curves on $C$ that minimize the local tension of the
strings.  Expressed mathematically, near a point $z_0$ on the curve
we choose a branch of the cover and define a local coordinate  by $w
= \int^z_{z_0} \lambda$.  The curve is then a straight line in the
$w$-plane.  BPS (half)-hypermultiplets are associated with curves
which begin and end on branch points of the cover $\Sigma \to C$,
while BPS vectormultiplets are associated with curves which are
closed. For reasons explained below, we call either of these
\emph{finite WKB curves}.  See Figures \ref{fig:bps-disc} and
\ref{fig:bps-annulus} for illustrations. The central charge of the
corresponding BPS state is $Z_\gamma = \pi^{-1} \oint_{\gamma}
\lambda$.  An important point below is that the phase of this
central charge is the angle $\vartheta$ between the straight line in
the $w$ plane and the ${\rm Re}(w)$-axis. When $K>2$ the analogous
description of BPS states is more involved, and makes use of string
webs on $C$.

As so often happens, our main result was in fact not the initial
goal of this work.  Rather, the original motivation came from a
recent construction \cite{Gaiotto:2008cd} of hyperk\"ahler metrics,
which in turn was motivated by the search for the physical
underpinnings of the Kontsevich-Soibelman wall-crossing formula
\cite{ks1}. In order to make the remainder of our summary
intelligible we must first recall here the most essential points of
\cite{Gaiotto:2008cd}. A more technical summary of
\cite{Gaiotto:2008cd} can be found in Section \ref{sec:review}.

The starting point of  \cite{Gaiotto:2008cd} is the compactification
of a general $d=4, \CN=2$ theory on a circle of radius $R$. As
mentioned above, at low energies the theory is a three-dimensional
sigma model whose target space   $\CM$  must carry a \hk\ metric.
The space $\CM$ has a fibration $\CM \to \CB$ where $\CB$ is the
moduli space of vacua of the four-dimensional theory  and the
generic fiber is a real torus of dimension $\dim \CB$.
  As $R\to
\infty$ the metric on $\CM$ becomes exponentially close to a simple
and explicit metric  which is \hk, but has singularities in real
codimension two. This metric is called the ``semiflat metric'' and
denoted
 $g^{\sf}$. It is easily derived by naive dimensional reduction
 along $S^1$ of the Seiberg-Witten effective Lagrangian.
Quantum corrections at finite values of $R$ smooth out $g^{\sf}$.
Moreover, these quantum corrections depend solely on the spectrum of
BPS states. Since that spectrum is itself a function of the
(four-dimensional) vacuum, the smoothness of the metric implies a
wall-crossing formula. In \cite{Gaiotto:2008cd} it is shown that
this is precisely the Kontsevich-Soibelman wall-crossing formula
(KSWCF).

From the purely mathematical viewpoint one can view
\cite{Gaiotto:2008cd} as giving a construction of \hk\ metrics from
the following three pieces of data:

\begin{itemize}

\item D1: A local system of lattices $\hat \Gamma\to \CB$ with an
integral antisymmetric form $\langle,\rangle$ (possibly degenerate)
on the fibers. Letting  $\Gamma$ be the  local system of symplectic
lattices obtained by dividing by the radical of $\hat \Gamma$, we
require $\CB$ to have real dimension equal to the rank of $\Gamma$.

\item D2:  A central charge function  $Z \in {\rm Hom}(\hat \Gamma,
\IC)$ such that $\langle dZ, dZ \rangle = 0$ where $d$ is the
differential along $\CB$.

\item D3: A piecewise constant function $\Omega: \hat \Gamma \to \IZ$
satisfying the KSWCF.

\end{itemize}

When a physical $\CN=2$ theory provides the data D1, D2, D3 the
moduli space $\CM$ of the physical problem can be identified with
the total space of the fibration
$\Gamma^* \otimes \IR/2\pi \IZ$ over $\CB$. Thus we expect that just given the
data D1, D2, D3 we can construct a \hk\ metric on
$\CM=\Gamma^* \otimes \IR/2\pi \IZ $, as indeed proves to be the
case. In fact, we construct a \emph{family} of \hk\ metrics on
$\CM$. The parameters of the family are described below.

An essential part of the construction of the metrics on $\CM$
involves the twistor description of \hk\ metrics. Exploiting the
fact that $\hat \Gamma^* \otimes \IR/2\pi \IZ$ has a fibration by
tori one reduces the construction of suitable holomorphic data on
twistor space   to the construction of a certain map $\CX^\RH: \CM
\times \IC^\times \to \hat \Gamma^* \otimes \IC^\times$,\footnote{In
\cite{Gaiotto:2008cd} this map was just called $\CX$.} where the
factor $\IC^\times$ in the domain is part of the twistor sphere. It
is very convenient to let $\CX^\RH_{\gamma}: \CM \to \IC^\times$ be
the contraction of $\CX^\RH$ with $\gamma\in \hat \Gamma$. We will
refer to these functions as \emph{Darboux coordinates}.
\footnote{There is an abuse of terminology here which we regret.
Once one chooses a basis $\{\gamma_i\}$ of $ \Gamma$ compatible with
a Lagrangian decomposition together with a lifting to $\hat \Gamma$
then $\log \CX_{\gamma_i}^\RH$ truly provide a system of Darboux
coordinates on the holomorphic symplectic manifold $\CM$. } Note
that $\CX^\RH_{\gamma}\CX^\RH_{\gamma'} = \CX^\RH_{\gamma+\gamma'}$.
The map $\CX^\RH$ is subject to a list of defining properties. The
full list of detailed properties is recalled in Section
\ref{sec:review} but three crucial properties must be stated here:

\begin{itemize}

\item  P1:  First,  the
Poisson structure is defined by equation (\ref{eq:poisson}):
\begin{equation}
\{ \CX^\RH_{\gamma}, \CX^\RH_{\gamma'} \} = \langle \gamma,\gamma'
\rangle \CX^\RH_{\gamma+\gamma'}.
\end{equation}

\item
P2: Second, the   $\CX^\RH_{\gamma}$ are asymptotic to the analogous
functions $\CX_{\gamma}^{\sf}$ associated with the semiflat metric
$g^{\sf}$.  Let $\zeta \in \IC^\times$ be an element of the twistor
sphere. The semiflat Darboux coordinates can be written very
explicitly as
\begin{equation}\label{eq:P2-sf}
\CX_{\gamma}^{\sf} := \exp \left( \pi R\zeta^{-1} Z_\gamma + i
\theta_\gamma + \pi R \zeta \bar Z_\gamma \right),
\end{equation}
where $\theta_\gamma: \hat \Gamma^*\otimes \IR/2\pi \IZ \to \IR/2\pi
\IZ$ are canonically defined by contraction. We   demand that
$\CX_{\gamma}^\RH \sim \CX_{\gamma}^{\rm sf}$
 \emph{both} for
$\zeta\to 0, \infty$ \emph{and} for $R\to \infty$.

\item
 P3: Third, the analytic structure of $\CX^\RH$ as a function of $\zeta$
 is constrained as follows. Define  the  \emph{BPS rays}  to be the rays
$\ell_{\gamma,u} := \{ \zeta: Z_\gamma(u)/\zeta \in \IR_-\}$. Then,
as $\zeta$ crosses a BPS ray $\ell_{\gamma_0,u}$ the
$\CX^\RH_{\gamma}$ are discontinuous by a Poisson transformation
$\CK_{\gamma_0}^{\Omega(\gamma_0;u)}$ where\footnote{Actually, we
should consider all multiples of $\gamma_0$, thus the correct
transformation to use is $\prod_{\gamma_0'' \parallel \gamma_0}
\CK_{\gamma_0''}^{\Omega(\gamma_0'';u)}$. In the examples we study
only a single charge will contribute to the discontinuity.}
\begin{equation}
\CK_{\gamma_0} : \CX^\RH_\gamma \to \CX^\RH_{\gamma} (1 \pm
\CX^\RH_{\gamma_0} )^{\langle \gamma, \gamma_0 \rangle}.
\end{equation}
The transformations $\CK_{\gamma_0}$ will be referred to as \emph{KS
transformations}. For the more precise equation (in particular the
choice of $\pm$ sign) see \eqref{eq:KS-Transfm} below. In addition
$\CX_\gamma^\RH$ must be \emph{holomorphic} (without any
singularities) as a function of $\zeta$ on the complement of the set
of BPS rays $\ell_{\gamma,u}$ with $\Omega(\gamma;u) \not=0$.
\end{itemize}

As explained in \cite{Gaiotto:2008cd}, from the functions
$\CX^\RH_{\gamma}$ one can recover the \hk metric on $\CM$. The
resulting metric smoothes out the real codimension two singularities
of $g^{\sf}$ (but some real codimension four singularities might
remain).

Returning to the physical viewpoint, the functions
$\CX^\RH_{\gamma}$ have nice interpretations in terms of line
operator expectation values as well as elements in a chiral ring in
a three-dimensional topological field theory. We hope to describe
these aspects of the $\CX^\RH_\gamma$ elsewhere (we again touch on
this point briefly in Remark 3 of Appendix \ref{app:MonodromyX}).

We said above that we obtain a family of \hk\ metrics. To understand
this first note that the  radical of $\hat \Gamma$ is, physically, a
lattice of flavor charges. We have
\begin{equation}
0 \rightarrow \Gamma_{\rm flavor} \rightarrow \hat \Gamma
\rightarrow \Gamma \rightarrow 0
\end{equation}
and the symplectic lattice $\Gamma$ is the lattice of electric and
magnetic gauge charges. The manifold $\hat \Gamma^*\otimes \IR/2\pi
\IZ$ is foliated by copies of $\CM$. Moreover, the
$\CX^\RH_{\gamma}$ for $\gamma \in \Gamma_{\rm flavor}$ take the
exact form (\ref{eq:P2-sf}), where $Z_\gamma$ encode hypermultiplet
masses and $\theta_\gamma$ encode flavor Wilson lines. These
parameters, together with $R$, parameterize the family of \hk\
metrics on $\CM$.

The last aspect of \cite{Gaiotto:2008cd} we must recall is the
explicit construction of $\CX^\RH$. This is done by a series of
maneuvers using the properties P2 and P3 to characterize $\CX^\RH$
as a solution of a Riemann-Hilbert problem that is in turn
equivalent to an integral equation. This integral equation,
incidentally, turns out to be a version of the Thermodynamic Bethe
Ansatz (TBA).\footnote{Another relation between four-dimensional
super Yang-Mills theory and the TBA has recently been discussed by
Nekrasov and Shatashvili \cite{Nekrasov:2009ui}.} This TBA equation
can then be solved by iteration \emph{provided} the radius $R$ is
large. The result is an explicit series expansion in terms of
multiple integrals whose integrands are small when
$\CX_{\gamma}^{\sf}$ is small.  Hence we obtain an explicit
construction of the \hk\ metric. We must stress that $R$ should be
large in order to justify the solution of the integral equation by
iteration. (An important fact used here is that $\CX_\gamma^{\sf}$
are exponentially small on $\ell_{\gamma,u}$ as $R\to \infty$.)

Having recalled the main features of \cite{Gaiotto:2008cd} we can at
last return to describing  the original goal in writing the present
paper: It is to give an alternative construction of the
$\CX^\RH_{\gamma}$ which does not rely on the integral equation and
is valid for all $R$. In this paper we will indeed give an
alternative construction of the functions $\CX^\RH_{\gamma}$ for the
theories in $\CS$  associated to $SU(2)$ Hitchin systems.  Our
definition is indeed sensible for \emph{all values of $R$}. Moreover
the new construction lends itself to elegant geometrical
verifications of the key defining properties P1, P2, and (part of)
P3.\footnote{The reader experienced with the Thermodynamic Bethe
Ansatz may view our results as a broad generalization of the work
\cite{Dorey:1999uk}, where solutions to TBA equations (in a
``conformal limit'') were basically constructed from the monodromy
data of a holomorphic connection.}

Before explaining the new construction we must confess at the outset
that one difficulty will remain unresolved. Concerning the behavior
of the Darboux coordinates at small $R$  there is some tension
between this paper and \cite{Gaiotto:2008cd}. In
\cite{Gaiotto:2008cd} we proposed that the TBA equation would have a
regular solution for all $R$.  This would yield a $\CX^\RH$ with no
poles in the $\zeta$-plane.\footnote{More precisely, $\CX^\RH$ is
holomorphic in the complement of the set of BPS rays with
$\Omega\not=0$.}   The results of this paper suggest that the truth
might be more complicated:  we indeed find a natural candidate for
$\CX^\RH$, and it is indeed defined for all $R$, but for small $R$,
we are not able to show that it is pole-free. On the other hand,  it
is hard to envision a scenario where there are two \ti{different}
$\CX^\RH$, one with poles and one without. So we see two reasonable
options. One option is that the $\CX^\RH_\gamma$ of this paper
actually do not have poles. They could then be identified with
solutions of the integral equation of \cite{Gaiotto:2008cd} for all
$R$.  The other option is that the $\CX^\RH_\gamma$ of this paper do
have poles.  In that case they are, strictly speaking, not solutions
of the integral equation of \cite{Gaiotto:2008cd} for small $R$. The
appearance of such ``extra poles'' as a parameter is varied is
well-known in the literature on the TBA (see for example
\cite{Dorey:1996re}), and can be dealt with in that context.
Clearly, these matters deserve further attention!

Now let us summarize the new  construction of the
$\CX^\RH_{\gamma}$. The key idea   begins with the fact that
Hitchin's moduli space $\CM$ is \emph{also} a moduli space of flat
connections with fixed monodromy\footnote{Since the Hitchin system
may have both regular and irregular singularities, the ``monodromy
data'' includes Stokes data.} around the punctures of $C$. Indeed,
given a solution of the Hitchin equations we can form a non-unitary
connection (equation \eqref{eq:def-CA} below)
\begin{equation}
\CA : = \frac{R}{\zeta} \varphi + A + R \zeta \bar \varphi,
\end{equation}
and the Hitchin equations imply this is a flat connection on $C$ for
any $\zeta \in \IC^\times$. Conversely, the flatness of such a
connection for all $\zeta$ implies that $(A,\varphi)$ solve the
Hitchin equations. The next observation is that Fock and Goncharov
have constructed a beautiful set of coordinates on the moduli space
of flat connections, using the data
of a triangulation of $C$ \cite{MR2233852}. (This useful set of
coordinates is available only if $C$ has at least one puncture and
hence we will not attempt to extend our construction of the
$\CX_\gamma$ beyond that case, even though the physical theory makes
sense when $C$ has no punctures.) We will use the Fock-Goncharov
coordinates to construct our functions $\CX_{\gamma}$. In outline
our program is the following: First, given an angle\footnote{The
periodicity of $\vartheta$ can be an integer multiple of $2\pi$, or
it might even live in the universal cover $\IR$.} $\vartheta$, we
define a distinguished triangulation which we call a \emph{WKB
triangulation}. Second, applying the Fock-Goncharov construction to
that triangulation we get a set of functions
$\CX_{\gamma}^{\vartheta}: \CM \times \IC^\times \to \IC^\times$. We
then use the $\CX_{\gamma}^{\vartheta}$ in turn to construct
$\CX^\RH_{\gamma}$, by specializing $\vartheta = {\rm arg}(\zeta)$.
Third,  we show that the resulting functions satisfy the defining
properties outlined in Section \ref{sec:review} (in particular
P1, P2, P3). Let us now sketch how  this program is accomplished in
slightly more detail.

In Section \ref{sec:fg-coordinates} we recall the construction of
Fock and Goncharov \cite{MR2233852}. We deviate from their
discussion in two ways. One rather minor difference is that we
prefer to use \emph{decorated triangulations}. In the case of
regular singular points (on which we mostly focus), these are ideal
triangulations whose vertices are the singular points $\CP_i$ of the
Hitchin system, but where we add an extra piece of data at each
point $\CP_i$.  Specifically, we consider flat sections $s$ solving
$(d+\CA)s=0$, and the decoration of $\CP_i$ consists in choosing a
flat section $s_i$, defined up to scale, in a neighborhood of
$\CP_i$. Such a flat section is necessarily an eigenvector of the
monodromy around $\CP_i$, so equivalently, the decoration is a
choice of one of the two eigenlines of that monodromy.  (An
analogous notion of decorated triangulation for irregular singular
points is explained in Section \ref{sec:irregular}.)  The second,
more important, deviation from the work of Fock and Goncharov is
that
 the existence of the vectormultiplets in the BPS spectrum forces us
 to extend the notion of triangulation to include more elaborate objects
which we call  ``limit triangulations.'' These are described in
Section \ref{sec:limit-triangulation}. The heart of the
Fock-Goncharov construction is to use ``overlaps'' of the (parallel
transport of the) flat sections $s_i$ to describe the monodromy of
the flat connection $\CA$. In Appendix \ref{app:MonodromyX} we
explain how this can be done. The procedure naturally leads to the
key definition of equation (\ref{eq:cross-ratio-wedges}).

That the  property P1 of the $\CX^\RH_\gamma$ will emerge
correctly can already be seen nicely at this stage. The Hitchin
moduli space has a natural holomorphic symplectic form, given in
(\ref{eq:symplectic}),
\begin{equation}
\varpi_\zeta = \frac{1}{2}\int_{C} {\Tr} \delta \CA \wedge \delta
\CA.
\end{equation}
The corresponding Poisson brackets of the Fock-Goncharov coordinates
then take the simple form (\ref{eq:fg-poisson}). To be
self-contained, we give an elementary derivation of these Poisson
brackets in Sections \ref{sec:ham-flows}-\ref{sec:poisson}, using
results from Appendix \ref{app:poorman}.

A second key defining property (P2 above)   of the coordinates
$\CX^\RH_{\gamma}$ is their asymptotic behavior for $\zeta \to
0,\infty $ and $R \to \infty$.  It is this property that motivates
our definition of a  WKB triangulation.  As described in Section
\ref{sec:WKB-triangulation}, we define WKB curves of phase
$\vartheta$ to satisfy $\langle \lambda, \p_t \rangle = e^{i
\vartheta}$. Of course, we have already met this condition above,
when discussing BPS states! It is equivalent to the assertion that
in the local coordinate $w = \int^z_{z_0} \lambda$, where $z_0$ is a
point on the curve, the curve is a straight line parallel to $e^{i
\vartheta}$. These WKB curves tend to be ``captured'' by the
singularities (as shown in the local analysis of Section
\ref{sec:wkb-local}) and hence the generic WKB curve begins and ends
on a singularity.  The WKB triangulation is then defined by choosing
a suitable finite set of ``topologically distinct'' generic WKB
curves using a procedure explained in Sections
\ref{subsec:GlobalBehaviorWKB} and \ref{subsec:DefiningWKB-Triang}.
Moreover, there is a canonical decoration given by choosing the flat
section which becomes exponentially small along a WKB curve
plummeting into a singularity. This choice of decorated
triangulation  is motivated by the WKB analysis  (with small
parameter $\zeta, 1/\zeta$ or $1/R$) of the equation $(d+\CA)s=0$
for the flat sections. Recall that in the WKB approximation,
exponentially small wavefunctions can be computed reliably, but
exponentially large wavefunctions are ambiguous by the addition of
an unknown exponentially small component. For this reason we must
take care when computing ``overlaps'' of flat sections $s_i$
transported from the different locations $\CP_i$ of $C$: We must
transport these sections along WKB curves.  In summary, given an
angle $\vartheta$ and a vacuum $u\in \CB$ -- or, better, a
Seiberg-Witten differential $\lambda$ -- we have a canonically
determined decorated triangulation. It will be denoted as
$T_\WKB(\vartheta,u)$ or as $T_\WKB(\vartheta, \lambda^2)$. The
second notation reflects the fact that the edges are unoriented, and
hence only depend on the quadratic differential $\lambda^2$.

Turning now to the third key property P3 we must consider how
different WKB triangulations are related as we vary $\vartheta$ at
fixed $\lambda$. Quite generally, different decorated triangulations
(not necessarily of WKB type)  can be turned into each other by a
series of elementary transformations which we refer to as the
\emph{flip}, \emph{juggle}, and \emph{pop}.   We may view the
decorated triangulations as objects in a groupoid, and the flips,
  \juggles , and pops are elementary morphisms which generate all other
morphisms in the groupoid.  A flip is simply the standard
transformation of flipping an edge within a quadrilateral formed by
two triangles, as in Figure \ref{fig:flip}. The decoration is
unchanged.  A pop, on the other hand, leaves the triangulation
unchanged but alters the choice of distinguished eigenline at a
specified vertex. When $\CA$ has structure group $SL(2,\IC)$ and the
singularity is a regular singularity with diagonalizable monodromy
the pop transformation simply exchanges the two eigenlines. The most
difficult transformation, the \juggle, relates different limit
triangulations. See Section \ref{sec:limit-triangulation}  for the
detailed discussion and Figure \ref{fig:vectormultiplet-twist} for
an illustration.  An important aspect of the Fock-Goncharov theory
is that under flips the coordinates undergo \emph{cluster
transformations}. These cluster transformations turn out to be
special cases of  the Kontsevich-Soibelman transformations
$\CK_{\gamma_0}$. The transformations under pops are explicitly
known, but in general are rather cumbersome. A significant point for
our main result is that, nevertheless, the  \emph{omnipop}, defined
to be the transformation that simultaneously pops all vertices, is a
simple and computable transformation ${\bf S}$, which we call the
\emph{spectrum generator} for reasons which will be clear below. The
omnipop transformation ${\bf S}$ is derived in Section
\ref{subsec:SuperPop}. The transformation under the \juggle\ is
described in Section \ref{sec:limit-triangulation}.

Having set up all the machinery in Sections \ref{sec:fg-coordinates}
and \ref{sec:WKB-triangulation} we finally give the crucial
definition of the functions $\CX^{\vartheta}_\gamma:\CM \times
\IC^\times \to \IC$ in equations \eqref{eq:canonical-darboux} and
\eqref{eq:CX-product}. Since the definition is given in terms of
Fock-Goncharov coordinates the Poisson brackets (property P1 above)
follow naturally. Moreover, our choice of decorated triangulation
$T_\WKB(\vartheta, \lambda^2)$ leads to a straightforward derivation
of the $\zeta \to 0, \infty$ and $R \to \infty$ asymptotics, as
indeed it was designed to do. Now, rather beautifully, as
$\vartheta$ varies the WKB triangulations undergo
flips, \juggles, and pops precisely when $\vartheta$ is the
inclination of some BPS ray $\ell_{\gamma,u}$. Indeed, this is quite
natural since, as we described above, the result of
\cite{Klemm:1996bj} simply states that BPS states are associated
with nongeneric \emph{finite} WKB curves. Recall these are closed or
begin and end on branch points of the covering $\Sigma \to C$.
The basic morphisms are illustrated in Figures
\ref{fig:hypermultiplet-flip} and \ref{fig:vectormultiplet-twist},
and in equations \eqref{eq:KS-flip} and \eqref{eq:twist-KS} we show
that the corresponding discontinuities in $\CX^{\vartheta}_{\gamma}$
are precisely those associated with KS transformations, with the
correct value of $\Omega$.

In Section \ref{sec:UniversalStokesMatrix}  we use the results of
Section \ref{sec:darboux-coordinates} to give our algorithm for
computing the BPS spectrum of theories with $K=2$.
Choose a half-plane $\CH(\vartheta,\vartheta+\pi)$ of the complex
$\zeta$ plane bounded by rays at angles $[\vartheta, \vartheta +
\pi]$  and consider evolving the triangulation
$T_\WKB(\vartheta,\lambda^2)$ as $e^{i\vartheta}$ rotates   to
$-e^{i \vartheta}$   in this half-plane. Remarkably, it turns out
that in this continuous evolution the pops always occur in special
circumstances\footnote{More technically: The pop occurs at the
center of a degenerate triangle such as in Figure
\ref{fig:degenerate-face}.} such that the corresponding
transformation of the $\CX^{\vartheta}_{\gamma}$ is the identity.
This surprising fact is shown in Section \ref{sec:no-flavor-jump}.
Hence the net transformation for evolving $\vartheta \to \vartheta +
\pi$ just involves a sequence of flips and juggles, and the effect on
the Darboux coordinates is the transformation ${\bf S} = \prod
\CK_{\gamma}^{\Omega(\gamma;u)}$ with the factors ordered by
  $\arg Z_\gamma$. Now, every BPS state (or its
antiparticle) has a BPS ray in the chosen half-plane
$\CH(\vartheta,\vartheta+\pi)$, so the product captures precisely
half the spectrum, while the other half are just the antiparticles.
On the other hand, the initial and final triangulations
$T_\WKB(\vartheta, \lambda^2)$ and $T_\WKB(\vartheta+\pi,\lambda^2)$
turn out to be simply related by an omnipop and, as we mentioned
above, that transformation can be computed explicitly (Section
\ref{sec:UniversalStokesMatrix}). Since the product decomposition
${\bf S} = \prod \CK_{\gamma}^{\Omega(\gamma;u)}$ is unique, (given an
ordering of BPS rays, which is in turn determined by $u$), we can
read off the spectrum from the aptly named spectrum generator ${\bf
S}$. It is worth asking how this algorithm improves upon the
prescription already given in \cite{Klemm:1996bj} for computing the
BPS spectrum of the $A_1$ theories of class $S$. The latter prescription
requires one to know the critical values of $\vartheta$ for which the
BPS states exist. The crucial point of the above algorithm is that
one need only choose a generic value of $\vartheta$, and no prior
knowledge of the phases of occupied BPS central charges is required.

In Sections \ref{section:Superconformal} and
\ref{section:SU2Examples} we work out a large number of examples of
our formalism.  In Section \ref{section:Superconformal} we show how
various limits of the linear $SU(2)$ quiver theories include all the
possible Argyres-Douglas (AD) superconformal theories. Already the
simplest examples of AD theories provide beautiful illustrations of
the KSWCF.  It turns out that all the wall-crossing identities in
these theories are consequences of a basic pentagon identity
(\ref{eq:simple-pentagon}). In Section \ref{section:SU2Examples} we
consider $SU(2)$ gauge theory with $N_f=0,1,2,3,4$ flavors of
fundamental hypermultiplets. In this case the BPS spectra are very
elaborate, and the wall-crossing typically involves infinite
products of KS transformations, generalizing the basic example which
appears for $N_f=0$, equation (\ref{eq:wcf-prod2}). It turns out
that all the wall-crossing formulae in these examples --- intricate
though they may be --- are obtained by successive use of this basic
identity and the pentagon. The most elaborate and beautiful
spectrum occurs in the $N_f=4$ case. We locate a particularly
interesting strong coupling region in which the finite spectrum
splits into two $N=4$ AD points and can therefore be described very
concretely. We expect this observation to be useful in some future
investigations.

Let us conclude this survey by returning to our original goal of
defining the functions $\CX^\RH_\gamma$ relevant to the construction
of \cite{Gaiotto:2008cd}. An important preliminary result is given
in Section \ref{sec:R-To-Infinity}, where we show that in the $R\to
\infty$ limit the $\CX_{\gamma}^{\vartheta}$ indeed are asymptotic
to the semiflat coordinates $\CX_{\gamma}^{\sf}$ as long as $\zeta $
is in the halfplane $\IH_{\vartheta}$ centered on $e^{i \vartheta}$.
The proof uses an interesting connection to the sinh-Gordon equation
(and a generalization thereof) on the Riemann surface
$C$.\footnote{Incidentally, this relation has also recently played a
useful role in the work of Alday and Maldacena \cite{Alday:2009yn},
and that connection allowed those authors to put some of our results
in this paper to good use.  We discuss this connection a bit more in
Section \ref{sec:pnut}.} From this we are able to deduce that the only singularities of
$\CX_{\gamma}^{\vartheta = \arg \zeta}$
 are the essential singularities at $\zeta = 0,\infty$ together with
discontinuities across those BPS rays with $\Omega\not=0$.  Given
the other results in Section \ref{sec:darboux-coordinates} it
follows that they satisfy the full set of defining properties in
Section \ref{sec:review}, and hence do indeed provide the desired
alternative construction we set out to find. Maddeningly, as we have
mentioned, this reasoning is valid for large enough $R$ but might
break down at some finite $R$, as explained in Section
\ref{sec:comparison}. This leaves the behavior as $R$ decreases to
zero as an important open question.

We close with a few comments and possible future directions for
research:

\begin{itemize}

\item First, and most obviously, the
generalization to higher rank theories ($K>2$) should be carried
out.
Since the first preprint version of this paper appeared,
we have made some progress in this direction.  The general story will
appear in \cite{gmn5-to-appear} and some important
examples in \cite{gmn6-to-appear}.

\item Our constructions
raise some tantalizing possible connections to the mathematics of
BPS state counting via Hall algebras (and perhaps from there to the
algebras of BPS states).  It seems likely that there should be a
category underlying this story, a sort of Fukaya category for
1-manifolds in a Riemann surface.  As we remark in Section
\ref{sec:bps-comments}, it is conceivable that this category is in
fact equivalent to a category of quiver representations. Moreover,
the geometry of the decompositions of $C$ we consider suggests a
method for realizing the Harder-Narasimhan filtration in this
category.  We briefly sketch these ideas in Section
\ref{sec:category}.

\item One consequence of our results is a new description of the \hk metrics
on certain moduli spaces of Higgs bundles.  Moduli spaces of Higgs bundles
play a prominent role in a new approach to the
geometric Langlands program initiated in \cite{geom-lang-n4},
and indeed the way these moduli spaces appear in this paper is not unrelated
to the way they appear in \cite{geom-lang-n4}.  It is thus natural to wonder
whether our results can be of any use for geometric Langlands.

\item Fock and Goncharov's construction was motivated in part by the desire to construct new
infinite-dimensional modular functors (associated with Liouville
theory and its higher rank ``Toda'' generalizations).  We believe
that some of the ideas of this paper, particularly the notion of
``limit triangulation,'' might provide some useful insights into
these new modular functors.

\item The story of this paper applies most directly to $\N=2$ theories which are not conformal.
However, there are some closely related conformal theories, discussed further in \cite{Gaiotto:2009we},
which could be obtained by adjusting the eigenvalues of the monodromies at the singular points in $C$ to zero.
Now, in the closely analogous case of
$\N=(2,2)$ theories in two dimensions, \cite{Cecotti:1993rm} exploited information about massive
deformations to get information about the conformal points (for example, their spectrum of conformal dimensions).
It is natural to wonder whether a similar trick would work here.

\item Finally, as mentioned above, it would be very interesting to understand
the analytic structure of the functions $\CX_\gamma$ constructed in this paper
at small $R$.

\end{itemize}

\section{Review} \label{sec:review}

Let us quickly recall the setup, notation and main proposal of
\cite{Gaiotto:2008cd}, to which we refer for more details.

\subsection{Setup}

We consider a $d=4$, $\CN=2$ supersymmetric gauge theory.  Call its Coulomb branch $\CB$.
At each point $u \in \CB$ the gauge
group is broken to a maximal torus $U(1)^r$.  There is a lattice $\hat\Gamma_u$
of charges, equipped with an antisymmetric integer-valued pairing $\inprod{,}$.
The radical of this pairing is the sublattice $(\Gamma_\flav)_u$ of flavor charges.
Dividing out by $(\Gamma_\flav)_u$ gives the quotient lattice $\Gamma_u$ of gauge charges.
$\Gamma_u$ has rank $2r$ and is equipped with a symplectic pairing.

The lattice $\hat\Gamma_u$ is
the fiber of a local system $\hat\Gamma$, with nontrivial monodromy
around the complex-codimension-1 singular loci in $\CB$,
where some BPS particles become massless.
There is a ``central charge'' homomorphism
\begin{equation}
Z: \hat\Gamma_u \to \IC
\end{equation}
varying holomorphically with $u$.  In particular, given any local section $\gamma$ of $\hat\Gamma$, there is
a corresponding locally-defined holomorphic function $Z_\gamma(u)$, the central charge
of a particle with charge $\gamma$.

We formulate the theory on $\IR^3 \times S^1$, with $S^1$ of
radius $R$.  At energies $\ll 1/R$ this theory looks
effectively three-dimensional.  Its moduli space is locally a
product of two \hk manifolds. One factor is the Higgs branch of
the $d=4$ theory, which we do not consider here. The other
factor is the $d=3$ Coulomb branch $\CM$. At generic,
non-singular points in $\CM$ the Higgs branch is actually
absent. $\CM$ is a fibration
\begin{equation}
\pi: \CM \to \CB
\end{equation}
with generic fiber a $2r$-torus.
The torus fibers appear because the gauge fields in $d=4$ give rise to scalars in $d=3$, namely the
holonomies of the gauge fields (both electric and magnetic) around $S^1$.
For each $\gamma \in \Gamma_u$ we have a corresponding circle-valued holonomy
$\theta_\gamma$, with
$\theta_{\gamma + \gamma'} = \theta_\gamma + \theta_{\gamma'}$.

Because of supersymmetry the metric $g$ on $\CM$ is \hk. A
first approximation $g^\sf$ to $g$ is obtained by naive
dimensional reduction.  To determine $g$ exactly, one must also
include instanton (and multi-instanton) effects, coming from
BPS particles of the $d=4$ theory winding around $S^1$. These
effects are weighted by the second helicity supertraces
$\Omega(\gamma;u)$ which count particles of charge $\gamma$.

\subsection{Darboux coordinates}

In \cite{Gaiotto:2008cd} we proposed an exact description of $g$.
The main idea is that to describe $g$ it is enough to describe
holomorphic Darboux coordinates for $\CM$ considered as a holomorphic symplectic manifold.

The construction is local over the base $\CB$.  Fix an open set $U
\subset \CB$ over which $\hat\Gamma$ is trivializable.
Also fix a choice of quadratic refinement $\sigma: \hat\Gamma \to \{ \pm 1 \}$
of the antisymmetric pairing mod 2.  Our holomorphic Darboux
coordinates are labeled by sections $\gamma$ of $\hat\Gamma$ over
$U$. They are functions $\CX_\gamma$ on $\pi^{-1}(U) \times
\IC^\times$ obeying
\begin{enumerate}[a)]
\item $\CX_{\gamma+\gamma'} = \CX_\gamma \CX_{\gamma'}$. \label{it:c1}
\item For any fixed $\zeta \in \IC^\times$, $\CX_\gamma(\cdot; \zeta)$ is valued in $\IC^\times$, and holomorphic in
complex structure $J^\pz$ on $\CM$. (Recall $\CM$ is \hk\ and hence
has a $\IC \IP^1$ of complex structures.) \label{it:c2}
\item The holomorphic Poisson brackets of the $\CX_\gamma$ are given by\label{it:c3}\footnote{For later
convenience we have rescaled the Poisson bracket by a factor $4
\pi^2 R$ relative to that in \cite{Gaiotto:2008cd}.  }
\begin{equation} \label{eq:poisson}
\left\{ \CX_\gamma, \CX_{\gamma'} \right\} = \inprod{\gamma, \gamma'} \CX_\gamma \CX_{\gamma'}.
\end{equation}
\newcounter{enumsaved2}
\setcounter{enumsaved2}{\value{enumi}}
\item For any fixed $(u, \theta) \in \CM$, $\CX_\gamma(u, \theta; \zeta)$ is holomorphic in $\zeta$.
Here $\theta$ is an angular coordinate on the fiber of $\CM \to \CB$
above $u$. \label{it:c4}
\item $\CX_{\gamma}(\cdot; \zeta) = \overline{\CX_{-\gamma}(\cdot; -1/\bar\zeta)}$.
\newcounter{enumsaved}
\setcounter{enumsaved}{\value{enumi}}
\end{enumerate}
Moreover, in \cite{Gaiotto:2008cd} it turned out to be particularly interesting to consider coordinate systems
subject to a further asymptotic condition, namely
\begin{enumerate}[a)]
\setcounter{enumi}{\value{enumsaved}}
\item $\lim_{\zeta \to 0} \CX_\gamma(u, \theta; \zeta) \exp \left[ - \zeta^{-1} \pi R Z_\gamma(u) \right]$ exists. \label{it:c6}
\end{enumerate}
However, there is a Stokes phenomenon in play here: for the $\CM$ of
interest, it turns out to be impossible to construct Darboux
coordinates which obey the conditions \ref{it:c1})-\ref{it:c6}). The
right thing to do is to ask for all desired properties to hold for
$\zeta$ in some half-plane, centered on a ray $e^{i \vartheta}
\IR_+$,\footnote{We emphasize that in some situations $\CX_\gamma^\vartheta$ actually depends on
$\vartheta \in \IR$, not just $\vartheta \in \IR / 2 \pi \IZ$.  We will
encounter this situation in Sections \ref{sec:irregular},
\ref{section:Superconformal}, and \ref{sec:UniversalStokesMatrix}
below.}
\begin{equation}
\IH_\vartheta := \left\{ \zeta: \vartheta-\frac{\pi}{2} < \arg \zeta < \vartheta+\frac{\pi}{2} \right\},
\end{equation}
and for a single $u_0 \in U \subset \CB$.
So we ask for a \ti{collection} of coordinate systems $\CX_\gamma^{\vartheta, u_0}$, each defined on $\pi^{-1}(U) \times \IH_\vartheta$.
Each one should obey \ref{it:c1})-\ref{it:c3}), and
\begin{enumerate}[a')]
\setcounter{enumi}{\value{enumsaved2}}
\item $\CX^{\vartheta, u_0}_\gamma(u, \theta; \zeta)$ is
    holomorphic in $\zeta$, for $\zeta \in
    \IH_\vartheta$.\label{it:c4p}
\item $\CX^{\vartheta, u_0}_{\gamma}(\cdot; \zeta) = \overline{\CX^{\vartheta+\pi, u_0}_{-\gamma}(\cdot; -1/\bar\zeta)}$. \label{it:c5p}
\item $\lim_{\zeta \to 0} \CX_\gamma^{\vartheta, u_0}(u_0, \theta; \zeta)
 \exp \left[ - \zeta^{-1} \pi R Z_\gamma(u_0) \right]$ exists, when $\zeta$
is restricted to $\IH_\vartheta$. \label{it:c6p}
\setcounter{enumsaved}{\value{enumi}}
\end{enumerate}
In this paper we will give a construction of functions
$\CX_\gamma^{\vartheta, u_0}$ obeying these conditions, for
sufficiently large $R$. These $\CX_\gamma^{\vartheta, u_0}$ really
do depend on $(\vartheta, u_0)$:  there is a real-codimension-1
subset in the space of $(\vartheta, u_0)$ where
$\CX_\gamma^{\vartheta, u_0}$ jumps.  From \ref{it:c2}),
\ref{it:c3}) it follows that these jumps are holomorphic Poisson
morphisms.

These jumps in the coordinates $\CX_\gamma^{\vartheta, u_0}$ are the most important part of the whole
story.  In particular, the jumps are determined by, and determine, the BPS degeneracies
of the $d=4$ field theory.  If we think of $u_0$ as fixed, then the jumps occur at specific values
of $\vartheta$, namely those $\vartheta$ which are the phases of central charges of BPS states
in the vacuum labeled by $u_0$.  Moreover the precise jumps are determined by the gauge charges
of the BPS states.
We state this more precisely as follows:
\begin{enumerate}[a)]
\setcounter{enumi}{\value{enumsaved}}
\item $\CX_\gamma^{\vartheta, u_0}$ is piecewise constant as a function of $(\vartheta, u_0)$,
with discontinuities  at pairs $(\vartheta, u_0)$ for which there
is some $\gamma_\BPS$ with $\arg -Z_{\gamma_\BPS}(u_0) = \vartheta$
and $\Omega(\gamma_\BPS;u_0) \neq 0$.\label{it:c7}

\item Fix $\vartheta_0 \in \IR / 2 \pi \IZ$, $u_0 \in \CB$, and define
\begin{equation} \label{eq:sdef}
{\bf S}_{\vartheta_0, u_0} := \prod_{\gamma_\BPS: \, \arg
-Z_{\gamma_\BPS}(u_0) = \vartheta_0}
\CK_{\gamma_\BPS}^{\Omega(\gamma_\BPS;u_0)},
\end{equation}
where $\CK_{\gamma_\BPS}$ is a holomorphic Poisson transformation of the $\CX_\gamma$ given by
\begin{equation}\label{eq:KS-Transfm}
\CK_{\gamma_\BPS}: \CX_{\gamma} \mapsto \CX_{\gamma} (1-\sigma({\gamma_\BPS}) \CX_{\gamma_\BPS})^{\langle \gamma,{\gamma_\BPS} \rangle}.
\end{equation}
Then\footnote{Our convention here differs by a sign from \cite{Gaiotto:2008cd}.}
\label{it:c8}
\begin{equation} \label{eq:ks-discontinuity}
\left( \lim_{\vartheta \to \vartheta_0^+} \CX^{\vartheta,
u_0}_\gamma \right) = {\bf S}_{\vartheta_0, u_0} \left(
\lim_{\vartheta \to \vartheta_0^-} \CX^{\vartheta, u_0}_\gamma
\right).
\end{equation}
(A genericity assumption is made here: The   charges with $\arg
-Z_{\gamma_\BPS}(u_0) = \vartheta_0$ are all proportional, so we
needn't order the product in ${\bf S}_{\vartheta_0,u_0}$. )
\setcounter{enumsaved}{\value{enumi}}

\end{enumerate}
A final important property is
\begin{enumerate}[a)]
\setcounter{enumi}{\value{enumsaved}}
\item When $\zeta \in \IH_\vartheta$,
we have the large $R$ asymptotics
\begin{equation}
\CX_{\gamma} \sim  \CX^\sf_\gamma
\end{equation}
and in fact the corrections are exponentially small, i.e. $\CX_\gamma = \CX^\sf_\gamma (1 + \CO(e^{-const \cdot R}))$
in regions bounded away from the singular points of $\CB$.
\label{it:c9}
\end{enumerate}

In the rest of this paper we will sometimes lighten the notation,
writing $\CX_\gamma^{\vartheta,u_0}$ just as $\CX_\gamma^\vartheta$
when we are not trying to emphasize the $u_0$ dependence, or even
just as $\CX_\gamma$.  (This $\CX_{\gamma}$ must \emph{not} be
confused with the $\CX_{\gamma}$ of \cite{Gaiotto:2008cd}. Those
functions are denoted by $\CX_{\gamma}^{\RH}$ in this paper. See
Section \ref{subsec:RH-problem} below.)

\subsection{Wall-crossing} \label{subsec:wallcrossing}

As we have just reviewed, at least for large enough $R$, the moduli space $\CM$ carries a family of local coordinate systems
$\CX^{\vartheta, u_0}_\gamma$, obeying the conditions
\ref{it:c1})-\ref{it:c3}), \ref{it:c4p}')-\ref{it:c6p}'), \ref{it:c7})-\ref{it:c9}).

The mere existence of these coordinates has a rather strong consequence.  Consider a point
$u \in \CB$ and two different phases $\vartheta_\pm$, with $\vartheta_+ - \vartheta_- < \pi$.
The coordinate systems $\CX^{\vartheta_\pm,u}_\gamma$
are generally not equal; to see how they are related, one must apply \eqref{eq:ks-discontinuity} to each
Stokes line which lies between $\vartheta_-$ and $\vartheta_+$.  This gives the relation as
\begin{equation}
\CX_{\gamma}^{\vartheta_+,u} = \bS(\vartheta_-, \vartheta_+;u) \CX_{\gamma}^{\vartheta_-,u},
\end{equation}
where
\begin{equation} \label{eq:ks-product}
\bS(\vartheta_-, \vartheta_+;u) = \prod^\ccwarrow_{\gamma_\BPS: \  \vartheta_- < \arg -Z_{\gamma_\BPS}(u) < \vartheta_+} \CK_{\gamma_\BPS}^{\Omega(\gamma_\BPS;u)},
\end{equation}
with the product taken in increasing order of $\arg -Z_{\gamma_\BPS}(u)$.

A key algebraic fact \cite{ks1}
is that product decompositions of the form \eqref{eq:ks-product} are
unique:  \eqref{eq:ks-product} actually \ti{determines} the $\Omega(\gamma_\BPS;u)$ for
$\arg -Z_{\gamma_\BPS}(u)$ between
$\vartheta_-$ and $\vartheta_+$ (with the exception of $\gamma_\BPS \in \Gamma_\flav$, which have
$\CK_{\gamma_\BPS} = \id$ and hence are invisible in $\bS(\vartheta_-, \vartheta_+;u)$.)
One can thus think of $\bS(\vartheta_-, \vartheta_+;u)$ as a kind of ``generating function''
for the $\Omega(\gamma_\BPS;u)$.

Now suppose we deform $u$ continuously to $u'$.  As long as no $\arg Z_{\gamma_\BPS}(u)$ crosses $\vartheta_+$
or $\vartheta_-$ in the process, it follows from \ref{it:c7})
 that $\CX_\gamma^{\vartheta_\pm,u} = \CX_\gamma^{\vartheta_\pm,u'}$,
and so
\begin{equation} \label{eq:wcf}
\bS(\vartheta_-, \vartheta_+;u) = \bS(\vartheta_-, \vartheta_+;u').
\end{equation}
The formula \eqref{eq:wcf} determines the $\Omega(\gamma_\BPS;u')$
given  $\Omega(\gamma_\BPS;u)$   and hence gives a complete solution
to the wall-crossing problem. In our context, it is a direct
consequence of the existence of the functions
$\CX^{\vartheta,u}_\gamma$.

Since the product \eqref{eq:ks-product} is generally infinite, we
should perhaps comment on how it is to be understood.  One begins by
finding some basis $\{\gamma_i\}$ of $\hat\Gamma$, such that all
$\gamma_\BPS$ which contribute to $\bS(\vartheta_-, \vartheta_+;u)$
are \ti{nonnegative} linear combinations of the $\gamma_i$.  This
basis depends on $(\vartheta_-, \vartheta_+, u)$.\footnote{Actually,
the existence of such a basis is not obvious.  In the mathematical
work of \cite{ks1} an additional technical condition was imposed
which guarantees it, and which one hopes would hold in all physical
examples.  In Section \ref{sec:bps-comments} of this paper we will
show that such a basis exists in the examples we consider.} Each
$\CK_\gamma$ defines a Poisson automorphism of the algebra $F :=
\IC[[x_{\gamma_1}, \dots, x_{\gamma_i}]]$. (Infinite Taylor series
arise because we expand denominators, $1/(1-x) = 1 + x + \cdots$.)
Now how about their product?  To understand that, we begin by
defining a collection of finite-dimensional unipotent groups
$G_N(\vartheta_-, \vartheta_+;u)$ ($N \ge 0$). We first use the
basis $\gamma_i$ to define the degree of $\gamma = \sum n_i \gamma_i
$ to be $\vert \gamma \vert := \sum_i n_i $ and informally
``truncate $F$ to Fourier modes of order less than or equal to
$N$.'' More precisely, there is a filtration of $F$ by ideals $I_N
\subset F$, generated by monomials with degree greater than $N$.
Since $\CK_\gamma$ maps $I_N$ to itself, the transformations with $
\vartheta_- < - \arg Z_{\gamma}(u) < \vartheta_+$
 generate
a group $G_N(\vartheta_-, \vartheta_+;u)$  of Poisson automorphisms
of $F_N := F / I_N$. Moreover, again because the $K_\gamma$ preserve
the filtration there is a projection map $G_N \to G_{N-1}$ and we
can use these to define a group $G(\vartheta_-, \vartheta_+;u)$ as
the inverse limit of the system of the $G_N(\vartheta_-,
\vartheta_+;u)$.  When projected to any $G_N(\vartheta_-,
\vartheta_+;u)$, the product \eqref{eq:ks-product} involves only
finitely many nontrivial factors and hence is well defined. Moreover
it behaves coherently with respect to  the projections $G_N \to
G_{N-1}$. This is sufficient to define it in $G(\vartheta_-,
\vartheta_+;u)$.

Let us also comment a bit more on the uniqueness of the product
decomposition which was claimed above. We have $\CK_\gamma
=e^{f_{\gamma}}$ where $f_\gamma := \sum_{n\geq 1}
\frac{\sigma(n\gamma)}{n^2} \{ \CX_{n \gamma},\cdot \} $. The
$f_{\gamma}$ with $ \vartheta_- < - \arg Z_\gamma(u) < \vartheta_+$
span the Lie algebra of $G(\vartheta_-, \vartheta_+;u)$. For the
unipotent groups $G_N$ the exponential map is bijective.  It follows
that the decomposition of a group element $g\in G(\vartheta_-,
\vartheta_+;u)$ into a product of the form (\ref{eq:ks-product}) is
indeed unique. From a more practical viewpoint, there is an
algorithm for extracting the $\Omega(\gamma_\BPS;u)$ from
$\bS(\vartheta_-, \vartheta_+;u)$, easily implemented on a computer
for any reasonably small $\gamma_\BPS$.  It amounts to considering
inductively the projections of $\bS(\vartheta_-, \vartheta_+;u)$ to
the successive subgroups $G_N(\vartheta_-, \vartheta_+;u)$. At the
$N$-th step one can determine the $\Omega(\gamma_\BPS;u)$ for
$\abs{\gamma_\BPS} \leq N$.

The formula (\ref{eq:wcf}) was first presented in \cite{ks1} in a
very general context.  It generalizes the primitive and
semi-primitive wall-crossing formulae, first derived in
\cite{Denef:2007vg} using Denef's multicentered BPS black hole
solutions of $\CN=2$ supergravity \cite{Denef:2000nb}.  The
arguments of \cite{Denef:2007vg} apply both to supergravity and to
its field theory limit.  Unfortunately, the constructions in
\cite{Gaiotto:2008cd} and in this paper are restricted to field
theory. Thus, an important open problem remains: Give a physical
derivation of \eqref{eq:wcf} for type II string theory on a
compact Calabi-Yau.

\subsection{Riemann-Hilbert problem}\label{subsec:RH-problem}

In \cite{Gaiotto:2008cd} we did not introduce the functions $\CX^{\vartheta}_\gamma$
explicitly.  Instead we formulated a Riemann-Hilbert problem, the solution of
which would lead to a \ti{single} Darboux coordinate system
$\CX^\RH_\gamma(\zeta)$, obeying \ref{it:c1})-\ref{it:c6}) for all
$\zeta$, \ti{except} that $\CX^\RH_\gamma(\zeta)$ is not holomorphic in $\zeta$, but
only piecewise holomorphic; it jumps by
$\CK_{\gamma_\BPS}^{\Omega({\gamma_\BPS};u)}$ along each ray $\zeta
\in Z_{\gamma_\BPS} \IR_-$.  We argued moreover that a solution
indeed exists for sufficiently large $R$.

There is a simple correspondence between such
$\CX^\RH_\gamma(\zeta)$ and $\CX_\gamma^{\vartheta}(\zeta)$ obeying
\ref{it:c1})-\ref{it:c3}), \ref{it:c4p}')-\ref{it:c6p}').  Given
$\CX^\RH_\gamma(\zeta)$, $\CX_\gamma^{\vartheta}(\zeta)$ can be
obtained as the analytic continuation of $\CX^\RH_\gamma(\zeta)$ in
$\zeta$ from the ray $\zeta \in e^{i \vartheta} \IR_+$. Conversely,
given $\CX_\gamma^\vartheta(\zeta)$, we divide the $\zeta$-plane
into slivers bounded by BPS  rays, and define
$\CX^\RH_\gamma(\zeta)$ to agree with $\CX_\gamma^\vartheta(\zeta)$
on the sliver containing the ray $\zeta \in e^{i \vartheta} \IR_+$.

One might naturally guess that if we apply this procedure to the $\CX^\vartheta_\gamma(\zeta)$
built in this paper, the $\CX^\RH_\gamma(\zeta)$ so obtained will give a solution to the Riemann-Hilbert
problem defined in \cite{Gaiotto:2008cd}, for all $R$.  However, we do not prove that in this paper.
The crucial problem is the possibility that the $\CX^\RH_\gamma(\zeta)$ we get could be only
piecewise \ti{meromorphic} away from $\zeta=0, \infty$.  This would be
incompatible with the Riemann-Hilbert problem, in which
$\CX^\RH_\gamma(\zeta)$ were required to be piecewise \ti{holomorphic}.
At sufficiently large $R$ this is not a problem:
the poles of the individual $\CX^\vartheta_\gamma$ lie outside the sector $\IH_\vartheta$, and
do not appear in $\CX^\RH_\gamma$.  The important question is whether as we go to small $R$ these poles
can move into the sliver around $\zeta \in e^{i \vartheta} \IR_+$.

Although our results on wall-crossing and BPS degeneracies are independent of this
question, it is relevant for the issue of describing the \hk metric using the methods
of \cite{Gaiotto:2008cd}.  We will discuss this matter at some length in Section
\ref{sec:comparison}.

\section{From brane constructions to Hitchin systems} \label{sec:physics}

In this section we describe a class of $d=4$, $\CN=2$ field
theories for which the \hk manifold $\CM$ is a moduli space of
solutions to Hitchin's equations.

In Section \ref{sec:scft} we discuss these theories in
terms of the six-dimensional $(2,0)$ theory reduced on a Riemann
surface $C$, with some number of real codimension-2 defects at
points of $C$, and explain why their further compactification
on a circle leads to Hitchin's equations.
In Section \ref{subsec:WittenConstruction} we give
a purely four-dimensional definition of a large class of
the theories we consider, by making
contact with the D4/D6/NS5-brane constructions
described in \cite{Witten:1997sc}.
Finally, in Section \ref{sec:iib} we briefly recall an alternative Type IIB
string theory construction which would lead to the same theories, involving a Calabi-Yau threefold
containing $C$ as a curve of ADE singularities.

\subsection{Compactifying the $(2,0)$ theory} \label{sec:scft}

Starting with the famous $(2,0)$ superconformal field theories
in $d=6$ and compactifying on
appropriate Riemann surfaces $C$, one can produce a very large
class of $d=4$, $\CN=2$ field theories.
As we will describe in Sections \ref{subsubsec:20AndCoulomb}-\ref{subsubsec:KineticTerms}
below, this construction also naturally realizes
the Seiberg-Witten curves in these theories as branched covers of
$C$ inside $T^*C$, as well as giving a direct handle on the BPS
spectrum, extending a picture described in \cite{Klemm:1996bj}.

Our main interest in this paper is in what happens when
we further compactify these four-dimensional theories on a circle.
That is, we consider the $(2,0)$ theory on $\IR^{1,2} \times S^1 \times
C$, where $S^1$ has radius $R$.  Then we find, at low energies, a
three-dimensional sigma model with \hk target space $\CM$.

What is $\CM$?  To answer that question, we describe the same theory
in a different way, by reversing the order in which we compactify.
If we first compactify the six-dimensional $(2,0)$ theory on $S^1$,
the low energy physics is described by five-dimensional
supersymmetric Yang-Mills theory. We can then consider further
compactification of this five-dimensional theory on $C$.  From this
viewpoint, the moduli space of 3-dimensional super-Poincar\'e
invariant vacua is the moduli space of solutions of certain BPS
equations for the gauge fields and scalars on $C$.  These BPS
equations turn out to be the Hitchin equations on $C$.  While these
two compactifications correspond to different limits, we do not
expect any phase transition in the low energy physics when we
exchange the relative length scales of $C$ and $S^1$.  The reason is
that the BPS-protected quantities that we study are insensitive to
the conformal scale of the metric on $C$, thanks to the topological
twist described below.  Therefore, we can identify the target space
$\CM$ of the three-dimensional sigma model as a moduli space of
solutions to Hitchin's equations on $C$. We explain this in Section
\ref{sec:hiteq}.

The analysis of $\CM$ which we carry out in the rest of this paper
can be applied only if $C$ carries some defect operators,
inserted at points $\CP_i$.  We make some comments about these
defects in Sections \ref{sec:defects}-\ref{sec:bc-intersection}.

\subsubsection{The $(2,0)$ theory on $\IR^{1,5}$ and its Coulomb
branch}\label{subsubsec:20AndCoulomb}

We begin with the $(2,0)$ superconformal theories in six-dimensional
spacetime
\cite{Witten:1995zh,Strominger:1996ac,Witten:1995em,Seiberg:1996vs,Seiberg:1997ax,
Seiberg:1997zk}.  These theories enjoy $osp(6,2\vert 4)$
superconformal invariance. Modulo topological subtleties, they are
obtained as products of two types of basic building block:
interacting theories, which have an ADE classification, and free
theories with an abelian ``gauge group.'' Our main interest in this
paper will be in the theories in the A series. From the point of
view of M-theory, the theory with ``gauge group'' $U(K)$ can be
described as a decoupling limit of a system of $K$ coincident
M5-branes \cite{Strominger:1996ac}.  We use this picture frequently
as a convenient shortcut for understanding properties of the theory.

Let us briefly recall some information about the chiral operators of
the $(2,0)$ theory; more detail can be found in
\cite{Aharony:1998an}, and see also  \cite{Bhattacharya:2008zy},
especially the table on page 31. There is a basis of operators transforming
in short representations of $osp(6,2\vert 4)$, labeled by the
Casimir operators of the ADE group $\fg$.  Label the Casimirs by
$k=1, \dots, r$. Within the $k$-th short multiplet we will focus on
the subspace $V_k$ of operators with lowest conformal weight.  $V_k$
is an irreducible representation of the $so(5)$ R-symmetry.  Its
conformal weight is twice the exponent $d_k$ of $\fg$.

The theory has a ``Coulomb branch'' parameterized by vacuum expectation values
of these chiral operators.  This branch is especially easy to understand in the $A_{K-1}$ theory:
it is just $(\IR^5)^K / S_K$, parameterizing configurations in which the $K$ M5-branes are
separated in the transverse $\IR^5$.

On the Coulomb branch the theory contains BPS strings,
geometrically described as the boundaries of M2-branes running between the separated M5-branes.
See Figure \ref{fig:string}.
\insfig{4in}{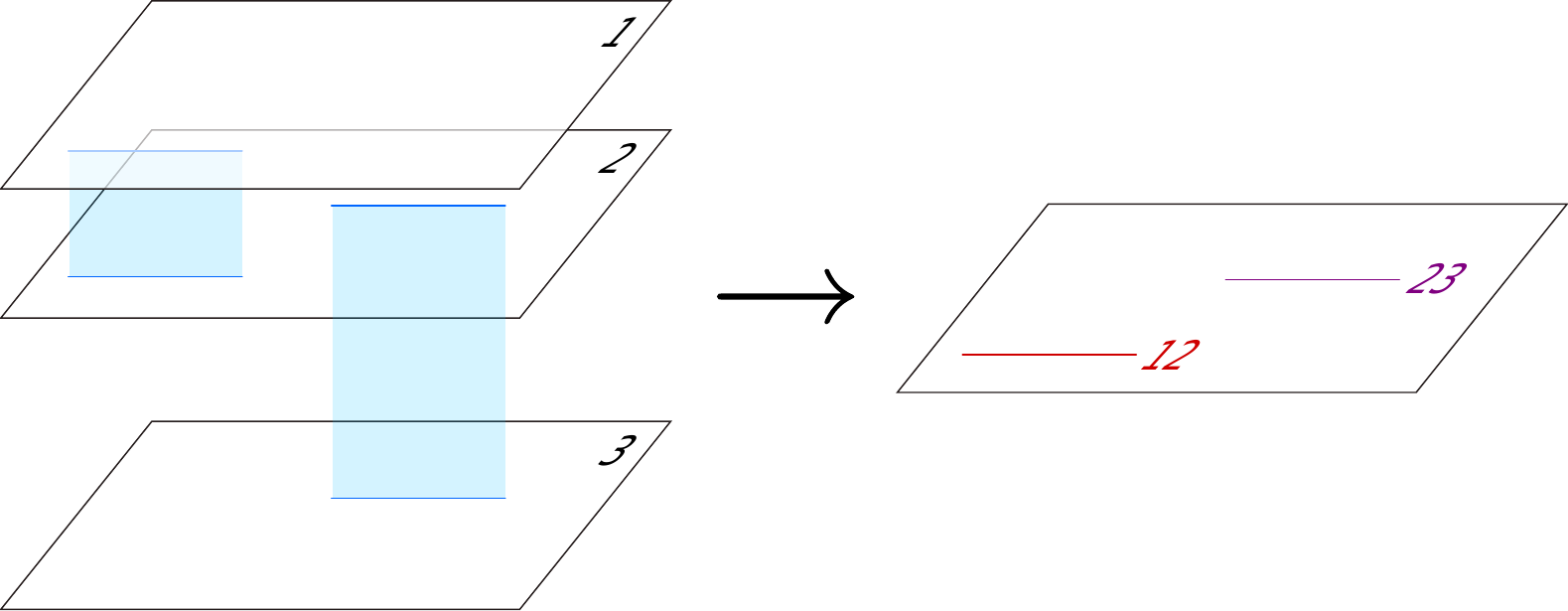}{Left: three separated M5-branes, including segments of two M2-branes stretching between them.
 Right: the corresponding picture in the $A_{K-1}$ $(2,0)$ theory with $K=3$.
 The two M2-brane segments have been projected down to string segments.}
Call the string that comes from an M2-brane running between brane
$i$ and brane $j$ an $ij$-string. These strings are oriented;
reversal of orientation exchanges $ij$-strings with $ji$-strings.
The BPS condition requires that the strings are straight lines in
$\IR^{5,1}$. The tension of a BPS $ij$-string can be calculated from
the M2-brane picture as\footnote{The tension of the M2-brane is
$2\pi/\ell^3$ and that of the electromagnetic dual M5-brane is
$2\pi/\ell^6$, where $\ell$ is the 11-dimensional Planck length.}
\begin{equation}
T_{ij} = \frac{2 \pi}{\ell^3} |x_i-x_j|
\end{equation}
where $|x_i-x_j|$ is the distance in $\IR^5$ (with dimensions of
length).   From BPS non-renormalization theorems one expects that
this geometric picture actually gives the exact tension, even in the
regime (most relevant for this paper) where the branes are separated
by distances of order $\ell$.

One can have supersymmetric ``junctions'' where strings of type
$ij$, $jk$, and $ki$  meet.
The appearance
of such junctions in the $(2,0)$ theory was noted in \cite{Lee:2006gqa}.
The supersymmetry preserved by a single
$ij$-string depends both on the slope of the string in the
worldvolume of the M5-branes, and on the direction of $x_i-x_j$ in
$\IR^5$. To find the condition for a quarter-BPS junction, we can
imitate the standard setup for $(p,q)$ string junctions
\cite{Sen:1997xi, Aharony:1997bh, Dasgupta:1997pu}: tie the relative
slopes of the three strings in the plane of the junction to the
relative slopes of the vectors $(x_i-x_j)$, $(x_j-x_k)$, $(x_k-x_i)$
in the plane in $\IR^5$ defined by the three points $x_i, x_j, x_k$.
Notice that this automatically gives mechanical equilibrium at the
junction, as the tension vectors of the three string segments  are
then linearly related to $(x_i-x_j)$, $(x_j-x_k)$, $(x_k-x_i)$, and
hence sum to zero.

\subsubsection{The $(2,0)$ theory on $\IR^{1,3} \times C$ and its Coulomb branch}

In order to compactify the $(2,0)$ theory on $C$ without
breaking four-dimensional supersymmetry, we need to consider a
partial twisting.  The super Poincar\'e subalgebra of
$osp(6,2\vert 4)$ has bosonic part $so(5,1) \oplus usp(4)\cong
so(5,1) \oplus so(5)$. The spinor representations  of
$Spin(1,5)$ and $Spin(5)$ are $\IH^2$, where $\IH$ is the
quaternions. Therefore the  Poincar\'e supercharges transform
in the $(\IC^4\otimes \IC^4)_{+}$ of $so(5,1) \oplus so(5)$,
where the subscript $+$ indicates a symplectic Majorana reality
constraint. Upon compactification on $C$ we preserve the
subalgebra $so(3,1)\oplus so(2)_C \oplus so(3) \oplus so(2)_R$,
under which the supercharges transform as
\begin{equation}
\left( \left((2,1)_{\half} \oplus (1,2)_{-\half} \right) \otimes
\left(2_{\half} \oplus 2_{-\half} \right) \right)_+
\end{equation}
The twisting consists of identifying the diagonal $so(2)$ of
$so(2)_C\oplus so(2)_R$ with the holonomy algebra of $C$, and
leaves us with supercharges transforming under $so(3,1) \oplus
so(3) \oplus so(2)_C'$ as
\begin{equation}\label{eq:supercharge-decomp}
(2,1;2)_{1} \oplus (2,1;2)_0 \oplus (1,2;2)_0 \oplus (1,2;2)_{-1}
\end{equation}
We can introduce a corresponding basis of supercharges:
\begin{equation}
Q_z^{\alpha A} ; Q^{\alpha A}; \bar Q^{\dot \alpha A}; \bar Q_{\bar z}^{\dot \alpha A}
\end{equation}
where $\alpha, \dot \alpha, A$ all run over $1,2$ and $z$ is a local
coordinate so that $dz$ has $so(2)_C'$ charge $+1$. Since the
supercharges $Q^{\alpha A}; \bar Q^{\dot \alpha A}$ of the middle
two summands of (\ref{eq:supercharge-decomp}) are uncharged under
$so(2)_C$ they are well-defined four dimensional supercharges, and
therefore we have preserved $\CN=2$ four-dimensional supersymmetry.
The commutator $\left\{ \bar Q^{\dot \alpha A}, \bar Q_{\bar
z}^{\dot \beta B}\right \} \sim \epsilon^{\dot \alpha \dot \beta}
\epsilon^{AB}\p_{\bar z}  $ will be quite useful below where we
consider   the chiral ring of the operators annihilated by $\bar
Q^{\dot \alpha A}$.

The moduli space of the four-dimensional theory is obtained
essentially by dimensional reduction from the Coulomb branch of the
six-dimensional theory. More precisely, choose a Cartan subalgebra
$so(2)_R \oplus so(2)$ of the $so(5)$ $R$-symmetry, and let $\CO_k$
be the operator in $V_k$ of weight $(d_k, 0)$. This operator has the
largest $so(2)_R$ charge in the multiplet and hence it must be
annihilated by any supercharge, such as $\bar Q^{\dot \alpha A}$,
with positive $so(2)_R$ charge. We reach the $d=4$ Coulomb branch by
giving vacuum expectation values to these chiral operators.

After the twisting, $\CO_k$ is a section of the bundle $K^{\otimes
d_k}$ over $C$, and in particular this is true of its vacuum
expectation value $\inprod{\CO_k}$. Since $\CO_k$ is annihilated by
$\bar Q^{\dot \alpha A}$, and since $\bar Q^{\dot \alpha A}$-exact
operators have vanishing vev's  \cite{Witten:1988ze},
$\inprod{\CO_k}$ must be annihilated by $\bar\partial$. This is the
only condition on $\inprod{\CO_k}$, so the $d=4$ Coulomb branch
which parameterizes these vevs is simply
\begin{equation}
\CB = \bigoplus_{k=1}^r H^0(C, K^{\otimes d_k}).
\end{equation}

In the $A_{K-1}$ case there is a nice geometric interpretation of
$\CB$. We are considering the low energy limit of a system of $K$
M5-branes which are wrapped on a holomorphic cycle $C$ inside a \hk
four-manifold $Q$.  To go to the Coulomb branch we separate the
branes so that they wrap some other cycle $\Sigma$ inside
$Q$.\footnote{One could also separate the branes in the 3 flat
transverse directions; this corresponds to moving onto the $d=4$
Higgs branch, which will not play a role in this paper.} Unlike the
flat space situation, we will consider cases where the branes cannot
be completely separated, so $\Sigma$ will be a \ti{connected}
divisor inside $Q$. By choosing holomorphic Darboux coordinates
$(x,z)$ for $Q$ as a holomorphic symplectic manifold we can identify
a neighborhood of $C$ with the holomorphic cotangent bundle $T^*C$.
Picking a point of $\CB$ just corresponds to picking the
coefficients $u_k \in H^0(C, K^{\otimes k})$, $k = 2, \dots, K$,
of the equation
\begin{equation} \label{eq:sigma-eq}
x^K + \sum_{k=2}^K u_k(z) x^{K-k} = 0
\end{equation}
defining $\Sigma \subset T^* C$.

We normalize the coordinates so that the holomorphic symplectic form is
\begin{equation}
\Omega = \frac{\ell^3}{2 \pi^2} dx \wedge dz.
\end{equation}
The projection map $T^* C \to C$ identifies $\Sigma$ as a $K$-fold
cover of $C$.  The distance between the $i$-th and $j$-th sheets is
a 1-form on $C$, which we call $\lambda_{ij}$.

We will see below that
$\Sigma$ should be identified with the Seiberg-Witten curve of the reduced theory.  The canonical one-form
\begin{equation}
\lambda = x\,dz,
\end{equation}
restricted to $\Sigma$, will be identified with the Seiberg-Witten differential.

\subsubsection{BPS strings on the Coulomb branch} \label{sec:bps-strings}

Our geometric picture of the Coulomb branch is a convenient way to
read off the properties of the BPS strings.  As in the flat space case, locally we have BPS
$ij$-strings labeled by pairs of sheets, and BPS junctions where three such strings meet.
The BPS tension of an $ij$-string is $\frac{1}{\pi} \abs{\lambda_{ij}}$;
this tension depends on the point of $C$,
unlike the flat space case.
There are also some special points on $C$, namely the $ij$-\ti{branch points}
where $\lambda_{ij} = 0$, i.e. the $i$-th and $j$-th sheets of $\Sigma$ come together.
An $ij$-string can end at an $ij$-branch point.  One quick way of deriving this fact is
to recall the description of these states in the M5-brane picture:  the $ij$-string is an M2-brane
foliated by segments connecting sheet $i$ and sheet $j$, and can end smoothly when these segments
shrink to zero size.  See Figure \ref{fig:m2-branch}.

\insfig{4in}{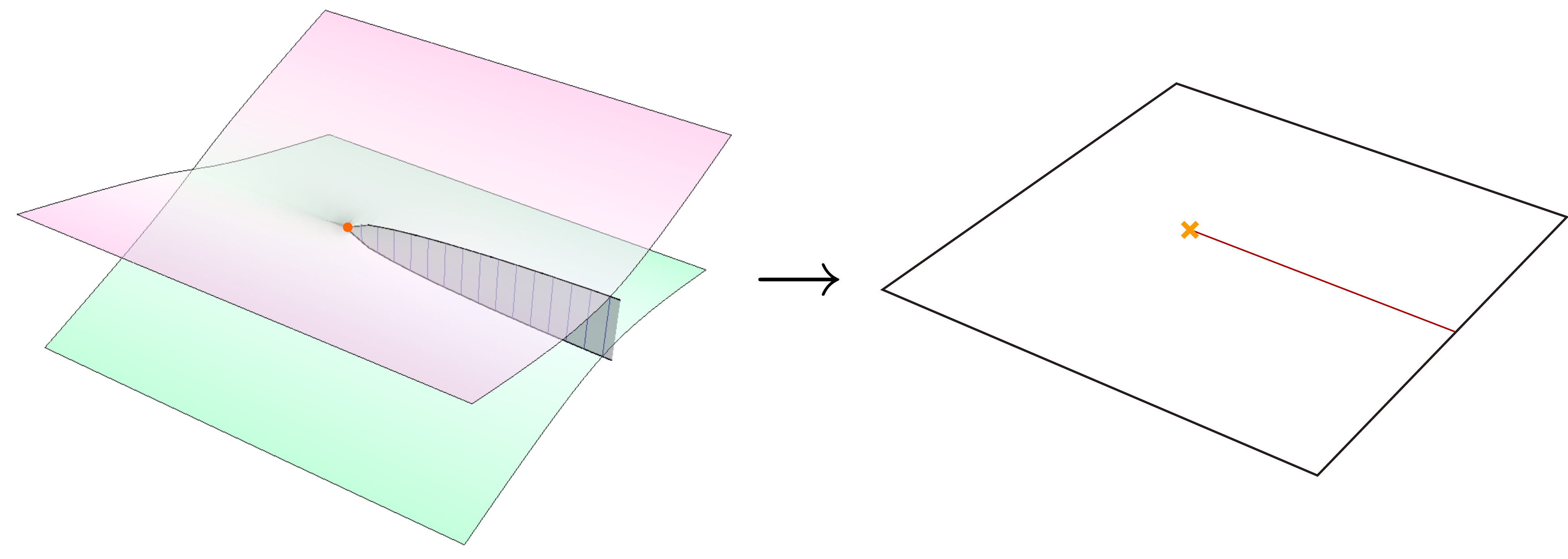}{Left: a portion of an M2-brane stretching between two sheets of an
M5-brane.  The M2-brane
is foliated by ``vertical'' segments, each of which lies in a single fiber of $T^* C$.  At the branch
point where the two sheets collide, the vertical segments shrink to zero length.  Right: the
projection of the M2-brane onto $C$ is a string in the $(2,0)$ theory which ends on the branch
point.}

The central charge of a segment of $ij$-string extended along the curve $c$ is given by
\begin{equation}\label{eq:Z-fragment}
Z = \frac{1}{\pi} \int_c \lambda_{ij}.
\end{equation}
The mass of the same segment on the other hand is just the integral of the tension,
\begin{equation} \label{eq:stringmass}
M = \frac{1}{\pi} \int_c \abs{\lambda_{ij}}.
\end{equation}
So the BPS bound $M \ge \abs{Z}$ is saturated if and only if
$\lambda_{ij}$ has the same phase $\vartheta$ everywhere along the
curve $c$, i.e., if $\partial_t$ denotes the tangent vector to $c$, we require
\begin{equation}\label{eq:BPS-WKB-curve}
\lambda_{ij} \cdot \partial_t \in e^{i \vartheta} \IR_+.
\end{equation}
In this case $Z = e^{i \vartheta} M$.

For a multi-string junction the phase $\vartheta$ must be the same for all strands.  This
condition is equivalent to the no-force condition at the junction:  indeed, the vector
representing the force exerted by the $ij$-strand on the junction can be expressed as the complex number
$e^{i \vartheta} \bar{\lambda}_{ij}$, so since the three $\lambda_{ij}$ sum to zero, the forces do as well.

The simplest BPS object is an $ij$-string stretched between two $ij$-branch points.  In the M5-brane picture it would be represented by a disc;
see Figure \ref{fig:bps-disc}.  Such an M2-brane has no moduli.  Its quantization yields
a single BPS hypermultiplet in four dimensions.

\insfig{5in}{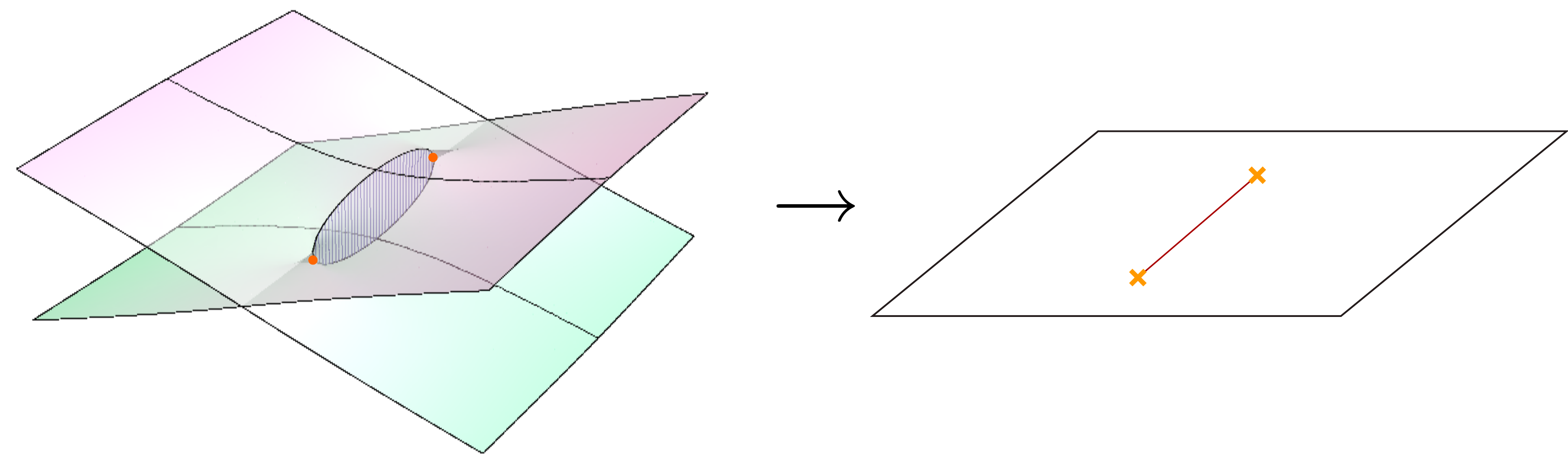}{Left: An M2-brane wrapped on a disc $D$, stretched between two sheets of the curve $\Sigma$ supporting the M5-brane.  Right: Under the projection $T^* C \to C$, the disc projects to
a string in the $(2,0)$ theory, which ends on the branch points.}

Similarly we can consider an $ij$-string stretched along a closed loop $c$ in $C$.  In the
M5-brane picture such a string is represented by a cylindrical M2-brane; see
Figure \ref{fig:bps-annulus}.  Such an M2-brane has a single
bosonic modulus and corresponding fermion zero mode.  Its quantization yields a BPS vectormultiplet.

\insfig{5in}{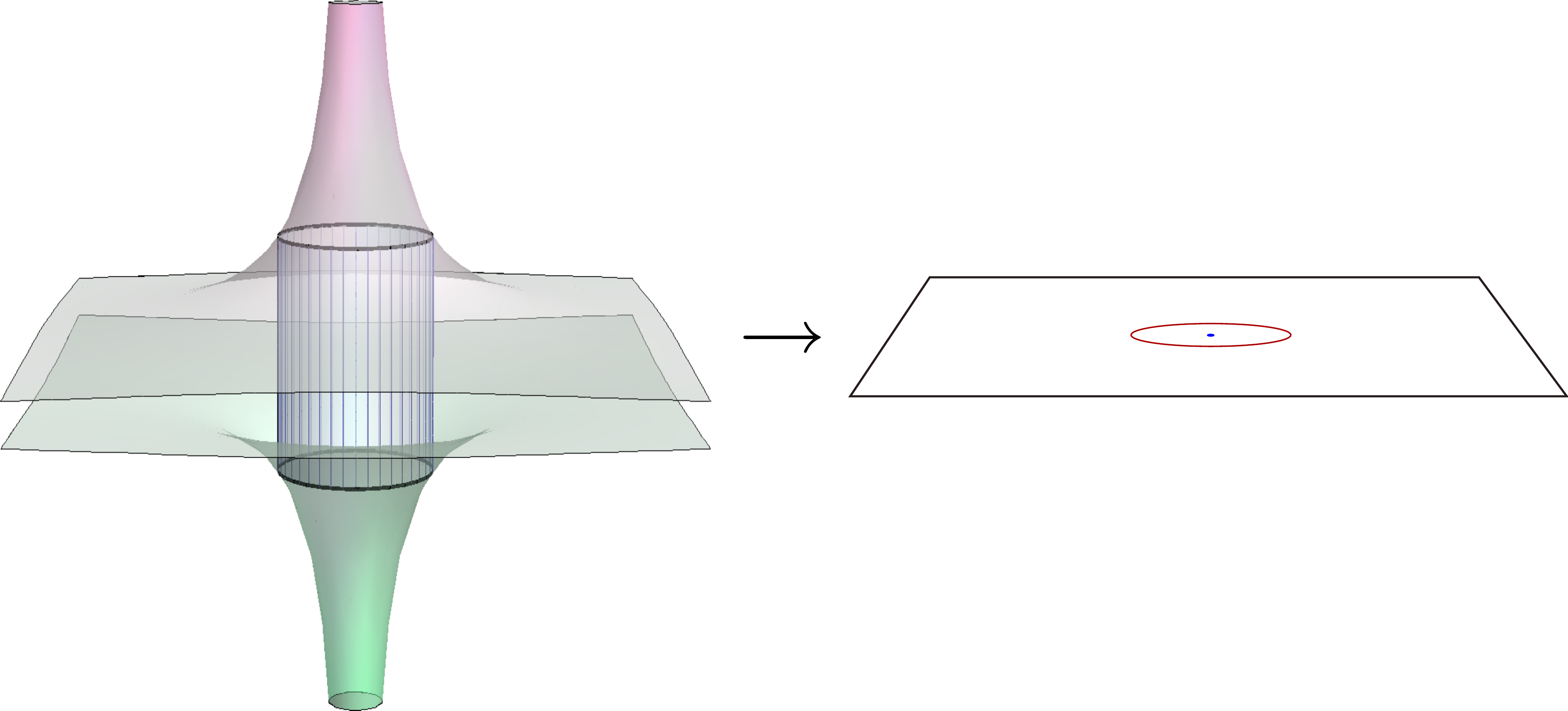}{Left: An M2-brane wrapped on a cylinder,
stretched between two sheets of the curve $\Sigma$ supporting the M5-brane.
Right: Under the projection $T^* C \to C$, the disc projects to
a closed string in the $(2,0)$ theory.}

If $K=2$ these cases exhaust the possibilities; this
description of the states was used in
\cite{Klemm:1996bj,Brandhuber:1996xk} to study the BPS
spectrum.  For $K>2$ one should also consider string webs.
Again lifting each strand $c$ to $c_i - c_j$, the sum of all
the lifts yields a closed cycle in $\Sigma$ which gives the
charge of the BPS particle.  The number of bosonic moduli is
equal to the number $\ell$ of loops in $c$.  Indeed, $\ell =
1+(j-b)/2$, where $b$ is the number of branch endpoints and $j$
the number of three-string junctions.  Each strand can be moved
perpendicular to itself, except that each strand attached to a
branch point loses this degree of freedom, and furthermore each
junction has only two degrees of freedom and hence imposes one
further constraint on the three strands ending on it.  The
number of strands is $(b+3j)/2$, so this naive count of the
bosonic moduli gives $(j-b)/2$.  With a bit of work it is
possible to see that one overall combination of the constraints
at the junctions is trivial, thus giving one extra modulus, for
a total of $\ell$ (see \cite{Aharony:1997bh}, equation 23). By
supersymmetry we should have $\ell$ fermionic moduli as well.
Taking into account these fermionic moduli we expect that
quantization should give a multiplet of spin $\half (\ell +
1)$.\footnote{As the moduli space has boundaries, depending on
the precise boundary conditions multiplets of lower spin might
possibly arise.}

Finally let us consider the central charges of these BPS states. The
oriented curve $c$ on $C$ representing a BPS $ij$-string can be
lifted to a pair of curves $c_i$, $c_j$ on $\Sigma$, namely, we take
the preimages of $c$ on the $i$-th and $j$-th sheets, where $c_i$
has the same orientation as $c$ and $c_j$ has the opposite
orientation.  Then \eqref{eq:Z-fragment} can be rewritten as
\begin{equation} \label{eq:central-lambda}
Z = \frac{1}{\pi} \int_c \lambda_{ij} = \frac{1}{\pi} \left( \int_{c_i} \lambda - \int_{c_j} \lambda \right).
\end{equation}
For any combination of $ij$-strings representing a BPS state,
the sum of the lifted curves is a closed cycle $\gamma$ on $\Sigma$; it is simply
the boundary of the M2-brane representing the BPS state.  What we have found is
\begin{equation} \label{eq:central-charge}
Z = \frac{1}{\pi} \oint_\gamma \lambda.
\end{equation}
This formula will be crucial in what follows.

Incidentally, one can also understand \eqref{eq:central-charge} as a
consequence of Stokes's theorem and the fact that the BPS condition
requires that the M2-brane is special Lagrangian in $Q$, i.e., when
restricted to the M2-brane the holomorphic symplectic form $\Omega =
\frac{\ell^3}{2 \pi^2} d \lambda$ is $e^{i \vartheta}$ times the
volume form. In the field theory limit, we are considering M5-branes
which lie close to the zero section $C$ in $T^* C \subset Q$; in
this limit the special Lagrangian M2-branes become ``vertical'' and
reduce to the string webs (see \cite{Mikhailov:1997jv} for related
discussion).

\subsubsection{Charge lattices}
\label{subsubsec:ChargeLattices}

This is a convenient moment to pause and consider the lattice $\hat\Gamma$ of
all charges (flavor and gauge) in the $d=4$ theory.
We said above that the charge of a BPS state is determined by a 1-cycle $\gamma$ in $\Sigma$,
the boundary of an M2-brane ending on the M5-brane.
So naively one might identify the charge lattice as $H_1(\Sigma; \IZ)$.  This is slightly
wrong, for two reasons.

First, not all classes in $H_1(\Sigma; \IZ)$ can support a BPS
state.  This is easiest to see in the case $K=2$; since all BPS
states come from membranes which connect the two sheets, they are
invariant under the operation of exchanging the two sheets and also
reversing the orientation.  It follows that all BPS charges lie in
the sublattice of $H_1(\Sigma; \IZ)$ which is invariant under this
combined operation.

Second, some classes in $H_1(\Sigma; \IZ)$ should be regarded as physically
equivalent, because it costs zero energy to move from one to the other.
This can happen for $K>2$ when there is some unbroken nonabelian gauge symmetry at a
defect inserted somewhere on $C$; then several sheets of the covering become identified near
the defect, and in particular, a loop around the defect can move freely between these sheets.
So to get the physical charge lattice we have to take a quotient.  Assuming there are no accidental degeneracies
one can describe this quotient operationally as dividing out by the kernel of $Z$, i.e. we
identify cycles $\gamma$ which have the same period $Z_\gamma$.
So altogether $\hat\Gamma$ is a subquotient (a quotient of a sublattice) of $H_1(\Sigma; \IZ)$.

The lattice $\Gamma$ of gauge charges is simpler to describe:  it is
just $H_1(\bar\Sigma, \IZ)$, where $\bar\Sigma$ is the compact Riemann surface
obtained by filling in the punctures of $\Sigma$. The flavor
lattice $\Gamma_\flav \subset \hat\Gamma$ is the radical of the
intersection pairing $\inprod{,}$ on $H_1(\Sigma;\IZ)$.  Any element
in $\Gamma_\flav$ can be represented by a linear combination of
small loops around the punctures.  $\hat\Gamma$ is an extension of
$\Gamma$ by $\Gamma_\flav$,
\begin{equation}\label{eq:Lattice-ExactSeq}
0 \to \Gamma_\flav \to \hat\Gamma \to \Gamma \to 0.
\end{equation}

As we move in the Coulomb branch $\CB$ the curve $\Sigma$ varies,
and in particular $H_1(\Sigma; \IZ)$ might have monodromy around loci
where $\Sigma$ degenerates.  Hence $\hat\Gamma$ and $\Gamma$ are not
fixed lattices, but rather local systems of lattices varying over
$\CB$.  In contrast $\Gamma_\flav$ is monodromy invariant.

\subsubsection{Kinetic terms}\label{subsubsec:KineticTerms}

For completeness, let us discuss in a bit more detail the
four-dimensional Lagrangian obtained by compactifying the $(2,0)$
theory on $C$, and verify that $\Sigma$ is indeed the Seiberg-Witten
curve and $\lambda$ is the Seiberg-Witten differential.

We begin by calculating the kinetic terms of the vectormultiplet
scalars in the IR 4d gauge theory, starting from the M5-brane
picture of the Coulomb branch. The scalar part of the action of the
M5-brane is just the DBI action, $2 \pi / \ell^6$ times the volume form.
A vacuum of the
four-dimensional theory is determined by some fixed holomorphic
curve $\Sigma_0 \subset Q$.  We consider configurations of the
M5-brane which approach $\IR^{3,1} \times \Sigma_0$ near infinity in
$\IR^{3,1}$, but may fluctuate in the interior of $\IR^{3,1}$. Such
fluctuations (at least sufficiently mild ones, which are all we need
here) are given by maps
\begin{equation}
f: \IR^{3,1} \times \Sigma_0 \to Q
\end{equation}
Let $y\in \IR^{3,1}$ and define $f_y:=f(y,\cdot): \Sigma_0 \to Q$.
We require that as $y\to \infty$, $f_y$ approaches the identity map.

In the low energy limit, we restrict attention to maps which have
zero potential energy, i.e. we require that $\Sigma_y :=
f_y(\Sigma_0)$ is a volume-minimizing cycle for all $y$.
Equivalently, we require that $\Sigma_y$ is holomorphic in $Q$ for
all $y$.  We can then define $\hat f: y\to \Sigma_y$ as a map into
the space $\CD$ of all holomorphic deformations of $\Sigma$,
\begin{equation}
\hat f: \IR^{3,1} \to \CD.
\end{equation}
Now we want to evaluate the kinetic energy of such a map. We work in
the approximation where $\hat f$ is \ti{slowly varying} over
$\IR^{3,1}$, so that we may truncate to second order in the
variations $\partial_\mu \hat f$.  These variations are elements of
the tangent space to $\hat f(y) \in \CD$, which in turn may be
identified with the space of holomorphic sections of the normal
bundle $T_{\hat f(y)}(\CD) \cong H^0(N(\Sigma_y))$. The expansion of
the DBI action to quadratic order in derivatives (with the constant
part subtracted) is
\begin{equation} \label{eq:KK-volume}
S = \frac{\pi}{\ell^6} \int_{\IR^{1,3}} d^4x \, \eta^{\mu\nu}
\inprod{\partial_\mu \hat f, \partial_\nu \hat f},
\end{equation}
where $\inprod{,}$ is the standard Hermitian metric on the space of
sections of $N(\Sigma_y)$, induced from the metric on $Q$. We are
interested in fluctuations for which $S$ is finite; to arrange this
we require that $\partial_\mu \hat f$ is \ti{normalizable} for each
$\mu$,
\begin{equation}
\norm{\partial_\mu \hat f}^2 < \infty.
\end{equation}
The normalizable modes span a subspace of $T\CD$ and define an
integrable  distribution on $\CD$. We let $\CB$ be the leaf of the
distribution passing through  $\Sigma_0$. We call $\CB$ the space of
``normalizable deformations'' of $\Sigma_0$. So we are restricting
attention to maps
\begin{equation}
\hat f: \IR^{3,1} \to \CB.
\end{equation}
Let $r$ denote the number of normalizable modes, i.e. the complex
dimension of $\CB$.

So far we have found that the scalar sector of our four-dimensional
theory is a sigma model into $\CB$. The six-dimensional tensor
multiplet also contains an abelian 2-form gauge potential with
self-dual field strength, and Kaluza-Klein reduction of this field
leads to abelian four-dimensional gauge fields. The number of
independent gauge fields is equal to $\half$ the dimension of the
space of normalizable harmonic $1$-forms on $\Sigma$, or
equivalently to the dimension of the space of normalizable
holomorphic $1$-forms on $\Sigma$.  In fact this is also equal to
$r$. To see this, let us briefly consider the metric on $Q$ in the
vicinity of $\Sigma$.  Let $z$ be a local coordinate on $\Sigma$.
Then we can choose a transverse coordinate $x'$ such that $\Sigma$
is the locus $x' = 0$ and  $\Omega = \frac{\ell^3}{2 \pi^2} dx'
\wedge dz$, and the \kahler form is of the form
\begin{equation}
\omega \vert_\Sigma = i g(z, \bar{z}) d z \wedge d \bar{z} + i \frac{\ell^6}{4 \pi^4} g^{-1}(z, \bar{z}) Dx' \wedge D\bar{x}',
\end{equation}
where $D$ is the covariant derivative in $N(\Sigma)$. Now, by the
adjunction formula there is an isomorphism $N(\Sigma) \cong
T^{*1,0}\Sigma$ most conveniently described by contraction with
$\Omega$. Moreover, there are hermitian products on $H^0(N(\Sigma))$
and $H^0(T^{*1,0}\Sigma)$ defined by
\begin{equation}
\langle v_1 \frac{\p}{\p x'}, v_2 \frac{\p}{\p x'} \rangle:=
\int_{\Sigma} \left( v_1 \bar v_2 \frac{\ell^6}{4\pi^4}
g^{-1}(z,\bar z) \right) g(z, \bar z) \frac{i}{2} dz d\bar z
\end{equation}
and
\begin{equation}
\langle \omega_1, \omega_2 \rangle:=  \frac{i}{2} \int_{\Sigma}
\omega_1 \overline{\omega_2}
\end{equation}
respectively. With these inner products the contraction with
$\Omega$ is an isometry:
\begin{equation} \label{eq:inprod-equiv}
\norm{v \partial_{x'}}^2 = \frac{i}{2}\int_{\Sigma} \left( g(z,
\bar{z}) dz \wedge d\bar{z} \right) \frac{\ell^6}{4 \pi^4} g^{-1}(z,
\bar{z}) \abs{v}^2 = \frac{i}{2} \frac{\ell^6}{4 \pi^4}
\int_{\Sigma} \abs{v dz}^2 = \frac{\ell^6}{4 \pi^4} \norm{v dz}^2.
\end{equation}
So in particular, $v dz$ is normalizable if and only if $v \partial_{x'}$ is.

Altogether we have obtained an abelian $\N=2$ supersymmetric gauge theory in four dimensions with $r$
vectormultiplets.  (Recall that the abelian vectormultiplet
contains one complex scalar and one vector field.)

We have not yet described this $\N=2$ theory in the standard way.
For this, suppose that $\Sigma$ lies close to $C$, and identify a
neighborhood of $C$ with $T^* C$ as we did above.  Now to obtain a good
basis for the space of normalizable holomorphic 1-forms, it is convenient to remember
that these are precisely the ones which extend holomorphically over
$\bar\Sigma$.  So choose a
basis of $A$ and $B$ cycles $\{A^I, B_I\}$ on $\bar\Sigma$,
$I=1,\dots, r$, and let $\{\alpha_I\}$ denote the normalizable holomorphic 1-forms dual to
the $A$ cycles,
\begin{equation}
\oint_{A^I} \alpha_J = \delta^I_J.
\end{equation}
Then we may also define the period matrix of $\bar\Sigma$ as usual by
\begin{equation}
\tau_{IJ} := \oint_{B_I} \alpha_J.
\end{equation}
The metric on normalizable holomorphic 1-forms on $\Sigma$ is the the same as that on holomorphic 1-forms
on $\bar\Sigma$:
\begin{equation} \label{eq:metric-1-forms}
\inprod{\alpha_I, \alpha_J} = (\im \tau)_{IJ}.
\end{equation}

Further define
\begin{equation}
a^I := \frac{1}{\pi} \oint_{A^I} \lambda.
\end{equation}
The $a^I$ give local coordinates on $\CB$. We want to describe the
scalar action \eqref{eq:KK-volume} in these coordinates. To do this
we need to understand the isometry of  $T\CB$ with
$H^0(N(\Sigma))_{\rm normalizable}$ more explicitly. We claim that
if $v_J dz = \pi \alpha_J$ then $\frac{\p}{\p a^J} \to v_J
\frac{\p}{\p x'}$. To see this note that the action of  $v
\frac{\p}{\p x'}$ shifts the surface so that
\begin{equation}
\pi \left( v \frac{\p}{\p x'}\right) a^I = \int_{A^{I'}}  \lambda
-\int_{A^{I}}   \lambda= \int_{W^I}   d\lambda = \int_{A^I} \iota( v
\frac{\p}{\p x'})  d \lambda =   \int_{A^I} v dz
\end{equation}
where $W^I$ is an infinitesimal bit  of surface given by pushing
$A^I$ along the vector field $ v \frac{\p}{\p x'}$.

Now, combining with the isometry (\ref{eq:inprod-equiv}) we have
\begin{equation}
\p_\mu \hat f \cong \p_\mu a^I \frac{\p}{\p a^I} \to \p_\mu a^I
\frac{\ell^3}{2\pi} \alpha_I
\end{equation}
and therefore
\begin{equation} S = \frac{1}{4\pi} \int_{\IR^{1,3}}
d^4x \, \eta^{\mu\nu} (\im \tau)_{IJ} \partial_\mu a^I \partial_\nu
\bar{a}^J.
\end{equation}
This is the standard form for the kinetic term in an abelian $\N=2$
theory, where we \ti{identify $\Sigma$ with the Seiberg-Witten
curve, and $\lambda$ with the Seiberg-Witten differential}.

\subsubsection{Compactifying the $(2,0)$ theory on $C \times S^1$:  Hitchin's equations} \label{sec:hiteq}

Now we are ready to consider the theory obtained by further dimensional reduction from $d=4$ to $d=3$ on $S^1$.

As we have mentioned, our approach to understanding this theory is to go back
to six dimensions and consider compactifying \ti{first} on $S^1$ and \ti{then} on $C$.
So we begin by compactifying the $(2,0)$ theory on $S^1$ of radius $R$.
This yields a theory which at low energies is five-dimensional supersymmetric Yang-Mills theory.
We are interested in this five-dimensional theory further compactified on $C$,
with an appropriate topological twist.

The moduli space $\CM$ of the resulting three-dimensional theory is just the space of
BPS configurations of the five-dimensional theory, which are moreover Poincare invariant in $\IR^3$.
What are these configurations?
Denote the adjoint scalars of the super Yang-Mills theory by
$Y^I$, $I=1, \dots, 5$, so that
\begin{equation} \label{eq:phiy}
\varphi := \half(Y^1 + i Y^2)
\end{equation}
has $so(2)_R$ charge $+1$, and $Y^{3,4,5}$ have charge zero.
In the twisted theory, $\varphi = \varphi_z dz$ is a $(1,0)$ form on $C$.
Then the BPS equations are simply the Hitchin equations for the gauge field $A = A_z dz +
A_{\bar z} d\bar z$ cotangent to $C$ and the adjoint scalar $\varphi$:\footnote{One efficient proof that the BPS equations
are the Hitchin equations goes as follows.  We are studying a stack of M5-branes on $S^1 \times C \times \IR^3$, or equivalently a stack
of D4-branes on $C \times \IR^3$, and looking for BPS configurations invariant under translations along $\IR^3$.
It is well known that the BPS configurations on a stack of D4-branes which are invariant under translations in
one direction are given by solutions
of the self-dual Yang-Mills equations.  The self-dual Yang-Mills equations, when evaluated on configurations which are invariant
under translations in two directions, become Hitchin's equations \cite{MR89a:32021}.}
\begin{align}
F + R^2 [\varphi , \bar\varphi] & = 0, \label{eq:hitchin-1} \\
 \bar\p_A \varphi := d\bar z\left( \p_{\bar z} \varphi  + [ A_{\bar z}  ,\varphi]\right) & = 0, \label{eq:hitchin-2} \\
 \p_A \bar\varphi := dz\left( \p_z \bar\varphi  + [ A_z,\bar\varphi] \right)& = 0. \label{eq:hitchin-3}
\end{align}

We briefly digress to explain the origin of the slightly unconventional factor
$R^2$ appearing in \eqref{eq:hitchin-1}.
First note that after reducing the $(2,0)$ theory on
a circle of radius $R$ one gets the five-dimensional SYM action in
the form\footnote{From the relation \eqref{eq:phiy} between $\varphi$ and $Y$,
and the relation \eqref{eq:central-charge} between $Z$ and $\varphi$,
it follows that when $Y^I = y^I \sigma^3$, an open membrane stretched
between the two M5-branes produces a string of tension $\frac{1}{2\pi} y^I$. Since the
tension of the membrane is $2\pi/\ell^3$, it follows that  the
physical distance in the transverse dimension to the brane is
$\frac{\ell^3}{4\pi^2} y^I$. A nice check on the relative
normalization of the two terms in \eqref{eq:D4-action} is obtained
by computing the mass of a W boson, i.e. a string running between two
displaced D4-branes, which comes out to be $Ry^I$.
The overall normalization can be obtained by reducing the 6-dimensional DBI action.}
\begin{equation}\label{eq:D4-action}
S =  \frac{ R}{8\pi^2} \int_{\IR^{1,2}\times C} {\rm Tr} \left(
\frac{1}{R^2} F \wedge \star F + DY^I \wedge \star D Y^I + \cdots
\right)
\end{equation}
We could rescale $Y^I = R^{-1} \hat Y^I$ to reach the standard
normalization:
\begin{equation}\label{eq:5deeAct}
S = \frac{1}{8\pi^2 R} \int_{\IR^{1,2}\times C} {\rm Tr} \left( F
\wedge \star F + D\hat Y^I \wedge \star D \hat Y^I + \cdots \right)
\end{equation}
It is in this frame that the BPS equations take the $R$-independent form
$F + [\hat \varphi, \overline{\hat \varphi}]\,dz\,d \bar z=0$.
This accounts for the factor of $R^2$ in \eqref{eq:hitchin-1}.
Our reason for preferring $\varphi$ over the rescaled $\hat \varphi$
is that the boundary conditions on the Higgs fields at the singular
points are $R$-independent when expressed in terms of $\varphi$,
as we will see later.

So we have found that the moduli space $\CM$ of the $d=3$ theory is the space of
solutions of the Hitchin equations on $C$.
In \cite{Cherkis:1997aa,Gukov:1998kt,Kapustin:1998xn,Cherkis:2000cj}
dualities were used to argue that the moduli space of $SU(K)$ $d=4$, $\CN=2$
supersymmetric field theory compactified on a circle should be
identified with a specific Hitchin system.
The present section generalizes their result.\footnote{We thank Edward Witten for suggesting that the
general Hitchin system could be realized in this way.}  The general relation between
M5-branes and Hitchin systems was already sketched by Witten in Section 2.3
of \cite{Witten:1997sc}.

It has been observed that the Seiberg-Witten solutions
of many $\CN=2$ theories can be understood in terms of known complex integrable systems, as discussed
e.g. in \cite{Gorsky:1995zq,Martinec:1995by,Donagi:1996cf,Itoyama:1995nv}.  Here we have constructed
a large general class of theories for which the relevant complex integrable system is a Hitchin system.

Let us say a few more words about the general structure of $\CM$.
On general field theory grounds \cite{Seiberg:1996nz} we expect
$\CM$ to be \hk and moreover to be a fibration over
the Coulomb branch $\CB$ of the $d=4$ theory, with generic fiber
a compact torus.  This torus fiber is moreover expected to be a complex submanifold
with respect to one ``distinguished'' complex structure on $\CM$.
How do we see this structure in our case?

The projection from
$\CM$ to $\CB$ is easy to describe:  $\CB$ is parameterized by the
vacuum expectation values $\inprod{\CO_k}$, which in the
five-dimensional Yang-Mills theory are identified with the
independent Casimirs of $\varphi$, so the projection is just
\begin{equation}
(A, \varphi) \mapsto \{ \text{Casimirs of } \varphi \}.
\end{equation}
This map is well known in the mathematics literature as the ``Hitchin fibration''
\cite{MR89a:32021}.  Its fiber over a generic $u \in \CB$ is indeed an abelian variety,
the Prym variety of the projection $\bar\Sigma_u \to
C$, defined as the kernel of a corresponding map of Jacobians
$J(\bar \Sigma_u) \to J(C)$.  In the important special case $C =
\IC\IP^1$, where $J(C)$ is trivial, this is simply $J(\bar\Sigma_u)$.

In the case of the $A_{K-1}$ theory, as we noted earlier, the $\inprod{\CO_k}$
determine the Seiberg-Witten curve $\Sigma \subset T^* C$.  Since we have identified
these with the Casimirs $\Tr \varphi^k$, $\Sigma$ given by \eqref{eq:sigma-eq} is nothing but
the \ti{spectral curve} determined by $\varphi$,
\begin{equation}\label{eq:speccurve}
\det (x\,dz - \varphi) = 0.
\end{equation}
In other words, the positions $x_i$ ($i = 1, \dots, K$) of the
various sheets of $\Sigma$ in the cotangent directions can be
interpreted as the various \ti{eigenvalues} of the matrix-valued
1-form field $\varphi$. (Thus the coefficients $u_k(z)$ in
\eqref{eq:sigma-eq} are elementary symmetric functions of the eigenvalues,
and can be written as polynomials in the $\langle \CO_k \rangle$.)

\subsubsection{Defects and geometric Langlands} \label{sec:defects}

So far we have been a bit vague about exactly what $C$ should be, and exactly
what boundary conditions we should put on the fields.
\ti{A priori} one might think that the
simplest possibility is just to choose $C$ a compact Riemann surface,
and require the fields to be regular everywhere.
As it turns out, the BPS spectra of the resulting $\N=2$ theories are difficult to analyze
using the methods of this paper.  However, there is a simple modification
which simplifies the story:  we
let $C$ be a compact Riemann surface with \ti{defects} inserted at finitely many
points.  In order to get some idea about
what kind of defects should be allowed, we now take a brief detour from
three dimensions down to two.

If we further compactify our three-dimensional sigma model
on a circle $\tilde S^1$ we obtain a two-dimensional $\CN = (4,4)$
nonlinear sigma model, with the same target space $\CM$.
This further compactification makes contact with the work of Witten et al on the
geometric Langlands program \cite{geom-lang-n4,Gukov:2006jk,Frenkel:2007tx}.

Indeed, altogether we have compactified the $(2,0)$ theory first on
$C$ and then on $S^1 \times \tilde S^1$. Suppose we do this in the
opposite order.  Reducing the ADE $(2,0)$ theory on $S^1 \times
\tilde S^1$ gives an ADE gauge theory with four-dimensional $\CN=4$
supersymmetry. As noted in \cite{Harvey:1995tg,Bershadsky:1995vm},
reduction of the $\CN=4$ theory on the Riemann surface $C$
leads to a sigma model into $\CM$. This theory was the starting point for
\cite{geom-lang-n4,Gukov:2006jk,Frenkel:2007tx}.

In the mathematics literature on the
geometric Langlands program there is a well-studied class of defects
(or ``ramifications'') one can introduce on $C$.  These defects are
defined by specifying certain singularities for the solutions of the
Hitchin equations.  Physically, they can be interpreted as coming from
surface operators in the four-dimensional super Yang-Mills theory.
Such operators have been investigated in \cite{Gukov:2006jk}.
Within field theory they can be defined either
by prescribing specific singular behavior for the fields in the path integral,
or by coupling a certain two-dimensional sigma model
to the four-dimensional gauge theory.  Alternatively, some elementary defect
operators can be defined by taking the field theory limit of intersecting-brane
configurations:  we intersect the stack of D3-branes which gives
rise to the four-dimensional gauge theory with some extra D3-branes along
a codimension-two locus.

Ultimately, these surface operators in the four-dimensional gauge theory should arise
from codimension-two defects in the $(2,0)$ theory.
However, the methods we use for defining the defect in four dimensions do not apply in six:
we do not have a path integral definition of the six-dimensional $(2,0)$ theory, and we
also do not know how to couple it to a four-dimensional theory living on a codimension-two locus.
Instead we characterize the defects mainly through the singularities which they induce in the
protected operators of the theory.
This will be adequate for us in this paper as we concern ourselves mostly with
the Hitchin system associated with the $\CN=2$ field theory, and this
depends only on the singularities in the protected operators.

We will assume that all the defects
used in \cite{Gukov:2006jk,Witten:2007td} to produce Hitchin
systems with ramification descend from
defects in the six-dimensional $(2,0)$ theory.  We can provide some partial evidence for this
assumption, as follows.
In Section \ref{sec:bc-intersection} we will describe some simple defects as the field
theory limit of certain M5-brane intersections.  Similar defects,
which induce the same type of singularities in the protected
operators, appear as boundary conditions ``at infinity'' on a non-compact $C$.
Given these simple defects, arbitrarily complicated ramification
in the Hitchin system can be produced from their collision, and by other limiting procedures,
which appear to have a simple physical meaning in the four-dimensional
$\CN=2$ field theory.  We will consider several examples in the text.

Although it is hard to give a precise six-dimensional definition of
the defect operators, it should be possible, upon compactification
on $S^1$, to give  at least a five-dimensional one.  Indeed, in the
infrared the five-dimensional gauge coupling goes to zero, and one
should be left with some $\CN=4$ three-dimensional SCFTs living at
the defect, weakly coupled to the five-dimensional gauge theory. It
would be interesting to identify them.

\subsubsection{Boundary conditions from fivebrane intersections} \label{sec:bc-intersection}

Let us now briefly discuss one specific kind of defect which is relatively straightforward to understand.
(We will encounter some more complicated defects in Section \ref{subsec:WittenConstruction} below.)

We consider again the special case of the $A_{K-1}$ theory, which we realized in terms
of $K$ M5-branes wrapped on $C$.
Now consider a simple transverse intersection between $C$ and
another curve $C_i$, supporting a single M5-brane.
Suppose that the two intersect at a point $\CP_i \in C$.  Then near
$C$, $C_i$ is just the fiber of $T^* C$ over $\CP_i$.  At a
generic point on the Coulomb branch, there is a (non-normalizable)
deformation of the theory which smoothes the intersection between $C_i$ and
$C$.  Namely, choosing a local coordinate $z$ for which $\CP_i$ is at $z = 0$,
we require that near $\CP_i$ one of the sheets of
$\Sigma$ looks like $xz = \rho_i$, for some $\rho_i \neq
0$.  More invariantly, recalling the identification between $x$ and an
eigenvalue of $\varphi$, we require that $\varphi$ is gauge
equivalent near $\CP_i$ to
\begin{equation}\label{eq:commonsing}
\varphi = \frac{ 1 }{z} \begin{pmatrix} \rho_i & & & & & \\ & 0 & & & & \\ & & & \ddots & & \\ & & & & & 0 \end{pmatrix}  + \cdots
\end{equation}
We must also choose boundary conditions on the gauge field; a
natural choice compatible with the Hitchin equations is
\begin{equation}\label{eq:Aboundryc}
A =\frac{1}{2i}  \left(\frac{d z}{z} -
\frac{d \bar z}{\bar z}\right) \begin{pmatrix} \alpha_i & & & & & \\ & 0 & & & & \\ & & & \ddots & & \\ & & & & & 0 \end{pmatrix} + \cdots
\end{equation}
where $\alpha_i$ is imaginary.

As we will see in examples later, $\rho_i$ has a natural
interpretation as a  mass parameter for a $U(1)$ flavor
symmetry in the four-dimensional gauge theory. Moreover, after
the reduction to three dimensions, the three real parameters
encoded in $\rho_i$, $\alpha_i$ can be interpreted as vevs of
three scalar fields in a vectormultiplet of a $U(1)$ flavor
symmetry.

Equation (\ref{eq:commonsing}) should be interpreted with care
in the limit $\rho_i \to 0$; we address this point further
below in Section \ref{sec:couplings}.

\subsection{Witten's construction}\label{subsec:WittenConstruction}

\def\a{\alpha}

An important special case of the construction we have discussed
in Section \ref{sec:scft} was considered
by Witten some time ago in \cite{Witten:1997sc}.
Indeed, suppose we consider an $\N=2$, $d=4$ theory defined by
some conformal or asymptotically free linear quiver of unitary
groups (perhaps with fundamental matter).
In \cite{Witten:1997sc} such quivers were realized in Type IIA string theory using certain
D4/D6/NS5-brane configurations.  Moreover, Witten observed that
upon lifting to M-theory these brane configurations are naturally
described in terms of a single M5-brane.

In this section we review this construction in some detail, and argue that it
can be viewed as an example of the general story of Section \ref{sec:scft}.
In particular, upon dimensional reduction to $d=3$, the moduli space $\CM$ of
the resulting theory should be a Hitchin system.
We thus give a rule for associating a specific ramified Hitchin
system on $\IC\IP^1$ to any conformal or asymptotically free linear quiver
of unitary groups.  This Hitchin system is not the most general possible ---
we will see that the defect operators which arise are of a restricted sort.

Actually, by taking appropriate further scaling/decoupling limits of the quiver gauge
theories, one could produce purely four-dimensional realizations of
$\CN=2$ field theories associated to more general Hitchin systems on
$\IC \IP^1$ or even higher genus curves $C$.
A four-dimensional construction of the $\CN=2$ field theories
associated to Hitchin systems with regular
singularities on a general Riemann surface has recently
appeared in \cite{Gaiotto:2009we}.

\subsubsection{Type IIA}

Following Witten, we begin in Type IIA
string theory. We consider a system of $n+1$ parallel NS5-branes
($n\geq 1$), labeled by $\alpha = 0, 1, \dots, n$.  The branes
are extended along the directions $x^{1,2,3,4,5}$, at common values of $x^{7,8,9}$, and separated
from one another in the $x^6$ direction.  We refer to the
interval
\begin{equation}
I_\alpha := \{ x^6: x^6_{\a-1} < x^6 < x^6_{\a} \}
\end{equation}
as the $\a$-th interval.
Next introduce a collection of D4-branes, at fixed values of $x^{4,5,7,8,9}$, and
extended over some intervals in $x^6$ ending on the NS5-branes.
The NS5- and D4-branes are all located at the same value of
$x^{7,8,9}$.  There are $k_0$ semi-infinite D4-branes on the left
($x^6 \to -\infty$), $k_\alpha$ D4-branes
in the interval $I_\alpha$ for $1 \leq \alpha\leq n$, and
$k_{n+1}$ semi-infinite D4-branes on the right ($x^6 \to
+\infty$).
Finally we may introduce D6-branes, extended along $x^{1,2,3,7,8,9}$, and at fixed values of
$x^{4,5,6}$.  An example of such a configuration is illustrated in
Figure \ref{fig:witten-3}.

\insfig{4.7in}{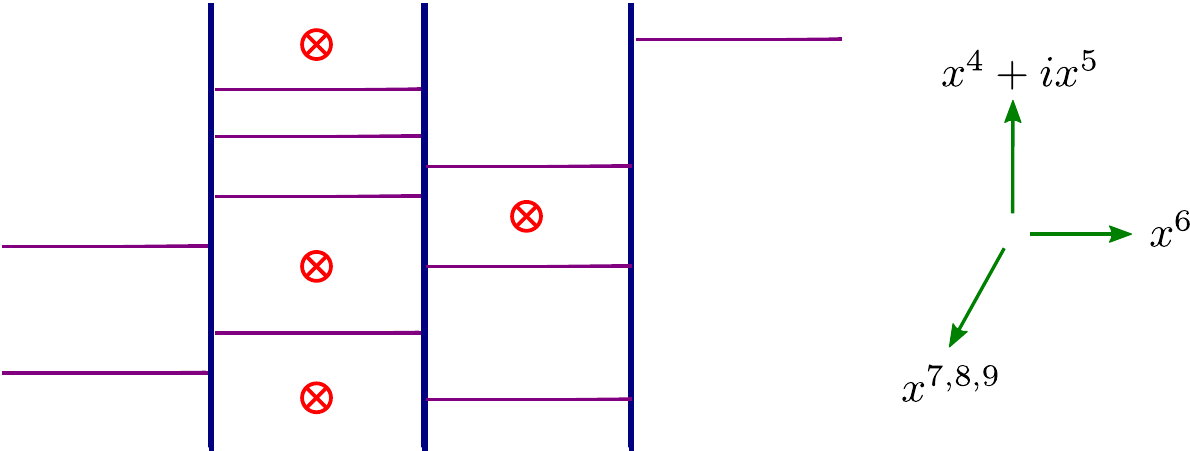}{A configuration of
Type IIA NS5-branes (blue), D4-branes (purple), and D6-branes (red circles with crosses).
We have chosen $n = 2$ and
$(k_0, k_1, k_2, k_3) = (2, 4, 3, 1)$.  The D6-branes are at definite values
of $x^{4,5,6}$.}

The ground state of this system preserves $8$ real
supercharges.  On length scales much larger
than the string length and the
distances between the NS5-branes, the fluctuations of the branes
are described by a $d=4, \CN=2$ gauge theory in
the spacetime coordinatized by $x^\mu = x^{0,1,2,3}$.
This gauge theory is a linear quiver with gauge group $U(k_1) \times\cdots
\times U(k_n)$; see Figure \ref{fig:witten-2}.

\insfig{3in}{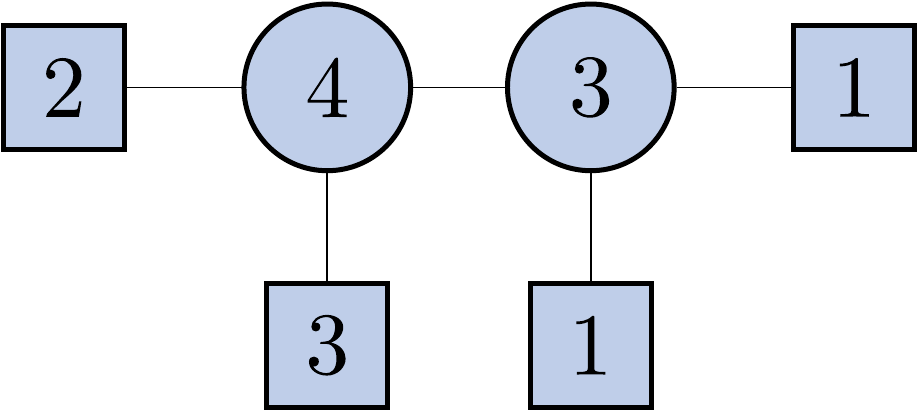}{A linear quiver, corresponding to the Type IIA configuration illustrated in
Figure \protect\ref{fig:witten-3}. Circular nodes correspond to the $U(k_\alpha)$ gauge groups.
Links between circular nodes correspond to bifundamental matter.
Links between circular and square nodes correspond to fundamental matter, with multiplicity given
by the number $d_\alpha$ in the square node.}

The matter content of this quiver theory consists of   $d_\a$ matter
fields transforming in the fundamental representation of each
$U(k_\a)$ factor.  These fundamental matter fields arise in two
ways.  First, they can come from strings stretching between the D4-
and D6-branes in the $\a$-th interval, in which case they are
charged under $U(k_\alpha)$.  Second, strings stretching between the
group of semi-infinite D4-branes at either end of the $x^6$ interval
and the adjacent group of D4-branes give $k_0$ fundamentals of
$U(k_1)$ and $k_{n+1}$ fundamentals of $U(k_n)$. When $n\geq 2$,
D4-D4 strings also give hypermultiplets in the bifundamental of
$U(k_\a) \times U(k_{\a+1})$ for $\a = 1, \dots, n-1$.

As explained in \cite{Witten:1997sc}, there is a
beautiful relation between the forces exerted by the D4-branes on
the NS5-branes and the beta functions of the gauge couplings of
the effective four-dimensional theory.  This relation shows that in
order to avoid large backreaction effects we must restrict attention
to asymptotically free or conformal theories. Therefore we
require the beta function coefficients to be nonnegative:
\begin{equation}\label{eq:NegBetas}
 b_\a:=  - 2k_\a  + k_{\a-1} +  k_{\a +1} +  d_\a \leq 0 , \qquad \a=1,\dots, n.
\end{equation}

The centers of the gauge groups $U(1)_\a \subset U(k_\a)$ are not involved in
the four-dimensional dynamics.  The relative $U(1)$ factors are Higgsed,
so up to a quotient by a finite group, the effective gauge group is $\prod_{\a} SU(k_\a)\times U(1)_d$,
where the diagonal $U(1)_d \subset \prod_{\a} U(1)_{\a}$ represents the
overall center-of-mass coordinate of the full collection of D4-branes.  This diagonal $U(1)_d$ is free
and decouples.

Let us now summarize the parameters of this theory. The UV
parameters are the gauge couplings or UV scales $\tau_\a$ or
$\Lambda_\a$,  together with the bifundamental masses $\mu_\a$  and
the fundamental masses $\hat \mu_{\a,j}$, $1 \leq j \leq d_\a$
(which enter the superpotential). There are also IR
``$u$-parameters'' specifying a point of the Coulomb branch: letting
$\Phi^{(\a)}$ denote the adjoint scalar field in the vectormultiplet
of $SU(k_\a)$, these parameters are the coefficients in the
characteristic polynomials $\langle \det(y - \Phi^{(\a)}) \rangle$
for each $\a$.

Finally we comment that this theory also has important flavor
symmetries.  Each bifundamental hypermultiplet has a $U(1)$
flavor symmetry (enhanced to $Sp(1)\sim SU(2)$ for
bifundamentals of the gauge groups $SU(2) \times SU(2)$), and
each set of $d_\a$ fundamental flavors has a $U(d_\a)$ flavor
symmetry (enhanced to $SO(2 d_\a)$ for fundamentals of the
gauge group $SU(2)$).  Not all of these flavor symmetries are
manifest in the brane construction.  Some, like the
enhancements for $SU(2)$ gauge groups, will be visible in the
six-dimensional $(2,0)$ theory setup, after the spurious $U(1)$
gauge group factors have been expunged.  Others emerge only
after the flow to the four-dimensional gauge theory. We will
see various examples throughout the paper.

\subsubsection{Lift to M-theory}

The second key element in Witten's construction is the lift of the
above Type IIA configuration to M-theory. Let $x^{10}$ denote a
periodic coordinate, of period $2\pi$, parameterizing the M-theory
circle. In the absence of D6-branes we take the 11-dimensional
M-theory metric to be
\begin{equation}\label{eq:MthryMetric}
ds^2 = \sum_{\mu=0,1,2,3} dx^\mu dx_\mu + \sum_{i=4,5,7,8,9}
(dx^i)^2 + R_{11}^2[(dx^6)^2 + (dx^{10})^2].
\end{equation}
(Note that $x^6$ and $x^{10}$ are dimensionless.)
In the presence of D6-branes we will replace a summand in
\eqref{eq:MthryMetric} by a multi-centered Taub-NUT manifold $Q$,
by letting the $x^{10}$ circle fiber nontrivially over the $\IR^3$
parameterized by $x^{4,5,6}$.

The NS5- and D4-branes both lift to M5-branes, and on the Coulomb
branch of the theory they are in general unified into a single
smooth M5-brane, which wraps a non-compact holomorphic curve
$\Sigma$ in $Q$. $\Sigma$ is constrained by the requirement that we
recover the IIA picture upon reducing along $x^{10}$.  This leads to
rules for the asymptotic shape of $\Sigma$, which are explained in
\cite{Witten:1997sc} and will be recalled in the next subsection.
The IR parameters of the theory are summarized in the coefficients
of the polynomial equation defining the curve $\Sigma$.  There is a
fairly elaborate map from the UV parameters $\tau_\a,\Lambda_\a,
\mu_\a, \hat \mu_{\a,j}, \langle \det(y^i - \Phi^{(\a)} ) \rangle$
to these IR parameters, which will be partially explained in the
next section.

We will need to take a limit of the M-theory system in which the
M-theory circle is large, i.e. $R_{11}/\ell \to \infty$ where
$\ell$ is the M-theory length scale. Thus the IIA string coupling
$g_s = (R_{11}/\ell)^{3/2}$ diverges. Nevertheless, we would also
like to use the low energy gauge dynamics of D-branes. This is
possible because, as explained in Section 2.3 of
\cite{Witten:1997sc}, marginal and relevant parameters of the gauge
theory are invariant under a simultaneous scaling of $g_s$ and
$x^6$.  In particular, we would like to hold the
four-dimensional gauge couplings fixed. Using the standard DBI action for
the D4-brane, we find the four-dimensional gauge coupling
$(g_{YM}^{(\a)})^{-2} \sim g_s^{-1} \frac{R_{11}\Delta x^6}{\ell_s}
\sim \left(\frac{\ell}{R_{11}}\right)^{1/2}(x^6_{\a} - x^6_{\a-1})$,
$1\leq \a \leq n$, up to numerical factors.  Thus, the NS5-branes
have separation $\Delta x^6 \to \infty$.  See \cite{Witten:1997sc}
for a detailed discussion of the regime of validity of the
construction.

Our aim is to reinterpret this brane setup so that
this construction can be matched to our general discussion about the
$(2,0)$ theory:  indeed it is equivalent
to considering an $A_{K-1}$ $(2,0)$ theory on the cylinder, with simple defects at various finite
points on the cylinder, and (possibly intricate) boundary conditions
at the two ends of the cylinder.  For clarity's sake, we proceed via
examples of increasing complexity.

\subsubsection{Conformal quivers}

The simplest setup involves $K$ infinite D4-branes
intersecting $n+1$ NS5-branes.  That is, we take $k_0 = k_1 = \cdots
= k_{n+1} = K$.  This corresponds to a linear quiver of $n$ $SU(K)$
gauge groups, with $K$ fundamentals for the first and final factors.

The lift to M-theory is straightforward:
the $K$ D4-branes lift to $K$ M5-branes, wrapping the cylinder
parameterized by $s := x^6 + i x^{10}$.  This cylinder is to be
identified with the Riemann surface $C$ of Section \ref{sec:scft}.
These $K$ M5-branes will give rise to an $A_{K-1}$ $(2,0)$ theory on
$C$.  The NS5-branes also lift to M5-branes,
intersecting the cylinder at distinct points $s = s_\a$, $\a=0,
\dots, n$.  Their worldvolumes fill the $x^4$-$x^5$ plane.  They
give rise to simple defects in the $A_{K-1}$ theory on $C$.

\insfig{3in}{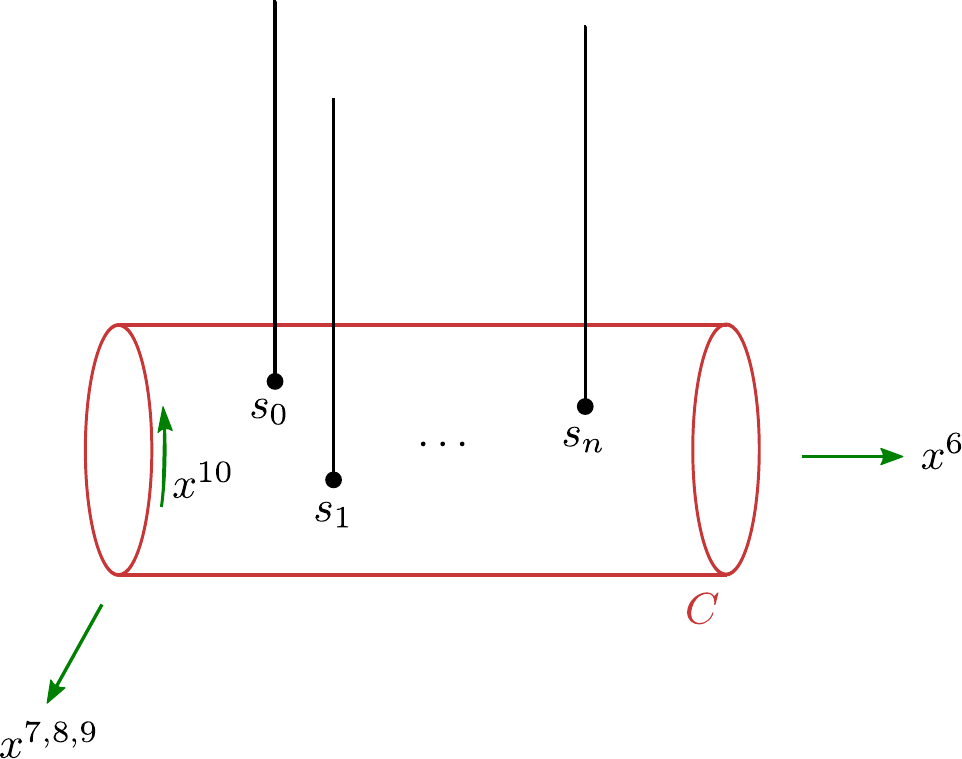}{$K$ M5-branes wrap a cylinder $C$.
There are $n+1$ transverse singly-wrapped fivebranes intersecting
$C$ at points $s_\a$.}

Let us introduce dimensionless coordinates $t := e^{-s} \in
\IC^\times$ and  $v := (x^4 + i x^5)/\ell \in \IC$.\footnote{The
algebraic relation between $t$ and $v$ together with the exponential
relation between $t$ and $s$ ultimately leads to the logarithmic RG
flow of couplings.} Figure \ref{fig:witten-1} suggests that the
curve $\Sigma$ wrapped by the M5-brane is simply the locus in
$\IC^\times \times \IC$,
\begin{equation}\label{eq:degencurve}
v^K \prod_{\a = 0 }^n (t-t_\a) = 0.
\end{equation}
As we will see when describing the four-dimensional
gauge theory interpretation, \eqref{eq:degencurve} actually corresponds to the conformal
point, with all masses and all vevs $\langle {\Tr} (\Phi^{\a})^s
\rangle $ ($1\leq \a \leq n$, $s=1,\dots, k_\a$) vanishing.
On the Coulomb branch, the various M5-branes
of the setup join into a single smooth Riemann surface,
defined by a polynomial equation in $(t,v)$ which
deforms (\ref{eq:degencurve}) by terms of lower order in $v$:
\begin{equation}\label{eq:swcurve1}
F(t,v) = v^K \prod_{\a =0 }^n (t-t_\a) + \sum_{i=1}^K p_i(t) v^{K-i} = 0.
\end{equation}
If $p_1(t_\a)\not=0$, then at $t = t_\a$ one and only one of
the roots $v(t)$ has a pole, with residue $\rho_\a:=-\frac{p_1(t_\a)}{\prod_\beta (t_\a-t_\beta)}$.
The branch of $\Sigma$ corresponding to this
divergent root at $t=t_\a$ is physically identified as the M-theory
lift of the $\a$-th NS5-brane in the IIA picture.

We also want to have $K$ fivebranes going to constant values of $v$
for very large and very small $t$. For very small $t$ the $K$ roots
of (\ref{eq:swcurve1}) represent the asymptotic coordinates of the $K$
D4-branes stretching to $x^6 \to +\infty$. If we wish
also to have $K$ distinct roots for $v(t)$ in the limit $t \to \infty$
then each $p_i$ should have degree at most $n+1$.   Thus, $F$ can
also be written as
\begin{equation} \label{eqn:sw1}
F(t,v)= \sum_{\a=0}^{n+1} q_\a(v) t^{n+1-\a} = 0.
\end{equation}
Here the $q_\a(v)$ are polynomials in $v$ of degree $\leq K$.
Moreover, $q_0(v)$ and $q_{n+1}(v)$ must be of degree $K$.  The roots
of $q_0$ and $q_{n+1}$ are the $k_0=K$ and $k_{n+1}=K$ asymptotic
values of the roots $v(t)$ for very large or very small $t$,
respectively. We denote these by $v^{(0)}_1, \dots, v^{(0)}_K$ and
$v^{(n+1)}_1, \dots, v^{(n+1)}_K$, respectively.

\subsubsection*{Weak coupling}

We would like to get some intuition about the physical meaning of
the coefficients of $F(t,v)$. This is easy if the four-dimensional
gauge theories are weakly coupled. The
gauge couplings of the four-dimensional $SU(K)$ theories are
determined by the dilaton, which sets the coupling of the
five-dimensional gauge theory living on the D4-branes, and by the
distance between NS5-branes in the $x^6$ direction. In the lift to
M-theory, this combination only depends on the coordinate $s$ of the
defects, and is independent of the overall scale $R_{11}$ of the
cylinder. The relation, including the theta angles, is
\begin{equation}
- i \pi \tau_\a = - i\pi \left(\frac{\theta_\a}{2 \pi} + \frac{4 \pi
i}{g_\a^2}\right) = s_{\a} - s_{\a-1}, \qquad\qquad \a = 1, \dots, n.
\end{equation}
This relation is based on the assumption that to read off the gauge
couplings, it makes sense first to descend from the $(2,0)$ theory
on the cylinder with defects to the 5d SYM worldvolume theory on the
D4-branes, and then to the four-dimensional low energy theory. This
is true only in the limit where the separation between the defects
is much larger than the radius of the cylinder, i.e. the gauge
couplings are weak. On the other hand, at strong coupling the
$\tau_\a$ actually lack a precise definition: even if the theories
are finite, there still is a scheme-dependent finite
renormalization. The simplest choice is  simply to take the $t_\a$
(up to an overall rescaling) as a convenient parameterization of the
space of marginal deformations of the ${\cal N}=2$ SCFT.

At weak coupling, we can take
\begin{equation}\label{eq:wkcpl}
\frac{\vert t_{\a}\vert }{\vert t_{\a-1} \vert}  = \epsilon_\a \to
0, \qquad\qquad 1\leq \a \leq n,
\end{equation}
so that in Type IIA language the NS5-branes are widely spaced.
Next, let $p_0(t) =\prod_{\a=0}^n (t-t_\a) = \sum_{s=0}^{n+1} c_{n+1-s}t^s $.
Then in this limit $\vert c_j\vert = \vert t_0 \cdots t_{j-1}\vert (1+ \CO(\epsilon_{j}) ) $,
$j=1,\dots, n+1$. Let us also set
\begin{equation}\label{eq:wkcpl2}
q_\a(v) = c_\a \tilde q_\a(v)
\end{equation}
The claim is that the roots of the monic polynomials $\tilde q_\a(v)$ parameterize
the positions of the $\a$-th group of D4-brane segments, i.e. the
Coulomb branch of the $\a$-th gauge group in the quiver. Indeed we can take $\vert t\vert $
in the range
\begin{equation}
\vert t_{\b} \vert \ll \vert t \vert \ll \vert t_{\b-1} \vert
\end{equation}
so that, in Type IIA language,
the $x^6$ positions of D4-branes lie well inside the
interval between the $\b^{th}$ and $(\b-1)^{th}$ NS5-branes. Then
\begin{equation}
\vert t^{n+1} \vert \ll \vert t^{n} t_0 \vert \ll \cdots \ll \vert
t^{n+1-\b} t_0 \cdots t_{\b-1} \vert
\end{equation}
\begin{equation}
\vert t^{n+1-\b} t_0 \cdots t_{\b-1} \vert \gg \vert t^{n-\b
}t_0\cdots t_{\b} \vert \gg \cdots \gg \vert t_0 \cdots t_{n+1}
\vert
\end{equation}
and hence in the equation $F(t,v) = 0$ the dominant term is
\begin{equation}
t^{n+1-\b} t_0 \cdots t_{\b-1} \tilde q_\b(v) =0
\end{equation}
and the transverse position of the M5-branes or D4-branes is close to the roots of $\tilde q_\b(v)$.
Again, at strong coupling this approximate statement becomes incorrect.  At the same time
scheme dependent finite renormalization and operator mixing in the field theory makes it a bit meaningless to talk about ``the Coulomb branch of the $\a$-th gauge group'', and try to identify the vevs  $\langle {\Tr} (\Phi^{\a})^s \rangle$ of specific operators in the gauge theory.

\subsubsection*{The Seiberg-Witten curve}

A beautiful insight of \cite{Witten:1997sc} is that the  curve
(\ref{eq:swcurve1})  should be identified with the Seiberg-Witten
curve for the four-dimensional linear quiver gauge theory.   From
our present perspective, this follows from the general discussion of
Section \ref{sec:scft}. The parameters of $F$ other than $t_\a$,
i.e., the coefficients of the polynomials $p_i(t)$, encode the vevs
of the adjoint scalars and the mass parameters.

The precise identification of the mass parameters is subtle, and
important. The key to making this identification is the observation
that the residues of the   Seiberg-Witten differential encode the
mass parameters. There is a simple canonical form for this
differential for the problem at hand, which was only implicit in
\cite{Witten:1997sc}, and was made explicit in
\cite{Fayyazuddin:1997by,Henningson:1997hy,Mikhailov:1997jv}.

The computation is a special case of the general discussion of
Section \ref{subsubsec:KineticTerms}, but is more straightforward in
this case, so we repeat it. Thus we consider a low energy
configuration of the M5-brane where the embedding into
$\IR^{1,3}\times \IC^\times \times \IC$ is given by $(x^\mu, t,
v(t;\xi_i(x^\mu) ))$. Here $\xi_i$ are a set of independent
parameters of the curve, say, the coefficients of the polynomial
$F(t,v)$.\footnote{Note that $t = e^{-s}$ is \emph{not} a time
coordinate; it is the coordinate along $C = \IC^\times$.}  The
moduli fields $\xi_i(x^\mu)$ are assumed to be slowly varying as
functions of the $x^\mu$.  The M5-brane action has a contribution
from the fluctuations of the normal bundle scalars coming from the
induced worldvolume metric. Using the metric of equation
(\ref{eq:MthryMetric}), $ds^2 = dx^\mu dx_\mu + \ell^2 \vert
dv\vert^2 + R_{11}^2 \vert \frac{dt}{t} \vert^2 + dx^2_{7,8,9}$ we
find the kinetic energy after subtracting the energy from the
tension of the M5-brane in the reference configuration:
\begin{equation}\label{eq:KEreduxii}
\frac{R_{11}^2}{\ell^2} \int_{\IR^{1,3}} dx^{0123}\int_{\Sigma}
\frac{dt d \bar t}{\vert t\vert^2} \frac{\p v }{\p \xi_i} \p_\mu
\xi_i \overline{ \frac{\p v }{\p \xi_j} \p_\mu \xi_j }.
\end{equation}
Now, since $\frac{\p v }{\p \xi_i}$ is a holomorphic function of $t$,
we can do the integral over $\Sigma$ as follows. First note that if
the one-form $\frac{dt}{t}\frac{\p v }{\p \xi_i}$ has a simple pole
at any value of $t$, including $t=0, \infty$, then the integral
diverges. Thus, the \emph{normalizable} variations of the parameters
of $F(t,v)$ are precisely those which do not alter the poles of
$\lambda = v \frac{dt}{t}$. We label these normalizable parameters by
$u_i$. Now, let $\bar \Sigma$ be the compact Riemann surface obtained
by embedding $\Sigma$ in projective space and filling in the punctures.  These
punctures correspond to the noncompact branches at $t_\a$ and the
$K$ branches above $t=0,\infty$. Introduce a basis $\gamma_a$ for
$H_1(\bar \Sigma;\IZ)$ and let $\CI^{ab}$ be the (dual) intersection
form in this basis.  Then we may compute
 \begin{equation}
 \int_{\Sigma} \frac{dt d \bar t}{\vert t\vert^2} \frac{\p v }{\p
 \xi_i}   \overline{ \frac{\p v }{\p \xi_j}  }= \int_{\bar\Sigma}
 \frac{dt d \bar t}{\vert t\vert^2} \frac{\p v }{\p \xi_i} \overline{
 \frac{\p v }{\p \xi_j}  }=\sum_{a,b} \CI^{ab} \oint_{\gamma_a}
\frac{\p   }{\p \xi_i}\lambda \oint_{\gamma_b} \overline{\frac{\p
}{\p \xi_j}\lambda }.
 \end{equation}
 In this way we derive the low energy effective action for the
 kinetic energy of the normal bundle scalars:
 \begin{equation}\label{eq:dualinvtke}
 \frac{R_{11}^2}{\ell^2} \int_{\IR^{1,3}} dx^{0123} \CI^{ab} \p_\mu
 \Pi_a \p_\mu \bar \Pi_b
 \end{equation}
 where $\Pi_a = \oint_{\gamma_a} v \frac{dt}{t}  $. If we choose a duality
 frame $(\gamma_I, \gamma^I)$ then (\ref{eq:dualinvtke}) becomes
 the standard expression $\int {\rm Im}(\tau_{IJ}) \p_\mu a^I \p_\mu \bar a^J$
 that one finds in the
$\CN=2$ effective action. Thus we conclude that the  Seiberg-Witten
differential is
\begin{equation}
\lambda_{SW} = \lambda = v \frac{dt}{t}.
\end{equation}

It is well-known that the residues of the Seiberg-Witten
differential depend affine-linearly on the mass parameters of the UV
Lagrangian. The differential $v \frac{dt}{t}$ has simple poles on
the $K$ branches covering $t=0$ and on the $K$ branches covering
$t=\infty$. The residues at these poles are the asymptotic positions
  in the $v$-plane of the semi-infinite D4-branes on the left and
right end of the quiver, respectively. We can verify that these
depend affine-linearly on the masses as follows:
 Consider the Type IIA picture and suppose that $k_1$ D4's are at
a common $v$-position $v_0$ in the first interval $I_1$. Classical
string theory shows that the
  mass of the $K$ fields in the fundamental
representation of $SU(k_1)$ are given by $\ell_s^{-1}\vert v_i^{(0)}
- v_0\vert $, where $v_i^{(0)}$ are the $K$ roots of $q_0(v)$.  By
holomorphy we see that the complex mass parameters $\hat
\mu^{(1)}_i$ satisfy $ v_i^{(0)} - v_j^{(0)} \propto \hat \mu_{1,i}
- \hat \mu_{1,j} $. A similar statement holds for the fundamental
fields associated with the branches stretching to $t\to +\infty$.
Thus, the differences of  mass parameters associated with the
$SU(K)$ flavor symmetry at each end are given by the differences
between the roots of $q_0(v)$ and between the roots of $q_{n+1}(v)$.

Finally, for generic parameters of the polynomial, as we have seen,
precisely one root $v(t)$ has a simple pole as $t\to t_\a$, leading
to a simple pole for the Seiberg-Witten differential with residue
$\rho_\a/t_\a$. In the weak-coupling regime described in equations
(\ref{eq:wkcpl}) et. seq. it is not difficult to show that if we
write:
\begin{equation}\label{eq:Paramqalpha}
q_{\a}(v) = c_\a (v^K- \mu_\a v^{K-1} - u_2^{(\a)} v^{K-2} - \cdots
- u_K^{(\a)} )
\end{equation}
then the residues at $t_\a$ are approximately $(\mu_\a -
\mu_{\a+1})$, for $0\leq \a \leq n$, and coincide either with
 the mass parameters for the bifundamental matter fields between the $\a$ and $\a+1$ nodes of the quiver, or with the overall $U(1)$ mass parameter of the fundamental matter fields at either end of the quiver for $\a=0$ or $\a=n$.

In the field theory limit the overall center of mass degree of
freedom of the M5-brane system decouples. We aim to arrive at a
description which involves only the degrees of freedom of the
$A_{K-1}$ $(2,0)$ theory.  A natural way ignore the shift of the
overall center of mass is to eliminate the coefficient of $v^{K-1}$
in $F(t,s)$.  To do this, we define
\begin{equation}
\tilde v := v - p(t)
\end{equation}
and make the choice  $p(t) = -\frac{p_1(t)}{K\prod_\alpha
(t-t_\alpha)}$. Note that the residue at $t=t_\alpha$ is
$\rho_\alpha/K$.

Next, recall that all physical quantities depend on integrals of the
Seiberg-Witten differential along closed paths in the Seiberg-Witten curve. In a given IR
theory the Seiberg-Witten differential $\lambda$ is not unique:  it is required to
satisfy (in any duality frame)
\begin{equation}
\tau_{IJ} = \frac{\p}{\p a^J} \oint_{\gamma_I} \lambda,
\end{equation}
but it may be modified by the addition of a one-form which does not
depend on the normalizable deformations. (Recall the notion of a
normalizable deformation is defined by the finiteness of the kinetic
energy (\ref{eq:KEreduxii})). In particular, we are free to change
the definition of the Seiberg-Witten differential by the shift $v
\to v + p(t)$, since $p(t)$ is a single-valued rational function of
$t$. The residues of $p(t) \frac{dt}{t}$ are linear combinations of
the mass parameters of the theory, and the the first derivatives
$\lambda_{u_i}$ are unchanged, and still coincide with the
holomorphic 1-forms on the Seiberg-Witten curve. The periods of
$\lambda$ are shifted by a certain linear combination of the mass
parameters. In order to write a central charge as a sum of
contributions from ``gauge charges'' and ``flavor charges'' we need
to choose a splitting of the sequence \eqref{eq:Lattice-ExactSeq}.
Once such a splitting has been made the shift of $\lambda$  amounts
to a shift of the flavor charges of BPS particles by multiples of
their gauge charges, i.e. to a legal redefinition of the flavor
currents. The shift of flavor charges is not simply harmless, it is
actually a useful improvement. For example, in the $K=2$ case, we
saw that the flavor symmetry groups are enhanced from $U(1)$ to
$SU(2)$ in the field theory limit. The shift in $\lambda$ gives a
charge assignment to BPS particles which is compatible with the
organization in irreps of the new $SU(2)$'s (see Section
\ref{sec:flavor} for more information).

The new Seiberg-Witten differential $\tilde v \frac{dt}{t}$ has
simple poles on the different sheets above $t=t_\a$  whose list of
residues is given by
\begin{equation}
\left( - (K-1) m_\a, m_\a, \dots, m_\a \right)
\end{equation}
where
\begin{equation}\label{eq:def-m-alpha}
m_\a = \frac{1}{K t_\alpha} \frac{p_1(t_\a)}{\prod_{\beta \not=\a}
(t_\alpha -t_\beta)}= - \frac{\rho_\a}{K t_\a}
\end{equation}
is identified with the mass parameter of a $U(1)$ flavor symmetry.

At $t=0$ (and similarly at $t=\infty$) the poles on different sheets
will be of the form
\begin{equation}
\left( - \sum m_{0,i}, m_{0,1}, \dots, m_{0,N-1} \right)
\end{equation}
where $m_{0,i}$ (and the corresponding $m_{\infty,i}$ at $t=\infty$)
are linked to the relative positions of the semi-infinite D4-branes
and are identified with the mass parameters of the $SU(K)$
flavor symmetry of either group of $K$ fundamentals. Notice that if
$K=2$ the two types of defect are identical, and indeed the $U(1)$'s are promoted to $Sp(1) \sim SU(2)$.

\subsubsection*{An example}

We close this section with an example. Let us consider $n=1$, so
that we have a simple gauge group $SU(K)$ with $2K$ fundamentals.
The standard Seiberg-Witten curve for this theory naively appears to be very different
from the one we consider here, especially when all the mass parameters are included \cite{Argyres:1995wt}:
\begin{eqnarray}
y^2 & = & P(w)^2 - (1-g^2) Q(w)\\
P(w) & = & \langle \det(w-\Phi)\rangle  = w^K - u_2 w^{K-2} -\cdots
- u_K \\
Q(w) &=&\prod_{I=1}^{2K} (w + g \mu + \mu_I)
\end{eqnarray}
Here the parameters $\mu, \mu_I$ determine the flavor masses by $\mu = \frac{1}{2K}
\sum_{I=1}^{2K} m_I $ and $\mu_I = m_I - \mu$. The parameter $g$
is a certain modular function of the coupling
$\tau$ (see eq. (5.15) of \cite{Argyres:1995wt})
which for weak coupling reduces to $g= 1 + \CO(e^{i \pi \tau})$.
The Seiberg-Witten differential is
\begin{equation}
\lambda^{\rm Standard} = \frac{w + (g-1) \mu}{2\pi i} d \log \left(
\frac{y-P}{y+P} \right).
\end{equation}

By a transformation of variables, we can bring the curve into the
form $F(t,v)=0$ where $F(t,v)$ is of the form (\ref{eq:swcurve1}).
First we introduce $\gamma$ obeying
\begin{equation}\label{eq:couplrel}
\frac{4}{(\gamma + \gamma^{-1})^2} = 1-g^2.
\end{equation}
Next, we set $v= w  + (g-1) \mu$; then introducing $v_I := m_I$ we have
\begin{equation}
Q(w) = \prod_{j=1}^{2K} (v-v_I).
\end{equation}
The symmetric group $S_{2K}$ acts on the set of roots $v_I$.
($S_{2K}$ is the Weyl group of the $U(2K)$ flavor symmetry group
which is broken to the Cartan by the masses.) We now explicitly
break the symmetry to $S_K \times S_K$ by choosing a set of $K$
roots $v_i$, $i=1,\dots, K$. We denote the remaining roots by
$\tilde v_i$, $i=1, \dots, K$.  Now we set
\begin{equation}
t=  \half(\gamma + \gamma^{-1}) \frac{y-P(w)}{\prod_{i=1}^K( v-v_i)}
\end{equation}
  Finally we take $ \tilde P(v) = P(w)$. In this way we
bring the curve to the form
\begin{equation}\label{eq:APStwo}
\prod_{i=1}^K (v-v_i) t^2 + (\gamma + \gamma^{-1}) \tilde P(v) t +
\prod_{i=1}^K (v-\tilde v_i) = 0
\end{equation}
which is of the form (\ref{eqn:sw1}). Note that if we put equation
(\ref{eq:APStwo}) into the form (\ref{eq:swcurve1}) we find that
$t_0 = \gamma, t_1 = \gamma^{-1}$, and thus from (\ref{eq:couplrel})
we confirm that $t_0/t_1 \sim e^{i \pi \tau}$ in accord with the
discussion of the weak-coupling limit.  Also, note that in this
presentation $p_1(t)\not=0$, so we have not yet fixed the center of
mass degree of freedom in the standard way.

It is worth noting that the standard Seiberg-Witten differential differs from
that naturally arising from our discussion:
\begin{equation}
i \pi \lambda^{\rm Standard} - \lambda^{\rm Hitchin} = \half v d
\log \left( \frac{\prod_{i=1}^K (v-v_i)}{\prod_{i=1}^K (v-\tilde
v_i)} \right)
\end{equation}
The residues of the difference, $v_i$, are just mass parameters, so
this corresponds again to a simple redefinition of the flavor
charges. Before we remove the center of mass piece, $\lambda^{\rm
Hitchin}$ has a pole on a single sheet at $t_0= \gamma$ or $t_1 =
\gamma^{-1}$. It takes some work to extract the residue through all
changes of variables, but the result is remarkably simple: the
residue at $t_0$ is $\sum \tilde v_i$, the overall mass parameter
for the $K$ fundamental flavors associated to the $K$ semi-infinite
branes on one side. The residue at $t_1$ is similarly $\sum v_i$.

The residues of $\lambda^{\rm Hitchin}$ at $t=0, \infty$ coincide
with the $v_i$ or $\tilde v_i$, but after we remove the center of
mass, the residues are modified to $v_i - \frac{1}{K} \sum v_i$ and
$\tilde v_i - \frac{1}{K} \sum \tilde v_i$, which are the mass
parameters of the $SU(K)$ flavor subgroups acting on each group of
$K$ fundamental flavors.

This splitting is natural: from the point of view of the $(2,0)$
theory the $U(2K)$ flavor symmetry is an accidental IR symmetry. The
six dimensional theory only has a $U(1)^2 \times SU(K)^2$ flavor
symmetry, and each factor is somehow associated to one of the four
``punctures'' at $t=0, t_0,t_1,\infty$. Notice that the punctures at
$0,\infty$ have different properties from the punctures at $t_0,
t_1$. We will see in the next subsection how the difference
manifests itself in the context of the Hitchin system.

\subsubsection{Mapping to a Hitchin system}\label{subsubsec:MapToHitchin}

We are now ready to return to the Hitchin system described in
Section \ref{sec:scft}.  We interpret the $K$
branches $\tilde v(t)$ of the solutions to $F(t,v)=0$ as the $K$
eigenvalues of the Higgs field $\varphi_s$, because these are the
positions of the M5-branes.  Our choice to fix the center of mass
to eliminate the $\tilde v^{K-1}$ term in the polynomial $F(t,v)$
guarantees that $\varphi_s$ is valued in $SU(K)$ and not $U(K)$. The
spectral curve of the Hitchin system on the cylinder is, by
definition, $\det (\tilde v + \varphi_s) =0 $. We want to identify
this spectral curve with the curve $F(t,v)=0$ wrapped by the IR
M5-brane. The only important point is to remember that $\varphi_s$
transforms as a one-form, $\varphi_t dt= \varphi_s ds$. Identifying
the spectral curve with the curve $F(t,v)=0$ and making the
definition $x := \frac{\tilde v}{t}$, the spectral curve in
$T^* \IC^\times$ has the form
\begin{equation}\label{eq:MatchHitchin}
\det (x-\varphi_t) = x^K + \sum_{i=2}^{K} \frac{\tilde p_i(t)}{ \left( t \prod_{\alpha=0}^{n} (t-t_\alpha) \right)^i } x^{K-i} = 0.
\end{equation}
(The $\tilde p_i$ differ from the $p_i$ which appeared in \eqref{eq:swcurve1} because of
the shift we made from $v$ to $\tilde v$.)
If we do not factor out the $U(1)$ degree of freedom we identify
$x=v/t$. In any case, this equation is to be identified with
\eqref{eq:speccurve}.  Note that the Seiberg-Witten
differential is $x\,dt$, as asserted in Section
\ref{sec:scft}.

In order to treat the singularities in a more symmetric way it is useful to
embed $\IC^\times \to \IC \IP^1$ via $t \to [t:1]$. Then we recognize that
the boundary conditions at the ends of the cylinder, i.e. at $t=0,
\infty$, state that $\varphi_t dt$ has a simple pole  with
residue given by the asymptotic values of $v$. Thus, we deduce
the boundary conditions on the Higgs field:
\begin{equation}\label{eq:SimpleAtZero}
\varphi_t dt  \to \frac{dt}{t} {\rm Diag}\{v^{(0)}_1, \dots, v^{(0)}_K
\}\qquad t\to 0
\end{equation}
\begin{equation}\label{eq:SimpleAtInfty}
\varphi_t dt  \to \frac{dt}{t} {\rm Diag}\{v^{(n+1)}_1, \dots, v^{(n+1)}_K
\}\qquad t \to \infty
\end{equation}

There are in addition simple poles at $t=t_\a$ with a residue  of
the form
\begin{equation}\label{eq:SimplePoleResidue}
{\rm Diag}\{-(K-1) m_\a, m_\a,m_\a,\cdots m_\a,m_\a \}  ,\qquad \a =
0, \dots, n
\end{equation}
if we factor out the $U(1)$ degree of freedom, and ${\rm
Diag}\{\rho_\a, 0,\dots, 0 \}$ if we do not. Recall that $m_\a$ is
given in equation (\ref{eq:def-m-alpha}). Note that the residues at
$t=0, \infty$ are generic semisimple elements of $su(K)$ and hence
correspond to regular singularities, but this is not at all true
when we consider the residues at $t=t_\a$, $\a = 0, \dots, n$, for
$K>2$, since then the residues are annihilated by all but the first
simple root (for the standard choice of simple roots).

We stress that the   conditions (\ref{eq:SimpleAtZero}),
(\ref{eq:SimpleAtInfty}) and (\ref{eq:SimplePoleResidue}) mean that
there is a local gauge in which the fields can be put in this form.

In the main part of the paper we will focus on the case $K=2$.  In that case, after
the center of mass $U(1)$ is removed as described above, there is no
longer any distinction between the two kinds of singularity.  Thus, the
Hitchin system turns out to be a general $SU(2)$ Hitchin
system on $\IC \IP^1$ with $n+3$ regular singularities. The equation
(\ref{eq:MatchHitchin}) is equivalent to
\begin{equation}\label{eq:su2exple}
{\rm tr} \varphi_t^2 = \half \frac{ (p_1(t))^2 - 4 p_2(t)
\prod_{\a=0}^n (t-t_\a) }{(t \prod_0^n (t-t_\a))^2}.
\end{equation}
Note that $\varphi_t dt$ has simple poles at $t = 0$ and $t =
\infty$ as well as at the $n+1$ points $t_a$. We will return to this
formula in Section \ref{section:Superconformal}.

\subsubsection{Linear conformal quivers with fundamental matter}
\label{subsubsec:LinConfQuivMatt}

It is straightforward to extend this analysis to a more general
linear conformal quiver with fundamental matter. This is a quiver of
$SU(k_\a)$ gauge groups, with $d_\a = 2 k_\a - k_{\a-1} - k_{\a +
1}$ fundamental fields at the $\a$-th node, $\a = 1, \dots, n$. The
extra fundamental fields are represented in the brane construction
by $d_\a$ D6-branes at fixed values of $x^{4,5,6}$ with the value of
$x^6$ in the interval $I_\a$. We will denote with $K$ the maximum of
the $k_\a$. When we include D6-branes the semi-infinite D4-branes on
the left and right do not change the analysis in any interesting
way, so in the remainder of this section we will omit them. As
before, for $\a =1,\dots, n$, $k_\a$ D4-branes are stretched between
the NS5-branes along the interval $I_\a$. We assume that initially
there are no D4-branes ending on D6-branes. In the weak IIA coupling
limit the D4-D6 strings provide the fundamental matter. When the
$k_\a$ D4-branes are coincident, the mass parameter of these
fundamental hypermultiplets is given by the difference of the $v$
coordinates for the D4-brane and the D6-brane.

The  lift to M-theory of the configuration we have just described is
obtained by lifting the configuration of D6-branes to a multi-center
Taub-NUT geometry $Q$ and then -- on the Coulomb branch of vacua --
lifting the D4- and NS5-branes to a single M5-brane with
worldvolume $\IR^{1,3} \times \tilde \Sigma \times \{\CP\}$,
where $\tilde \Sigma \subset Q$ is a holomorphic surface. Our goal in this
section is to make contact once again with a Hitchin system on a
cylinder, and show how the fundamental matter changes the boundary
conditions on the Higgs field.

It turns out that there are several distinct ways to do this,
as the same brane system can be subject to certain deformations which lead to the same IR
four dimensional field theory, but to different intermediate six-dimensional setups.
The detailed analysis is a bit intricate, so we will first anticipate the main result.

For linear conformal quivers there is a canonical choice, which still leads to a $SU(K)$ Hitchin system on the cylinder with regular singularities. The boundary conditions at $t=0$ and $t=\infty$,
\eqref{eq:SimpleAtZero} and \eqref{eq:SimpleAtInfty}, are modified slightly, as the
$v^{(0)}_i$ and the $v^{(n+1)}_j$ are not generic anymore. Instead, the $K$ $v^{(0)}_i$
are partitioned in blocks of identical values, so that each block of $\a$ identical $v^{(0)}_i$
corresponds to a fundamental flavor
at the $\a$-th  node of the quiver, and each block of $\a$ identical $v^{(n+1)}_i$
corresponds to a fundamental flavor
at the $(n+1-\a)$-th  node of the quiver.
For example, if $K=3$ we have quivers of $SU(3)$ gauge nodes, possibly
with a single $SU(2)$ gauge group at either or both ends. If, say, $k_1=2$ and $k_2=k_3=3$,
 then $d_1=d_2=1$, $d_3=3$, and there is a single fundamental at the first node,
and another at the second node. Then the residue of the regular singularity
at $t=\infty$ will have two identical eigenvalues.

This canonical Hitchin system is characterized by the
fact that the $U(1)$ R-symmetry of the 4d theory is identified with the $U(1)$ symmetry rotating $v$.
Other choices lead to Hitchin systems (of rank higher than $K$) on the cylinder,
with no punctures away from $t=0, \infty$, and with wild ramification either at
$t=0$ or $t=\infty$. The $U(1)$ R-symmetry of the 4d
theory is identified with a combination of the $U(1)$ symmetries
rotating $v$ and rotating $t$. It is surprising that the same 4d field theory,
and hence the same \hk moduli space should be described by different Hitchin systems.
The matter is discussed further in Section \ref{sec:surprise}.

To begin, let us recall the metric on $Q$.  The metric on $Q$ is
determined by the positions $\vec r_a = (r_{1,a}, r_{2,a}, r_{3,a})
\in \IR^3$ where $a$ runs over some index set. If we describe $Q$ as
a circle fibration over $\IR^{3}$, then the metric is
\begin{equation}
ds^2_{TN} = V^{-1} (d\psi +A)^2 + V d\vec r^2
\end{equation}
where $\psi \sim \psi + 4\pi$,
\begin{equation}
V =1 + \sum_{a} \frac{1}{\vert \vec r - \vec r_a\vert}
\end{equation}
and $dA = *dV$. The full M-theory metric is
\begin{equation}
ds^2 = dx^\mu dx_\mu + \frac{R_{11}^2}{4} ds^2_{TN} + dx^2_{7,8,9}
\end{equation}
We identify $s = x^6 + i x^{10} = \half ( r_3 + i \psi)$ and $r_1+i
r_2 = 2 \frac{\ell}{R_{11} } v$.   The Taub-NUT space $Q$ carries a
hyperk\"ahler structure, but there is a distinguished complex
structure in which the coordinate $v $ is holomorphic. In this
complex structure the manifold $Q$, as a complex manifold, has
equation
\begin{equation}\label{eq:MultiCenterTNa}
UW  = \prod_a (v-v_a)
\end{equation}
The parameters $v_a$ are the complex structure parameters, while the
 $x^6_a$ are the K\"ahler parameters of the hyperkahler metric on
 $Q$.

Now we turn to a description of the holomorphic  curve $\tilde
\Sigma$   in the complex manifold (\ref{eq:MultiCenterTNa}). The
curve can   be described as a polynomial equation $\tilde F(W,v)=0$.
Alternatively, in a different chart, using $U=W^{-1} \prod_a
(v-v_a)$ this equation can be also be rearranged as $\tilde
G(U,v)=0$ for a polynomial $\tilde G$. In \cite{Witten:1997sc} it is
shown that the constraint that there are no semi-infinite D4-branes
on the left or the right leads to the following structure for
$\tilde F(W,v)$: To each interval $I_\a := \{ x^6: Re(s_{\a -1}) <
x^6 < Re(s_\a)\}$ we associate a polynomial of degree $d_\a$:
\begin{equation}
J_\a(v):= \prod_{x^6_a\in I_\a} (v-v_a), \qquad\qquad \a=1,\dots, n.
\end{equation}
Now introduce
\begin{eqnarray}\label{eq:Aalphdef}
A_0(v) &= & c_0\\
 A_\a(v) &= & c_\a g_\a(v) \prod_{\beta =1}^{\a }
J_\beta^{\a-\beta}(v), \qquad\qquad \a = 1, \dots, n \\
A_{n+1}(v) & =& c_{n+1}   \prod_{\beta =1}^{n}
J_\beta^{n+1-\beta}(v),
\end{eqnarray}
where $c_\a$, $\a = 0, \dots, n+1$,  are nonzero constants and
$g_{\a}(v)$ are  \emph{monic} polynomials in $v$ of degree $k_\a$,
$\a = 1, \dots, n$. (It is also convenient to define $g_0 =
g_{n+1}=1$.) Then we have the curve (\cite{Witten:1997sc}, eq.
(3.23))
\begin{equation}\label{eq:TNCurveii}
\tilde F(W,v) =  \sum_{\a=0}^{n+1} A_\a(v) W^{n+1-\a}=0
\end{equation}
or equivalently
\begin{equation}\label{eq:TNCurveiii}
\tilde G(U,v) =  c_{n+1}U^{n+1} + c_n g_n(v) U^n + \sum_{\a=0}^{n-1}
g_{\a}(v)\left( \prod_{\b = \a+1}^n J_\b(v)^{\b-\a}\right) U^\a =0
\end{equation}
Moreover, the coefficients of $g_{\a}(v)$ are interpreted in
\cite{Witten:1997sc} as the usual order parameters $\langle
{\Tr}(\Phi^{(\a)})^s\rangle$ of the $SU(k_\a)$ gauge group. The
constants $c_\a$ can of course be rescaled by a common nonzero
factor and hence should be viewed as a point in projective space.
They encode  the gauge couplings. (In the weak coupling region the
$c_\a$ can be given in terms of the $\a^{th}$ elementary symmetric
function of the $t_\a$. )

Naively the Taub-NUT setup does not seem to lead to a situation
where a Hitchin system is useful, as it looks quite different from
our previous picture of  $T^*C$ with $K$ fivebranes wrapping $C$
together with  transverse defects. However, as we now explain, we
can take a limit of the Taub-NUT geometry in which we recover the Hitchin
description.

The curve $\tilde \Sigma$ will again be interpreted as the
Seiberg-Witten curve. Recall  that the polynomial $\tilde F(W,v)$ is
independent of the K\"ahler parameters $x^6_a$, as is the complex
structure of $Q$.  As we saw in Section \ref{subsubsec:KineticTerms}
and again in \eqref{eq:KEreduxii} et. seq., the kinetic terms and
Seiberg-Witten differential only depend on the complex structure,
and hence are unaffected by changes in $x^6_a$: two theories which
differ only by translation of the D6-branes in the $x^6$ direction
are described by the same IR fixed point. On the other hand, in the
limit where the D6-branes are brought far to the left or far to the
right of the NS5-brane system, the setup strongly resembles the
setup without D6-branes, since evidently $V\to 1$ if $\vert
x^6_a\vert \to \infty$ at finite values of $\vec r$.

Now, let us divide the D6-branes into two disjoint subsets
$\CL \amalg \CR$ and consider a limit where the D6-branes with $a \in
\CR$ move far to the right while the D6-branes with $a\in \CL$ move far
to the left. In Appendix \ref{app:HoloTN} we show that, when
properly normalized, the limiting value of $W$ is
\begin{equation}\label{eq:W-t-limit}
W \to t \prod_{a\in\CL}(v-v_a).
\end{equation}
We substitute this into \eqref{eq:TNCurveii} and simplify. To do
this note that each polynomial $J_\a(v)$ factorizes according to the
D6-branes which move to the left and right respectively: $J_\a(v) =
J_{\a,L}(v) J_{\a,R}(v)$. After factoring out $\prod_{\a}
J_{\a,L}^{n+1-\a}$ we are left with
\begin{equation}\label{eq:Fhatcurve}
\hat F(t,v) = \sum_{\a=0}^{n+1} \hat g_{\a}(v) t^{n+1-\a},
\end{equation}
where
\begin{eqnarray}\label{eq:hatgee}
\hat g_0(v) & =&  c_0 \prod_{\beta=1}^n J_{\beta,L}^\beta \\
\hat g_{\a}(v) & =& c_{\a} g_{\a}(v) \prod_{\beta=1}^{\a} J_{
\beta,R}^{\a-\beta} \prod_{\beta=\alpha+1}^{n}
J_{\beta,L}^{\beta-\a}\qquad\qquad \a =1, \dots, n-1\\
\hat g_n(v) & = & c_{n} g_{n}(v) \prod_{\beta=1}^{n} J_{
\beta,R}^{n-\beta}\\
\hat g_{n+1}(v) &= & c_{n+1} \prod_{\beta=1}^{n}
J_{\beta,R}^{n+1-\beta}.
\end{eqnarray}

There is an elegant interpretation of the factors in $\hat
g_{\a}(v)$ based on the Hanany-Witten effect \cite{Hanany:1996ie}.
In our context, the Hanany-Witten effect states that for each
D6-brane the ``linking number''
\begin{equation}
 (L-R) - \half (l -r )
\end{equation}
is constant. Here $l$, ($r$) is the number of NS5-branes to the left
(right) of the D6-brane and $L$, ($R$) is the number of D4-branes
ending on the D6 from the left (right).  In particular, if a D6-brane
moves in the $x^6$ direction at constant $v$ across an NS5-brane, a
D4-brane stretched between them is created.   After we have
moved the D6-branes with $a\in \CR$ far to the right, and those with
$a\in \CL$ far to the left each interval consists of ``free'' and
``frozen'' D4-branes.\footnote{The adjective ``frozen'' refers to
the so-called $S$-rule discussed in \cite{Hanany:1996ie}. The
D4-branes in the interval $I_\a$ which are created  from the
horizontal motion of the D6-branes cannot move independently in
the $v$-direction while preserving supersymmetry.} The factor
$g_{\a}(v)$ in $\hat g_{\a}$ accounts for the ``free'' D4-branes.
The coefficients of $g_{\a}(v)$ correspond to the normalizable
deformations in the usual way. The factor $\prod_{\beta=1}^{\a}
J_{R,\beta}^{\a-\beta} $ accounts for the ``frozen'' D4-branes in
the interval $I_\a$ which have been created by the motion of
D6-branes  to the right, and the factor $\prod_{\beta=\alpha+1}^{n}
J_{\beta,L}^{\beta-\a}$ accounts for the ``frozen'' D4-branes in the
interval $I_\a$ which have been created from the motion of D6-branes
to the left.

It is of some interest to compute the order $\hat k_\a$ of the
polynomials $\hat g_{\a}$. To do this note that, after a D6-brane
has passed through the interval $I_\a$ in either direction the net
value of $-2k_\a + k_{\a -1} + k_{\a +1} + d_\a$ remains constant.
Therefore, since our initial configuration has all these values set
to zero (for conformality) it follows that after all the D6-branes
have been moved to the far left or far right we have $-2\hat k_\a +
\hat k_{\a+1} + \hat k_{\a-1}=0$, $1\leq \a \leq n$. Here $\hat k_0$
is the number of D4-branes ending on the D6-branes on the left,
i.e.,   the order of $\prod J_{\beta,L}^\beta$ while similarly $\hat
k_{n+1}$ is the number of D4-branes ending on the D6-branes on
the right, i.e. the order of $\prod J_{\beta,R}^{n+1-\beta}$. Using
the results of Appendix \ref{app:ConvexKay} we see that $\frac{
(\hat k_{n+1}  - \hat k_0 )}{n+1}=r$ is a nonnegative integer and
\begin{equation}\label{eq:solvekhat}
\hat k_\a = \hat k_0 + \a r.
\end{equation}

There are many different ways in which we can move D6-branes to the
left and to the right. At one extreme, we could move them all to the
right, so that $\hat k_{0}=0$ and $\hat k_\a$ increases linearly as
a function of $\a$.   Of course, the other extreme has all D6-branes
on the left, and then $\hat k_\a$ decreases linearly as a function
of $\a$. Amongst the different ways of moving the D6-branes to the
left and the right there is a canonical choice which we will refer
to as the ``balanced case.''  To describe this choice we appeal
again to Appendix \ref{app:ConvexKay} to note that the $k_\a$ grow
as a function of $\a$ up to some maximum $K$ attained at some
$\a=\a_{-}$, then $k_\a=K$ is constant up to some $\a=\a_+\geq
\a_-$, and then $k_\a$ decreases monotonically for $\a > \a_+$. This
suggests a canonical movement of the D6-branes:  we bring all those
in the intervals $\a < \a_-$ to the left and all those in the
intervals $\a > \a_+$ to the right. In the interval $I_{\a_-}$ we
bring $k_{\a_-}-k_{\a_--1}$ branes to the left and in the interval
$I_{\a_+}$ we bring $k_{\a_+}-k_{\a_++1}$ branes to the right. It is
not difficult to show that this leads to $\hat k_{\a_--1} = \hat
k_{\a_-} = \cdots = \hat k_{\a_+} = \hat k_{\a_++1}$. Since the
$\hat k_\a$ grow linearly it follows that $r=0$ and hence all $\hat
k_\a$ are equal to some common value $\hat K$.

This ``balanced'' motion of the D6-branes maps the system to one very
similar to that of the previous section.  We will consider the
other possible distributions of D6-branes to the left and right in the next
section, but for the remainder of this section we focus on the
canonical choice.  The curve \eqref{eq:Fhatcurve} is then a special
case of \eqref{eqn:sw1}, where $q_\a(v)$ are of the form $\hat
g_\a(v)$. As before we can rearrange $\hat F(t,v)$ to be a
polynomial in $v$ of order $\hat K$, of the form
\eqref{eq:swcurve1}:
\begin{equation}
\hat F(t,v) = v^{\hat K} \prod_{\a=0}^n(t-t_\a) + \sum_{i=1}^{\hat
K} v^{\hat K-i} p_i(t) \end{equation}
Dividing by $\prod_{\a=0}^n(t-t_\a)$ we produce a monic polynomial
in $v$ with   coefficients $R_i(t) =
\frac{p_i(t)}{\prod_{\a=0}^n(t-t_\a)}$ which are rational functions
of $t$. We now interpret this as the equation for a   spectral curve
\begin{equation}
\det(v- t \varphi_t) =  v^{\hat K}   + \sum_{i=1}^{\hat K} v^{\hat
K-i} R_i(t)
\end{equation}
where the Higgs field $\varphi_t$ is in $u(\hat K)$. Our goal is now
to understand what the special structure of the coefficients $\hat
g_\a(v)$ implies about the boundary conditions on the Higgs field
$\varphi_t$ at the defects.

Accordingly, let us  analyze the behavior of the $\hat K$ roots
$v_i(t)$ of (\ref{eq:Fhatcurve}). Since there are $n+1$ roots $t(v)$
for $t$ as a function of $v$, and since $v$ can freely go to
infinity,  there must be $(n+1)$ values $t_\a$ at which $v\to
\infty$. Generically, $v$ will have a simple pole at $t_\a$. These
simple poles imply that, (after   shifting away the center of mass),
$\varphi_t$ has a first order pole with residue exactly as in
(\ref{eq:SimplePoleResidue}). This is just the situation we had
before.

On the other hand,  the behavior of the roots when $t\to 0$ or $t\to
\infty$ requires a refinement of our earlier analysis. For $t\to 0$
the roots tend to the roots of $\hat g_{n+1}(v)$ while for $t\to
\infty$ they tend to those of $\hat g_0(v)$. These roots are all at
finite values of $v$. However,  the factors of $J_{\beta,L}$ with
$\beta>1$ in $\hat g_0$ (and those of $J_{\beta,R}$ with $\beta <
n+1$ in $\hat g_{n+1}$) lead to multiple roots. In general, the
existence of multiple roots of the coefficient $q_0(v)$ (or
$q_{n+1}(v)$) in (\ref{eqn:sw1}) means that several D4-branes end on
the same D6-brane. Now, there is an important distinction between
the multiple roots obtained by moving several D6-branes attached to
a single D4-brane to the same $v$-coordinate, and the multiple roots
resulting from the Hanany-Witten effect. In the former case, the
roots of $q_0(v)$ are in general different from the roots of
$q_\a(v)$ for $\a>1$. In the latter case, the structure of
(\ref{eq:hatgee}) et. seq.  shows that the multiple roots from the
factor $J_{\beta,L}$ are also (multiple) roots of $\hat g_{\a}$ for
$\a <  \b$. Thus, in general, as $t \to \infty$, if $q_0(v)$ has a
root $v_*$  of order $\beta$, and $v_*$ is not a root of $q_1(v)$,
then $\beta$ roots $v_i(t)$ behave like
\begin{equation}\label{eq:FracRootLimit}
v_i(t) \to v_* + \frac{\xi_i}{t^{1/\beta}} + \cdots
\end{equation}
as $t\to\infty$, where $\xi_i$ are constants independent of $t,v$.
On the other hand, in the case of multiple roots arising from the
factors $J_{\beta,L}$ in (\ref{eq:hatgee}) the analogous set of
$\beta$ roots $v_i(t)$ behave like
\begin{equation}\label{eq:WholeRootLimit}
v_i(t) \to v_* + \frac{\xi_i}{t}+ \CO(t^{-2})
\end{equation}
Analogous statements hold for the behavior of the roots associated
with D6-branes on the right, for $t\to 0$.

The behavior of the roots $v(t)$ for $t\to \infty$ we have just
described have implications for the boundary conditions of the Higgs
field $\varphi_t dt$ as $t\to \infty$. This Higgs field will behave
like
\begin{equation}
\varphi_t \to \frac{R}{t} + \frac{R_2}{t^2} + \cdots
\end{equation}
Equating $\hat F(t,v) = \hat g_0(v) t^{n+1} + \cdots $  with
$\prod_0^n (t-t_\a) \det(v-t\varphi_t)$, and taking $t\to \infty$,
we see that the characteristic polynomial of the residue $R = t
\varphi_t\vert_{t=\infty}$ is just $\det(v-R) = \hat g_0(v)$. Since
this has multiple roots we must consider the possibility that $R$
has nontrivial Jordan form.

We now claim that a nontrivial Jordan form leads to roots behaving
like (\ref{eq:FracRootLimit}) while if $R$ is semisimple then the
roots will behave like (\ref{eq:WholeRootLimit}). We may prove this
as follows. For simplicity suppose that the characteristic
polynomial $\det(v-R) = v^\beta$ for some integer $\beta > 1$. Now
consider the perturbation $R(t) = R + C/t $ where $t$ is large and
$C$ is \emph{generic}. If $R$ is semisimple, then $R=0$ and the
eigenvalues of $R(t)$ are $c_i/t$ where $c_i$ are the distinct
  eigenvalues of $C$. At the other extreme, suppose $R$
is a Jordan block of size $\beta$, which we denote as $N_\beta$.
\footnote{i.e $N_\beta = e_{1,2} + e_{2,3} + \cdots +
e_{\beta-1,\beta}$ in terms of matrix units.}  By a gauge
transformation with $g= \exp[\epsilon/t]$ we can bring $R(t)$ to the
form
\begin{equation}
R(t) \to N_\beta + \frac{1}{t}\left( \sum_{i=1}^\beta \mu_i
e_{\beta,i} \right) + \CO(1/t^2)
\end{equation}
for some constants $\mu_i$. Here $e_{i,j}$ is the matrix unit, that
is, the matrix whose only nonzero entry is $1$ in the $i^{th}$ row
and $j^{th}$ column.  On the other hand,
\begin{equation}\label{eq:UsefulDet}
\det\left[ v - N_\beta -\frac{1}{t}\left( \sum_{i=1}^\beta \mu_i
e_{\beta,i} \right)\right]= v^{\beta} +
\frac{\mu_{\beta}}{t}v^{\beta-1} + \frac{\mu_{\beta-1}}{t}
v^{\beta-2} + \cdots + \frac{\mu_1}{t}
\end{equation}
From the standard relation between the elementary symmetric
functions and the power sum functions it follows that the roots $v_i
\sim c_i t^{-1/\beta} + \CO(t^{-2/\beta}) $. In the intermediate
cases, when $R$ has several Jordan blocks the roots will fall off
with a fractional power of $t$ with the fractional power governed by
the largest Jordan block of $R$.

The upshot is that the boundary condition for the Higgs field at
$t\to \infty$ has the block diagonal form:
\begin{equation}\label{eq:RegSingBalanced}
\varphi_t dt \to  \frac{dt}{t} {\rm Diag}\{ \cdots  v^{(\beta)}_1
1_{\beta},\dots,v^{(\beta)}_{d_{\beta,L}} 1_{\beta}, \cdots \} +
\CO(t^{-2})
\end{equation}
where $1_{\beta}$ is the diagonal matrix and $v^{(\beta)}_1
 ,\dots,v^{(\beta)}_{d_{\beta,L}}$ are the roots of $J_{\beta,L}$
(assumed distinct, for simplicity). Put more simply, the boundary
conditions preserve a subgroup
\begin{equation}
\prod_{\beta=1}^n (U(\beta))^{d_{\beta,L}}
\end{equation}
of the $U(\hat K)$ gauge group. Entirely parallel remarks apply to
the limit $t\to 0$ and the group of branes on the right.

In conclusion, the conformal linear quivers of unitary groups give
rise to Hitchin systems on $\IC \IP^1$ with two regular
singularities of a generic type, labeled by two partitions of $K$,
$K = \sum \beta d_{\beta,L}$ and $K = \sum \beta d_{\beta,R}$, and
an arbitrary number of ``basic'' singularities, associated to the
partition $K = (1) + (K-1)$.

\subsubsection{Brane-bending and irregular
singularities}\label{subsubsec:Brane-bending}

We continue to study the conformally invariant linear quiver of the
previous subsubsection but now consider the case where we move the
D6-branes in an ``unbalanced'' way so that $\hat k_{n+1} - \hat
k_0$ is nonzero.

The behavior of the roots is now dramatically different. Recall that
the order of $\hat g_{\a} $ is $\hat k_\a = \hat k_0 + \alpha r$.
For simplicity assume that $r>0$,   so that, by (\ref{eq:solvekhat})
  $\hat g_{n+1}= c_{n+1} v^{\hat k_{n+1}} + \cdots $ has the highest
 power of $v$. Then, since  $\hat g_{n+1} $ dominates all the other coefficients
$\hat g_\a(v)$ for $v\to \infty$  it follows that none of the roots
$v(t)$ go to infinity at finite values of $t$.

The fact that there are only two singular points in $t$  can also be
seen more physically by considering the $U(1)_R$ symmetry of the
theory. This symmetry rotates $v \to e^{i \theta} v$, and hence
changes the masses $v_a \to e^{i \theta} v_a$. However, it must not
change the coupling constants, and therefore $c_\a \to c_\a$.  In
the balanced case the curve $\hat F(t,v)=0$ is invariant under this
scaling with fixed $t$. On the other hand, in the unbalanced case
this is not true and we must rescale $(t,v) \to (e^{i r\theta} t,
e^{i \theta} v)$. Then, a singularity at a finite point $t_\a\in
\IC^\times$ would be incompatible with the equation.

Next, let us consider the roots for $t\to 0$. Here $v_i(t)$ simply
approaches the $\hat k_{n+1}$  roots of $\hat g_{n+1}(v)$ and as we
have discussed, they behave like $v_i(t) = v_i + \xi_{i,s} t + \cdots$.

On the other hand, the behavior of the roots at $t \to \infty$ is
more complex. There are $\hat k_0$ roots behaving like $v_i(t) \to
v_i + \frac{\xi_{i,a}}{t}+ \CO(1/t^2)$ where $v_i$ are the roots of
$\hat g_0(v)$. Since $\hat k_{n+1} > \hat k_0$  this does not account
for all the  roots.  In addition, there are $(n+1)r$ roots where $t,
v$ both go to infinity.  In this case, keeping the leading order
terms in $\hat F(t,v)$, we see that the roots with $(t,v)$ both going
to infinity must asymptote to the roots of
\begin{equation}\label{eq:BigRoots}
0 = c_{n+1} v^{(n+1)r} + c_n t v^{nr} + \cdots + c_1 t^n v^r + c_{0}
t^{n+1} := c_{n+1} \prod_{i=1}^{n+1} (v^r - \nu_i t).
\end{equation}

\insfig{3.3in}{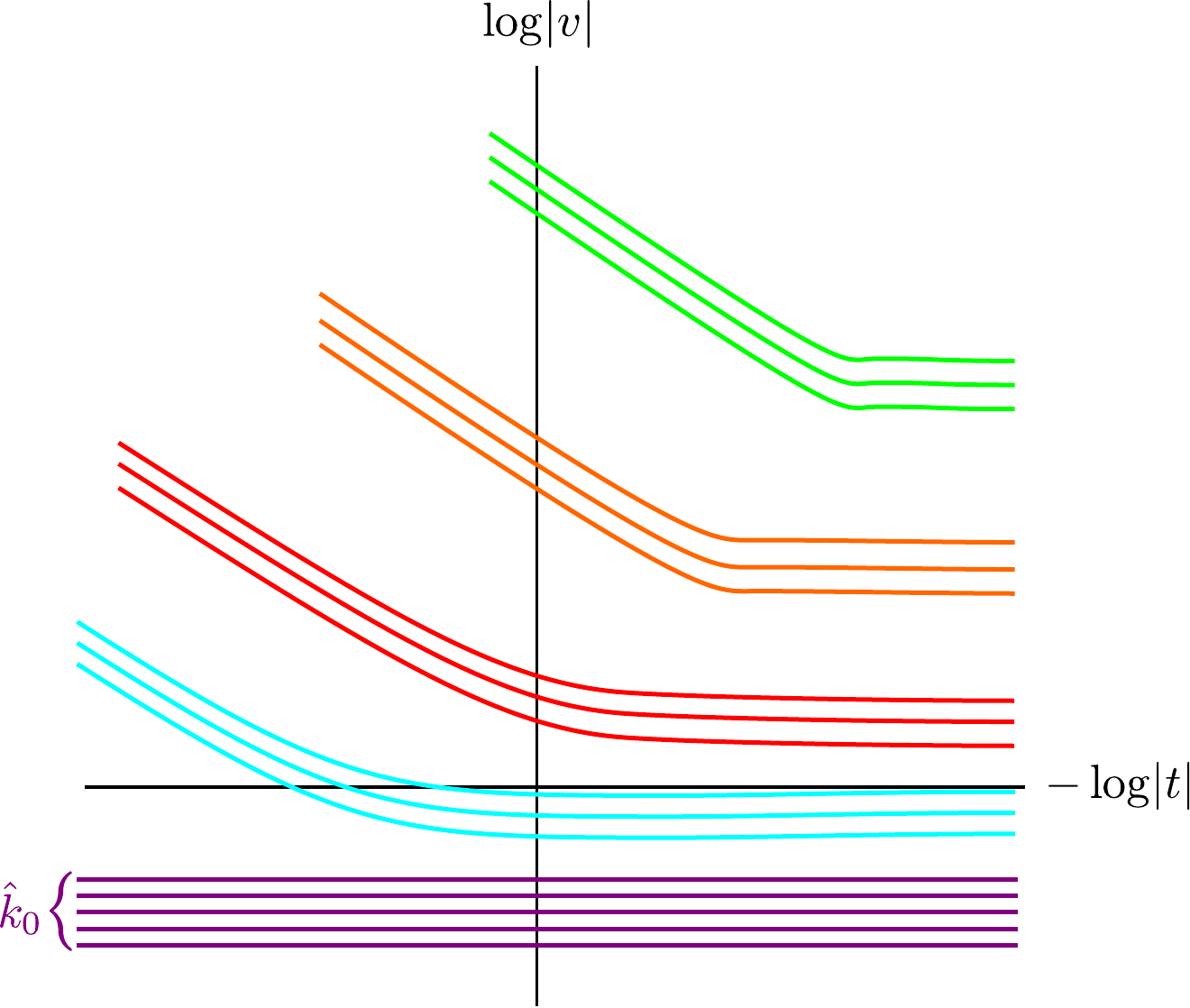}{Brane bending in the conformal but
unbalanced case when $\hat k_{n+1} - \hat k_0 = (n+1) r > 0$. Note that
$n+1$ bundles of $r$ branes go to infinity as $t\to \infty$. Here
$r=3$ and $n+1=4$. There are $\hat k_{n+1}=17$ horizontal branes at
$t\to 0$.}

Note that we have now found the phenomenon of ``brane bending'':
some of the roots $v(t)$ go to infinity as $t\to \infty$.
To picture the situation note that the projection
of $\Sigma$ into the $(-\log\vert t\vert, \log \vert v \vert)$
plane asymptotes to the form shown in Figure \ref{fig:brane-bending}.

Now let us consider the mapping to a Hitchin system.  In this case
we identify the characteristic polynomial of the Higgs field
according to
\begin{equation}
\det(v - t \varphi_t) = \frac{1}{c_{n+1}} \left( \hat g_{n+1}(v) + t
\hat g_n(v) + \cdots + t^{n+1} \hat g_0(v) \right)
\end{equation}
where $\varphi_t \in u(\hat k_{n+1})$.  From our discussion of the
roots above,  all the singularities of the Higgs field must lie at
$t=0$ or $t=\infty$. Our previous discussion applies to the roots at
$t\to 0$ and hence $\varphi_t$ has a regular singularity at $t \to
0$ as in (\ref{eq:RegSingBalanced}).

By contrast, for $t\to \infty $ we find that $\varphi_t$ has a
block diagonal form. There is a $\hat k_0 \times \hat k_0$ block
corresponding to a regular singularity with residues given by the
roots of $\hat g_0(v)$. We interpret \eqref{eq:BigRoots} to imply
that in addition there are $(n+1)$ blocks, labeled by $i=1,\dots,
n+1$ with limiting behavior
\begin{equation}
\det(v-t\varphi_t) \sim v^r - \nu_i t.
\end{equation}
(The matrices are understood to be restricted to the $i^{th}$ block.
We do not indicate this to avoid cluttering the notation.)  Thus, in
the unbalanced case, the boundary condition at $t\to \infty$
corresponds to an \emph{irregular singularity},
with $(n+1)r$ eigenvalues of $t\varphi_t$ behaving like
\begin{equation}
\omega^s (\nu_i t)^{\frac{1}{r}} (1 + \CO(1/t)) \qquad\qquad
s=1,\dots, r;\qquad  i = 1, \dots, n+1,
\end{equation}
where $\omega$ is a primitive $r^{th}$ root of unity.

We can also consider the opposite case $\hat k_0 > \hat k_{n+1}$.
For this case we reverse the sign of $r$ and take $r = (\hat
k_0-\hat k_{n+1})/(n+1)>0$. In this case $t \varphi_t$ has a regular
singularity at $t\to \infty$ and an irregular one at $t \to 0$.  The
irregular singularity involves $(n+1)$ blocks, each with eigenvalues
\begin{equation}
\omega^s (\frac{\nu_i}{t})^{\frac{1}{r}} (1 + \CO(t)) \qquad\qquad
s=1,\dots, r;\qquad  i = 1, \dots, n+1
\end{equation}
where $\omega$ is a primitive $r^{th}$ root of unity.

Let us now discuss what these conditions mean for the behavior of
the Higgs field $\varphi_t$ at $t\to 0,\infty$. We focus on $t\to 0$
and take $r = (\hat k_0-\hat k_{n+1})/(n+1)>0$. Using
(\ref{eq:UsefulDet}) we see that by a \emph{complex} gauge
transformation we can put $\varphi_t$ in the form
\begin{equation} \label{eq:phi-formal}
\varphi_t \to \frac{1}{t} \left( \frac{\nu_i}{t} e_{r1} + N_r +
\CO(t) \right) \qquad t \to 0 \end{equation}
However, to define the Hitchin system properly we must specify an
actual solution to the Hitchin equations at $t\to 0$. We can do this
block by block.  For the blocks corresponding to the irregular
singularity it  is easy to give such a solution in the diagonal
gauge,
\begin{equation}
\varphi_t = \frac{1}{t} \left(\frac{\nu_i}{t}\right)^{1/r} {\rm
Diag} \{ \omega, \omega^2, \dots, \omega^r \}
\end{equation}
The problem is that this is not single-valued.  We can make
$\varphi_t$ single-valued by a singular gauge transformation
\begin{equation}
g = \eta S \eta^{-1}
\end{equation}
with $S_{ab} = \frac{1}{\sqrt{r}} e^{\frac{2\pi i }{r} ab}$ and
$\eta$ is given by
\begin{equation}
\eta = \left(\frac{t}{\bar t} \right)^{\frac{r-1}{4r}} {\rm Diag}\{
1,\left(\frac{\bar t}{ t} \right)^{\frac{1}{2r}},\left(\frac{\bar
t}{ t} \right)^{\frac{2}{2r}},\dots, \left(\frac{\bar t}{ t}
\right)^{\frac{r-1}{2r}} \}
\end{equation}
In order to keep the gauge field single valued we take
\begin{equation}
A = - d\eta \eta^{-1}
\end{equation}
in diagonal gauge and the upshot is that our single-valued
asymptotic solution of the Hitchin equations (for the $i^{th}$
block) is
\begin{equation}\label{eq:ISPatInfty}
\varphi_t \to \frac{\nu_i^{1/r}}{t} \frac{1}{\vert t\vert^{1/r}}
\left( \frac{\bar t}{\vert t\vert } e_{r1} + N_r + \CO(t) \right)
\end{equation}
with
\begin{equation}\label{eq:GaugeFldInfty}
A  =  \left( \sum_{j=1}^r \frac{2(j-1)-(r-1)}{4r} e_{jj} \right)
\left( \frac{dt}{t} - \frac{d\bar t}{\bar t}\right).
\end{equation}

\subsubsection{The asymptotically free case}

Let us finally turn to the general asymptotically free quiver with
fundamental matter. For such a quiver the $\beta$-function
coefficients $b_{\a} \leq 0$, $1\leq \a \leq n$, and $b_\a<0$ for at
least one value of $\a$. Physically, such quiver gauge theories can
be obtained by decoupling fundamental fields in a conformal linear
quiver by taking the masses of some of the fundamental fields to
infinity. They then decouple, leaving an asymptotically free theory.

The decoupling procedure can be elegantly carried out at the level
of Seiberg-Witten curves. In each interval $I_\a$ we choose a set of
centers $v_a$, $a\in S_\a$   which we will send to infinity. We then
scale $c_\a \to 0$ in such a way as to leave the coefficients
$A_\a(v)$ in eq. (\ref{eq:Aalphdef}) finite.   However, since we are
most interested here in the application to Hitchin systems we
consider the decoupling limit \emph{after} we have moved the
D6-branes to large values of $\vert x_a^6\vert$. Therefore, we split
the set $S_\a = S_{\a,L} \amalg S_{\a,R}$ according to whether the
D6 to be decoupled  has first moved to the far left or the far
right. In this limit we have
\begin{equation}
J_{\a,L}(v) \to \c J_{\a,L}(v) \prod_{a\in S_{\a,L}} (-v_a) (1 +
\CO(v/v_a)),
\end{equation}
where $\c J_{\a,L}(v)$ are the factors for the D6-branes not taken
to infinity, and similarly for $J_{\a,R}$. Thus, we send $c_\a$ to
zero in such a way that
\begin{equation}
\c c_{\a} = c_\a \prod_{\b=1}^\a \left( \prod_{a\in
S_{\b,R}}(-v_a)\right)^{\a -\b}\prod_{\b=\a+1}^n \left( \prod_{a\in
S_{\b,L}}(-v_a)\right)^{\b -\a}
\end{equation}
remains finite.  We now have the analogs of eqs.
(\ref{eq:Fhatcurve}) and (\ref{eq:hatgee}):
\begin{equation}\label{eq:Fhatcurvechck}
\c F(t,v) = \sum_{\a=0}^{n+1} \c g_{\a}(v) t^{n+1-\a}
\end{equation}
where
\begin{eqnarray}\label{eq:hatgeechck}
\c g_0(v)&=& \c c_0 \prod_{\beta=1}^n \c J_{\beta,L}^\beta  \\
\c g_{\a}(v) &=& \c c_{\a} g_{\a}(v) \prod_{\beta=1}^{\a} \c J_{
\beta,R}^{\a-\beta} \prod_{\beta=\alpha+1}^{n} \c
J_{\beta,L}^{\beta-\a}\qquad\qquad \a =1, \dots, n-1, \\
\c g_{n}(v) &=& \c c_{\a} g_{\a}(v) \prod_{\beta=1}^{n} \c J_{
\beta,R}^{\a-\beta}, \\
\c g_{n+1}(v) & =&  \c c_{n+1} \prod_{\beta=1}^{n} \c
J_{R,\beta}^{n+1-\beta}.
\end{eqnarray}
The only difference from the previous case is that now
$\c b_\a = -2\c k_\a + \c k_{\a + 1} + \c k_{\a -1}$ is $\leq 0$
for all $1\leq \a \leq n$ and can be strictly less than zero.
Here $\c k_\a$ is the order of $\c g_{\a}$.

Again referring to Appendix \ref{app:ConvexKay} the  points $(\a, \c
k_\a)$ define a convex polygonal curve. The $\c
k_\a$ are strictly increasing for $0 \leq \a \leq \a_-$. The maximal
value $\c K := \c k_{\a_-}$ is attained for $\a_- \leq \a \leq \a_+$.
Thereafter the $\c k_\a$ are strictly decreasing for $\a \geq
\a_+$. If
$\c k_1 - \c k_0 =0 $ then $\a_-=0$ and if
$\c k_1 - \c k_0 < 0 $ then $\a_+=0$. For simplicity, in what
follows
we will just describe the case $\c k_1 - \c k_0 > 0 $.

\insfig{4.7in}{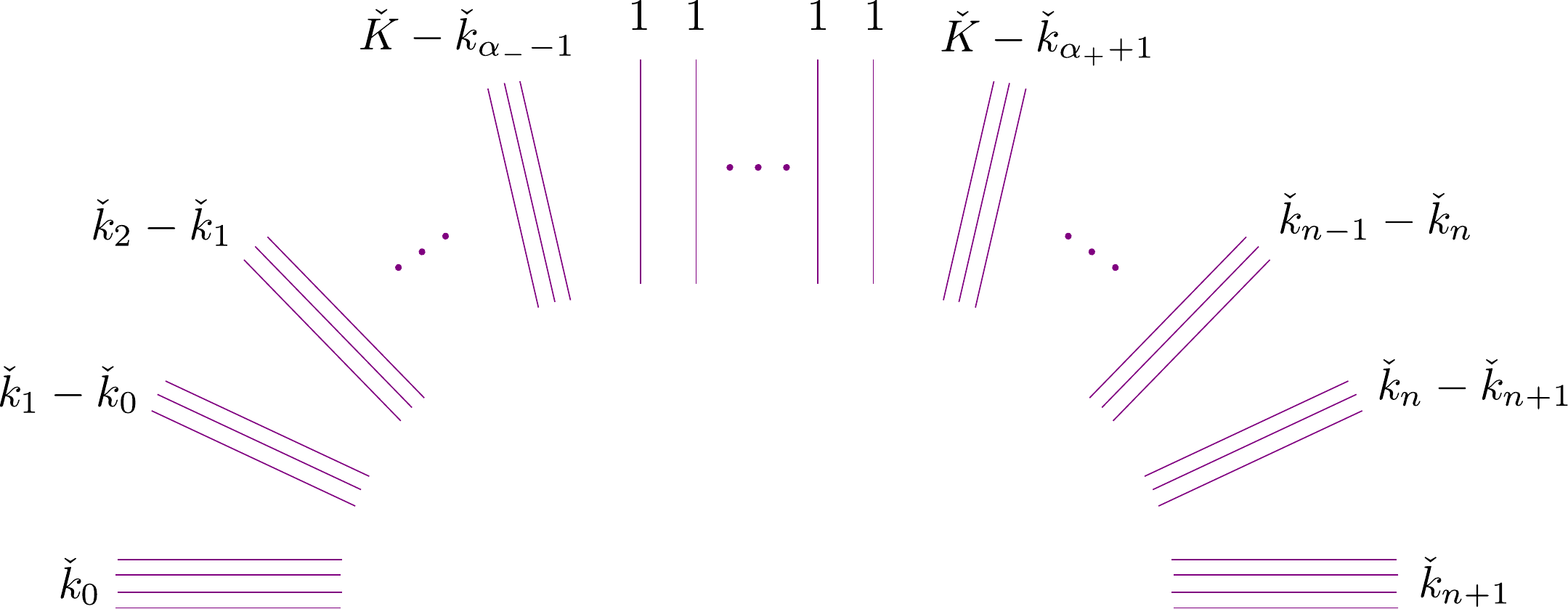}{The boundary of the Newton polygon of
$\c F(t,v)$ dictates the asymptotic behavior of the M5-brane curve.
Here it is projected on the $(-\log\vert t \vert, \log \vert v \vert
)$ plane.}

Now the polygonal path can be viewed as the boundary of the Newton
polytope for $\c F(t,v)$. Accordingly, we can extract the
asymptotic
behavior of the roots of the equation $\c F(t,v)=0$. To do this we
balance   terms on the boundary faces of the polytope. Assuming,
for simplicity, that a face consists of a single segment we have:
\begin{equation}
\c c_\a t^{n+1-\a} v^{\c k_\a} + \c c_{\a +1} t^{n-\a} v^{\c k_{\a +1} }
\cong 0
\end{equation}
for $0 \leq \a \leq \a_--1$ or $\a_+\leq \a \leq n$. Each value of
$\a$ gives $\vert \c k_{\a+1} - \c k_{\a} \vert$ roots $v(t)$
asymptotic to
\begin{equation}
t \cong - \frac{\c c_{\a +1}}{\c c_\a} v^{(\c k_{\a+1} - \c k_{\a}
)}
\end{equation}
For these roots $v(t)\to \infty$ with $t\to \infty$ for $0\leq \a \leq
\a_--1$ and $t\to 0$ for $\a_+\leq \a \leq n$, respectively. The
remaining roots at $t\to \infty $ and $t\to 0$ are the roots of
$\c g_0(v)$ and $\c g_{n+1}(v)$, respectively. These latter roots
remain finite. In addition, if $\a_+>\a_-$ then at the  roots $t\in \IC^\times$ of
\begin{equation}\label{eq:intermed}
\sum_{\a=\a_-}^{\a_+} \c c_{\a} t^{\a_+-\a} =0
\end{equation}
we have $\a_+-\a_-$ roots $v(t)$ going to infinity. Generically,
these will be simple poles $v(t) \sim \frac{\rho_\a}{t-t_\a}$.\footnote{The
$U(1)_R$ symmetry forbidding these roots in the conformal case is anomalous here.}
In this way the boundary of the Newton polygon dictates the
asymptotic shape of the M5-branes. This is illustrated in Figure
\ref{fig:brane-bending-2}.

The physical interpretation of the constants $\c c_\a$ can now be
deduced from the above pictures. Consecutive branches of the M5
surface measured at the same value of $v$ have corresponding values
of $t$ related by
\begin{equation}
\frac{t^{(\a+1)}}{t^{(\a)}} \cong \frac{ \c c_{\a+1} \c c_{\a
-1}}{\c c_\a^2} v^{\c b_\a}
\end{equation}
In the weak coupling limit we have seen that
$\log\vert\frac{t^{(\a)}}{t^{(\a +1)}}\vert$ should be interpreted
as an inverse coupling constant.  Thus, for $\c b_\a<0$ we see that
it is logarithmically running with scale $\vert v \vert$, and
moreover the UV cutoff scale is
\begin{equation}
\Lambda^{-\c b_\a} =\frac{ \c c_{\a+1} \c c_{\a -1}}{\c c_\a^2}
\end{equation}
while for $\c b_\a=0$ the combination $\frac{ \c c_{\a+1} \c c_{\a
-1}}{\c c_\a^2}$ encodes the UV couplings,  as before.

Now we are ready to describe the relevant Hitchin system. This is a
$U(\c K)$ system with singularities at $t=0,\infty$, as well as
singularities at the roots of (\ref{eq:intermed}). At the latter
singularities $\varphi_t$ has a regular singularity. At $t\to
\infty$, $\varphi_t$ has block diagonal form. There is a regular
singular block with residues given by the roots of of $\c g_0(v)$.
In addition there are $\a_-$ blocks of irregular singularities of
the form (\ref{eq:ISPatInfty}) where we should substitute  $r= \c
k_{\a +1}-\c k_{\a}$ and $\nu_i = - \c c_{\a}/\c c_{\a +1}$, $0\leq
\a \leq \a_--1$. Similarly, at $t =\infty$ there is a block of
regular singular points governed by the roots of $\c g_{n+1}(v)$ and
$n+1-\a_+$ blocks of irregular singular points of the form
(\ref{eq:ISPatInfty}) with $r= \c k_{\a}-\c k_{\a +1}$ and $\nu_i =
- \c c_{\a}/\c c_{\a +1}$, $\a_+\leq \a \leq n$.

The large mass limit leading to irregular singularities can be
carried out directly at the level of the Hitchin system. For
example, consider the $\epsilon\to 0$ limit of the Higgs field:
\begin{equation}
\frac{1}{t} \begin{pmatrix} 0 & 1/2 \\ - \rho/\epsilon & 0 \\
\end{pmatrix} +\frac{1}{t-\epsilon} \begin{pmatrix} 0 & 1/2 \\ \rho/\epsilon & 0 \\
\end{pmatrix}
\end{equation}
For fixed $\epsilon$ there are two regular singular points. However,
for $\epsilon\to 0$  the Higgs field develops an irregular singular
point of the type we have described.

The simplest example of the above constructions is $n=1$, with $\c
K=2$. This is a $U(2)$ theory with $d_1 = N_f$ fundamental flavors.
If $N_f=0$, then $\c k_0=0, \c k_1 = 2, \c k_2 =0$, so $\c b_1 =
-4$. The curve (\ref{eq:Fhatcurvechck}) is $\c c_0 t^2 + \c c_1 t
(v^2-u) + \c c_2=0$ and,  after a rescaling of $t$, the curve
becomes the standard $SU(2)$ Seiberg-Witten curve
\begin{equation}\label{eq:Standard-SU2-curve}
v^2 - u - t - \frac{\Lambda^4}{t} = 0,
\end{equation}
and hence $t^2 \lambda^2 =\left( u + t + \frac{\Lambda^4}{t}\right)
(dt)^2$ where $\pm \lambda$ are the eigenvalues of $\varphi_t dt$.
The boundary conditions on the Higgs field state that, up to gauge
equivalence,
\begin{equation}
\varphi_t dt \to \frac{\Lambda^2}{\vert t \vert^{1/2}}
\begin{pmatrix} 0 & 1 \\ e^{-i \theta} & 0 \\ \end{pmatrix}
\frac{dt}{t} \qquad\qquad t\to 0
\end{equation}
\begin{equation}
\varphi_t dt \to -  \vert t \vert^{1/2}
\begin{pmatrix} 0 & 1 \\ e^{i \theta} & 0 \\ \end{pmatrix}
\frac{dt}{t} \qquad\qquad t\to \infty
\end{equation}
where $t = \vert t \vert e^{i \theta}$.  We will discuss this
example in great detail in Section \ref{section:SU2Examples}.

\subsubsection{A surprising isomorphism}\label{sec:surprise}

We can make a surprising mathematical prediction based on
our physical setup. Since the IR fixed point is independent of the
motion of the D6-branes, the different Hitchin moduli spaces
obtained from different distributions of the D6-branes to the
left and the right must be isomorphic!

Put more precisely, begin with the polynomial $\tilde F(W,v)$ of
equation (\ref{eq:TNCurveii}). Then, as described in equations
(\ref{eq:W-t-limit}) et. seq., consider the different polynomials
$\hat F(t,v)$ arising from the different movements of the D6-branes
to left and right.  The resulting polynomials can be interpreted as
spectral curves for a $U(\hat K)$ Hitchin system where $\hat K =
{\rm max}[\hat k_0, \hat k_{n+1}]$, as we have explained.  From
these polynomials we can read off the boundary conditions for the
Higgs field on $\IC^\times$. Physics predicts that the resulting
moduli spaces are isomorphic as hyperk\"ahler manifolds.  We will
see one explicit example in Section \ref{section:SU2Examples}.
Examples of this phenomenon have very recently appeared in the
mathematical literature as well \cite{irkm}.

Our methods offer a strategy for proving
the isomorphism between these \hk moduli spaces:  the
spectral curves of the different Hitchin systems are isomorphic
(although not as fibrations $\Sigma \to C$), so the semiflat limits $g^\sf$
of the \hk metrics coincide.  As soon as the BPS spectra are also found
to be the same, the Riemann-Hilbert problems of
\cite{Gaiotto:2008cd} coincide, and hence so do the hyperk\"ahler metrics.

\subsubsection{Non-Abelian flavor symmetries and
punctures} \label{sec:flavor}

The relation with the linear quivers has suggested that each
puncture on $C$, i.e. each defect in the six-dimensional $(2,0)$
theory, is associated with a certain flavor (sub)group of the
resulting four-dimensional $\CN=2$ theory. The defects which lead to
the most general regular singularity for the $SU(K)$ Hitchin system
are associated to the $K-1$ mass parameters in the Cartan of an
$SU(K)$ flavor group.  The simplest defects which break the $SU(K)$
gauge symmetries to $S(U(1) \times U(K-1))$ are instead associated
to the mass parameters of some $U(1)$ flavor group (unless $K=2$, in
which case the two types of singularity are identical, and indeed
the $U(1)$ is enhanced to $SU(2)$).  We already met examples of
regular singularities with a generic pattern of gauge symmetry
breaking $S(\prod U(\beta)^{d_\beta})$ labeled by a partition of
$K$, $d_\beta$, in equation \eqref{eq:RegSingBalanced}.  They have a
single mass parameter for each $U(\beta)$ gauge factor. It can be
shown easily \cite{Gaiotto:2009we} that a non-Abelian flavor
symmetry $S(\prod U(d_\beta))$ is associated with these
singularities.

In later sections of this paper we mostly focus on the $K = 2$ case,
but here we briefly digress to see how the non-Abelian $SU(K)$ flavor
symmetries manifest themselves in the spectrum of BPS string webs
whenever the mass parameters at the corresponding singularity in $C$
go to zero.

Consider first the case $K = 2$. Then we can take a local
coordinate $z$ near the singularity and
\begin{equation}
\lambda^2 \sim \left (\frac{m^2}{z^2} + \frac{u}{z} + \cdots \right) dz^2.
\end{equation}
As we tune $m \to 0$ a zero of $\lambda^2$ must be coming close to
the double pole, to reduce it to a single pole.  What is the
behavior of BPS strings near such a ``molecule'' made up of a
singularity and a zero?  We know that the BPS strings follow curves
of constant phase $\vartheta$ for the 1-form $\lambda_{12} = \lambda
- (-\lambda) = 2 \lambda$.  For a given value of $\vartheta$, three
such curves emanate from the branch point (this important general
fact will be discussed at length in Section
\ref{sec:WKB-triangulation}). One plunges into the singularity,
while the other two wind around the singularity in opposite
directions and escape together.

\insfig{4.5in}{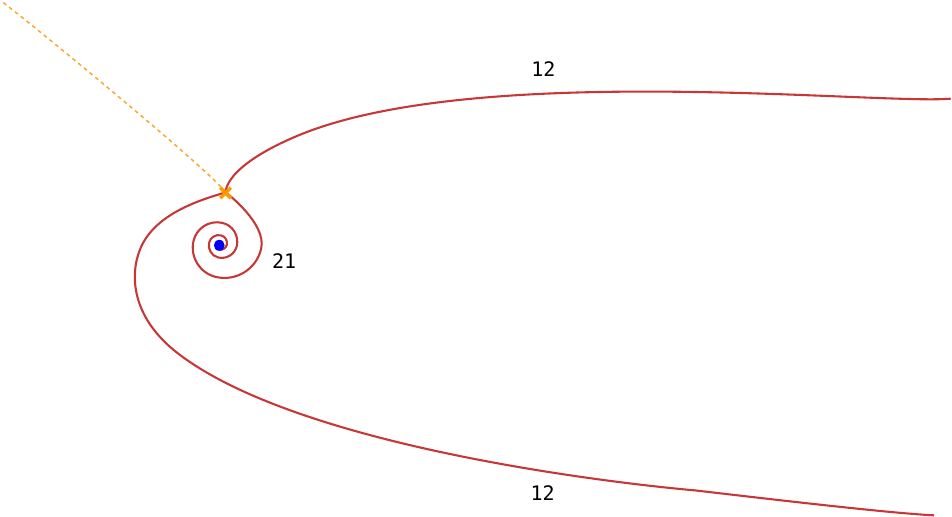}{The pattern of flow lines (red) around a molecule consisting of
a nearly-coincident double pole (blue dot) and zero (orange cross) of
$\lambda^2$.  Three lines emanate from the zero for each value of the
phase $\vartheta$.  If we choose the (orange dotted) branch cut in $\lambda$ as
depicted, the lines flowing from the zero correspond to BPS strings
of type $12$, $21$, as labeled.}

\insfig{4.5in}{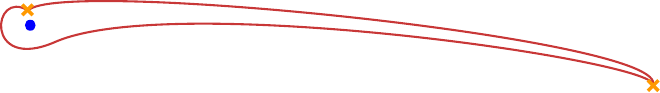}{For
two nearby values of the phase $\vartheta$ the two flow
lines which escape from the molecule may hit another turning point, giving rise to a doublet
of BPS hypermultiplets.}

As we vary the phase $\vartheta$ of the central charge, this doublet
of escaping lines will swipe across other zeroes of $\lambda^2$.  This gives rise to a
pair of BPS strings, which join the same two branch points, but pass
on opposite sides of the singularity, as shown in Figure \ref{fig:doublet-b}.
The central charges of the BPS particles differ by the period of $2 \lambda$ around the
singularity, i.e. by $2m$.  The fact that for small enough $m$ BPS
states always come in doublets of flavor charge $\pm 1$ is a clear
symptom of the presence of an $SU(2)$ flavor symmetry associated to
each regular singularity in the $K=2$ case, and predicts that
the BPS states transform in doublets of these important
$SU(2)$ flavor (sub)groups.  We will see in our examples of $SU(2)$
gauge theories how the $SU(2)$ groups are always embedded in the
full flavor symmetry group in such a way that this condition is
satisfied.

In the cases with $K > 2$ it is important to consider string
webs.  $K=3$ will be sufficient to illustrate this point.  Locally we
have
\begin{equation}
\lambda^3 \sim \left (\frac{\sum m_i m_j}{z^2} + \frac{u}{z} + \cdots \right) dz^2 \lambda+ \left (\frac{\prod m_i}{z^3} + \frac{v}{z^2} + \cdots \right) dz^3
\end{equation}
Different sheets of $\lambda$ meet at the zeroes of the discriminant
of this equation. The discriminant has a degree $6$ pole, which
reduces to $4$ when the masses are turned off, hence we expect to
see two branch points come close to the singularity. It is easy to see
they must be branch points of different type, say for the sheets
$12$ and $13$. For a given value of $\vartheta$, three lines
emanate out of each branch point.  One line emerging from each branch point flows into the
singularity.  A second goes around the
singularity and the other branch point, and turns into a $23$ line.
These two $23$ lines then escape together. The third pair of
lines will intersect, and we can set a string junction at the
intersection, from which a third $23$ line escapes, close to the
other two.
\insfig{4.5in}{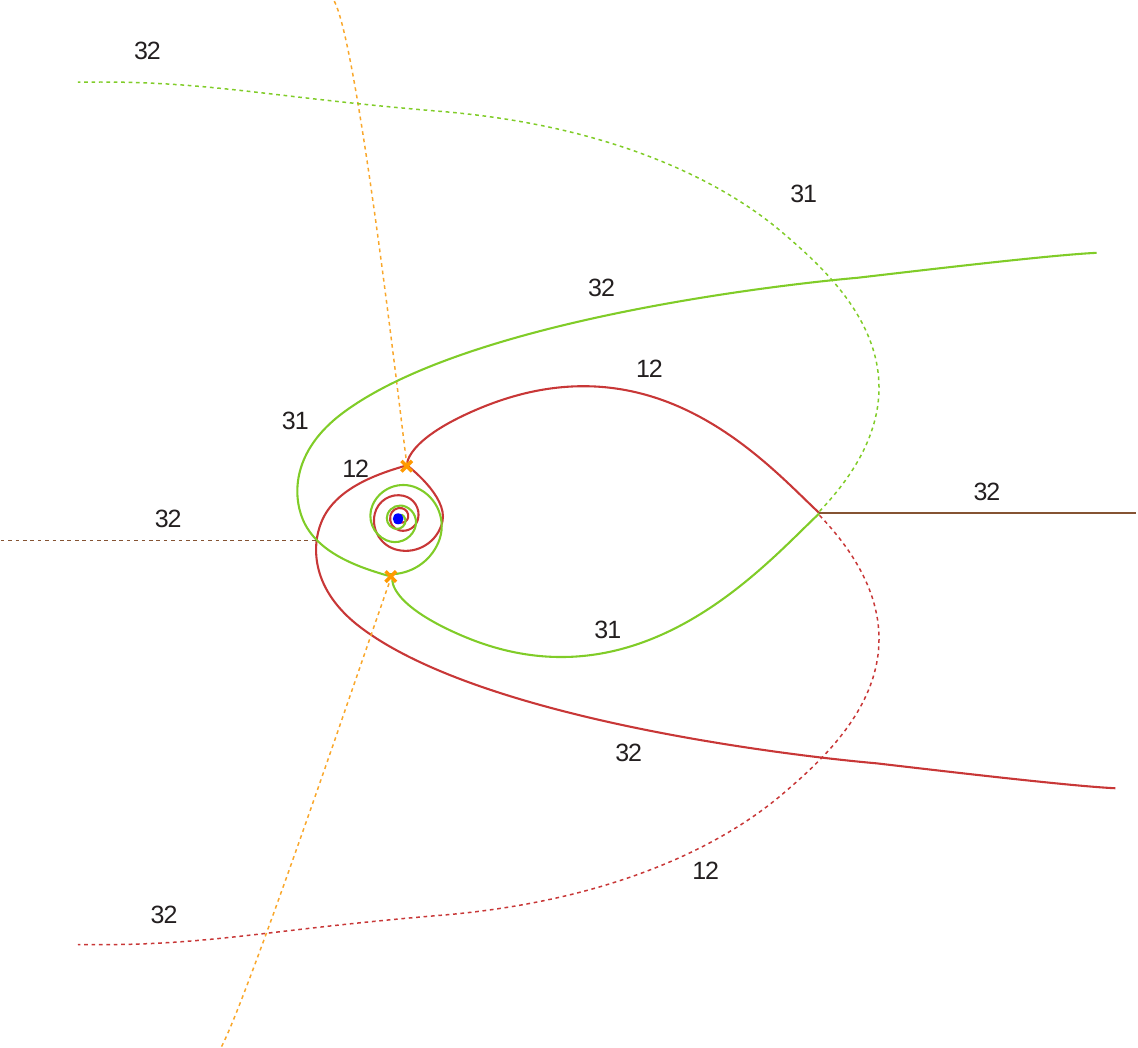}{The pattern of flow lines
(red, green, continuous or dotted) in a region where a singularity
(blue dot) and two zeroes of the discriminant (orange dots) come
close. Three lines emanate from each zero for every value of the
phase $\vartheta$. We take the zeroes to be of the type $12$ and
$13$. If we put the (orange dotted) cuts in $\lambda$ as depicted,
the lines flowing from the zeros correspond to BPS strings of the
indicated type. When two lines of compatible type intersect, we
allow for a possible web junction. All in all, for every value of
$\vartheta$, two groups of three lines of the same type flow away
from the molecule.}

\insfig{4.5in}{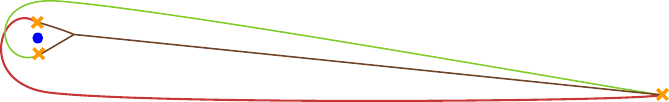}{For nearby
values of the phase $\vartheta$, a group of three flow lines escaping from
the molecule may hit another turning point, giving rise to a triplet of BPS
hypermultiplets.}

Again, as we vary the phase $\vartheta$ of the central charge, this
triplet of lines may swipe across a branch point of the $23$ type
and give rise to a triplet of BPS strings (or, more generally, may be
connected through junctions to give a triplet of string
webs).  This triplet of states carry the same gauge charges, while
their central charges differ pairwise by $m_1-m_2$, $m_1-m_3$, $m_2-m_3$
because they wind in different ways around the singularity.  These states thus
form a triplet for the $SU(3)$ flavor symmetry associated with the
singularity.

\subsection{Type IIB construction} \label{sec:iib}

The M-theory realization of our theories which we use in Section \ref{sec:scft} gives a
convenient geometric way of understanding many properties of the theory quickly, but may be
unfamiliar for the reader who is motivated by Donaldson-Thomas invariants or categories of D-branes
on Calabi-Yau threefolds.  In this section we briefly recall an alternative perspective on our
construction.

The essential point is that the $(2,0)$ theory arises in
Type IIB string theory on an ADE singularity \cite{Witten:1995zh}. Therefore, to
compactify this theory on $C$ we can consider Type IIB on a
Calabi-Yau threefold which in an appropriate scaling limit
develops a curve $C$ of ADE singularities. The scaling limit
decouples gravity and leaves us with the desired $\CN=2$ field
theory. (See \cite{Katz:1997eq} for a detailed construction.)
The BPS states of this theory arise from D3-branes wrapping
special Lagrangian cycles. In the scaling limit such cycles are
obtained as fibrations of the vanishing spheres of the ADE
singularity over string webs in $C$, which are identified with
the BPS webs we considered above.

For example, in the $A$ case, the integral of the holomorphic three
form over the $ij$-th vanishing sphere is identified with the
differential $\lambda_{ij}$ on $C$.  A simple $ij$-string stretched
between two turning points where $\lambda_{ij} = 0$ lifts to a special
Lagrangian with the topology $S^3$: an $S^2$ fibered over a
segment, shrinking at the endpoints. A closed $ij$-string wrapping a
non-trivial cycle of $C$ lifts to a special Lagrangian with the
topology $S^1 \times S^2$.

It would be interesting to take inspiration from this
correspondence, and develop a mathematical definition of some
sort of Fukaya category of string webs on a Riemann
surface $C$, with stability conditions specified by a spectral
curve $\Sigma \subset T^*C$. This would allow a more direct
connection to the work of \cite{ks1}.  We make some tentative
comments in this direction in Section \ref{sec:category}.

\section{Hitchin systems} \label{sec:hitchin}

In Section \ref{sec:physics} we have given an extended review and
discussion of the physical motivations for studying a certain class
of Hitchin systems.  In this section we summarize the mathematics
problem motivated by this discussion, and fill in some standard facts
about the \hk geometry of the relevant moduli spaces.

\subsection{Moduli space}

Let $G = U(K)$ or $SU(K)$ for some $K$.\footnote{One can also
consider quotients of $SU(K)$ by subgroups of its center. The
different forms of the gauge group lead to Hitchin moduli spaces
which are related but can differ in global structure. When the genus
of $C$ is zero these considerations lead, for example, to changes in
the periodicity of $m^{(3)}$. When $C$ has positive genus further
issues arise which are only briefly addressed in Appendix
\ref{app:MonodromyX}.} We have a complex curve $C$ with a
topologically trivial $G$-bundle $V$ on it. We are considering
connections $D =
\partial + A$ in $V$, and Higgs fields $\varphi \in
\Omega^{1,0}(\End V)$.

There are finitely many points $\CP_i \in C$ where the pair
$(A, \varphi)$ are required to be singular.  Let us focus on a
single such point, and choose a local coordinate $z$ which
vanishes there. A rigorous discussion of boundary conditions
for Hitchin's equations and the construction of their \hk
moduli spaces has been given in \cite{MR1206652,MR1397988} for
regular singularities, and \cite{wnh} for irregular ones.  Here
we give a schematic account of the boundary conditions we
encountered in Section \ref{sec:physics}, which will be
adequate for our purposes.

Consider the regular case first. Fix simultaneously diagonalizable
elements $\alpha \in \fg$ (skew-Hermitian) and $\rho \in \fg_\IC$,
and write
\begin{align}
\varphi_0 & = \frac{\rho}{2} \frac{dz}{z}, \label{eq:reg-sing-bc-1} \\
A_0 & = \frac{\alpha}{2i} \left( \frac{dz}{z} - \frac{d\bar z}{\bar
z} \label{eq:reg-sing-bc-2}
\right).
\end{align}
Near $z = 0$ we require the pair $(A, \varphi)$ to be close to these fiducial ones,
i.e.
\begin{align}\label{eq:SingConnBc1}
\varphi &= \varphi_0 + {\rm regular}, \\
A &= A_0 + {\rm regular}. \label{eq:SingConnBc2}
\end{align}
We let $\CN$ denote the space of all $(A, \varphi)$ obeying this
condition at each singularity. So $\CN$ depends on the data
$(\rho,\alpha)$ at each singularity, as well as on $C$ and the
points $\CP_i$; we do not write this dependence explicitly.

There is a natural action of $G$-valued gauge transformations on
$(A, \varphi)$.  This gives an action on $\CN$ as well, provided we
consider only gauge transformations which preserve the singularity
conditions.  Roughly this means we consider gauge transformations
which, at each singularity, are restricted to lie in the subgroup $H
\subset G$ commuting with the pair $(\rho, \alpha)$. Let $\CG$ denote the group
of gauge transformations so restricted.

In most of our examples we also need to allow a wilder kind of singularity.
The most elementary example of such a boundary condition was given
in \eqref{eq:ISPatInfty}, \eqref{eq:GaugeFldInfty}. In those
equations $t$ denoted a local coordinate on $C$ vanishing at the
singularity; here we called that coordinate $z$, so those equations
become
\begin{align} \label{eq:irreg-sing-bc-1}
\varphi_0 &= \frac{\nu^{1/K}}{\vert z\vert^{1/K}}
\left( \frac{\bar z}{\vert z \vert } e_{K1} + N_K \right) \frac{dz}{z}, \\
A_0  &=  \left( \sum_{a=1}^K \frac{2(a-1)-(K-1)}{4K} e_{aa} \right) \label{eq:irreg-sing-bc-2}
\left( \frac{dz}{z} - \frac{d\bar z}{\bar z}\right).
\end{align}
As before we require that near $z = 0$ the pair $(A, \varphi)$ are close to these fiducial ones.

Generally we
will have some block structure at the singular points, with each
block of $(A_0, \varphi_0)$ either of the ``regular'' form
\eqref{eq:reg-sing-bc-1}, \eqref{eq:reg-sing-bc-2} or the
``irregular'' form \eqref{eq:irreg-sing-bc-1},
\eqref{eq:irreg-sing-bc-2}.  As in the case of regular singularities,
we let $\CN$ denote the space of all $(A, \varphi)$ obeying our
singularity conditions.  Again, $\CN$ is acted on by an appropriate group $\CG$
of gauge transformations.

More general irregular singularities can also arise if we take some
scaling limits of the parameters of the theory:  we do not discuss
this situation now, but we will meet it in Section
\ref{section:Superconformal}.

Having specified our boundary conditions,
our desired moduli space $\CM$ is then the
subspace of $\CN$ consisting of solutions of Hitchin's equations,
\begin{equation}\label{eq:HitchinEqs-redux}
\begin{split}
F + R^2 [\varphi , \bar\varphi ]  & = 0,\\
 \bar\partial_A \varphi:=
 \left( \p_{\bar z} \varphi_z  + [ A_{\bar z}  ,\varphi_z] \right) d\bar{z} \wedge dz &  = 0,\\
 \partial_A \bar\varphi:=
 \left( \p_z \bar\varphi_{\bar z}  + [ A_z,\bar\varphi_{\bar z}] \right) dz \wedge d\bar{z} & = 0,
\end{split}
\end{equation}
modulo gauge transformations.  (Here by $\bar\varphi$ we mean the Hermitian conjugate
of $\varphi$.)  The arguments of Section \ref{sec:physics} identify this as the physically defined
moduli space of an appropriate $\CN=2$ field theory reduced on $S^1$ of radius $R$.

\subsection{Hyperk\"ahler structure}

As we have already emphasized, $\CM$ is equipped with a natural \hk structure.
In particular, this implies that it has a $\IC\IP^1$ worth of complex structures $J^\pz$,
and a complex symplectic form $\varpi_\zeta$, which is holomorphic at each fixed $\zeta$.
In this subsection we explain how these complex structures arise and what can be quickly said
about $(\CM, J^\pz)$ from the point of view of complex geometry.  The most crucial point for later sections
will be the identification of $(\CM, J^\pz)$ as a moduli space of flat connections when $\zeta \in \IC^\times$.

A convenient way to think about the \hk structure on $\CM$ begins
from the observation that $\CN$ is an infinite-dimensional affine
space, which is \hk in a very simple way, basically as the cotangent
bundle to the affine space of complex-valued connections.
The action of the gauge group on $\CN$ preserves the \hk
structure, and moreover admits an \hk moment map $\vec{\mu}$. The
components of this moment map (in one natural basis) are precisely
what appear on the left side of the Hitchin equations
\eqref{eq:HitchinEqs-redux}. So the procedure of imposing Hitchin's
equations and then dividing out by the gauge group $\CG$ is precisely the
usual notion of \hk quotient \cite{Hitchin:1986ea}, i.e. $\CM = \CN \hkq \CG$.

There is another way to view the \hk quotient, which in fact explains why
it induces an \hk structure on $\CM$.  Upon choosing a
$\zeta$, we can divide $\vec{\mu}$ into a real-valued moment map
$\mu_\IR$ and a complex-valued moment map $\mu_\IC$.  Then instead
of imposing $\vec{\mu} = 0$ we can impose only $\mu_\IC = 0$, and
divide out by the action of a complexification $\CG_\IC$ of the gauge group.
In favorable circumstances (where each $\CG_\IC$-orbit is ``stable,'' i.e.
contains a unique $\CG$-orbit consisting of solutions of $\mu_\IR = 0$),
this procedure gives exactly $\CM$.
Moreover $\mu_\IC$ is
holomorphic on $\CN$, and the complexified gauge group
acts holomorphically, so this procedure induces a complex structure
on $\CM$, which is $J^\pz$.

The way this works out for the $\CM$ we are considering depends drastically on whether $\zeta \in \IC^\times$ or $\zeta \in \{0,\infty\}$.
We now describe these two cases in turn.

\subsubsection{Flat connections}

We begin with the case $\zeta \in \IC^\times$.
Use $A$ and $\varphi$ to form a complex-valued connection,
\begin{equation} \label{eq:def-CA}
\CA : = \frac{R}{\zeta} \varphi + A + R \zeta \bar \varphi.
\end{equation}
At each of our marked points $\CA$ is singular, with leading
behavior determined by the singular parts of $\varphi$ and $A$.
For example, in the case of a regular singularity this amounts to
\begin{equation} \label{eq:A-sing-reg}
\CA_0 = \left( \frac{R}{\zeta} \frac{\rho}{2}  + \frac{\alpha}{2i}
\right) \frac{dz}{z} +  \left( R \zeta \frac{\bar \rho}{2} -
\frac{\alpha}{2i} \right) \frac{d\bar{z}}{\bar{z}}.
\end{equation}

All of the holomorphic information in complex structure $J^\pz$ is naturally expressed in terms of $\CA$.
For example, the holomorphic symplectic form on $\CN$ is simply
\begin{equation} \label{eq:symplectic}
\varpi_\zeta = \frac{1}{2} \int_{C} {\Tr}\,\delta \CA \wedge \delta \CA.
\end{equation}
Note that although $\CA$ is singular, $\delta \CA$ is regular, so the integral defining $\varpi_\zeta$ is well defined.
Considered as a function of $\zeta$, $\varpi_\zeta$ has simple poles at $\zeta=0$ and $\zeta = \infty$; this
is a standard expectation from \hk geometry (sometimes expressed as the statement that $\varpi$ is twisted
by $\CO(2)$ over the twistor sphere).

Now the equation $\mu_\IC = 0$ simply says that $\partial + \CA$ is flat.
Dividing out by the complexified gauge
group we thus identify $(\CM, J^\pz)$ as a moduli space of flat, $G_\IC$-valued connections on $C$.
Using a theorem of \cite{hbnc} (for regular singularities) and \cite{wnh} (more generally) we can describe
this moduli space more precisely:  it consists of all flat $G_\IC$-connections with the requisite fixed boundary conditions,
subject to a certain stability condition.

\subsubsection{Higgs bundles} \label{sec:higgs}

At $\zeta = 0$ or $\zeta = \infty$ the story is rather different.  Take for example $\zeta = 0$.
In this case the equation $\mu_\IC = 0$ just says that
\begin{equation}
\partial_{\bar z} \varphi + [A_{\bar z}, \varphi] = 0.
\end{equation}
In other words, we have a holomorphic structure on $V$ (determined by the operator
$\bar\partial := (\partial_{\bar z} + A_{\bar z}) d\bar{z}$)
and a holomorphic 1-form $\varphi$ valued in $\End V$, with appropriate singularities at the marked points.
The triplet $(V, \varphi, \bar\partial)$ is called a \ti{Higgs bundle}.
Dividing out by the complexified gauge transformations we thus identify
$(\CM, J^{(\zeta = 0)})$ as a moduli space of Higgs bundles, with appropriate boundary conditions on $\varphi$
at the punctures.
As above, the theorems of \cite{hbnc,wnh} tell us that in fact we get all Higgs bundles in this way (subject to
a certain stability condition which will not play much role in this paper.)
Note that this space, considered as a complex manifold, is actually independent of $R$.

Given a Higgs bundle there is a simple way of extracting gauge invariant information:  consider the
characteristic polynomial of $\varphi$, i.e. write
\begin{equation}
\det (x - \varphi) = x^N + \sum_{i=1}^N p_{i} x^{N-i}
\end{equation}
(where $p_1 = 0$ if $G = SU(K)$ rather than $U(K)$).  Since $\varphi$
is a 1-form the coefficients $p_i$ of its characteristic polynomial are forms of degree $i$ on $C$.  These forms are meromorphic, with some specific conditions on their singular behavior near the marked points, dictated by the type
of singularity we have fixed.  We do not write these conditions in general but just note that the space of forms
$p_i$ obeying them is an affine space $\CB$, with dimension half that of $\CM$.  This $\CB$ is to be identified
with the moduli space of the 4-dimensional gauge theory which we discussed in Section \ref{sec:physics}.

The map $\CM \to \CB$ just discussed is sometimes called the Hitchin fibration.  Its generic fiber is a compact torus,
which is moreover a complex Lagrangian submanifold in complex structure $J^{(\zeta = 0)}$.
The $p_i$ can be thought of as a maximal set of algebraically independent commuting Hamiltonians, which make
$\CM$ into an integrable system.

\subsection{Another viewpoint on defects} \label{sec:couplings}

So far we have considered the singular behavior of $( A,\varphi)$ as
a fixed ``boundary condition'' which we introduced by hand.  There
is another viewpoint which is sometimes handy to keep in mind: at
least for regular singularities, the poles in $(\varphi, A)$ can be
interpreted as arising from sources in the Hitchin equations (see
\cite{Kapustin:1998pb} and also section 3 of \cite{Gukov:2008sn}).
In other words, we deform the equations to
\begin{align}
F + R^2 [\varphi , \bar\varphi] & =  2 \pi \mu_{\IR} \delta^{(2)}(z-z_i), \label{eq:shitchin-1} \\
 \bar\p_A \varphi := d\bar z\left( \p_{\bar z} \varphi  + [ A_{\bar z}  ,\varphi]\right) & = \pi \mu_{\IC} \delta^{(2)}(z-z_i), \label{eq:shitchin-2} \\
 \p_A \bar\varphi := dz\left( \p_z \bar\varphi  + [ A_z,\bar\varphi] \right)& =  \pi \bar \mu_{\IC} \delta^{(2)}(z-z_i). \label{eq:shitchin-3}
\end{align}
The residues of $(A,\varphi)$ obeying these deformed equations then
turn out to be proportional to $\mu_{\IR}, \mu_{\IC}, \bar
\mu_{\IC}$.  (If there are multiple singularities we will have a sum on the
right-hand side.)

The usefulness of this point of view arises from interpreting the sources $\vec\mu$
as moment maps for an action of $SU(K)$ on a coadjoint orbit $O_i$ of $SL(K,\IC)$.
In other words, the modified equations \eqref{eq:shitchin-1}-\eqref{eq:shitchin-3}
are obtained by an \hk quotient $(\CN \times O_i) \hkq {\mathcal G}$, where a gauge
transformation $g(z)$ acts on $\CN$ as usual and on $O_i$ by the coadjoint action of $g(z_i)$.
Indeed, this is the natural supersymmetric coupling of the 5-dimensional
super Yang-Mills theory to degrees of freedom living at the defects $z = z_i$.

The simplest example is obtained by taking $O_i$ to be a minimal
orbit. This is exactly the example we considered in Section
\ref{sec:bc-intersection}. In the complex structure at $\zeta = 0$,
the parameter $m_i$ determines the orbit as a complex manifold,
while $m_i^{(3)}$ enters only into its metric. This viewpoint is
particularly helpful for understanding how to take the limit $m_i
\to 0$: the limit of the minimal semisimple orbit is not the zero
orbit but rather the minimal nilpotent orbit. So in the holomorphic gauge
the natural limiting boundary condition on $\varphi$ is
actually (up to conjugation as usual)
\begin{equation}\label{eq:commonsingii}
\varphi_0 = \frac{ 1 }{z} \begin{pmatrix} 0 & 1 & & & & \\ & 0 & &
& & \\ & & & \ddots & & \\ & & & & & 0 \end{pmatrix}.
\end{equation}
The corresponding solution of Hitchin's equations, related to \eqref{eq:commonsingii} by
a complex gauge transformation, is similar to \eqref{eq:irreg-sing-bc-1}, and has a milder
singularity (of order $z^{-1/2}$) for $\varphi_0$.

The appearance of the minimal orbit can also be understood more
directly. Recall that in Section \ref{sec:bc-intersection} we were
considering an M-theory setup involving a set of $K$ M5-branes on
$C$ intersecting a single transverse M5-brane. Reducing on the $S^1$
discussed in Section \ref{sec:scft} to Type IIA so that all of these
M5-branes become D4-branes  one would generally expect to get a
fundamental hypermultiplet of $SU(K)$ at the intersection.  The
minimal semisimple orbit of $SU(K)$ is very close to that:  if we
start from the $K$ hypermultiplets coupled to $U(K)$ and Higgs the
overall $U(1)$, the resulting hyperk\"ahler quotient yields the
minimal orbit coupled to $SU(K)$.

There is a similar story for more general singularities where
the residue of $\varphi$ lies in some
non-minimal semisimple orbit, i.e. it is conjugate to a diagonal
matrix with a different pattern of eigenvalues.  The eigenvalues
still play the role of mass parameters.  In the limit as the mass parameters are
turned off, the semisimple orbit smoothly approaches some
nilpotent orbit, so the residue of $\varphi$ becomes a nilpotent
matrix with a specific Jordan form.

\section{Fock-Goncharov coordinates} \label{sec:fg-coordinates}

For the next few sections we specialize to a simple case:  let $\CM$
be the moduli space of solutions of Hitchin's equations on $C$ with
gauge group $G = SU(2)$, with $\nsing$ regular singularities with
semisimple residues, at points $\CP_1, \dots, \CP_\nsing$.  We always
assume $\nsing \ge 1$, and if $C$ has genus zero we assume $\nsing \ge
3$.\footnote{The case $\nsing = 3$, $C = \IC\IP^1$ is somewhat degenerate
since in this case $\CM$ is zero-dimensional.}
We will remove the restriction to regular singularities in Section
\ref{sec:irregular}.

Fix some $\zeta \in \IC^\times$.  Then we can identify $\CM$
(considered as a holomorphic symplectic manifold, in complex structure $J^\pz$) with
a moduli space of flat $SL(2,\IC)$-connections $\CA$.
In this section we will use the approach of Fock and Goncharov
\cite{MR2233852} to define a useful collection of holomorphic Darboux coordinate
systems on $\CM$.
Each coordinate system $\CX^T$ will be associated to a
``decorated triangulation'' $T$, a certain combinatorial object to
be defined momentarily.  In the following sections we will explain
how to build the desired functions $\CX_\gamma$ on
$\CM$ from the coordinate systems $\CX^T$.

\subsection{Defining the Fock-Goncharov coordinates}

By a \ti{triangulation} we will always mean a triangulation of $C$,  with all vertices at the
singularities $\CP_i$, and at least one edge incident on each
vertex.  At each $\CP_i$ we have the operator $M_i$ giving the clockwise monodromy of $\CA$-flat sections,
which is $SL(2,\IC)$-valued and
generically has two distinct eigenlines.  We define a \ti{decoration} at $\CP_i$ to be a choice of
one of these two eigenlines, and a \ti{decorated triangulation} $T$ to be a triangulation plus a decoration
at each vertex.  Let $\mu^T_i$ denote the corresponding monodromy eigenvalue.

To avoid confusion, it is useful to
observe a slight difference between our setup and that of
\cite{MR2233852}. Their point of view was to include the
choice of decoration in the moduli space, so they really built
coordinate systems on a moduli space of ``decorated flat
connections.'' In our situation, where the conjugacy classes of
the $M_i$ are fixed once and for all, we instead include the
choice of decoration as part of the discrete datum $T$.

\insfig{3in}{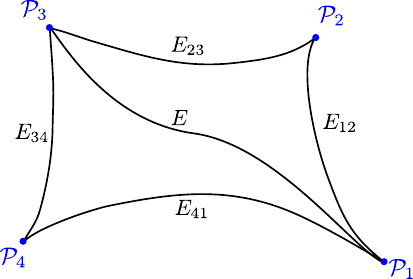}{The quadrilateral $Q_E$ associated to an
edge $E$ in the triangulation $T$.}

Now fix a decorated triangulation $T$.
For each edge $E$ of $T$, we define a coordinate function $\CX^T_E$,
as follows. The two triangles bounding $E$ make up a quadrilateral
$Q_E$.  Number its vertices $\CP_i$, $i = 1, 2, 3, 4$, in
counterclockwise order (using the standard orientation on $C$),
and with $E$ running between vertices $1$
and  $3$. The edges are unoriented, so such a labeling
is determined only up to the simultaneous exchange $1
\leftrightarrow 3$ and $2 \leftrightarrow 4$. The $\CX^T_E$
constructed below will be independent of this ambiguity.
See Figure \ref{fig:quadrilateral}.

Over $Q_E$ we now choose four sections $s_i$ of $V$, obeying the flatness equation
\begin{equation}
(d + \CA) s_i = 0,
\end{equation}
and with each $s_i$ an eigenvector of $M_i$, with eigenvalue $\mu^T_i$.
The $s_i$ cannot be made globally single-valued and smooth on $C$ (the monodromy would
require introducing a branch cut somewhere)
but we emphasize that we do choose them to be single-valued and smooth on $Q_E$.
Each $s_i$ is uniquely determined up to complex rescaling.

The Fock-Goncharov coordinate is constructed from the $s_i$:
\begin{equation} \label{eq:cross-ratio-wedges}
\CX^T_E := - \frac{(s_1 \wedge s_2)(s_3 \wedge s_4)}{(s_2 \wedge
s_3)(s_4 \wedge s_1)},
\end{equation}
where all $s_i$ are evaluated at any common point $\CP_* \in Q_E$.  Because the
connection $\CA$ is valued in $sl(2,\IC)$, this quantity is
independent of $\CP_*$. Moreover, the ambiguity of each $s_i$
by a complex rescaling cancels out in $\CX^T_E$.

The functions $\CX_E^T$ go to $0$ or $\infty$ only when $s_i \wedge
s_j = 0$ for two adjacent vertices $\CP_i$, $\CP_j$, which happens
on a codimension-$1$ subvariety of $\CM$.  Thus the $\CX_E^T$ are
well defined in a Zariski open patch $\CU_T \subset \CM$. In fact,
the $\CX_E^T$ give a coordinate system on this patch
\cite{MR2233852}: in Appendix \ref{app:MonodromyX} we show how to
reconstruct the connection $\CA$ modulo gauge equivalence (i.e. the
monodromy representation of $\CA$) from the $\CX_E^T$.

As an aside we note that one can think of $\CX^T_E$ as a
cross-ratio, in the following sense. After choosing some fiducial
basis in the $2$-dimensional space of $\CA$-flat sections of $V$ over $Q_E$,
the four 1-dimensional
subspaces $\{ \lambda s_i: \lambda \in \IC^\times\} \subset \IC^2$
give four points $x_i \in \IC\IP^1$. Then, one can show that
\eqref{eq:cross-ratio-wedges} is just
\begin{equation} \label{eq:cross-ratio}
\CX^T_E = - \frac{(x_1 - x_2)(x_3 - x_4)}{(x_2 - x_3)(x_4 - x_1)}.
\end{equation}

\subsection{Monodromies} \label{sec:monodromies}

Certain combinations of the $\CX^T_E$ have a simple interpretation.
Indeed, consider any singular point $\CP_i$, and consider the product
of $\CX^T_E$ over all edges $E$ which meet $\CP_i$. Without accounting
for the monodromy of $s_i$ one formally finds that there is a
telescoping cancellation of terms in the product, leaving $1$.
However, in defining $\CX_E^T$ one must take care to define $s_i$ to
be single-valued and continuous in the whole quadrilateral $Q_E$.
Such a choice may be made for each $E$, but we cannot construct a
single $s_i$ satisfying this condition everywhere; we will have to
include a branch cut somewhere.  If we choose our cut to run through
one of the triangles meeting $\CP_i$, then precisely two
quadrilaterals are affected.  They are each multiplied by a factor
of $\mu^T$, so we find that
\begin{equation} \label{eq:monodromy-product}
\prod_{E\,\textrm{meeting}\,\CP_i} \CX^T_{E} = (\mu^T_i)^{2}.
\end{equation}
(We have assumed implicitly that $\CP_i$ meets at least two edges; it is actually possible that it
meets only one, but \eqref{eq:monodromy-product} will contine to hold in that case, for which
see Section \ref{sec:degenerate} below.)

\subsection{Counting the coordinates}

As we have remarked, the functions $\CX^T_E$ give a local coordinate system on
$\CM$.  As a simple check we show that the dimension counting works
out correctly.

The dimension of $\CM$ can be obtained by considering the monodromy
data.  Let $g$ be the genus of $C$, and recall that our local system
has $\nsing$ regular singularities $\CP_i$ on $C$.
There are $3(\nsing+2g)$ degrees
of freedom in the $SL(2,\IC)$-valued monodromy matrices $M_i$, $A_j$, $B_j$ (around the
singularities, $A$ and $B$ cycles respectively), minus
$3$ for the usual constraint $\prod M_i = \prod A_j B_j A_j^{-1} B_j^{-1}$,
minus another $3$ for the $SL(2,\IC)$ gauge symmetry, minus $\nsing$
because the monodromy eigenvalues are fixed.  Altogether this gives
\begin{equation}
\dim \CM = 2\nsing + 6g - 6.
\end{equation}
How many coordinates $\CX^T_E$ do we get?
Using Euler's formula
\begin{equation}
\nsing + (\# F) - (\# E) = 2 - 2g
\end{equation}
and the fact that all of our faces are triangles
\begin{equation}
2 (\# E) = 3 (\# F)
\end{equation}
we obtain
\begin{equation} \label{eq:counts}
\# E = 3\nsing + 6g - 6, \quad \# F = 2\nsing + 4g - 4.
\end{equation}
As we argued above, $\nsing$ combinations of the $\CX^T_E$ give the
monodromy eigenvalues $(\mu^T_i)^2$, which are fixed up to a discrete choice;
the other $2\nsing + 6g - 6$ are just enough to give coordinates on
$\CM$ as desired.

\subsection{Hamiltonian flows} \label{sec:ham-flows}

As we noted in Section \ref{sec:hitchin}, $\CM$ has a natural
holomorphic symplectic form.  We now describe the Hamiltonian flow
generated by $\log \CX^T_E$.  This flow turns out to be very simple.

Given a connection $\CA \in \CM$, we may describe its image $\CA^t$
under the time-$t$ flow generated by $\log\CX^T_E$,
 as follows.  We
consider the connection $\CA$ on $C$ as divided into two
pieces, namely the restrictions to the quadrilateral $Q_E$
(``inside'') and its complement $C \setminus Q_E$
(``outside'').
On the common boundary of the two pieces we have an identification
between boundary values of sections of $V$.  Gluing the two pieces
back together using this identification one would recover the original $V$ with its connection
$\CA$.  Now we consider gluing them together using a different identification.
Namely, on edge $E_{ij}$ (with $i$, $j$ cyclically ordered)
we glue $s_i$ outside to $e^{\pm t/2} s_i$
inside, and $s_j$ outside to $e^{\mp t /2} s_j$ inside, where the
sign $\pm$ is $+$ for $i = 2,4$ and $-$ for $i = 1,3$.
This gluing is still $SL(2,\IC)$-valued since
it preserves $s_i \wedge s_j$, and preserves the flatness of the
connection since we have glued flat sections to flat sections.  It
defines the new connection $\CA^t$.

Note that the monodromy of the section $s_i$ around $\CP_i$ is the
same for $\CA^t$ as it was for $\CA$ (since the transformations of
$s_i$ coming from the two edges which meet $\CP_i$ cancel one
another).  It follows that $\CA^t$ and $\CA$ have the same monodromy
eigenvalues, so this flow really preserves our
moduli space $\CM$ as it should.

The derivation of this flow from the symplectic structure
\eqref{eq:symplectic} is basically straightforward but notationally
a bit awkward, so we have sequestered it in Appendix
\ref{app:poorman}.

\subsection{Poisson bracket} \label{sec:poisson}

Given two edges $E$ and $E'$ of the triangulation $T$, we define
$\inprod{E, E'}$ to be the number of faces $E$
and $E'$ have in common, counted with a sign $+1$ ($-1$) if $E$
comes immediately before $E'$ in counterclockwise (clockwise) order
going around the common face.
In this section we show that the Poisson brackets of the $\CX^T_E$
are determined by this pairing,
\begin{equation} \label{eq:fg-poisson}
\{\CX^T_E, \CX^T_{E'}\} = \inprod{E, E'} \CX^T_E \CX^T_{E'}.
\end{equation}

To check \eqref{eq:fg-poisson} we consider the action of the
Hamiltonian flow generated by $\log \CX^T_E$ on $\CX^T_{E'}$.
Recall from Section \ref{sec:ham-flows} that this flow involves
cutting and gluing along the sides of the quadrilateral $Q_E$.
If $\inprod{E,E'} = 0$, then one can compute $\CX^T_{E'}$ using only
the connection $\CA$ outside $Q_E$, and hence $\CX^T_{E'}$ is
invariant under the flow, in agreement with \eqref{eq:fg-poisson}.
Now suppose $\inprod{E,E'} = +1$. The definition
\eqref{eq:cross-ratio-wedges} of $\CX^T_{E'}$ requires us to
transport the sections $s_i$ from the vertices $\CP_{1,2,3,4}$ of
$Q_{E'}$ to a common point $\CP_* \in Q_{E'}$; let us choose that
point to lie also in $Q_E$. See Figure \ref{fig:poisson-bracket} to
fix the labeling of the vertices.
\insfig{3.5in}{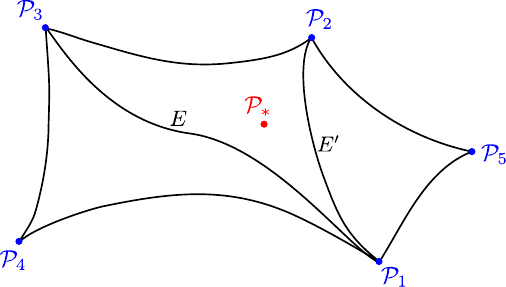}{Fixing notation for calculating the
effect of the flow generated by $\log \CX^T_{E}$ on the function
$\CX^T_{E'}$, in the case $\inprod{E,E'} = +1$.} The transport requires us to bring $s_5$
from outside $Q_E$ to inside, across the edge $E' = E_{12}$. It
follows that the $s_5$ appearing in the definition of $\CX^T_{E'}$
is transformed by the flow. Expanding $s_5$ in the convenient basis
of flat sections
\begin{equation}
s_5 = a s_1 + b s_2
\end{equation}
the action of the flow replaces $a \mapsto a e^{-t/2}$ and $b
\mapsto b e^{t/2}$, and hence takes
\begin{align}
s_1 \wedge s_5 &\to e^{t/2} s_1 \wedge s_2, \\
s_5 \wedge s_2 &\to e^{-t/2} s_5 \wedge s_2.
\end{align}
So the flow generated by $\log \CX^T_E$ multiplies $\CX^T_{E'}$ by
$e^{t}$.  The desired Poisson bracket \eqref{eq:fg-poisson} follows
directly.

It is also possible to have $\inprod{E,E'} = +2$, if the two edges
share two faces.
In this case a similar argument shows that the flow generated by $\log
\CX^T_E$ multiplies $\CX^T_{E'}$ by $e^{2t}$, and the desired
\eqref{eq:fg-poisson} still holds.

\subsection{Coordinate transformations and the groupoid of decorated triangulations}

Now we come to a crucial point.  The
coordinate system $\CX^T$ depends on the choice of triangulation $T$.
We would like to know the coordinate transformation
$S_{T,T'}$ relating $\CX^T$ to $\CX^{T'}$.

There is a procedure to compute $S_{T,T'}$, which is straightforward
in principle, but cumbersome in practice.  The cross ratios
$\CX^T_E$ contain enough information to parallel-transport flat
sections $s$ along paths in $C$, as shown in Appendix
\ref{app:MonodromyX}.
Therefore, if $T$ and $T'$ have the same decoration, one can first
use the $\CX^T_E$ to determine $s_i$ and $s_j$ along the edges
$E'_{ij}$ of $T'$, and then use those to compute the
$\CX^{T'}_{E'_{ij}}$.  If the decorations are different, one also
needs to determine the new monodromy eigensections $s'_i$, by
transporting reference sections around the vertices of the
triangulation $T$.  The explicit expressions for $S_{T,T'}$ obtained
in this way are generally complicated rational transformations, and
will be of little use to us (except for one crucial exception, which
we will meet in Section \ref{sec:UniversalStokesMatrix}).

Instead, it is more useful to study the coordinate transformations
corresponding to simple local changes of the triangulation $T$.
Any two triangulations $T$, $T'$ are related to one another by a
sequence of such simple moves, which we call ``flips at edges'' and ``pops
at vertices'':

\begin{itemize}
\item Given an edge $E$ of $T$, we define the flip $\sigma_E$ as
follows:  the original $T$ and the flipped $T'$ are identical except
that in the quadrilateral $Q_E$ the edge $E = E_{13}$ in $T$ is
replaced by an edge $E' = E_{24}$ in $T'$, as illustrated in Figure
\ref{fig:flip}.  (We assume here that $E$ is distinct from all
edges of the quadrilateral $Q_E$.  This can fail in the presence of
degenerate triangles with two coincident edges, to be discussed below.
We will not need to define a flip for such edges.)
\insfig{5.5in}{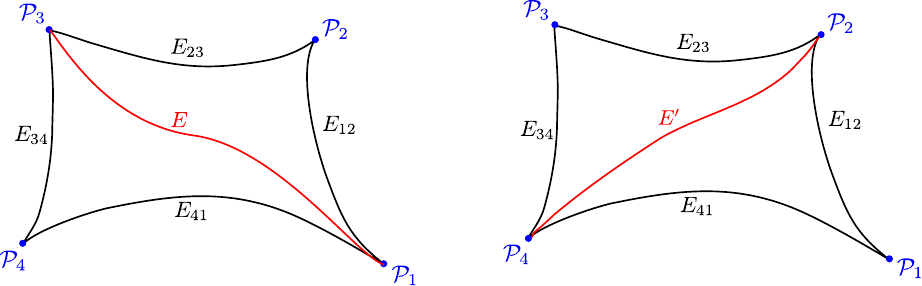}{Flipping a triangulation $T$ to $T'$.}

\item Given a vertex $\CP_i$, we define the pop $\pi_i$ as follows:  the original $T$ and the popped
$T'$ differ only by reversing the choice of monodromy eigenvalue (decoration)
at $\CP_i$.
\end{itemize}
The coordinate transformation $S_{T,T'}$ is then usefully described as
the composition of the coordinate transformations associated to a
sequence of flips and pops which takes $T$ to $T'$.

It is convenient to rephrase the above in the language of
groupoids.\footnote{A ``groupoid'' is a category all of whose
morphisms are invertible. In plain English this simply means we have
a system of points (``objects'') and a collection of arrows between
points (``morphisms''). There is an associative composition law on
arrows, a unit arrow on every point, and every arrow has an inverse
arrow.} First we contemplate a groupoid $\CT$ which has as objects
the various decorated triangulations, and a unique morphism betwen
any two decorated triangulations. (That is, the objects have no
automorphisms.)  It turns out that $\CT$ is freely generated by the
flips and pops, subject to four kinds of relations:
\begin{itemize}
\item $\sigma_E \sigma_{E'} = \sigma_{E'} \sigma_E$
when $\inprod{E, E'} = 0$.

\item When $\inprod{E,E'} = 1$, $\sigma_E \sigma_{E'} \neq \sigma_{E'} \sigma_E$,
but $\sigma_E \sigma_{E'}$ can be rewritten as a product of \ti{three} flips, as
shown in Figure \ref{fig:pentagon-flips}.\footnote{When $\inprod{E,E'} = 2$
there is no relation between $\sigma_E$ and $\sigma_{E'}$, but nevertheless
this case will be important in Section \ref{sec:limit-triangulation} below.}

\item Each $\pi_i$ commutes with everything else.

\item $\pi_i^2 = \id$.

\end{itemize}

\insfig{4in}{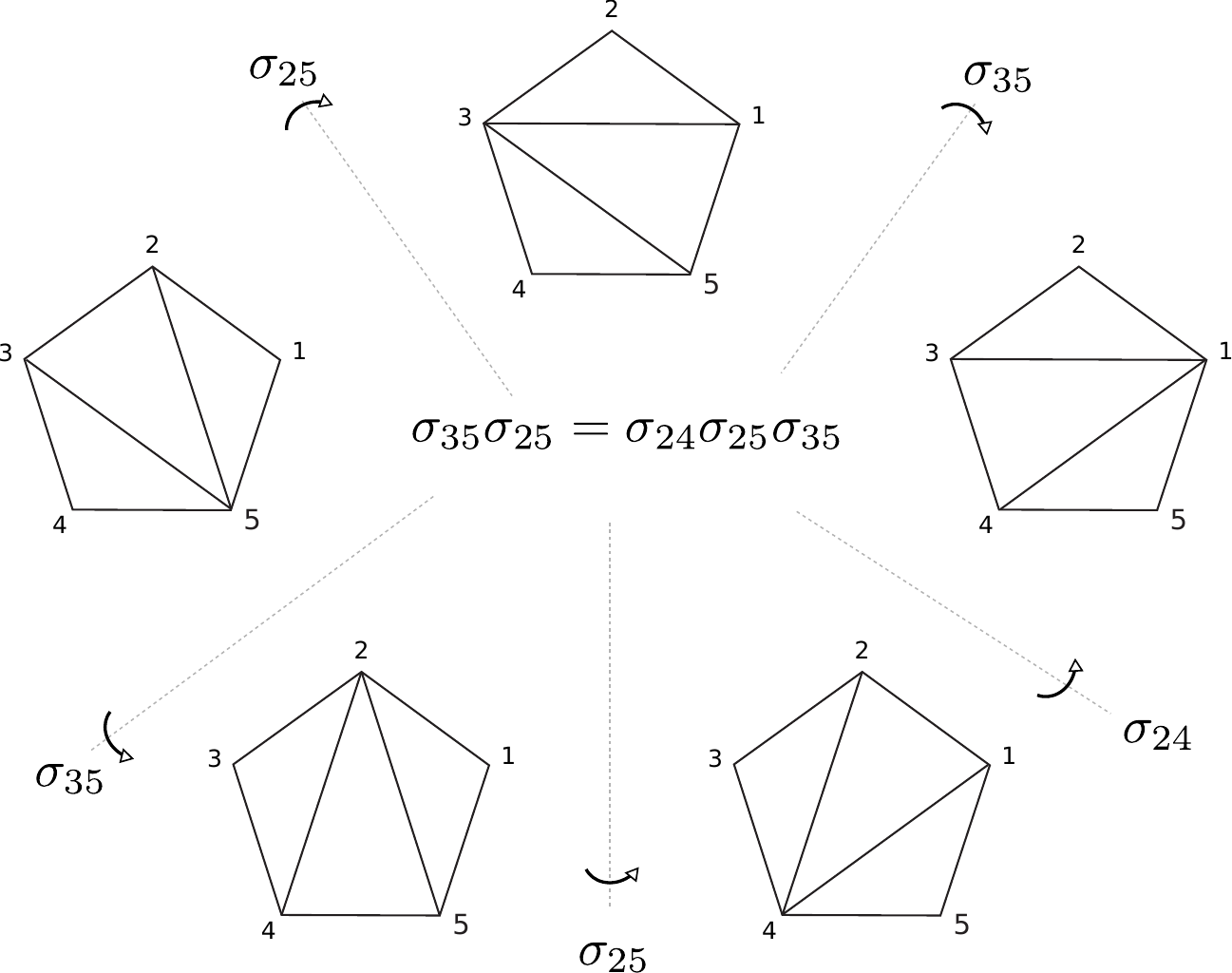}{The ``pentagon relation'': two
different sequences of flips which relate a pair of triangulations.}

Second, we define the groupoid $\CC$ of Darboux coordinate systems
on $\CM$.  To define this groupoid first define a \ti{Poisson torus}
to be a space isomorphic to $(\IC^\times)^{\dim \CM}$, equipped with
a Poisson structure such that the bracket of the standard coordinate
functions $x_i$ is of the form $\{x_i, x_j\} = a_{ij} x_i x_j$. Then
a \ti{Darboux coordinate system} is an injective Poisson map from a
Zariski-open subset of $\CM$ to a Poisson torus.  The groupoid $\CC$ has as
objects the Darboux coordinate systems, and a unique morphism
between any two objects, which may be identified with the unique
Darboux coordinate transformation between them (a Poisson bijection
between the two images of the overlap in $\CM$).

One can then consider \eqref{eq:cross-ratio-wedges} as defining a functor,
\begin{align}
\CT & \to \CC \\
T & \mapsto \CX^T.
\end{align}
Under this functor the morphisms of triangulations get mapped to corresponding coordinate
transformations, so relations among morphisms
imply relations among coordinate transformations.  As we will see momentarily, these relations
can be surprising when written explicitly.

\subsection{Transformation under flips}\label{subsec:TransfmFlips}

The effect of a flip $\sigma_E$ on the coordinates is rather simple.
First, referring to Figure \ref{fig:flip} it is easy to see that the
coordinates attached to $E$ in $T$ and $E'$ in $T'$ are trivially
related,
\begin{equation} \label{eq:fg-trans-1}
\CX^T_E = (\CX^{T'}_{E'})^{-1}.
\end{equation}

\insfig{5in}{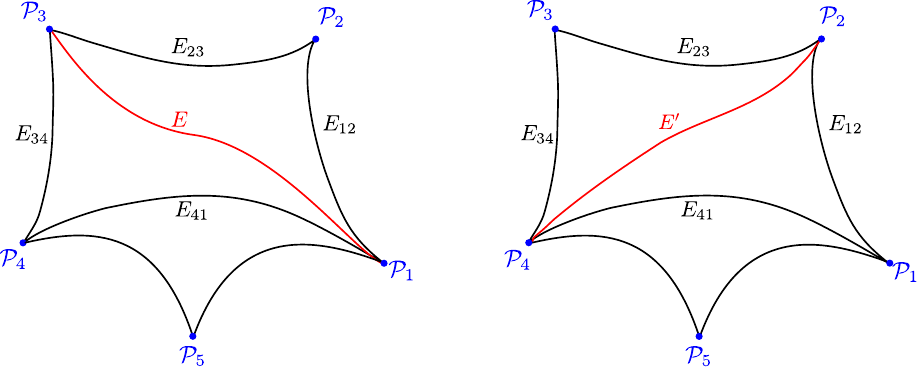}{The effect of the flip on the quadrilaterals used to compute $\CX^T_{E_{41}}$ and $\CX^{T'}_{E_{41}}$.}

The flip also changes the coordinates attached to the four edges of $Q_E$.
For example, consider the coordinate $\CX^T_{E_{41}}$ in Figure \ref{fig:flip-edge-coordinate}.
It is the cross-ratio of $s_1$, $s_3$, $s_4$, $s_5$ in that order, which we denote as
$\CX^T_{E_{41}} = r(1, 3, 4, 5)$.  After the flip, we have instead $\CX^{T'}_{E_{41}} = r(1, 2, 4, 5)$.
On the other hand, we also have $\CX^T_E = r(1, 2, 3, 4)$.  Since any five points on $\IC\IP^1$ have only two independent
cross-ratios, it follows that there must be an algebraic relation between $\CX^T_{E_{41}}$, $\CX^{T'}_{E_{41}}$ and $\CX^T_E$.
Indeed a direct computation gives this relation, and similar ones involving the other edges of $Q_E$:
\begin{align}
\CX^{T'}_{E_{12}} &= \CX^T_{E_{12}} (1 + \CX^T_E), \label{eq:fg-trans-2} \\
\CX^{T'}_{E_{23}} &= \CX^T_{E_{23}} (1 + (\CX^T_E)^{-1})^{-1}, \label{eq:fg-trans-3} \\
\CX^{T'}_{E_{34}} &= \CX^T_{E_{34}} (1 + \CX^T_E), \label{eq:fg-trans-4} \\
\CX^{T'}_{E_{41}} &= \CX^T_{E_{41}} (1 + (\CX^T_E)^{-1})^{-1}. \label{eq:fg-trans-5}
\end{align}

(Above we assumed that
each $\inprod{E_{ij}, E} = \pm 1$, i.e. the only common face is the one appearing in the figure.
If, say, $\inprod{E, E_{41}} = 2$ instead, then the transformation is
instead $\CX^{T'}_{E_{41}}  = \CX^T_{E_{41}} (1 + (\CX^T_E)^{-1})^{-2}$,
and similarly for the other edges.)

These transformations of course preserve the Poisson bracket \eqref{eq:fg-poisson}
(in a non-trivial fashion, since the intersection matrix $\inprod{E,E'}$ changes under
the flip.)  So they define a morphism of Darboux coordinate systems.
Moreover, note that this morphism strongly resembles the KS transformations $\CK_\gamma$ we reviewed in
Section \ref{sec:review}.  This is the first hint of the connection
between the Fock-Goncharov coordinates and the Kontsevich-Soibelman
wall-crossing formula.  The transformation laws are still not quite the same, though.  The
identification will require one more important step, to be described below.

Finally we comment that the pentagon identity in the groupoid $\CT$ of decorated
triangulations implies a corresponding pentagon identity among
the morphisms in $\CC$.  How does it arise concretely?  Consider
again Figure \ref{fig:pentagon-flips}, and call the two coordinate
functions attached to the two interior edges of the $n^{th}$ pentagon $(x_n, y_n)$, with $n$ considered
modulo $5$.  The flip relating adjacent pentagons gives the coordinate transformation
\begin{align}
y_{n+1} & = x_n^{-1}, \\
x_{n+1} & = y_n(1 + x_n).
\end{align}
This sequence of coordinate transformations indeed has period $5$.
In particular, by eliminating $y_n$ we get the beautifully
simple period-5 relation $x_{n+1} x_{n-1} = 1+x_n$.

\subsection{Degenerate triangulations and transformation under pops} \label{sec:degenerate}

The effect of popping a vertex $\CP$ of a generic triangulation $T$ is
in general somewhat intricate. To construct the new flat section at
$\CP$ in terms of the old one, we would need to use the parallel transport all the way
around $\CP$.  While in principle this is determined by the $\CX^T_E$ for edges
incident on $\CP$, in practice the result is generally a complicated rational
function.  In this subsection we will meet a special class of
triangulations with only a \ti{single} edge incident on $\CP$,
for which the effect of the pop becomes very simple.
These are triangulations which include degenerate
faces in which two of the edges are identified, as pictured in
Figure \ref{fig:degenerate-face}.

With a degenerate face included, our rules for constructing the
coordinates $\CX^T_E$ have to be amended slightly. We ``resolve''
the face by passing to a covering $\tilde{U}$ of a neighborhood $U
\subset C$, ramified only at the center vertex.
The flat connection $\CA$ on $U$ pulls back to a flat connection on
$\tilde{U}$, with regular singularities at the preimages of the
singularities on $U$, and $T\vert_U$
lifts to a triangulation $\tilde{T}$.  We choose the covering so
that $\tilde{T}$ is not degenerate, so we can use
our standard rule to define coordinates
$\CX^{\tilde{T}}_{\tilde{E}}$. (In order to
resolve the face completely the covering should have at least three
sheets, as in Figure \ref{fig:degenerate-quadrilateral}.)
We then define $\CX^T_E := \CX^{\tilde{T}}_{\tilde{E}}$
where $\tilde{E}$ is any preimage of $E$. Since the connection
is pulled back, it is invariant under the
automorphisms of the covering, so $\CX^T_E$ is independent of the choice of preimage
$\tilde{E}$.  It is similarly independent of the precise choice of covering.

\insfig{1.25in}{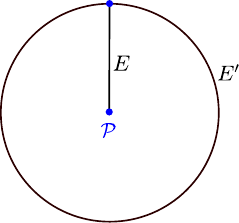}{A degenerate face:  it can be thought of as a triangle
whose three edges are $E'$, $E$ and $E$.}

\insfig{4.5in}{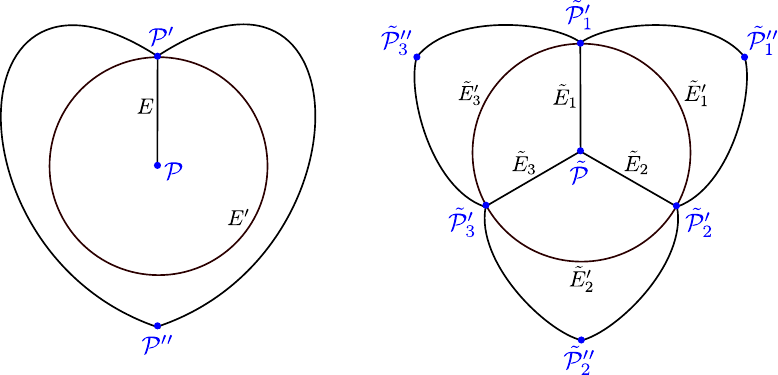}{A neighborhood
of a degenerate face, and its resolution by passing to a threefold cover.}

For example, we compute  $\CX^T_E$ using the quadrilateral
$Q_{\tilde E_1}$ as follows.  Let $s$ be the decoration at the preimage $\tilde{\CP}$ of $\CP$, which we
may choose to be the pullback of a decoration at $\CP$.
Let $s'_i$ be the decorations at the various preimages
$\tilde{\CP}'_i$ of $\CP'$; we may choose all of the $s'_i$ to be pullbacks of a single decoration
at $\CP'$.
Now \eqref{eq:cross-ratio-wedges} becomes
\begin{equation}
\CX^T_E = \CX^{\tilde{T}}_{\tilde{E}_1} = - \frac{(s'_3 \wedge s'_1) (s'_2 \wedge s)}{(s \wedge s'_3) (s'_1 \wedge s'_2)},
\end{equation}
which may be further simplified as follows.  Let $M$ denote the clockwise monodromy around $\CP$ in the
original degenerate face.
After transporting all $s'_i$ to a common point in $Q_{\tilde{E}_1}$ we will have $s'_1 = M^{-1} s'_3 = M s'_2$, so
\begin{equation} \label{eq:degenerate-monodromy}
\CX^T_E = - \frac{(M s'_1 \wedge s'_1) (M^{-1} s'_1 \wedge s)}{(s \wedge M s'_1) (s'_1 \wedge M^{-1} s'_1)}
= - \frac{(M s'_1 \wedge s'_1) (s'_1 \wedge M s)}{(M^{-1} s \wedge s'_1) (M s'_1 \wedge s'_1)} = (\mu^T)^{2},
\end{equation}
where we recall that $\mu^T$ is the eigenvalue of $s$ under $M$,
and we used the fact that $M \in SL(2,\IC)$.

For $\CX^T_{E'}$ we get
\begin{equation}
\CX^T_{E'} = \CX^{\tilde{T}}_{\tilde{E}'_3} = - \frac{(s \wedge s'_3)(s''_3 \wedge s'_1)}{(s'_1 \wedge s)(s'_3 \wedge s''_3)}
\end{equation}
and again after transporting to a common point in $Q_{\tilde{E}'_3}$ we have $s'_3 = M s'_1$, so
\begin{equation} \label{eq:degen-loop-coord}
\CX^T_{E'} = - \frac{(s \wedge M s'_1)(s''_3 \wedge s'_1)}{(s'_1 \wedge s)(M s'_1 \wedge s''_3)} =
- \frac{M^{-1} s \wedge s'_1}{s'_1 \wedge s} \frac{s''_3 \wedge s'_1}{M s'_1 \wedge s''_3} = (\mu^T)^{-1} \frac{M
s'_1 \wedge s''_3}{s''_3 \wedge s'_1}.
\end{equation}
Note that $s$ enters this result only through its monodromy
eigenvalue $\mu^T$. It follows that the only effect of a pop at the
vertex $\CP$ is through the relation $\mu^{T'} = (\mu^T)^{-1}$,
which gives using \eqref{eq:degenerate-monodromy},
\eqref{eq:degen-loop-coord} the transformation $\pi_{\CP}$:
\begin{align}
\CX^{T'}_E &= (\CX^{T}_E)^{-1}, \\
\CX^{T'}_{E'} &= \CX^T_E \CX^T_{E'}.
\end{align}

So the effect of the pop at the degenerate vertex $\CP$ is very simple.
On the other hand, the effect of the pop at $\CP$ in a generic triangulation $T$ can
be determined by first flipping all edges incident on $\CP$ but one to
reach a degenerate triangulation with a single edge incident on $\CP$,
then popping at $\CP$, and finally flipping the edges back in the opposite order. Since
pops and flips commute, this is the same as popping
$T$ at $\CP$.
More generally, one way to compute $S_{T,T'}$ for a generic pair of
triangulations $T$, $T'$ is to decompose this morphism into  a
sequence of flips which change the undecorated triangulation underlying $T$ into
the one underlying $T'$, and which pass through degenerate triangulations
where the effect of the required pops is simple.

\subsection{Limits of triangulations and the \juggle} \label{sec:limit-triangulation}

So far we have considered the transformation of the coordinates $\CX^T_E$ under flips and pops.
For our purposes it will actually be necessary also to consider a third, more complicated kind of
transformation, which is not quite a relation between two triangulations, but rather a relation
between two ``infinitely twisted limits'' of triangulations.  In this section we introduce these
limits.

We will consider triangulations $T$ containing an annular region
$W$, with a single vertex $\CP$ on the outer ring and $\CP'$ on the inner
ring.  Any such $T$ has two interior edges on $W$, with both vertices in common.
Suppose we hold the part of $T$ outside $W$ fixed, and consider varying the part on $W$.
There are various choices of such $T$, differing from one another in how many
times the edges wind around the annulus.
See Figure \ref{fig:annulus} for some examples.

In order to parameterize the possible $T$ we begin by choosing two fixed paths
$E_\pm$ from $\CP$ to $\CP'$, such that $E_+ - E_-$ winds once around $W$ counterclockwise.
Then we define $T_0$ to be the triangulation with interior edges
$E_{0+} := E_+$, $E_{0-} := E_-$.

Performing a flip on $E_{0-}$ we obtain another triangulation
$T_1$. We label its edges as $E_{1\pm}$, again with $E_{1+}$ differing
from $E_{1-}$ by one unit of counterclockwise winding:  so $E_{1+}$ is
the edge created by the flip, and $E_{1-} = E_{0+}$.
By flipping $E_{1-}$ we obtain a new triangulation
$T_2$. See Figure \ref{fig:annulus}. Repeating this process we
obtain a sequence of triangulations $T_m$ for $m \in \IZ_+$.
Flipping $E_{m-}$ takes us from $T_m$ to $T_{m+1}$. Conversely,
flipping $E_{m+}$ takes us from $T_m$ to $T_{m-1}$.

\insfig{13cm}{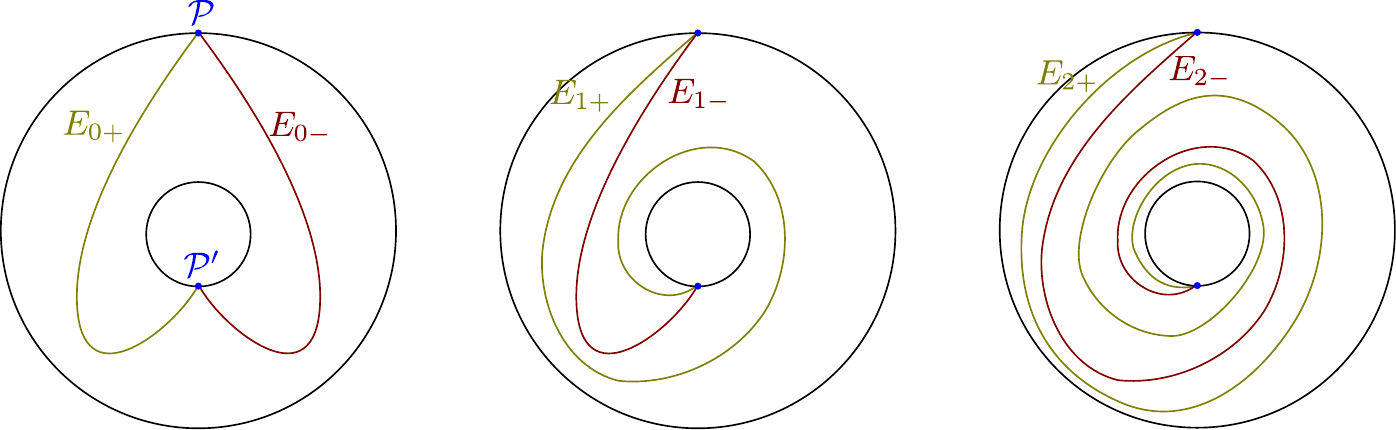}{An annulus in triangulation $T_0$,
and the triangulations $T_1$, $T_2$ obtained by flipping $E_{0-}$, $E_{1-}$.}

What are the Fock-Goncharov coordinates for the triangulation $T_m$?
To minimize confusion we pass to an infinite covering of the annulus,
like the coverings we used in our discussion of degenerate triangulations,
and choose specific preimages $\tilde{E}_\pm$ of $E_\pm$.
Define $s$, $s'$ to be the decorations at the two ends $\tilde{\CP}$, $\tilde{\CP}'$ of $\tilde{E}_+$,
and introduce the notation
\begin{equation}
K = - (s \wedge M s) (s' \wedge M s'), \quad c_k = (s \wedge M^k s')^2.
\end{equation}
The definition \eqref{eq:cross-ratio-wedges} becomes (see Figure \ref{fig:log-cover-Tm})
\begin{equation}
\CX_{E_{m+}}^{T_m} = \frac{K}{c_{1-m}}, \quad \CX_{E_{m-}}^{T_m} = \frac{c_{-m}}{K}.
\end{equation}
\insfig{11cm}{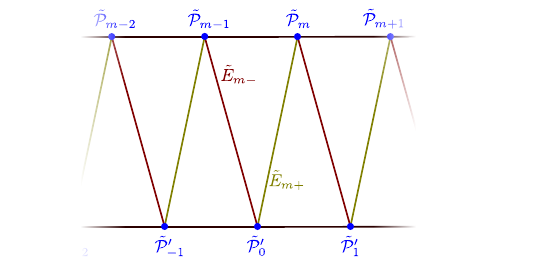}{Our infinite covering of the annulus, with the triangulation $T_m$ marked.  We defined
$\tilde\CP_0 := \tilde\CP$ and $\tilde\CP'_0 := \tilde\CP'$, and then call their shifts $\tilde\CP_m$ and
$\tilde\CP'_m$ respectively.}

Our main interest here is in the $m \to \infty$ limit. Define two
``limit coordinates'' $\CX^{T_{+\infty}}_{A,B}$, as follows. Denote
the eigenvalues of $M$ as $\xi_\pm$, where $\abs{\xi_+} > 1$. (We
assume that we are in the generic situation so that $\abs{\xi_\pm}
\neq 1$. In the physical application this corresponds to assuming
generic masses for vectormultiplets.) One of our coordinates is
simply
\begin{equation} \label{eq:A-positive}
\CX^{T_{+\infty}}_A := \xi_+^{2}.
\end{equation}
To define the other coordinate introduce the projection operators $P_\pm$ on the two monodromy eigenspaces.
Then define
\begin{equation} \label{eq:B-positive}
\CX^{T_{+\infty}}_B := -\frac{(s \wedge P_- s')^2}{(s \wedge M s)(s' \wedge M s')}.
\end{equation}
A direct computation shows that these are ``limiting'' coordinates in the sense that
\begin{align}
\CX^{T_{+\infty}}_A &= \lim_{m \to \infty} \CX^{T_m}_{E_{m+}} \CX^{T_m}_{E_{m-}}, \label{eq:limiting-1} \\
\CX^{T_{+\infty}}_B &= \lim_{m \to \infty} (\CX^{T_m}_{E_{m+}})^{-m} (\CX^{T_m}_{E_{m-}})^{1-m}. \label{eq:limiting-2}
\end{align}
Letting $E$ denote one of the boundaries of the annulus, the coordinates $\CX^{T_m}_E$
have well defined $m \to \infty$ limits, which we call $\CX^{T_{+\infty}}_E$.
The coordinates for all edges outside the annulus are just $m$-independent,
so letting $E$ be one of these edges we simply define $\CX^{T_{+\infty}}_E = \CX^{T_m}_E$ for any $m$.
We can then define a new coordinate system $\CX^{T_{+\infty}}$
consisting of $\CX^{T_{+\infty}}_A$, $\CX^{T_{+\infty}}_B$, and all the other $\CX^{T_{+\infty}}_E$.

The coordinate system $\CX^{T_{+\infty}}$ we obtained here depended on the choice we
made at the beginning, of which paths to call $E_\pm$.  $\CX^{T_{+\infty}}_A$ is independent of this choice, but
$\CX^{T_{+\infty}}_B$ for different choices differ by integral powers of $\xi_+$.

We can also consider the opposite kind of limit.  Beginning again with $T$ (and choosing the
same paths $E_\pm$), flipping $E_{m+}$ repeatedly we get a sequence of triangulations
$T_{m}$ with $m < 0$, with edges winding around the annulus counterclockwise.  Then define,
similarly to the above,
\begin{equation} \label{eq:A-negative}
\CX^{T_{-\infty}}_A := \xi_-^2
\end{equation}
and
\begin{equation} \label{eq:B-negative}
\CX^{T_{-\infty}}_B := -\frac{(s \wedge P_+ s')^2}{(s \wedge M s)(s' \wedge M s')}.
\end{equation}
These are also limiting coordinates, in the sense that
\begin{align}
\CX^{T_{-\infty}}_A &= \lim_{m \to -\infty} \CX^{T_m}_{E_{m+}} \CX^{T_m}_{E_{m-}}, \label{eq:limiting-3} \\
\CX^{T_{-\infty}}_B &= \lim_{m \to -\infty} (\CX^{T_m}_{E_{m+}})^{-m} (\CX^{T_m}_{E_{m-}})^{1-m}. \label{eq:limiting-4}
\end{align}

From the point of view of the groupoid $\CT$, we are introducing a
countable family of new objects:  for each of the countably many
possible choices of the fiducial paths $E_\pm$, we add two ``limit
triangulations'' $T_{\pm \infty}$. Relative to some fixed choice of
$E_\pm$ we can label the limit triangulations more concretely as
$T_{\pm \infty}^{[k]}$ for $k \in \IZ$. There is a morphism from
$T_m$ to any $T_{\pm \infty}^{[k]}$, representing the limit of an
infinite sequence of flips. There is a relatively trivial morphism
which changes $k$: at the level of coordinate systems, it leaves
$\CX_A$ invariant and changes $\CX_B$ by $(\xi_+)^{k'-k}$.  We also
introduce more interesting morphisms, which we call ``\juggles'':
there is one \juggle morphism from any $T^{[k]}_{+\infty}$ to any
$T^{[k']}_{-\infty}$, or vice versa. Morally speaking, introducing
the \juggles ``closes a loop'' in the groupoid of triangulations:
starting from any $T_m$ we can imagine flipping infinitely many
times to reach some $T_{+\infty}$, then \juggleing to reach some
$T_{-\infty}$, then flipping again infinitely many times to get back
to $T_m$.  We will abuse notation, using the symbol $\CT$ for the
groupoid of decorated triangulations augmented by these limit
triangulations and extra morphisms.

Now how does the \juggle act on coordinate systems, i.e. what is its
image under our functor $\CT \to \CC$? For the $A$ coordinates this
is simple:  from \eqref{eq:A-positive} and \eqref{eq:A-negative} we
just have
\begin{equation} \label{eq:A-relation}
\CX^{T_{-\infty}}_A = (\CX^{T_{\infty}}_A)^{-1}.
\end{equation}
What about the $B$ coordinates?
The relation here depends on the choice of paths $E_\pm$ we make
in defining $T_{+\infty}$ and $T_{-\infty}$.  Suppose that we choose the same $E_\pm$ for both.
Then a simple linear algebra calculation shows that
\begin{equation}
\frac{(s \wedge Ms)(s' \wedge Ms')}{(s \wedge P_+ s') (s \wedge P_- s')} = - (\xi_+ - \xi_-)^2.
\end{equation}
Combining this with the definitions \eqref{eq:B-positive}, \eqref{eq:B-negative} we find
\begin{equation} \label{eq:B-relation}
\CX^{T_{-\infty}}_B = (\CX^{T_{+\infty}}_B)^{-1} (\xi_+ - \xi_-)^{-4}.
\end{equation}

\section{The WKB triangulation}\label{sec:WKB-triangulation}

In the last section we reviewed some basic properties of the Fock-Goncharov coordinates $\CX^T_E$.
Now how can these be related to the coordinates $\CX^\vartheta_\gamma$ we want?
Upon trying to relate the two, an issue immediately presents itself:
$\CX^T_E$ depends on the triangulation $T$, while $\CX^\vartheta_\gamma$ does not depend on any
triangulation, but depends instead on the angle $\vartheta$.  So if we want to identify the two
we need a way of choosing a canonical triangulation $T$ depending on $\vartheta$.
We are not free to make this choice arbitrarily:  ultimately we want to engineer $\CX^\vartheta_\gamma$
to obey the properties we listed in Section \ref{sec:review}.  In particular, we want to control the
asymptotic behavior of $\CX_\gamma^\vartheta$ in the limit $\zeta \to 0$.

How can these asymptotics be determined?  The basic idea is very simple.  As $\zeta \to 0$, $\CA$ becomes
dominated by its leading term $R \varphi / \zeta$.  At any point of $C$,
$\varphi$ has two eigenvalues $\pm \lambda$, where
$\lambda$ is a multivalued one-form on $C$, or a
single-valued one on the spectral curve $\Sigma$.
The ``WKB approximation'' roughly says that, in a gauge where
\begin{equation}
\varphi = \begin{pmatrix} \lambda & 0 \\ 0 & -\lambda \end{pmatrix},
\end{equation}
there are two independent  approximate $\CA$-flat sections of the
form
\begin{equation} \label{eq:wkb-rough}
\psi^{(1)} \sim \begin{pmatrix} e^{- \frac{R}{\zeta} \int^z \lambda
} \\ 0 \end{pmatrix}, \quad \psi^{(2)} \sim \begin{pmatrix} 0 \\
e^{\frac{R}{\zeta} \int^z \lambda } \end{pmatrix}.
\end{equation}
So one might expect that computing the parallel transport in the
$\zeta \to 0$ limit will reduce to computing periods of the 1-form
$\lambda$ on $\Sigma$.

Working this idea out in detail, it turns out that not all
triangulations are created equal.  Indeed, if we fix the quadratic differential
$\lambda^2$ and choose an angle $\vartheta$, there is a unique ``WKB triangulation''
$T_\WKB(\vartheta, \lambda^2)$, for which the WKB approximation gives the correct
asymptotics when $\zeta$ lies in the half-plane $\IH_\vartheta$
centered on $e^{i \vartheta}$.  In this section we define this
triangulation and describe some of its basic properties.

Throughout this section we assume $\lambda^2$ to be held fixed and generic, meaning that it has
only simple zeroes.

\subsection{WKB curves}

Fix any $\vartheta \in \IR / 2 \pi \IZ$.  We define a \ti{WKB curve
with angle $\vartheta$} to be a curve in $C$, with tangent
vector $\partial_t$, such that
\begin{equation} \label{eq:wkb-curve}
\lambda \cdot \partial_t \in e^{i \vartheta} \IR^\times
\end{equation}
everywhere along the curve.  ($\lambda$
is defined on $C$ only up to a sign, but that ambiguity is
immaterial for this definition.)
These curves define the \ti{WKB foliation with angle $\vartheta$}.

(Note that the parameter $t$ in \eqref{eq:wkb-curve} has nothing to
do with the coordinate $t$ used in Sections \ref{sec:physics} and
\ref{section:Superconformal}. Moreover, when we speak of
``exponential growth'' of sections along WKB curves, we are choosing
a parametrization in which $\lambda \cdot \partial_t = \pm e^{i
\vartheta}$. )

\subsection{Local behavior of WKB curves} \label{sec:wkb-local}

Around a generic point of $C$, the local behavior of the WKB foliation is trivial:  in terms of the local
coordinate $w = \int \lambda$, it is just the foliation by straight lines $\Im e^{-i \vartheta} w = const$.

Now let us consider the behavior near one of the singular
points $\CP$.  We choose coordinates so that it is at $z = 0$, and
take $\rho = 2m \sigma^3$, $m \in \IC$, in
\eqref{eq:reg-sing-bc-1}. Fixing a choice of branch for
$\lambda$, near $z = 0$ we have approximately $\lambda =
\frac{m dz}{z}$.

Let
\begin{equation} \label{eq:defeps}
\varepsilon(\vartheta) = \sgn \left(\re e^{-i \vartheta} m \right)
\end{equation}
(with the convention that $\sgn 0 = 0$). If $\varepsilon(\vartheta)
= 0$ then there are no WKB curves going into the singularity
(instead the WKB curves nearby are circles running around it); we
assume for a while that we are not in this degenerate situation, but
we will return to this point in Section \ref{sec:no-flavor-jump}. So
long as $\varepsilon(\vartheta) \neq 0$ the most general WKB curve
is a logarithmic spiral, which we may parameterize so that
\begin{equation} \label{eq:log-spiral}
z(t) = z_0 e^{- \varepsilon(\vartheta) \frac{e^{i \vartheta}}{m} t}.
\end{equation}
This curve goes into the singularity as $t \to \infty$. Note that
$-\varepsilon(\vartheta)$ is the sign of $e^{- i \vartheta} \lambda
\cdot \partial_t$.  In particular,
this sign is the same for all WKB curves going into the same
singularity.

\insfig{5in}{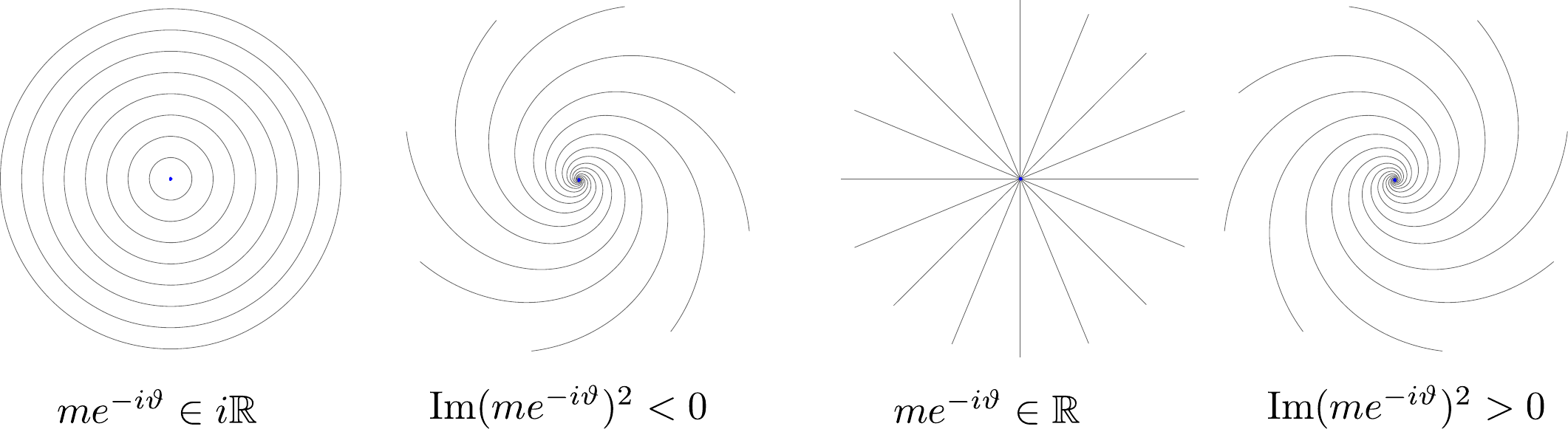}{Behavior of WKB curves near a
singularity.}

We will also be interested in the behavior of the WKB foliation
near a simple zero of $\lambda^2$, also known as a \ti{turning point}.
At such a point the foliation becomes singular.  Three separating WKB curves
emanate from the turning point, as shown in Figure \ref{fig:foliation-zero}.
\insfig{2.05in}{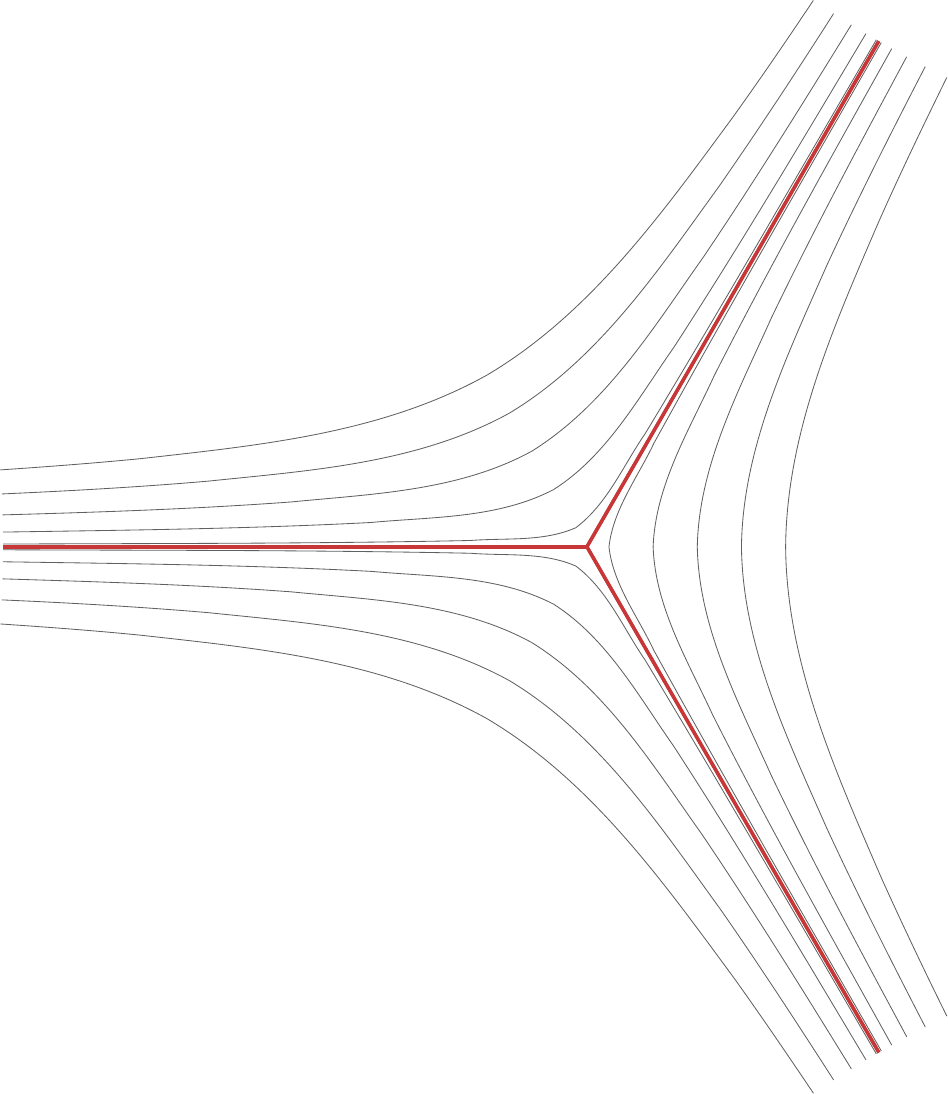}{Behavior of the WKB foliation near a
turning point.  Generic WKB curves are shown as thin black curves,
separating ones as thicker red curves.}
We will discuss these curves further below.

\subsection{Global behavior of WKB
curves}\label{subsec:GlobalBehaviorWKB}

Now let us discuss the global behavior of the WKB foliation.
For a general foliation of a Riemann surface, the behavior of the leaves can be quite wild.
Fortunately, the WKB foliation is much easier to control.
We divide up the WKB curves as follows:
\begin{itemize}
\item A \ti{generic} WKB curve is asymptotic in both directions to a singular point (possibly the same one).
\item A \ti{separating} WKB curve is asymptotic in one direction to a turning point and in the other direction to a singular point.
\item A \ti{finite} WKB curve is asymptotic in both directions to a turning point (possibly the same one), or closed.
\item A \ti{divergent} WKB curve is not closed and does not approach any limit in one or both directions.
\end{itemize}
In what follows, the values of $\vartheta$ for which a finite WKB curve exists
will play a very special role.  They correspond to places where a BPS state appears and the WKB triangulation jumps.
So for the moment let us assume that there are no finite WKB curves.
In that case, as we now show, there are also no divergent curves.

Suppose (for a contradiction) that $\alpha$ is a divergent WKB
curve.  In \cite{MR743423} it is shown that such a curve is actually
recurrent: defining $A$ to be the closure of $\alpha$, $\alpha$
comes arbitrarily close to every point of $A$ infinitely many times.
Moreover, the interior of $A$ is a nonempty connected domain, and
the boundary of $A$ consists of finite WKB curves.  Since we have
stipulated that there are no finite WKB curves, the only possibility
is that the boundary of $A$ is empty:  in other words $A$ fills up
$C$. In particular, $\alpha$ comes arbitrarily close to the singular
points. But recall from Section \ref{sec:small-flat} that the nearby
WKB curves are logarithmic spirals going into the singular point
(unless $e^{- i \vartheta} m \in i\IR$, in which case they are
closed BPS curves, but we have already excluded this case).  So
$\alpha$ cannot pass too close to the singular point, else it would
fall in.  This gives the desired contradiction. (Note that this
argument depends crucially on the fact that we assumed $\nsing \ge 1$,
i.e. we have some singular points on $C$.  If $\lambda^2$ were
regular the WKB foliation could be considerably more
complicated.\footnote{This is one indication that when $C$ has no
defect operators the spectrum of BPS states is qualitatively
different.  Indeed, in the absence of singularities on $C$ we would
expect that there are arbitrarily long closed WKB curves.
Correspondingly, when there are no defect operators we expect that
there is an infinite spectrum of BPS vectormultiplets with
arbitrarily large masses.}) Hence our foliation consists only of
generic and separating WKB curves. The generic WKB curves fall into
1-parameter families; each family sweeps out a ``cell'' bounded by a
union of separating WKB curves \cite{MR743423}.  In the $w$
coordinate such a cell just looks like a   strip at angle
$\vartheta$. Again using the assumption that there are no finite WKB
curves, the two boundaries of this strip each can contain only a
single turning point. One subtlety is that these two turning points
may actually be identified.  It follows that there are two possible
topologies for the closure of the cell, shown in Figure
\ref{fig:diamond-disc}.\footnote{If we had only $\nsing=2$ singular
points and $C = \IC\IP^1$, there would have been a third possible
kind of cell, consisting of the whole of $C$; but we excluded this
case by considering only $\nsing \ge 3$.}

\insfig{4in}{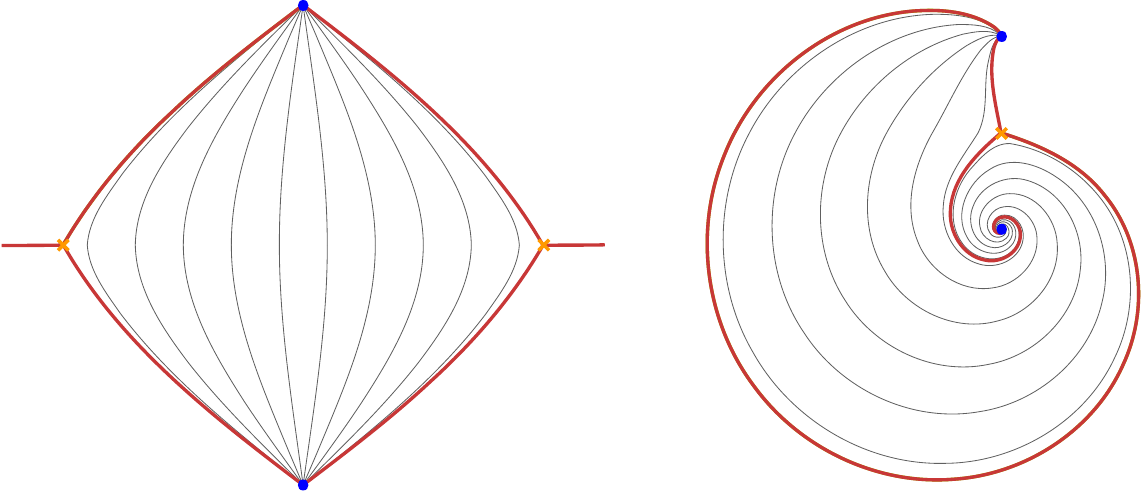}{The two possible cells
swept out by generic WKB curves:  a diamond and a disc. The disc can
be considered as a degenerate diamond, obtained by identifying the
two turning points.  Generic WKB curves are indicated by thin black curves,
separatrices by thick red ones.}

\subsection{Defining the WKB triangulation}\label{subsec:DefiningWKB-Triang}

Now we will define the WKB triangulation.  It is roughly ``dual'' to the cell decomposition we have just seen.

We assume that all turning points are \ti{simple} zeroes of $\lambda^2$.
Choose one generic WKB curve $E_i$ in each family.  The $E_i$ divide $C$ into
faces.  Each face must contain a turning point (this follows from the fact that the edges belong to different cells and
the boundary of each cell is made up of separating WKB curves.)  So let us focus on a single turning point $\CT$.
The WKB curves running near $\CT$ are the edges of a face $F$.
Looking at Figure \ref{fig:foliation-zero}, we see that $\CT$ is on the boundary of at most three cells.
If there are exactly three then $F$ is a triangle.  It can also happen that two of the three classes
of curve pictured in Figure \ref{fig:foliation-zero} actually belong to a single cell.  In this case $F$
is a degenerate triangle.

\insfig{4.5in}{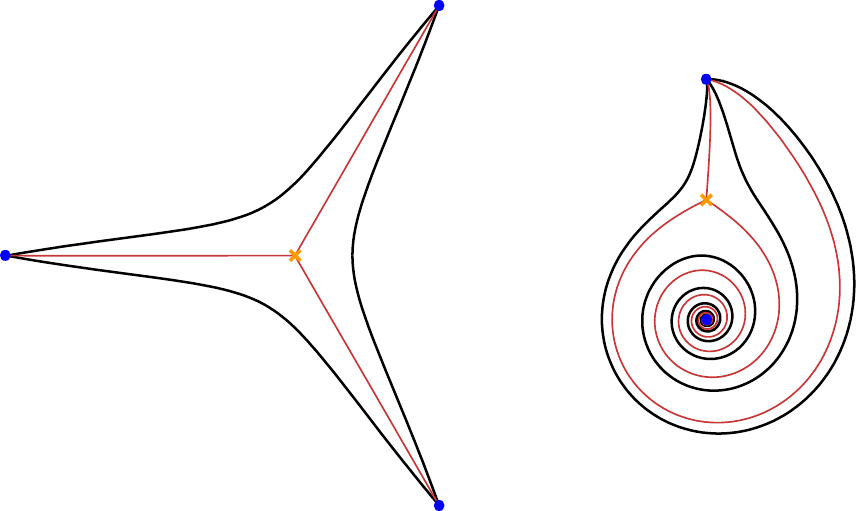}{The two types of faces occurring in the WKB triangulation.  On the
left is an honest triangle.  On the right is a degenerate triangle, with only two distinct edges and two vertices
(compare Figure \protect\ref{fig:degenerate-face}).  Each face contains a single turning point.
Edges of the WKB triangulation are indicated by thick black curves, separatrices by thin red ones.}

We conclude that the $E_i$ are the edges of a triangulation of
$C$, with faces of two types, pictured in Figure
\ref{fig:optimal-faces}. The topology of the triangulation does not
depend on which generic $E_i$ we choose within a cell, and therefore
we identify two triangulations that differ by such a choice.
 That is, by
``triangulation'' we really mean an isotopy class of triangulations.
With this identification understood, we have defined a triangulation
of $C$.

\subsection{The WKB decoration: small flat sections} \label{sec:small-flat}

So far we have used the WKB foliation to define an undecorated triangulation.
To construct $\CX^\vartheta_\gamma$, we also need some canonical way of
choosing a decoration.

We first consider the behavior of the flat sections around one
of the singular points, which for convenience we put at $z =
0$. The behavior of the connection around this point is given
by \eqref{eq:A-sing-reg}, determined by the residues $\rho$ and
$\alpha$ of $\varphi$ and $A$ respectively.  We take $\rho = 2
m \sigma^3$, where $m \in \IC$, and $\alpha = - 2 i m^{(3)}
\sigma^3$, where $m^{(3)} \in \IR$.

Then by the standard Frobenius analysis of the behavior around a
regular singular point (slightly modified here since we consider a
$C^\infty$ connection rather than a holomorphic one) there are two
flat sections of the form
\begin{align}
s^{(1)} &= z^{-R \zeta^{-1} m + m^{(3)}} \bar{z}^{-R \zeta \bar{m} -  m^{(3)}} \begin{pmatrix} 1 + O(\abs{z}) \\ O(\abs{z}) \end{pmatrix}, \\
s^{(2)} &= z^{R \zeta^{-1} m - m^{(3)}} \bar{z}^{R \zeta \bar{m} + m^{(3)}} \begin{pmatrix} O(\abs{z}) \\ 1 + O(\abs{z}) \end{pmatrix}.
\end{align}
They have clockwise monodromy eigenvalues $\mu^{(1)} = e^{2 \pi i
\nu}$, $\mu^{(2)} = e^{-2 \pi i \nu}$, where
\begin{equation}
\nu = R \zeta^{-1} m - 2 m^{(3)} - R \zeta \bar{m}.
\end{equation}
Let us evaluate their behavior along a WKB curve going into the
singularity. Using \eqref{eq:log-spiral} we obtain
\begin{equation}
s^{(1)} \sim \exp \left[ \varepsilon(\vartheta) \left\{ R (e^{i
\vartheta}\zeta^{-1}+ e^{-i\vartheta}\zeta) + m^{(3)}
\left(\frac{e^{-i\vartheta}}{\bar m} - \frac{e^{i\vartheta}}{
m}\right) \right\} t \right]
\end{equation}
where $\varepsilon(\vartheta)$ was defined in \eqref{eq:defeps}.
The piece multiplying $m^{(3)}$ is a pure phase and does not affect
the norm of $s^{(1)}$.  Looking at the remaining piece we see that
$\norm{s^{(1)}}$ is exponentially small as $t \to \infty$ if $\varepsilon(\vartheta)\,
\Re (e^{i \vartheta} \zeta^{-1} + e^{-i\vartheta} \zeta)<0$.
Similarly, $\norm{s^{(2)}}$ is exponentially small as $t \to \infty$ if
$\varepsilon(\vartheta)\,\Re
(e^{i \vartheta} \zeta^{-1} + e^{-i\vartheta} \zeta)>0$.

Thus, if $\zeta \in \IH_{\vartheta}$, the eigensection with clockwise
monodromy
\begin{equation} \label{eq:monodromy-formula}
\mu = e^{(- \varepsilon(\vartheta)) 2 \pi i \nu} = \exp \left[(- \varepsilon(\vartheta)) 2 \pi i \left( R \zeta^{-1} m - 2 m^{(3)} - R \zeta \bar{m} \right) \right]
\end{equation}
is asymptotically smaller in norm than all other flat sections along
a WKB curve going into the singularity.  We call it the ``small flat
section.'' We will see in Section \ref{sec:asymptotics} that if
$\zeta \to 0$ while remaining in $\IH_{\vartheta}$ the WKB
approximation gives us good control over the evolution of this
section along a WKB curve.  With this motivation, we choose this
small flat section as our canonical decoration at the singularity.

We have now finished defining a decorated triangulation
$T_\WKB(\vartheta, \lambda^2)$ for each $(\vartheta, \lambda^2)$. We
call it the \ti{WKB triangulation}.  To lighten notation we will
sometimes write it as $T_\WKB(\vartheta)$ or
even just $T_\WKB$.  Recalling that the quadratic differentials $\lambda^2$
correspond to points $u \in \CB$ we also sometimes write $T_\WKB(\vartheta, u)$.

\subsection{Jumps of the WKB
triangulation}\label{subsec:Jumps-WKB-triang}

As we vary the parameter $\vartheta$ the WKB foliation changes, and
correspondingly $T_\WKB(\vartheta)$ changes.  For generic
$\vartheta$, $T_\WKB(\vartheta)$ just changes by a continuous
homotopy of the edges, and hence the isotopy class of the
triangulation is constant. However, there are some special values
$\vartheta = \vartheta_c$ at which $T_\WKB(\vartheta)$ jumps
discontinuously.

These jumps are of crucial importance for us:  they are related to
the existence of BPS states. Indeed, the topology of $T_\WKB$ is
completely determined by the behavior of the separating WKB curves,
and this behavior jumps exactly when a separatrix degenerates to
include a finite WKB curve.\footnote{Consider a separating WKB curve
starting from $z_0$. In the $w$-plane, where $w= \int_{z_0}^z
\lambda$ is defined in some neighborhood of the separating curve,
the curve is a straight line. It varies continuously with
$\vartheta$.  This only fails when there is no open neighborhood of
the separating WKB curve that does not contain another turning
point. That is, it only fails when the separating WKB curve contains
a finite WKB curve.}  Now comparing \eqref{eq:BPS-WKB-curve} with
\eqref{eq:wkb-curve} we see that BPS strings and WKB curves obey
exactly the same equation; a finite WKB curve is just the same thing
as a BPS string with finite total mass. Thus, the values
$\vartheta_c$ at which the WKB foliation changes topology are the
phases of BPS states.

We now examine the three kinds of topology changes
of $T_\WKB(\vartheta)$ which can occur as $\vartheta$ varies.

\subsubsection{A jump from a BPS hypermultiplet} \label{sec:hyper-jump}

The fundamental example of a topology change occurs at  $\vartheta =
\vartheta_c$ for which a WKB curve appears connecting two turning
points.  As we described in Section \ref{sec:bps-strings}, this
finite WKB curve represents a BPS hypermultiplet of the $d=4$
theory.

As $\vartheta$ crosses $\vartheta_c$, the WKB foliation undergoes
a topology change; see Figure \ref{fig:hypermultiplet-flip}.
In particular, the generic WKB curve running from northwest to
southeast is replaced by one running from northeast
to southwest.  Since this generic WKB curve represents one of the
edges $E$ of $T_\WKB$, we see that this triangulation
undergoes a flip as $\vartheta$ crosses $\vartheta_c$.
\insfig{6in}{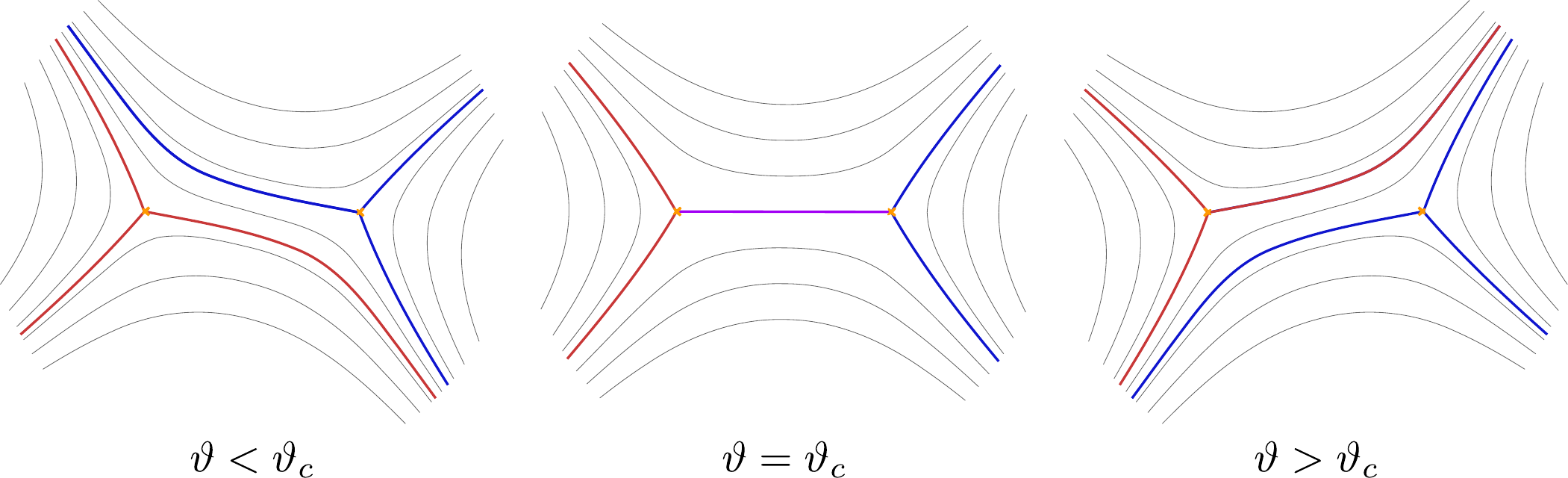}{The jump of the WKB foliation as
$\vartheta$ crosses a critical $\vartheta_c$ at which a finite WKB
curve appears, corresponding to a BPS hypermultiplet.}

\subsubsection{A jump when $m e^{-i \vartheta} \in i\IR$} \label{sec:flavor-jump}

A more intricate jumping behavior occurs at phases $\vartheta = \vartheta_c$ for which closed WKB curves exist.  As we described in Section \ref{sec:bps-strings}, these closed WKB curves
always appear in one-parameter families.

The simplest possibility is a family of closed WKB curves surrounding a single singular point.
As we already saw
in Section \ref{sec:wkb-local}, such a family appears whenever
$m e^{-i \vartheta} \in i\IR$.  The behavior of the WKB foliation near this value
of $\vartheta$ is shown in Figure \ref{fig:spiral-jump}.

\insfig{5.6in}{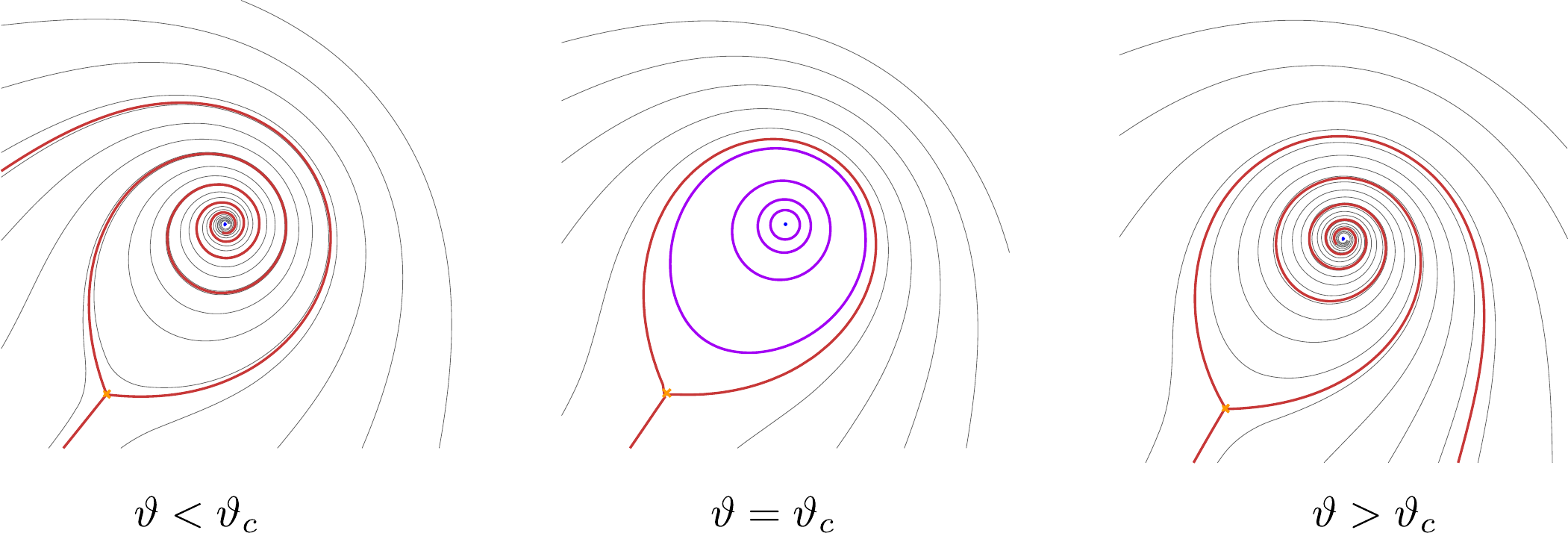}{The jump of the WKB foliation as $\vartheta$
crosses a critical $\vartheta_c$ at which a family of closed WKB curves appears surrounding a single singular point.}

In Section \ref{sec:small-flat} we defined the decoration of the WKB triangulation,
which is determined by the sign $\varepsilon(\vartheta)$ of
$\Re (e^{- i \vartheta} m)$.  This sign changes
sign at $\vartheta_c$.  It follows that at $\vartheta_c$ the WKB triangulation undergoes a pop.

\subsubsection{A jump from a BPS vectormultiplet} \label{sec:vector-jump}

Another possibility is a family of closed WKB curves which does not contract onto a single
singular point.
This corresponds to a BPS vectormultiplet which carries some nonzero gauge charge.

Generically, when such a vectormultiplet appears it is accompanied by two infinite
families of hypermultiplets:
if we look at a narrow enough interval $\vartheta_- < \vartheta < \vartheta_+$
containing $\vartheta_c$, then the WKB triangulation varies smoothly with $\vartheta$ except in one annular region of $C$, and
in this annular region the WKB triangulation undergoes an infinite sequence of flips.  The situation is precisely
the one we considered in Section \ref{sec:limit-triangulation}.
As $\vartheta$ decreases starting from $\vartheta_+$,
$T_\WKB(\vartheta, \lambda^2)$ runs through a sequence of triangulations
$T_m$ with $m > 0$, and $m \to \infty$ as $\vartheta \to \vartheta_c^+$.
On the other hand, if we instead start from $\vartheta_-$,
then as $\vartheta$ increases $T_\WKB(\vartheta, \lambda^2)$ runs through all of
the $T_{m}$ with $m<0$, and $m \to -\infty$ as $\vartheta \to \vartheta_c^-$.

\insfig{5.4in}{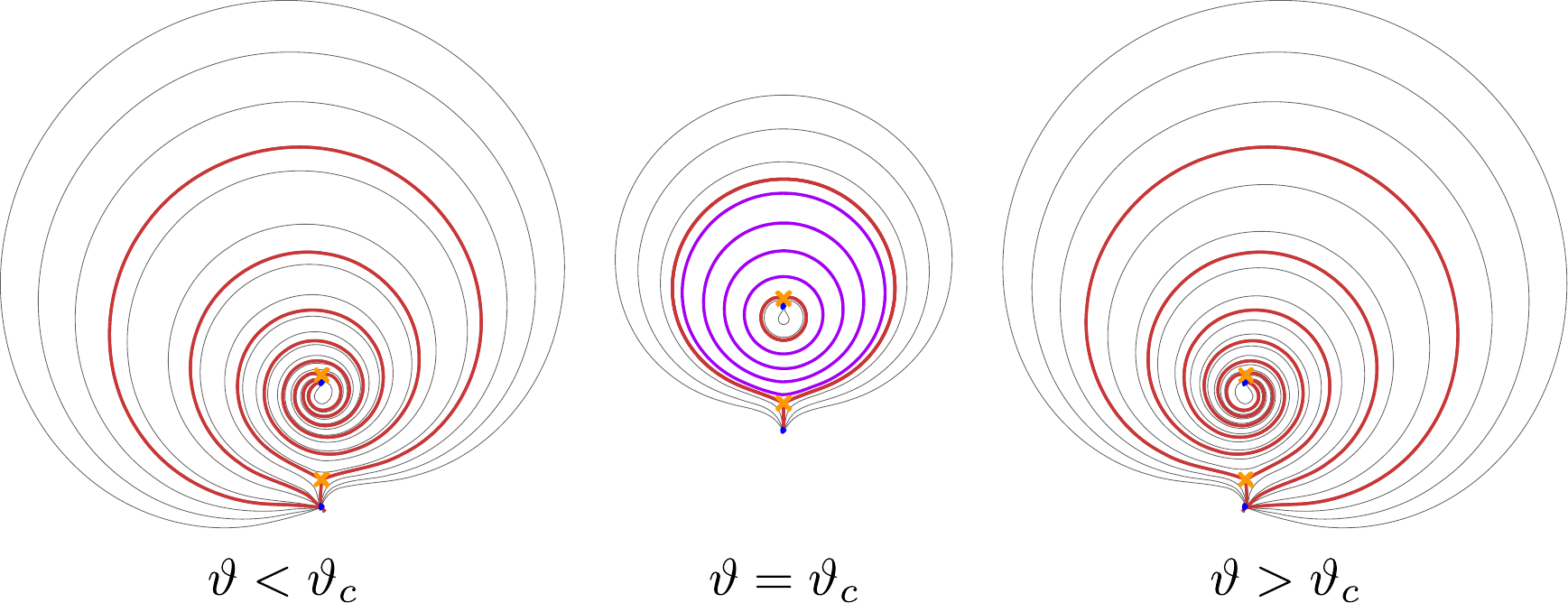}{An annular region of the WKB foliation, near a critical $\vartheta = \vartheta_c$ at which
a family of closed WKB curves representing a BPS vectormultiplet appears.}

Roughly speaking, after $m$ flips a typical WKB curve runs around the annulus $m$ times,
either clockwise or counterclockwise depending whether $\vartheta > \vartheta_c$ or $\vartheta < \vartheta_c$.
Exactly at $\vartheta = \vartheta_c$ the WKB curves become closed curves foliating an open region inside the annulus.
See Figure \ref{fig:vectormultiplet-twist}.

\subsubsection{Other degenerations are nongeneric}

Let us now briefly justify our
claim that for \emph{generic} quadratic differentials $\lambda^2$ only the above
three degenerations of the WKB triangulation occur as we vary $\vartheta$.
(For non-generic moduli, other
degenerations of $T_\WKB(\vartheta, \lambda^2)$ can and do occur. We will see an
example in Section \ref{sec:nf1} below.)

A critical value $\vartheta_c$ occurs when a
separating WKB curve degenerates to include a finite WKB curve,
which begins and ends on a turning point. There are only two
possibilities:  either these two turning points are the same or they are
different.  If the two turning points are the different, then the finite
WKB curve has no moduli, and the behavior near $\vartheta_c$
must be the hypermultiplet degeneration of Section
\ref{sec:hyper-jump}.  If the two turning points are the same, then
the WKB curve going through an infinitesimally displaced point (in
the cell of the foliation bounded locally by the finite WKB curve)
must be closed. These closed WKB curves come in a
one-parameter family. This family could terminate when the closed
WKB curve hits a singular point, shrinks to zero size, or hits a turning point.  Examining the
local behavior near singular points (for generic $m$) we see that
the family of closed WKB curves cannot end by hitting one.  If the closed WKB curves
shrink to zero size then
we are in the situation of Section \ref{sec:flavor-jump}.  The last possibility
is for the closed WKB curve to hit another turning point.  In the generic case
it will hit exactly one turning point.
Thus, our family describes precisely the situation encountered in
the degeneration of Section \ref{sec:vector-jump}.

\section{The canonical coordinates} \label{sec:darboux-coordinates}

In this section, we will finally define the functions
$\CX_\gamma$ and check that they have all of the
properties we promised in Section \ref{sec:review}.

\subsection{Labeling by homology} \label{sec:homology}

As we saw in the last section, there is a canonical choice of
triangulation $T_\WKB$ determined by the WKB foliation.  We will use
this triangulation. However, there is one more crucial issue to deal
with before we can identify $\CX_\gamma$ with $\CX^{T_\WKB}_E$.  The
coordinates $\CX^{T_\WKB}_E$ are labeled by the edges $E$ of
$T_\WKB$, but we want our $\CX_\gamma$ to be labeled by classes
$\gamma \in \hat\Gamma$. So we need to specify a map from the set of
edges of $T_\WKB$ to $\hat\Gamma$ and we will do so by defining
a homology class $\gamma_E \in H_1(\Sigma;\IZ)$ associated to each
edge $E$ of $T_\WKB$.\footnote{Recall from Section
\ref{subsubsec:ChargeLattices} that in the ``$K = 2$'' case we are
considering, $\hat \Gamma$ is simply the sublattice (not
subquotient!) of $H_1(\Sigma;\IZ)$ which is odd under the deck
transformation exchanging the two sheets.}

Since $T_\WKB = T_\WKB(\vartheta)$ depends on $\vartheta$, we will
sometimes write this map as $E \mapsto \gamma_E^\vartheta$.
Throughout this section we assume $\vartheta$ is generic.

\subsubsection{For ordinary edges}

\insfig{3.5in}{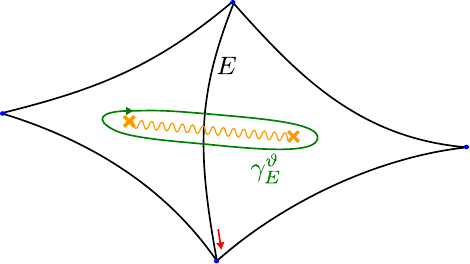}{The construction of $\gamma^\vartheta_E
\in H_1(\Sigma; \IZ)$.  $\Sigma$ is a double cover of the base
$C$. To draw the picture we choose definite branch cuts
for this double cover.  Having done so, we can speak about the
two sheets of the covering.  We call the two sheets ``upper''
and ``lower'', arbitrarily.  A solid green line indicates a
curve on the upper sheet. A red arrow next to a segment of an
edge indicates the orientation of the lift of this segment to
the upper sheet.  (Specifying this orientation for one segment
determines it for all segments.)} Given an edge $E$ of the WKB
triangulation, the quadrilateral $Q_E$ contains two turning
points.  Draw a loop in $Q_E$ which surrounds these two turning
points. We aim to define a connected lift of this loop to the
spectral curve to obtain a class $\gamma^\vartheta_E \in
H_1(\Sigma; \IZ)$. There are two ambiguous choices here:  the
orientation of the loop and which of the two sheets we lift it
to.  Clearly, reversing the orientation changes the sign of
$\gamma^\vartheta_E$.  Moreover, one can easily see that either
connected lift of the loop is odd under the deck transformation
exchanging the two sheets. So these two ambiguous choices just
affect the overall sign of $\gamma^\vartheta_E$.

We fix this sign as follows.  On $\Sigma$, $\lambda$ is a
single-valued 1-form.  Thus, the two possible lifts $\hat{E}$ of any
edge $E$ of the WKB triangulation each have an orientation, defined
by the condition that the positively oriented tangent vector
$\partial_t$ to $\hat{E}$ obeys $e^{- i \vartheta} \lambda \cdot
\partial_t > 0$. Note that $\hat E$ defines a cycle in the relative
homology group $H_1(\Sigma, \{ \CP_i \};\IZ)$. This has a well-defined pairing with $H_1(\Sigma;\IZ)$ and
we demand that the intersection $\langle \gamma^\vartheta_E, \hat{E}
\rangle = 1$. (This is independent of which of the two possible
lifts $\hat{E}$ we chose.) See Figure \ref{fig:loop}.\footnote{Our
convention is that in the standard orientation for the $xy$ plane,
$\langle x\mathrm{-axis}, y\mathrm{-axis} \rangle = +1$.}

The intersection pairing on these cycles agrees with the pairing on
edges we defined above. That is, for all $E,E'$:
\begin{equation}\label{eq:intersection-pairing}
\inprod{\gamma^\vartheta_E, \gamma^\vartheta_{E'}} = \inprod{E, E'},
\end{equation}
as illustrated in Figure
\ref{fig:loop-intersection}.
\insfig{3.5in}{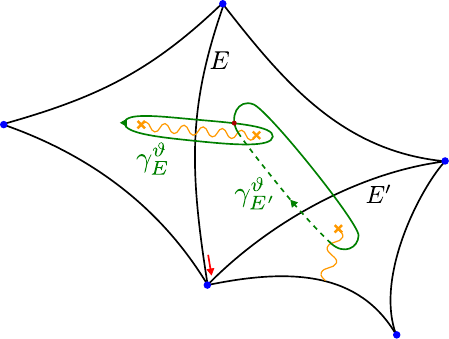}{A
pair of edges with $\inprod{E, E'} = 1$ also have
$\inprod{\gamma^\vartheta_E, \gamma^\vartheta_{E'}} = 1$.}

\subsubsection{For degenerate edges} \label{sec:homology-degenerate}

We can also consider faces of the WKB triangulation which are degenerate in the sense of Section \ref{sec:degenerate}.
In this case our rule for defining the cycles $\gamma_E$,
$\gamma_{E'}$ cannot be straightforwardly applied.  To get around this difficulty we pass
to a covering surface, just as we did to define the Fock-Goncharov coordinates in Section \ref{sec:degenerate}.
The edges $E$, $E'$ have multiple preimages
on the covering surface.  We choose any preimages $\tilde{E}$,
$\tilde{E}'$ and then construct $\gamma^\vartheta_{\tilde{E}}$ and
$\gamma^\vartheta_{\tilde{E}'}$ as above.  See Figure \ref{fig:yin-yang}.
These cycles then descend to the desired $\gamma^\vartheta_E$
and $\gamma^\vartheta_{E'}$, pictured in Figure \ref{fig:degenerate-loops},
which are independent of the choice of preimage.
\instwofigs{2.5in}{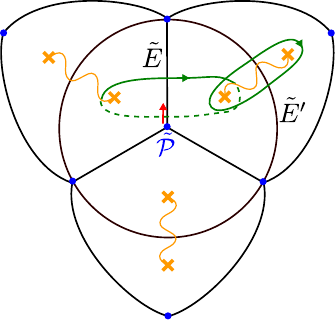}{The covering
surface which we use to resolve a degenerate face.}{2.25in}{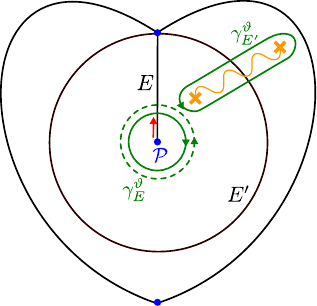}{Cycles $\gamma^\vartheta_{E'}$ and
$\gamma^\vartheta_E$ associated to a degenerate face.}

\subsubsection{Cycles around singular points} \label{sec:punctures}

To every singular point $\CP_i$ there is a corresponding privileged cycle $C_i$ on $\Sigma$,
consisting of two small loops running around $\CP_i$ in opposite directions on the two sheets,
oriented so that $\inprod{C_i, \hat{E}} = 1$ when $E$ is any edge incident on $\CP_i$ (recall that the lifted
edges $\hat{E}$ have natural orientations).  See Figure \ref{fig:C-cycle}.
\insfig{1.4in}{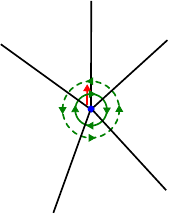}{The cycle $C_i$ associated to the puncture $\CP_i$.}
The cycle $C_i$ has a simple expression in terms of the $\gamma_E$:
\begin{equation}
C_i = \sum_{E \textrm{ meeting } \CP_i} \gamma_E.
\end{equation}
This fact is illustrated in Figure \ref{fig:vertex-sum} for $\CP_i$ a generic vertex (with
two or more edges incident on it).
\insfig{3.7in}{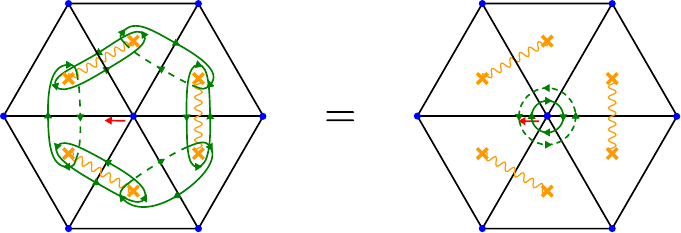}{The sum of $\gamma_E$ over all $E$ meeting the vertex $\CP_i$ gives $C_i$.}
If $\CP_i$ is a degenerate vertex (with only one edge $E$ incident on it) then
Figure \ref{fig:degenerate-loops} shows directly that $C_i = \gamma_E$.

\subsubsection{Lattice generated by $\{ \gamma_E\}$ } \label{sec:lattice-of-charges}

The   vectors $\{ \gamma_E\}$  for $E$ running over the edges of
$T_\WKB$ generate a sublattice of $\hat \Gamma$, and we now show
that they in fact generate the entire lattice $\hat \Gamma$.

First, we show that the lattice generated by $\{ \gamma_E\}$  has
the correct rank. When $C$ has genus $g_C$ the quadratic
differential $\lambda^2$ has $4g_C-4+2\nsing$ zeroes. Since these are
simple zeroes,   by the Riemann-Hurwitz formula the genus of
$\bar\Sigma$ is $g_{\bar \Sigma} = 4 g_C +\nsing-3$. Using the Lefshetz fixed point
formula the rank of the anti-invariant sublattice of
$H_1(\bar\Sigma;\IZ)$ is $6g_C +2\nsing-6$.  Introducing the punctures we find
that the rank of $\hat\Gamma$, the anti-invariant part of $H_1(\Sigma; \IZ)$,
is $6g_C + 3\nsing - 6$, precisely correct for
the lattice generated by $\{ \gamma_E\}$. (Recall equation
\eqref{eq:counts}.) Hence the rational vector space $\hat \Gamma
\otimes \IQ$ is generated by $\{ \gamma_E\}$. It follows that any
primitive vector $V \in \hat \Gamma$ can be written as $\sum c_E
\gamma_E$ with $c_E \in \IQ$. Now, the relative homology
$H_1(\Sigma, \{ \CP_i \};\IZ)$ is Poincar\'e dual to
$H_1(\Sigma;\IZ)$ so $\langle V, \hat E' \rangle \in \IZ$. Recall
that  $\langle \gamma_E, \hat E' \rangle = \delta_{E,E'}$ so we see
that the coefficients $c_E$ must in fact be integral.

\subsection{Defining the canonical coordinates}

We define $\CX^{\vartheta,u_0}_\gamma$ by the properties
\begin{equation} \label{eq:canonical-darboux}
\CX^{\vartheta,u_0}_{\gamma^\vartheta_E} :=
\CX^{T_\WKB(\vartheta,u_0)}_E
\end{equation}
for all edges $E$ in the triangulation $T_{\WKB}(\vartheta,u_0)$ and
\begin{equation} \label{eq:CX-product}
\CX^{\vartheta,u_0}_{\gamma + \gamma'} = \CX^{\vartheta,u_0}_\gamma
\CX^{\vartheta,u_0}_{\gamma'}.
\end{equation}
for all $\gamma, \gamma'$.  Since the $\gamma^\vartheta_E$ are a
basis of $\hat\Gamma$, these two properties define
$\CX^\vartheta_\gamma$.

We will mostly emphasize the $\vartheta$-dependence of these
functions at fixed $u_0$, and hence we almost always denote them as
$\CX_\gamma^\vartheta$.

\subsection{Some easy properties}

Let us note a few easy properties of these canonical coordinates.
First, using \eqref{eq:fg-poisson}, \eqref{eq:intersection-pairing},
and \eqref{eq:CX-product} we see that in terms of the homology
labeling the Poisson structure is simply
\begin{equation} \label{eq:homology-poisson}
\{ \CX^\vartheta_\gamma, \CX^\vartheta_{\gamma'} \} = \inprod{\gamma, \gamma'}
\CX^\vartheta_{\gamma + \gamma'} \qquad\qquad \forall \gamma, \gamma' \in
\hat\Gamma.
\end{equation}

Second,
in \eqref{eq:monodromy-product}
we noted that the product of the $\CX^T_E$ for all $E$ meeting the singular point $\CP_i$
is $\mu_i^{2}$.  On the other hand we saw in Section \ref{sec:punctures} that
the sum of the $\gamma^\vartheta_E$ over all these $E$ is the cycle $C_i$.
Combining these two statements we arrive at the simple rule that
\begin{equation} \label{eq:monodromy}
\CX^\vartheta_{C_i} = \mu_i^{2}.
\end{equation}

Finally we want to establish the reality condition obeyed by the
$\CX^\vartheta_\gamma$. Note that by the definition
\eqref{eq:def-CA} of $\CA$ we have
\begin{equation}
\CA(\zeta) = - \overline{\CA(-1/\bar\zeta)}
\end{equation}
from which it follows that, if $s$ is a flat section for
$\CA(\zeta)$, then $\bar{s} / \norm{s}^2$ is a flat section for
$\CA(-1 / \bar\zeta)$. In particular, if $s_i$ is the small flat
section for $\CA(\zeta)$ at some singularity $\CP_i$ and at angle
$\vartheta$, then $\bar{s}_i / \norm{s_i}^2$ is the \ti{large} flat
section at $\CP_i$ for $\CA(-1/\bar\zeta)$ at angle $\vartheta$, or
equivalently, it is the small flat section at $\CP_i$ for
$\CA(-1/\bar\zeta)$ at angle $\vartheta + \pi$. Comparing the
definitions \eqref{eq:cross-ratio-wedges} at $\zeta$ and $-1 /
\bar\zeta$, we obtain directly
\begin{equation} \label{eq:real-edges}
\CX^{T_\WKB(\vartheta)}_E (\zeta) = \overline{\CX^{T_\WKB(\vartheta + \pi)}_E (- 1 / \bar\zeta)}.
\end{equation}
We also have as usual
\begin{equation} \label{eq:comp-theta}
\gamma^\vartheta_E = - \gamma^{\vartheta+\pi}_E.
\end{equation}
Combining \eqref{eq:real-edges} and \eqref{eq:comp-theta} gives the
desired reality condition
\begin{equation}
\CX^{\vartheta}_{\gamma} (\zeta) = \overline{\CX^{\vartheta+\pi}_{-\gamma} (- 1 / \bar\zeta)}.
\end{equation}

\subsection{Asymptotic behavior} \label{sec:asymptotics}

Now we come to the main motivation of our definition of the WKB triangulation and our labeling of the coordinates
by cycles in $H_1(\Sigma; \IZ)$.  We claim that, as $\zeta \to 0$ within the half-plane $\IH_\vartheta$,
the asymptotics of $\CX_\gamma$ are simply
\begin{equation} \label{eq:wkb-asymp}
\CX^\vartheta_\gamma \sim c_\gamma \exp(\zeta^{-1} \pi R Z_\gamma),
\end{equation}
where $c_\gamma$ is some   function on $\CM$  which is independent
of $\zeta$.

To obtain the asymptotics \eqref{eq:wkb-asymp} we use directly the
definition of $\CX_\gamma$. Suppose $\gamma = \gamma^\vartheta_E$
for some nondegenerate edge $E$ as in Figure
\ref{fig:quadrilateral}.
 Then we will apply the WKB approximation
for the parallel transport of the small flat sections (decorations)
along the edges. Choose a gauge along the edges such that
\begin{equation}
\varphi = \begin{pmatrix} \lambda & 0 \\ 0 & -\lambda \end{pmatrix},
\end{equation}
with the sign chosen so that $e^{-i \vartheta} \lambda \cdot \partial_t < 0$ for $\partial_t$
along $E_{12}$ oriented away from $z_1$, toward $z_2$.

Let $I_1(z)$ be an antiderivative of $\lambda$, defined on a neighborhood of the two edges
$E_{12}$ and $E_{41}$ (here by ``edge'' we mean an open curve, excluding its endpoints).
One would expect by the WKB approximation that
one can choose the flat section $s_1(z, \zeta)$ such that along $E_{12}$ we have the $\zeta \to 0$ asymptotics
\begin{equation}
s_1(z, \zeta) \sim c_1(z) \begin{pmatrix} e^{- \frac{R}{\zeta} I_{1}(z)} \\ 0 \end{pmatrix}.
\end{equation}
In Appendix \ref{app:wkb-error} we argue that this is indeed the
case.  The argument depends crucially on the fact that $E_{12}$ is a
WKB curve and on our choice of the WKB decoration:  the point is
that these choices ensure that the errors introduced in using the
WKB approximation to $s_1$ are exponentially smaller than $s_1$
itself, and that they remain so as we evolve along $E_{12}$.
Similarly let $I_2(z)$ be an antiderivative of $\lambda$, defined on
a neighborhood of the two edges $E_{12}$ and $E_{23}$. Again using
Appendix \ref{app:wkb-error} one can choose $s_2(z, \zeta)$ such
that along $E_{12}$ we have the $\zeta \to 0$ asymptotics
\begin{equation}
s_2(z, \zeta) \sim c_2(z) \begin{pmatrix} 0 \\ e^{\frac{R}{\zeta} I_2(z)} \end{pmatrix}.
\end{equation}
Evaluating both at some general point $z_{12}$ of $E_{12}$ we get
\begin{equation}
s_1 \wedge s_2 \sim c_{12} \exp \left( \frac{R}{\zeta} (I_2(z_{12}) - I_1(z_{12})) \right).
\end{equation}
Similar arguments with the indices $1234$ permuted give
\begin{align}
s_2 \wedge s_3 &\sim c_{23} \exp \left( -\frac{R}{\zeta} (I_3(z_{23}) - I_2(z_{23})) \right), \\
s_3 \wedge s_4 &\sim c_{34} \exp \left( \frac{R}{\zeta} (I_4(z_{34}) - I_3(z_{34})) \right), \\
s_4 \wedge s_1 &\sim c_{41} \exp \left( -\frac{R}{\zeta} (I_1(z_{41}) - I_4(z_{41})) \right),
\end{align}
where we have made the obvious extensions of our notation, and we have chosen $\lambda$ to be single-valued
in a neighborhood of the union of all the edges.  Combining these gives
\begin{multline} \label{eq:asint}
\CX^\vartheta_\gamma \sim c_\gamma
\exp \bigg( \frac{R}{\zeta} \big(I_1(z_{41})-I_1(z_{12})+I_2(z_{12})-I_2(z_{23})+ \\  I_3(z_{23})-I_3(z_{34})+I_4(z_{34})-I_4(z_{41}) \big) \bigg).
\end{multline}
If $\lambda$ were single-valued on the whole $Q_E$ we could have taken all $I_i$ to be the same function, in which case
they would cancel out in \eqref{eq:asint}.  Because of the two branch points in the interior of $Q_E$ this cancellation does not occur.
Instead, using $I_i(z') - I_i(z) = \int^{z'}_z \lambda$, \eqref{eq:asint} becomes
\begin{equation} \label{eq:wkb-period}
\CX^\vartheta_\gamma \sim c_\gamma \exp \left(\frac{R}{\zeta} \oint_\gamma \lambda \right).
\end{equation}
This is the key result:  the period integral over $\Sigma$ has emerged naturally from the WKB approximation!

If $\gamma = \gamma^\vartheta_E$ with $E$ a \ti{degenerate} edge we obtain the same result in a slightly different way.
In Section \ref{sec:homology-degenerate} this $\gamma$ was defined as the sum of two loops
running in opposite directions around the two lifts of the vertex.
Then the period of $\lambda$ is just determined by the residue of $\varphi$
at the vertex, which was fixed by our boundary conditions:  we obtain
\begin{equation} \label{eq:degenerate-period}
\oint_{\gamma} \lambda = -4 \pi i m \varepsilon(\vartheta).
\end{equation}
On the other hand we know from \eqref{eq:degenerate-monodromy} that $\CX^T_E = (\mu^T)^{2}$, with $\mu^T$
the eigenvalue of the clockwise monodromy around the degenerate vertex, in turn given by \eqref{eq:monodromy-formula}:
\begin{equation} \label{eq:degenerate-asymptotics}
\CX^T_E \sim \exp \left[- \varepsilon(\vartheta) 4 \pi i \left( \frac{R}{\zeta} m - 2 m^{(3)} \right) \right].
\end{equation}
Comparing \eqref{eq:degenerate-period} and
\eqref{eq:degenerate-asymptotics} we see that we obtained the
expected \eqref{eq:wkb-period} just as for nondegenerate edges.
(We also got a small bonus in this case:  the constant
$c_\gamma$ which gives the finite part of the asymptotics is
just $c_\gamma = e^{8 \pi i \varepsilon(\vartheta) m^{(3)}}$.)

Having established \eqref{eq:wkb-period} for all $\gamma^\vartheta_E$, it holds for
all $\gamma \in \hat\Gamma$ by multiplicativity.
Then finally recalling that $Z_\gamma = \frac{1}{\pi} \oint_\gamma \lambda$, \eqref{eq:wkb-period} becomes
the desired \eqref{eq:wkb-asymp}.

\subsection{From triangulations to KS symplectomorphisms}

Let us briefly take stock of where we are.  We have just seen from
the WKB analysis that the asymptotic properties of the coordinate
systems $\CX^{T_\WKB}$ are captured well by certain simple
properties of the corresponding decorated triangulation $T_\WKB$. To
any sector $\CV$ around the origin in the $\zeta$-plane, with
angular opening $\pi$ or smaller, we can associate a nice
subgroupoid ${\cal T}_\CV$ of decorated triangulations.  A
triangulation is in $\CT_\CV$ if homotopy representatives of the
edges can be picked with two basic properties:  a) for all $\zeta
\in \CV$, the decorations are exponentially small going along the
edges into the singularities, and b) along each edge,
$\lambda \cdot
\partial_t$ lies in $\CV$ (for some choice of the sign of $\lambda$).
The WKB analysis tells us that the functor defined in Section
\ref{sec:fg-coordinates} maps each triangulation in ${\cal T}_\CV$
to a Darboux coordinate system with ``good asymptotics'' in the
sector $\CV$:  by this we mean that each function is naturally
labeled by a cycle in $\hat\Gamma$, and $\lim_{\zeta \to 0}
\CX_\gamma(\zeta) \exp \left[ - \zeta^{-1} \pi R Z_\gamma \right]$
is finite if $\zeta \in \CV$.

Because of this labeling, we can consider each coordinate system as
a map to an abstract complex torus with a system of Fourier modes
$X_\gamma$, and each coordinate transformation simply as a
symplectomorphism of that torus. For any $\vartheta_+, \vartheta_-$,
$\vartheta_+ -\vartheta_- \leq \pi$, the transformation of
coordinates relating the coordinate systems
$\CX^{\vartheta_\pm}_\gamma$ is an interesting symplectomorphism
$\bS(\vartheta_+, \vartheta_-;u)$. Anticipating our results, we use
the same nomenclature as in Section \ref{sec:review}. By
construction, if $\vartheta_- < \vartheta < \vartheta_+$, then
$\bS(\vartheta_+, \vartheta_-;u) = \bS(\vartheta_+, \vartheta;u)
\bS(\vartheta, \vartheta_-;u)$.

The ``Stokes factors'' mentioned in Section \ref{sec:review} emerge
in the limit where $\vartheta_+$ and $\vartheta_-$ approach the same
value $\vartheta_0=\mathrm{arg}\, Z_{-\gamma_0}$ from the left or from
the right. Then $\bS(\vartheta_+, \vartheta_-;u) \to
\bS_{\vartheta_0,u}$. Clearly, these Stokes factors are captured by
the comparison between the WKB triangulations $T_\WKB(\vartheta_+)$
and $T_\WKB(\vartheta_-)$. Unless the triangulation jumps at
$\vartheta_0 = \mathrm{arg}\, Z_{-\gamma_0}$ the Stokes factor will be
the identity. If the triangulation jumps, we can compute the
corresponding symplectomorphism. We will do that in the next few
subsections.

What we expect based on the comparison to Section \ref{sec:review}
--- but what is far from obvious at this stage --- is that
$\bS_{\vartheta_0,u}$ involves only symplectomorphisms generated by
functions of $X_{\gamma_0}$.  As we will see momentarily, this is
indeed the case: $\bS_{\vartheta_0,u}$ is a product of KS factors
$\CK_{n \gamma_0}$, with the expected multiplicities $\Omega(n
\gamma_0)$. The wall-crossing formula, which was written in Section
\ref{sec:review} as the invariance of $S(\vartheta_+,
\vartheta_-;u)$ under small changes of $\lambda$, thus follows
directly from the invariance of the decorated triangulations
$T_\WKB(\vartheta_+, \lambda^2)$ and $T_\WKB(\vartheta_-,
\lambda^2)$.

\subsection{Jumps at special $\vartheta$}

\subsubsection{Symplectomorphism from a BPS hypermultiplet}

We have seen that when $\vartheta$ crosses a critical value
$\vartheta_c$ corresponding to a BPS hypermultiplet the WKB
triangulation jumps. Now we would like to know how
$\CX^\vartheta_\gamma$ jumps at this $\vartheta_c$.

Let us state the problem a bit more precisely.
For $\vartheta$ in a small interval $\vartheta_- \le \vartheta < \vartheta_c$, the homotopy class of the WKB triangulation is constant;
call it $T_-$.
Similarly, considering $\vartheta$ in a small interval $\vartheta_c < \vartheta \le \vartheta_+$, we have
a triangulation $T_+$.
We want to compare $\CX^{\vartheta_-}_\gamma$ and $\CX^{\vartheta_+}_\gamma$.
To lighten the notation we will sometimes replace $\vartheta_\pm$ by simply $\pm$ below.

\insfig{5.75in}{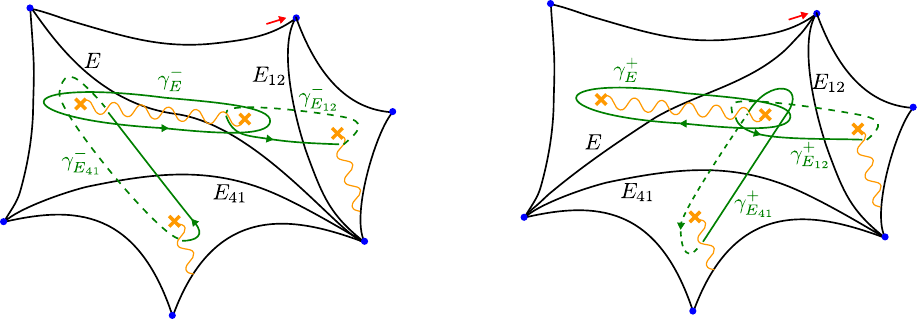}{The cycles attached to the triangulations $T_\WKB(\vartheta_\pm)$.}

In Figure \ref{fig:flip-cycles} we depict some of the cycles $\gamma^{\pm}_E$.
From this figure we can read off the relations among them:
\begin{align}
\gamma_E^{-} &= - \gamma_E^{+}, \label{eq:flip-c1} \\
\gamma_{E_{12}}^{-} &= \gamma_{E_{12}}^{+}, \label{eq:flip-c2} \\
\gamma_{E_{41}}^{-} &= \gamma_{E_{41}}^{+} - \gamma_E^{-}. \label{eq:flip-c3}
\end{align}
The last two equations can be neatly summarized as the transformation
$\gamma_{E_i}^+ = \gamma_{E_i}^- + \langle \gamma_{E_i}^-, \gamma_E^-\rangle_{{}_+} \gamma_E^-$,
where the subscript $+$ on the intersection means we only take the positive part.

Now we can describe the relations between $\CX_\gamma$ before and after the flip.
First, from \eqref{eq:fg-trans-1} we see that
\begin{equation}
\CX^+_E = 1/{\CX^-_E}.
\end{equation}
Combining this with \eqref{eq:flip-c1}, and using freely the definition of $\CX_\gamma$ from the $\CX_E$, gives
\begin{equation}
\CX^+_{\gamma_E^+} = (\CX^-_{\gamma_E^-})^{-1} = \CX^-_{\gamma_E^+}.
\end{equation}
In other words, $\CX_{\gamma_E^+}$ is continuous across the ray $\vartheta = \vartheta_c$.

Next let us consider the edge $E_{12}$.  From \eqref{eq:fg-trans-2} we have
\begin{equation}
\CX^+_{E_{12}} = \CX^-_{E_{12}} (1 + \CX^-_E).
\end{equation}
Combining this with \eqref{eq:flip-c2} gives
\begin{equation}
\CX^+_{\gamma^+_{E_{12}}} = \CX^-_{\gamma^+_{E_{12}}} (1 + \CX^-_{\gamma^-_E}).
\end{equation}
The same holds for $\CX^\pm_{\gamma_{E_{34}}}$, just by replacing $1 \to 3$ and $2 \to 4$ above.

The story for $E_{41}$ is slightly more complicated.
From \eqref{eq:fg-trans-5} we have
\begin{equation}
\CX^+_{E_{41}} = \CX^-_{E_{41}} (1 + (\CX^-_E)^{-1})^{-1}.
\end{equation}
Combining this with \eqref{eq:flip-c3} we have
\begin{equation}
\CX^+_{\gamma^+_{E_{41}}} = \CX^-_{\gamma^-_{E_{41}}} (1 + (\CX^-_E)^{-1})^{-1} = \CX^-_{\gamma^+_{E_{41}}} (\CX^-_E)^{-1} (1 + (\CX^-_E)^{-1})^{-1} = \CX^-_{\gamma^+_{E_{41}}} (1 + \CX^-_E)^{-1}.
\end{equation}
The same holds for $\CX^\pm_{\gamma_{E_{23}}}$, just by replacing $4 \to 2$ and $1 \to 3$.

To summarize our results, let us define
\begin{equation}
\gamma_{\hyper} := \gamma^-_E.
\end{equation}
This is the charge of the BPS hypermultiplet represented by the finite WKB curve.
What we have found is
\begin{equation} \label{eq:KS-flip}
\CX^+_\gamma = \CX^-_\gamma (1 + \CX^-_{\gamma_\hyper})^{\inprod{\gamma, \gamma_\hyper}}.
\end{equation}
This is exactly the expected transformation property \eqref{eq:ks-discontinuity}, if we put
$\Omega(\gamma_\hyper) = 1$ --- precisely agreeing with the fact that the finite WKB curve
represents a single BPS
hypermultiplet of charge $\gamma_\hyper$ --- and also put $\sigma(\gamma_\hyper) = -1$,
and all $\Omega(n \gamma_\hyper) = 0$ for $n > 1$.

\subsubsection{Symplectomorphism from a BPS vectormultiplet}

Let us now consider what happens at $\vartheta_c$ corresponding to a vectormultiplet.
Then as we discussed in Section \ref{sec:vector-jump}, the WKB triangulation near $\vartheta_c$ contains an annulus $W$,
triangulated by edges which undergo an infinite sequence of flips as $\vartheta \to \vartheta_c$ from either direction,
while exactly at $\vartheta_c$ we have an annulus foliated by closed WKB curves.

In what follows we will use some statements about the WKB foliation which were determined
by computer experimentation.  We believe that the picture we describe is correct at least
in the case when $C$ has genus zero; however, after the first preprint version of this paper appeared,
Ivan Smith pointed out to us that the picture may be more complicated if $C$ has genus $g>0$
and the annulus $W$ cuts off a component which contains a handle and contains no punctures.
In \cite{gmn5-to-appear} we will give a different way of studying that situation.
We will find that the main outcome of our analysis, \eqref{eq:twist-KS} below, continues to
hold even in that case.

First, to rigidify the picture it is convenient to consider some auxiliary
objects, namely WKB curves with phase $\vartheta+\frac{\pi}{2}$
instead of $\vartheta$; call these \ti{anti-WKB curves}. Generically
when there is a BPS state with phase $\vartheta_c$ there is no BPS
state with phase $\vartheta_c + \frac{\pi}{2}$, so the anti-WKB
curves vary smoothly near $\vartheta = \vartheta_c$, in contrast to
the WKB curves which are undergoing violent changes there. In
particular, for $\vartheta$ on either side of $\vartheta_c$, the
anti-separatrices give a convenient division of a region containing
the annulus into simply connected cells. Let $\alpha$ and
$\beta$ denote two anti-WKB curves belonging to two of these cells, as
shown in Figure \ref{fig:anti-cells}, and $\hat\alpha$ and $\hat\beta$
lifts to $\Sigma$ (with the anti-WKB orientation). Using this
division we can also give a canonical choice of the fiducial paths
$E_\pm$, also shown in Figure \ref{fig:anti-cells}.
\insfig{3.3in}{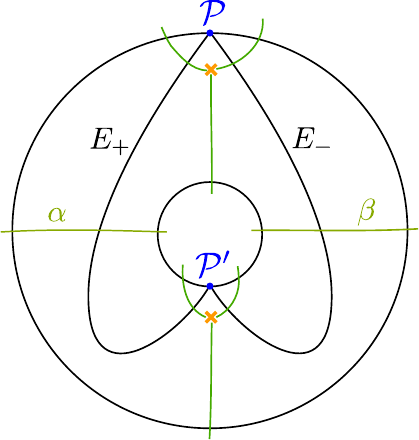}{Topology of the annulus, with anti-WKB
curves marked in green, and two fiducial paths $E_\pm$.}

As in Section \ref{sec:limit-triangulation}, we can label the various
possible triangulations of the annulus as $T_m$, for $m \in \IZ$.
As $\vartheta \to \vartheta_c$ from above, the WKB
triangulation runs through an infinite sequence of triangulations.
These triangulations can be identified as the $T_m$ for $m = m_+, m_+ + 1, \dots$, with some $m_+$.
Similarly, as $\vartheta \to \vartheta_c$ from below
the WKB triangulation runs through a different sequence of triangulations $T_m$,
with $m = m_-, m_- - 1, \dots$.
So the $T_m$ for sufficiently large or sufficiently small $m$ all occur as WKB triangulations,
but the $T_m$ for intermediate values of $m$ need not.
For notational convenience below, we look at a narrow range of $\vartheta$ so that
$m_+ > m_-$ (so each $T_m$ occurs at most once) and for each $T_m$ that does occur,
choose a $\vartheta_m$ for which $T_\WKB(\vartheta_m) = T_m$.

For $m>m_+$, $E_{m+}$ emerges from the singularity $\CP'$,
passes on the right of a turning point, turns right and winds clockwise around the annulus crossing $2m$ anti-separatrices, then
passes near a second turning point before reaching $\CP$.  $E_{m-}$ is similar but crosses only $2m-2$ anti-separatrices.
See Figure \ref{fig:wkb-m1} for the case $m=1$.
\instwofigs{2in}{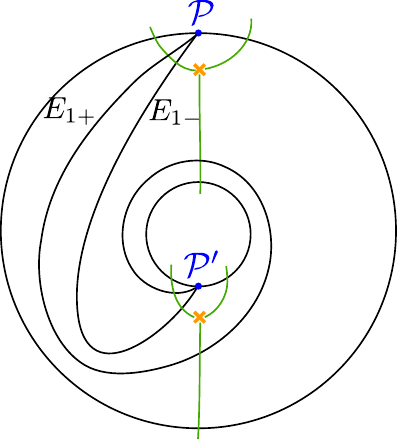}{The topology of $T_1 = T_\WKB(\vartheta_1)$ on the annulus.}
{2in}{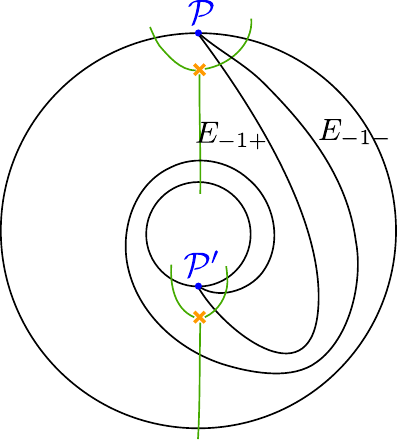}{The topology of $T_{-1} = T_\WKB(\vartheta_{-1})$ on the annulus.}
From this it follows that the corresponding cycles have
\begin{align}
\inprod{\gamma_{E_{m+}}^{\vartheta_m}, \hat\alpha} &= 1-m, \label{eq:int1} \\
\inprod{\gamma_{E_{m+}}^{\vartheta_m}, \hat\beta} &= 2-m,  \\
\inprod{\gamma_{E_{m-}}^{\vartheta_m}, \hat\alpha} &= m, \\
\inprod{\gamma_{E_{m-}}^{\vartheta_m}, \hat\beta} &= m-1. \label{eq:int4}
\end{align}

For $m<m_-$, the situation is very similar to the above, with the crucial difference that the word ``right''
is replaced by ``left'' at several points.  So $E_{m+}$ emerges from $\CP'$,
passes on the left of a turning point, turns left and winds
counterclockwise around the annulus crossing $-2m-2$
anti-separatrices, then passes near a second turning point before
reaching $\CP$.  $E_{m-}$ is similar but crosses
$-2m$ anti-separatrices. See Figure \ref{fig:wkb-m-1} for the case
$m=-1$. The relations \eqref{eq:int1}-\eqref{eq:int4} in this case
are replaced by
\begin{align}
\inprod{\gamma_{E_{m+}}^{\vartheta_m}, \hat\alpha} &= 1+m, \label{eq:int5} \\
\inprod{\gamma_{E_{m+}}^{\vartheta_m}, \hat\beta} &= m,  \\
\inprod{\gamma_{E_{m-}}^{\vartheta_m}, \hat\alpha} &= -2-m, \\
\inprod{\gamma_{E_{m-}}^{\vartheta_m}, \hat\beta} &= -1-m. \label{eq:int8}
\end{align}
We now define four cycles $\gamma^\pm_{A,B}$, shown in Figure
\ref{fig:limit-homology-annulus}, with
\insfig{5in}{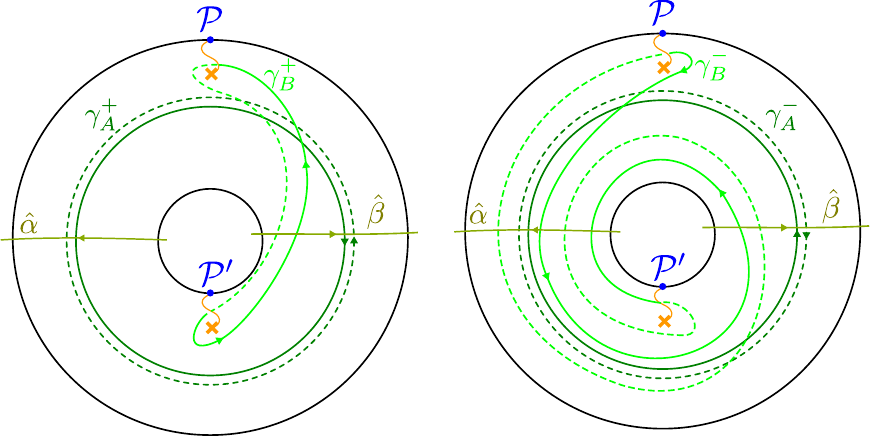}{Two bases $\gamma^+_{A,B}$ and
$\gamma^-_{A,B}$ for the part of $\hat\Gamma$ supported over the
annulus.}
\begin{align}
\inprod{\gamma^+_A, \hat\alpha} = 1, \quad & \inprod{\gamma^-_A, \hat\alpha} = -1, \\
\inprod{\gamma^+_A, \hat\beta} = 1,  \quad & \inprod{\gamma^-_A, \hat\beta} = -1, \\
\inprod{\gamma^+_B, \hat\alpha} = 0, \quad & \inprod{\gamma^-_B, \hat\alpha} = -2, \\
\inprod{\gamma^+_B, \hat\beta} = -1, \quad & \inprod{\gamma^-_B, \hat\beta} = -1.
\end{align}
Note that it follows that
\begin{align}
\gamma^-_A &= - \gamma^+_A, \label{eq:compare-a} \\
\gamma^-_B &= - \gamma^+_B + 2 \gamma^-_A. \label{eq:compare-b}
\end{align}
On the other hand, comparing the intersection numbers we see that these cycles
are related to the ones attached to the nearby WKB triangulations by the uniform formula
\begin{align}
\gamma^\pm_A & = \gamma^{\vartheta_m}_{E_{m-}} + \gamma^{\vartheta_m}_{E_{m+}}, \label{eq:gamA} \\
\gamma^\pm_B & = (1-m) \gamma^{\vartheta_m}_{E_{m-}} - m \gamma^{\vartheta_m}_{E_{m+}}. \label{eq:gamB}
\end{align}
where on the left we choose the sign $+$ when $m > m_+$ and $-$ when $m < m_-$.

In Section \ref{sec:limit-triangulation} we defined a ``limit
coordinate system'' $\CX^{T_{+\infty}}$.  Now, in the present case where
$T_{+\infty}$ arises as $T_\WKB(\vartheta_c)$, we can define a
corresponding labeling by homology:  namely we define
$\CX^+_\gamma$ by our usual rule \eqref{eq:canonical-darboux}, where $E$ can denote
either an edge of $T_{+\infty}$ away from the annulus, or one of
the special symbols $A$, $B$.
Using \eqref{eq:limiting-1},
\eqref{eq:limiting-2}, \eqref{eq:gamA}, \eqref{eq:gamB} we see that these coordinates indeed deserve
to be called limit coordinates:
\begin{equation}
\lim_{\vartheta \to \vartheta_c^+} \CX^{\vartheta}_\gamma = \CX^+_\gamma. \label{eq:limit-1}
\end{equation}
We may similarly define $\CX^-_\gamma$, and
using \eqref{eq:limiting-3}, \eqref{eq:limiting-4}, \eqref{eq:gamA}, \eqref{eq:gamB} we see that these similarly obey
\begin{equation}
\lim_{\vartheta \to \vartheta_c^-} \CX^{\vartheta}_\gamma = \CX^-_\gamma. \label{eq:limit-2}
\end{equation}

We are finally ready to compare $\CX^+_\gamma$ to $\CX^-_\gamma$.
For the $A$ cycles this is fairly straightforward:  using \eqref{eq:compare-a} and \eqref{eq:A-relation} we have
\begin{equation} \label{eq:twist-A}
\CX^+_{\gamma^+_A} = (\CX^-_{\gamma^-_A})^{-1} = \CX^-_{\gamma^+_A}.
\end{equation}
For the $B$ cycles it is a bit more complicated.  From \eqref{eq:compare-b} and \eqref{eq:B-relation} we find that
\begin{equation}
\CX^+_{\gamma^+_B} = (\CX^-_{\gamma^-_B})^{-1} (\xi_+ - \xi_-)^{-4}
= \CX^-_{\gamma^+_B} (\CX^-_{\gamma^-_A})^{-2} (\xi_+ - \xi_-)^{-4} =
\CX^-_{\gamma^+_B} \xi_-^{-4} (\xi_+ - \xi_-)^{-4}
\end{equation}
so finally
\begin{equation} \label{eq:twist-B}
\CX^+_{\gamma^+_B} = \CX^-_{\gamma^+_B} (1 - \xi_-^2)^{-4}.
\end{equation}

Define
\begin{equation}
\gamma_\vect := - \gamma^+_A.
\end{equation}
This is the charge of the BPS vectormultiplet represented by the closed WKB curves.
Combining our results \eqref{eq:twist-A}, \eqref{eq:twist-B} gives the simple transformation law
\begin{equation} \label{eq:twist-KS}
\CX^+_\gamma = \CX^-_\gamma (1 - \CX^-_{\gamma_\vect})^{-2 \inprod{\gamma, \gamma_\vect}}.
\end{equation}
Note that we can read off $\langle \gamma_B^+, \gamma_A^+ \rangle
=-2$ from Figure \ref{fig:limit-homology-annulus}.

Altogether we see that the two coordinate systems $\CX^\pm_\gamma$
are related by the symplectomorphism $\CK^{-2}_{\gamma_\vect}$.  So
again, we obtain exactly the expected transformation property
\eqref{eq:ks-discontinuity} for $\CX_\gamma(\zeta)$, if we put
$\Omega(\gamma_\vect) = -2$ --- precisely agreeing with the fact
that the closed WKB curves represent a single BPS vectormultiplet
of charge $\gamma_\vect$ --- and also put $\sigma(\gamma_\vect) =
+1$, and all $\Omega(n \gamma_\vect) = 0$ for $n > 1$.

\subsubsection{No symplectomorphism when $m e^{-i \vartheta} \in i \IR$} \label{sec:no-flavor-jump}

Now we dispose of a tricky point.  Recall that at each singularity $\CP_i$ we have a parameter $m_i \in \IC$
which controls the residue of $\varphi$.
What happens at the angles $\vartheta = \vartheta_c$ where some $m_i e^{-i \vartheta_c} \in i\IR$?

As we described in Section \ref{sec:flavor-jump}, the triangulation
$T_\WKB(\vartheta)$ jumps rather violently near the singular point $\CP_i$
as $\vartheta$ crosses $\vartheta_c$, and at the same time we have a pop
which changes the flat section $s_i$.
On the other hand, the family of closed WKB curves which appear at this $\vartheta$ do not correspond to a charged BPS state:
if they are a BPS state at all, it is one carrying only flavor charge (the one associated to the puncture $\CP_i$, of course).
Said differently, if we define $\gamma_\BPS$ by lifting these closed WKB curves in our usual way, while it is indeed true that
$Z(\gamma_\BPS) \in e^{i \vartheta_c} \IR$, this charge $\gamma_\BPS$ is in the radical of $\inprod{,}$.  So
the symplectomorphism $\CK_{\gamma_\BPS}$ is actually trivial.  Hence it seems that to be consistent with
\eqref{eq:ks-discontinuity} we should expect that the $\CX_\gamma^\vartheta$ do \ti{not} jump at $\vartheta = \vartheta_c$,
despite the jump of $T_\WKB$.  In this section we verify that this is indeed the case.

First, let us consider the coordinate $\CX_{C^+}$.
Recall from \eqref{eq:monodromy} that
\begin{equation}
\CX^\pm_{C^\pm} = \mu_\pm^{2}.
\end{equation}
On the other hand, we have
\begin{equation} \label{eq:ctrans}
C^+ = - C^-
\end{equation}
and
\begin{equation}
\mu_+ = 1/\mu_-.
\end{equation}
Hence
\begin{equation}
\CX_{C^+}^+ = (\mu_+)^{2} = (\mu_-)^{-2} = (\CX_{C^-}^-)^{-1} = \CX_{C^+}^-.
\end{equation}

\insfig{4.7in}{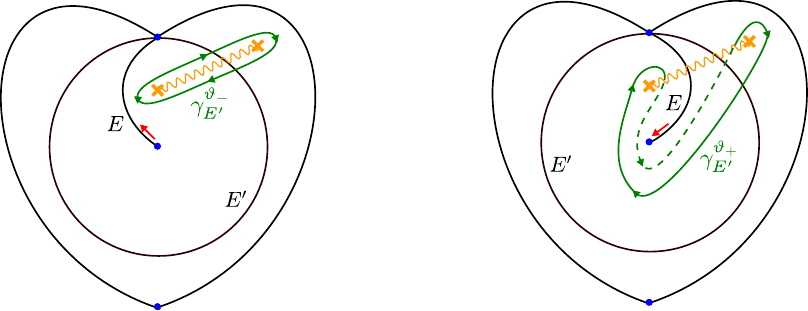}{An unwinding of Figure \protect\ref{fig:spiral-jump},
with the homology cycles corresponding to edges marked.}
There is one other $\CX_\gamma$ we have to worry about.
Let $E'$ be the ``loop'' of the degenerate face and define
\begin{equation}
\gamma := \gamma^{-}_{E'}.
\end{equation}
The equality $\CX^+_\gamma = \CX^-_\gamma$ is a consequence of two different effects which cancel one another.
As we showed in \eqref{eq:degen-loop-coord},
\begin{equation} \label{eq:deg-1}
\CX^{\pm}_{\gamma^\pm_{E'}} = (\mu_\pm)^{-1} S,
\end{equation}
where $S$ is continuous across the critical locus (in particular it does
not involve the section $s$).  Unwinding Figure
\ref{fig:spiral-jump} to see its topological content more easily, we
arrive at Figure \ref{fig:degenerate-combining-loops}, from which we
see that
\begin{equation} \label{eq:degenerate-homology-jump}
\gamma^{-}_{E'} = \gamma^{+}_{E'} + C^+.
\end{equation}
Combining all this we get
\begin{equation}
\CX^+_{\gamma^{-}_{E'}} = \CX^{+}_{\gamma^{+}_{E'}} \CX^{+}_{C^+} = (\mu_+)^{-1} S \mu_+^{2} = (\mu_-)^{-1} S = \CX^-_{\gamma^-_{E'}}
\end{equation}
as desired.

So indeed the $\CX^\vartheta_\gamma$ are continuous across this $\vartheta = \vartheta_c$.

\subsection{Quadratic refinement} \label{sec:quadratic-refinement}

Now we come to another pesky detail:  the sign $\sigma(\gamma)$
which occurs in the transformation \eqref{eq:KS-Transfm}.  In the
general story described in \cite{Gaiotto:2008cd} one expects that
$\sigma: \hat\Gamma \to \IZ_2$ is a quadratic refinement of the mod
2 intersection pairing $\inprod{,}$ on $\hat\Gamma$. By definition
this means that
\begin{equation}
\sigma(\gamma) \sigma(\gamma') = \sigma(\gamma + \gamma')\sigma(0)
(-1)^{\inprod{\gamma, \gamma'}}.
\end{equation}
In general, one would not expect such a quadratic refinement to exist globally over $\CB$; instead
one would have to pick different refinements in different local patches, and in gluing the patches
together one would have to keep track of some sign changes in $\CX^\vartheta_\gamma$, coming from the
fact that the refinements are not the same.

On the other hand, in this this paper we seemed to find a simpler
story.  The gluing laws for our functions $\CX^\vartheta_\gamma$ do
not involve any tricky signs.  Moreover, the transformations we
found for the $\CX^\vartheta_\gamma$ agree with
\eqref{eq:KS-Transfm}, provided that we have
\begin{align} \label{eq:dessigma}
\sigma(\gamma_\hyper) &= -1, \\
\sigma(\gamma_\vect) &= +1. \label{eq:dessigma2}
\end{align}
How can this be consistent?  It would be consistent if there exists a \ti{single} quadratic refinement $\sigma$
which obeys \eqref{eq:dessigma}, \eqref{eq:dessigma2}
for all hypermultiplets and vectormultiplets which appear in the spectrum at any $u \in \CB$.

We can easily construct such a $\sigma$ for any fixed $(\vartheta, u)$:  it is determined by requiring
$\sigma(0) = +1$, $\sigma(\gamma_E) = -1$ for all nondegenerate edges $E$ of $T_\WKB(\vartheta, u)$,
and $\sigma(\gamma_E) = +1$ for degenerate edges.
From this definition it is straightforward to see that $\sigma(\gamma_\hyper) = -1$ and $\sigma(\gamma_\vect) = +1$
for any BPS states which have phase $\vartheta$.
So from this perspective the trouble is to show that the $\sigma$ so defined is
actually independent of $(\vartheta, u)$.  This amounts to checking that this formula
for $\sigma$ is consistent with the transformations of the $\gamma_E$ when the triangulation
$T_\WKB$ jumps.  Fortunately this is indeed the case.

Indeed, as we have discussed, as we vary $(\vartheta, u)$ the WKB
triangulation undergoes three types of transformation.  The first
type (hypermultiplet) is given by
\eqref{eq:flip-c1}-\eqref{eq:flip-c3} and one checks directly that
it is consistent with our proposal for $\sigma$.  The second type
(vectormultiplet) is given in terms of the auxiliary cycles
$\gamma_{A,B}$ defined by \eqref{eq:gamA}, \eqref{eq:gamB}.  Using
those equations and our proposal for $\sigma$ gives
$\sigma(\gamma_A) = +1$ and $\sigma(\gamma_B) = -1$.  This $\sigma$
is indeed consistent with the transformations \eqref{eq:compare-a},
\eqref{eq:compare-b}.  Finally, the transformation of cycles from
the third type of jump ($m / e^{i \vartheta} \in i \IR$) is given by
\eqref{eq:ctrans} and \eqref{eq:degenerate-homology-jump}, once
again consistent with $\sigma$ (recalling that $C_+$ corresponds to
a degenerate edge and so $\sigma(C_+) = +1$.)

There is another, more intrinsic, way of describing this quadratic
refinement.  Given any homology class $\gamma \in H_1(\Sigma; \IZ)$
we first represent $\gamma$ by a disjoint union of oriented closed
curves which avoid the zeroes of $\lambda$.  On each of these closed
curves we have the \ti{phase function} defined as the phase of
$\lambda \cdot \partial_t$.  Letting $n_w$ denote the number of
times this phase winds around the circle as we go around the curve,
$\sigma(\gamma)$ is the product of $(-1)^{n_w+1}$ over all
components of our representative for $\gamma$.
It is straightforward to check that this is indeed well
defined as a function on homology, and that it gives a quadratic
refinement.  Moreover, from the fact that $\gamma_\hyper$ is
represented by a single closed constant phase curve, and
$\gamma_\vect$ is represented by a disjoint union of two such curves,
one easily sees that $\sigma$ has the desired
properties \eqref{eq:dessigma}, \eqref{eq:dessigma2}.

\subsection{Some comments on the full BPS spectrum} \label{sec:bps-comments}

At this point we have constructed the functions $\CX^\vartheta_\gamma$ and verified
that they jump by KS transformations as we vary $\vartheta$, in the class of
examples related to $SU(2)$ Hitchin systems with regular singularities.
As we explained above and in Section \ref{sec:review}, these functions are the
key ingredient in our explanation of the wall-crossing formula.
But that is not all:  we have also learned a strategy for
determining the BPS spectrum.  Indeed, if we pick two phases $\vartheta_\pm$
and compute the WKB triangulations $T_\WKB(\vartheta_\pm, u)$,
we can then reconstruct the coordinate transformation
$\bS(\vartheta_+, \vartheta_- ; u)$, decompose it uniquely into
a product of properly ordered KS transformations as described
in Section \ref{subsec:wallcrossing}, and read
off the spectrum of BPS states with phases in the sector
$[\vartheta_-, \vartheta_+]$.

This strategy is particularly potent if we choose $\vartheta_+ = \vartheta_- +\pi$.\footnote{To
be precise, we include only one of the two boundary rays in the sector.}
In that case the sector is a whole half-plane, and thus big enough to determine
the whole BPS spectrum (as the BPS states with phase outside this half-plane
are the antiparticles of ones inside it).  We therefore call
$\bS(\vartheta, \vartheta + \pi ; u)$ the ``spectrum generator.''
Luckily, it turns out that, unlike a random $\bS(\vartheta_+, \vartheta_- ; u)$,
this spectrum generator is actually computable!
The essential reason for this computability is that
$T_\WKB(\vartheta)$ and $T_\WKB(\vartheta+\pi)$ only differ in the
decoration: to go from one to the other we just have to pop at all of the
singularities.  We defer the computation of the spectrum
generator to Section \ref{sec:UniversalStokesMatrix}, but we note now
that the result is quite simple, and only depends on the
combinatorial data of $T_\WKB(\vartheta, u)$.

It would be very interesting to understand physically \ti{why} $T_\WKB(\vartheta, u)$
can capture the whole BPS spectrum.  At least one part of this story is easy to understand:
for each edge $E$ of $T_\WKB(\vartheta, u)$, there is a BPS hypermultiplet of charge $\gamma_E$
in the vacuum $u$.  Indeed, the corresponding cell of
the WKB foliation is mapped to a strip by the coordinate
transformation $z \mapsto w = \int^z \lambda$, with the two turning
points at opposite boundaries of the strip, and the
preimage in the $z$-plane of a straight segment running between
the two turning points in the $w$-plane yields a BPS string of
charge $\gamma_E$.\footnote{Similar statements hold for
limit triangulations:  in that case each annulus foliated by WKB curves
gives a family of closed BPS strings (a vectormultiplet), along with
an infinite tower of hypermultiplets, namely the
inverse images of straight paths in the $w$ coordinate between
the turning points, with various windings on the annulus.}

The charges $\gamma_E$ of these BPS particles form a basis of the
charge lattice, as shown in Section \ref{sec:lattice-of-charges}. In
fact, more is true: this basis has an important positivity property,
reminiscent of the relation between roots and simple roots in a
simple Lie algebra. Recall that a BPS state corresponds to a finite
WKB curve, giving a straight line segment in the $w$-plane. The
slope of this segment is the phase of the central charge of the BPS
state.  In particular, if this phase lies between $\vartheta$ and
$\vartheta+\pi$, then the BPS string has \emph{positive}
intersection with the WKB curves of phase $\vartheta$. On the other
hand, by the definition of the homology labeling given in Section
\ref{sec:homology}, $\langle \gamma_E^\vartheta, \hat E'\rangle =
\delta_{E,E'}$. It follows that, if a homology cycle $\gamma_{\BPS}$
supports a BPS state with phase between $\vartheta$ and
$\vartheta+\pi$, we have
\begin{equation}
\gamma_{\BPS} = \sum_E c_E \gamma_E^{\vartheta}
\end{equation}
with all $c_E\geq 0$.

The above facts suggest some natural speculations.  First, the
positive decomposition is a hint that all BPS states can be viewed
as bound states of a set of ``simple'' BPS states, which
are in correspondence with the edges $E$ of $T_\WKB(\vartheta, u)$.
Further evidence for this conjecture comes from the fact that
the spectrum generator $\bS$ can be computed purely from the
combinatorial data of $T_\WKB$.

Note that even for a fixed vacuum $u$, one can obtain various bases of
simple BPS states by changing $\vartheta$.  This is reminiscent of the
oft-employed description of BPS states in terms of quiver quantum mechanics;
in that story one sometimes describes the same quantum mechanics using various
different quivers, corresponding to different ``exceptional collections'',
which are related by mutations.  Indeed, in our context
there is a natural quiver around, with nodes labeled by the edges $E_i$ of the
triangulation, and the number of arrows determined by $\langle E_i, E_j \rangle$.  It is
possible that with appropriate FI terms and superpotentials the
BPS states of the quiver quantum mechanics would be in 1-1
correspondence with the full BPS spectrum of the theory with phases
in the sector $[\theta, \theta+\pi]$.
Relations between mutations of quivers and ``cluster transformations,'' closely
related to the KS transformations we encounter in this paper,
have been considered extensively in the mathematics literature
(see e.g. \cite{cluster-intro}).  (For further developments in this direction see
e.g. \cite{Gaiotto:2010be,Cecotti:2010fi,Cecotti:2011rv}.)

\section{Irregular singularities}\label{sec:irregular}

While we have focused on the case of regular singular points in the
past three sections, the constructions can also be adapted to the
case of irregular singularities.  The story is quite similar to the
regular case, with the following modifications:

\begin{enumerate}

\item We begin by defining the appropriate notion of triangulation when
irregular singularities are included. Suppose that $\CP_*$ is an
irregular singular point where $\lambda^2$ has a pole of order
$L+2$, with $L \geq 1$ integral. (The boundary conditions of Section
\ref{sec:physics} for rank two Hitchin systems involve irregular
singular points with $L = 1$, while in Section
\ref{section:Superconformal} we will meet singular points with $L =
N+2$, $N\geq 0$.) We draw a circle $S^1(\CP_*)$ around $\CP_*$,
bounding a disc $D(\CP_*)$, to be considered as infinitesimally
small.  On this circle we mark $L$ points $Q_i$, $i=1, \dots, L$.
These points are cyclically ordered by saying that $\dots, Q_i,
Q_{i+1}, \dots$ are going clockwise around $\CP_*$.  Then our
triangulations are really triangulations of the surface $C'$
obtained by cutting all the discs $D(\CP_*)$.  The vertices are the
marked points $Q_i$ around all of the irregular singularities, as
well as all of the regular singularities $\CP_i$.  The edges
necessarily include the segments on the circles $S^1(\CP_*)$ joining
consecutive points $Q_i$.  We call these segments \emph{boundary
edges}; they will have a special status below.

\item In order to define a \emph{decorated} triangulation, we
need to choose a flat section for the connection $\CA$ (up to scale)
near each vertex.  For irregular
singularities this means choosing a flat section near each point $Q_i$ on
$S^1(\CP_*)$.  As in the regular case we would like to narrow this
down to a discrete choice.  To this end we observe that in the case of an
irregular singular point, in addition to a possible monodromy there is also
Stokes phenomenon.\footnote{In what follows we are assuming that the standard
Stokes theory for \ti{meromorphic} connections on a complex curve can be
extended in the most obvious way to apply to the connection $\CA$, which is flat
but not meromorphic.  We have not found any literature on this precise
situation, although somewhat related constructions appear in \cite{MR1864833}.}
There are $L$ rays (``Stokes rays'') emerging from $\CP_*$, bounding sectors
of opening angle $2\pi/L$.  A flat section which is asymptotically
exponentially small as $z \to \CP_*$ along a
ray going into $\CP_*$ on one side of a Stokes ray becomes
exponentially large on the other side of the Stokes ray.\footnote{In the literature
on Stokes phenomenon there are two kinds of rays, named \emph{Stokes} and
\emph{anti-Stokes}, each of which plays an important role in the systematic
development of the theory. Regrettably, the terminology is not consistently
applied by various authors on the subject. In our convention, the standard Airy
function, $Ai(x)$, which has real exponential decay along the positive real axis
and power law decay with an oscillating envelope along the negative real axis,
has Stokes rays along the negative real axis and along $\vert {\rm arg}(x) \vert
= \pi/3$.\label{stokes-footnote}}
We
define a decoration to be a choice of flat section near $Q_i$ (modulo overall rescaling)
which, after analytic continuation around $\CP_*$, is exponentially small as
$z \to \CP_*$ in \emph{some} sector bounded by Stokes rays.

\item We can define an infinite sequence of such
sections (up to rescaling),
\begin{equation}\label{eq:bisequenceStokes}
\cdots, s_{-3}, s_{-2}, s_{-1}, s_0, s_1, s_2, s_{3}, \dots
\end{equation}
where the ordering is determined by saying that if $s_n$ is the
small solution in sector $\CS$, then $s_{n+1}$ is the small solution
in the next sector in the clockwise direction.
(If the monodromy is trivial, we can choose the scales of the $s_n$ so that
the sequence will have period $L$.)  Moreover, we further restrict
our choice of decoration so that
if we choose, say $s_n$ at point $Q_i$ then at point $Q_{i+1}$ we
must also choose $s_{n+1}$, and so on. Thus, the choice of
decoration at an irregular singularity boils down to a \emph{single}
choice of flat section at one marked point, rather than
an independent choice at each point.
The set of possible choices of decoration near an irregular singular point thus forms a
$\IZ$-torsor. If the monodromy is trivial, it can be reduced to
a $\IZ_L$-torsor.

\item The definition of the Fock-Goncharov coordinates $\CX_E^T$ can now be given just as
before, with the important caveat that we define $\CX_E^T = 0$ if $E$ is a boundary edge.

\item The local behavior of the
WKB foliation around an irregular singularity is rather different
from that around a regular singularity.  Rather than spiraling isotropically into
the singularity, each WKB curve is
asymptotically tangent to one of $L$ rays.  These rays, which we will call
\emph{WKB rays} (with phase $\vartheta$), determine points $Q_i$,
$i=1, \dots, L$, on an infinitesimal circle $S^1(\CP_*)$ around $\CP_*$; these
are the marked points we will use in defining $T_\WKB$.
If $\vartheta = \arg \zeta$, then the WKB rays with phase $\vartheta$ are
the same as the anti-Stokes rays for the connection $\CA$.
\insfig{2.7in}{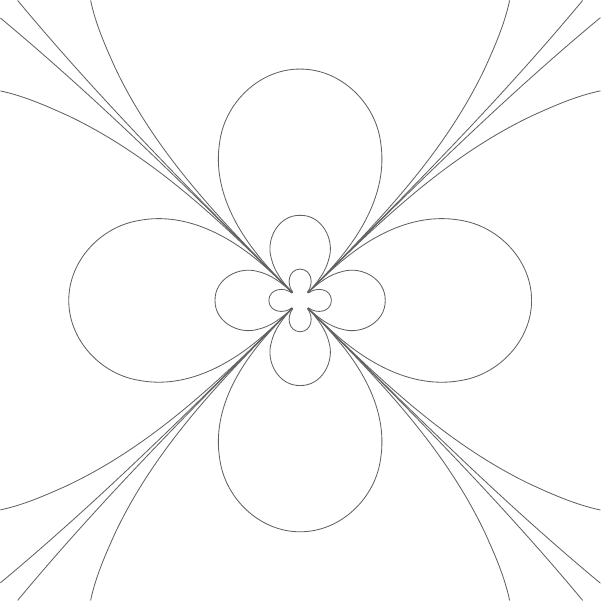}{Behavior of the WKB foliation near an
irregular singular point with $L = 4$.  The WKB curves cluster onto
the $4$ WKB rays, separated from one another by arcs of $\pi/2$
radians. WKB curves in a small neighborhood of the singularity look like flower petals which
connect adjacent WKB rays.}

\item
The definition of $T_\WKB$ proceeds essentially as
before.  If a WKB curve asymptotes to a WKB ray ending on
an irregular singularity $\CP_*$, we regard it as ending on the corresponding point $Q_i$ on
$S^1(\CP_*)$.  Then as usual, the separating WKB curves divide
$C'$ into cells foliated by generic WKB curves, and the edges of $T_\WKB$ consist of one
generic WKB curve from each cell.  Note that there are generic WKB curves which
sit entirely in an arbitrarily small neighborhood of the irregular singularity, and connect adjacent
WKB rays, as shown in Figure \ref{fig:petals}.  Among
the edges of $T_\WKB$ there are $L$ such curves, to be identified with the boundary edges
mentioned above which connect adjacent $Q_i$.
Each such edge bounds a petal-shaped region touching $\CP_*$;
the disc $D(\CP_*)$ is identified with the union of these $L$ petals.

\item The decoration of $T_\WKB$ at a vertex $Q_j$ is obtained by choosing
the flat section which becomes exponentially small when following the WKB ray through $Q_j$ going
into $\CP_*$.  Let us describe this section a bit more precisely.
Put $\CP_*$ at $z = \infty$ and suppose $\lambda^2 \sim z^{L-2} dz^2$ there.
The $L$ WKB rays are located at
\begin{equation}\label{eq:PolyRays}
r_j := \left\{ \arg(z) = \frac{2}{L} \vartheta + \frac{2\pi
j}{L} \right\}, \quad j=1, \dots, L.
\end{equation}
The formal asymptotics of flat sections are of the form
\begin{equation}
\exp \left[ \pm \frac{2}{L}\frac{R}{\zeta} z^{\half L} +
\cdots \right] s_{const}
\end{equation}
as $z \to \infty$.  In particular their norm is controlled by the sign of
the real part of the exponential, which changes across the Stokes rays
\begin{equation}
{\rm arg}(z) = \frac{2}{L}\vartheta_\zeta + \frac{(2j+1)\pi}{L},
\qquad j = 0, \dots, L-1,
\end{equation}
where we defined $\vartheta_\zeta = \arg \zeta$.  So long as
$e^{i\vartheta} \not=- e^{i\vartheta_\zeta}$, each of the sectors
bounded by these Stokes rays contains a unique WKB ray $r_j$.
Along this WKB ray the norm of a general flat section is asymptotic to
\begin{equation} \label{eq:norm-sign}
\exp\left[ \pm \frac{ 2 R}{L} \frac{\vert z\vert^{L/2}}{\vert
\zeta\vert} \Re \left(e^{i (\vartheta + \pi j - \vartheta_\zeta)}
\right) + \cdots \right].
\end{equation}
If we choose the $\pm$ sign in \eqref{eq:norm-sign} opposite to the
sign of ${\Re}\left(e^{i (\vartheta + \pi j - \vartheta_\zeta)}
\right)$, then this norm is exponentially small as $z \to \infty$
along $r_j$, for any $\zeta \in \IH_\vartheta$. So far we have just
discussed the formal asymptotics, but it is an important principle
that there exists a unique flat section (called the ``small
section'') whose norm indeed has this exponentially small asymptotic
behavior along $r_j$, for $\zeta \in \IH_\vartheta$.  We choose this
small section to be the decoration of $T_\WKB(\vartheta)$ at the WKB
ray $r_j$.  It is canonically determined by $\vartheta$ and $j$.  A
subtle point is that this small section generally has nontrivial
monodromy as $\zeta$ goes around $0$; it follows that it really
depends on $\vartheta \in \IR$, not just $\vartheta \in \IR / 2 \pi
\IZ$.

\item The definitions of $\gamma^\vartheta_E$ and $\CX^\vartheta_{\gamma}$ proceed precisely
as for the case with only regular singular points. We do not define
this cycle when $E$ is a boundary edge.

\item We define a pop at an irregular singular point $\CP_*$
to be the action by $1$ on the $\IZ$-torsor of decorations at
$\CP_*$. To fix conventions, if  $\dots, Q_j, Q_{j+1}, \dots$ are ordered
clockwise and the decoration associates to them the sections $\dots, s_n,
s_{n+1}, \dots$ then after the pop we associate to them the sections $\dots,
s_{n-1}, s_{n}, \dots$ .  If we replace
$\vartheta \to \vartheta + \pi$, then the decoration of $T_\WKB(\vartheta)$
at each irregular singular point undergoes a pop.

\item It can happen that a sequence of flips produces a new
triangulation which differs from the original one only by a rotation of
one of the boundary circles by $2\pi/L$, or, equivalently, a cyclic
permutation of the $Q_i$.  This is identical to the effect of a single
pop at the irregular vertex.  This reflects a relation among morphisms
in the groupoid of decorated triangulations.

\end{enumerate}

\section{Scaling limits of linear $SU(2)$ quivers:  the case of
one irregular singular point}\label{section:Superconformal}

\insfig{3.5in}{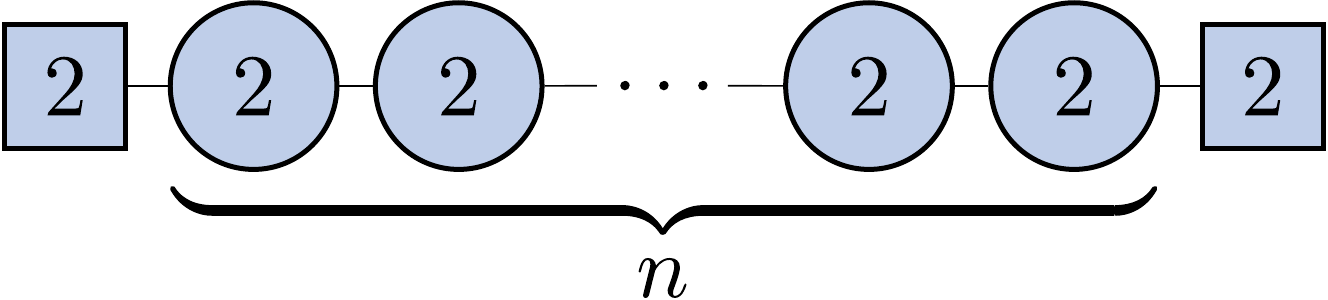}{Linear quiver for the theories
considered in Section \protect\ref{section:Superconformal}.}

In this section we illustrate some of the considerations of Sections
\ref{sec:fg-coordinates}-\ref{sec:irregular} for some particularly simple
theories.  These theories are obtained as certain scaling limits of linear quivers of $n$ $SU(2)$
gauge groups, with two fundamental hypermultiplets for each of the first and last gauge groups,
as shown in Figure \ref{fig:quiver-of-twos}.
The corresponding Hitchin systems have gauge group $SU(2)$ and (after the scaling limit)
only one singularity, an irregular one, on $C = \IC\IP^1$.

\subsection{Linear $SU(2)$ quivers and their parameter spaces}

This theory has $3n+3$ physical parameters:  the UV Lagrangian involves
$n$ gauge couplings, $n-1$ bifundamental masses and $4$ fundamental
masses, while the Coulomb branch is parameterized by $n$ vevs $\langle {\Tr}
(\Phi^{(\a)})^2 \rangle$.

Recall from Section \ref{sec:physics} that the corresponding Seiberg-Witten curve is $F(t, v) = 0$, where
\begin{equation}
F(t,v) = \sum_{\a=0}^{n+1} q_\a (v) t^{n+1-\a} = p_0(t) v^2 + p_1(t)
v + p_2(t) \end{equation} Moreover, we saw that one can parameterize
(with some redundancy)
\begin{equation}
q_\alpha(v) = c_\alpha (v^2 - \mu_\alpha v - u_\alpha).
\end{equation}
In the weak-coupling regime the couplings are determined from the
$c_\alpha$ and the masses from $\mu_\alpha$ and $u_\alpha$, while the
Coulomb branch is parameterized by the $u_\alpha$.
Finally, as we saw in \eqref{eq:su2exple}, after
factoring out the center-of-mass degree of freedom the Seiberg-Witten curve
becomes the spectral curve for an $SU(2)$ Hitchin system with
\begin{equation}
\lambda^2 = \half \frac{p_1(t)^2 - 4 p_2(t) p_0(t) }{(tp_0(t))^2} dt^2.
\end{equation}
This quadratic differential has double poles at $t = 0$ and $t = \infty$ as well as
at the $n+1$ zeroes of $p_0(t)$.

The above description somewhat obscures the S-duality properties of
the problem \cite{Witten:1997sc,Gaiotto:2009we}.  The physical gauge
couplings really only depend on a point in the moduli space of
spheres with $n+3$ marked points. Moreover, there is no S-duality
invariant distinction between the bifundamental and fundamental mass
parameters. These facts suggest that we should treat the $n+3$
singularities and mass parameters more democratically. This can be
achieved by introducing some redundancy into the description, making
a general fractional linear transformation from $[t:1] \in \IC
\IP^1$ to a new coordinate $z$. After making such a transformation
we have (subscripts on polynomials indicate their degree)
\begin{equation}\label{eq:quad-diff-1}
\lambda^2 = \frac{Q_{2n+2}(z)}{(D_{n+3}(z))^2} (dz)^2.
\end{equation}
After subtracting out 3 parameters for the $SL(2,\IC)$ action and 1 parameter for simultaneous rescaling
of $Q$ and $D$, there are still $3n + 7 - 4 = 3n+3$ physical parameters.

As we have mentioned, the physical parameters are not on an equal
footing. The couplings and masses specify the UV theory while
the Coulomb branch parameters specify the
vacuum.  To bring out this distinction it is useful to parameterize $\lambda^2$
in a slightly different way.

The $n+3$ mass parameters can be characterized as the residues of the $n+3$ poles
in $\lambda^2$.  (We will not consider the possibility that some of these poles collide,
corresponding to a strong coupling singularity of the physical theory.)
If the poles are located at $z_a$ for $a = 1, \dots, n+3$, and we assume all $z_a \neq \infty$,
we can write $\lambda^2$ in the form
\begin{equation}\label{eq:part-frac}
\lambda^2 = \sum_{a=1}^{n+3} \left(
\frac{m_a^2}{(z-z_a)^2} + \frac{c_a}{z-z_a}\right) dz^{2}.
\end{equation}
Requiring that $\lambda^2$ is regular at $z = \infty$ gives
three conditions on the $c_a$,
\begin{eqnarray}\label{eq:c-conds}
\begin{aligned}
\sum_{a=1}^{n+3} c_a & = 0, \\
\sum_{a=1}^{n+3} z_a c_a & =   - \sum_{a=1}^{n+3} m_a^2, \\
\sum_{a=1}^{n+3} z_a^2 c_a & =   - 2\sum_{a=1}^{n+3} m_a^2z_a.
\end{aligned}
\end{eqnarray}
The Coulomb branch $\CB$ is the space of $c_a$ solving
\eqref{eq:c-conds}. Because \eqref{eq:c-conds} is an inhomogeneous
linear equation for the $c_a$, it is an affine space of dimension
$n$.

(Another viewpoint is that, once we have specified the couplings
through $z_a$, the space of theories with arbitrary masses and vacua
is a linear space --- the space of polynomials $Q_{2n+2}(z)$ --- which can also be
thought of as $H^0(\IC \IP^1, K^{\otimes 2} \otimes \CO(2
\CP_1 + \cdots + 2\CP_{n+3})  )$. Fixing the masses then fixes an
affine subspace of this space.)

\subsection{Scaling limit}

The singularities of the Coulomb branch for generic masses occur
when two or more roots of $Q_{2n+2}(z)$ coincide. At these points
the metric on the Coulomb branch is singular and fluctuations around
this locus have infinite action. The reason is, of course, that at
least one BPS state becomes massless on this locus. The zeroes of
$Q_{2n+2}(z)$ occur on the discriminant locus $\CD$, which has
strata $\CD_k$ where precisely $k$ zeroes coincide. We will now
focus on a stratum $\CD_N$. Near this locus some number of BPS
states become light, and we would like to focus on the low-energy
physics of these states.

Accordingly, we consider the following scaling limit. Choose a point
$\CP\in C$ and a local coordinate $z$ with $z(\CP)=0$. The relevant
neighborhood in the space of $Q$'s can be parameterized by
$Q_{2n+2}(z) = \prod_{i=1}^N(z-\epsilon \theta_i) \tilde
Q_{2n+2-N}(z)$, with $\tilde Q_{2n+2-N}(0)\not=0$.  At $\epsilon \to
0$, $N$ zeroes of $Q$ collide at $z = 0$. The mass of a typical BPS
state associated with a string connecting two of these colliding
zeroes is of order $\epsilon^{(N+2)/2}$. We thus consider physics at
energies of this scale and below. We also define a scaling region of
the Riemann surface $C$ (i.e. a scaling region on the M5
worldvolume) by $z = \epsilon \tilde z$.  We will only be
considering fluctuations localized in this region.  After a suitable
rescaling of the Seiberg-Witten differential $\lambda$ and $\tilde
z$, we are therefore considering a theory with
\begin{equation}\label{eq:PolyDegen}
\lambda^2 = P_N(z) dz^2,
\end{equation}
where $P_N$ is monic of order $N$, and our scaling patch of the
M5-brane worldvolume has been blown up to the entire complex plane.

Many of the deformations of the original theory become
non-normalizable in the scaling limit; however, some deformations remain normalizable.
These are the polynomial deformations $P_N \to P_N + \delta P_N$
such that
\begin{equation}
\int_{\IC} \vert \delta \lambda \vert^2 = \int_{\IC} \biggl\vert
\frac{\delta P_N(z) }{2 \sqrt{P_N(z)}} \biggr\vert^2 d^2 z < \infty.
\end{equation}
Since the zeroes of $P_N(z)$ are simple, a divergence could only
come from $z \to \infty$, and hence the normalizability condition is
\begin{equation}
\delta P_N = \sum_{i < \half(N-2)} \delta p_i z^i.
\end{equation}
There is another useful point of view on this normalizability condition,
namely, we fix the singular part of the expansion of
$\sqrt{P_N(z)}$ around $z = \infty$:
\begin{equation}\label{eq:AsympOfP}
\sqrt{P_N(z)} = \Delta(z) + o(z^{-1})
\end{equation}
for some fixed $\Delta(z)$ (determined by the UV theory which we have mostly
discarded).

There is an important difference between even and odd $N$.
When $N$ is even, $\Delta(z)$ is an expansion in integer powers, ending with a simple pole; we
denote its residue by $m$.  Varying $m$ is a ``log-normalizable'' deformation,
with $i = \half(N-2)$.   As we will see
presently when we pass from Hitchin systems to flat connections, $m$ specifies the
formal monodromy at $\infty$.  In the case of $N$ odd there is no such log-normalizable
deformation.

After the scaling limit there are still singular loci on the Coulomb branch where
some $\theta_i = \theta_j$, i.e. where $P_N(z)$ has multiple zeroes.  These
loci are intersections between higher-dimensional strata of $\CD$ and the
scaling region around $\CD_N$.  At these loci some BPS states become massless.
Moreover, in some cases the BPS states which become massless are mutually nonlocal
ones, leading to the kinds of theories studied in \cite{Argyres:1995jj,Argyres:1995xn}.

To reach the most extreme case, we could tune the nonnormalizable
parameters so that we are considering normalizable
deformations of $\lambda^2 = z^N (dz)^2$.
The deep IR physics at this point in moduli
space is described by a superconformal field theory
\cite{Argyres:1995jj,Argyres:1995xn}, and one can define a UV
complete quantum field theory with a finite BPS spectrum by perturbing away from this
theory with the normalizable deformations identified above.  We
are then studying the Seiberg-Witten curves and BPS states of those theories.
From the point of view of the original UV theory
specified by the linear quiver, we are
focusing on a low-energy subsector in a region of
moduli space where a number of BPS
states are becoming parameterically light.
As we will see below, the number of
such states is bounded by $\half N(N-1)$
\cite{Shapere:1999xr}.

\subsection{Hitchin system} \label{subsec:HitchinForAD}

In the previous subsection we have described a degeneration of
the spectral curve of our Hitchin system.  To complete the
discussion we now explain the corresponding degeneration of the
Hitchin system.  The new Hitchin system is defined on
$\IC\IP^1$ and has a single irregular singularity at
$z=\infty$.  To be specific, when $N$ is even we have
\begin{equation}
A_0 = \begin{pmatrix} -m^{(3)} & 0 \\ 0 & m^{(3)}
\end{pmatrix} \left( \frac{dz}{z} - \frac{d \bar z}{\bar z} \right)
\end{equation}
and
\begin{equation}
\varphi_0 = \begin{pmatrix} \Delta(z) & 0 \\  0 & - \Delta(z)
\end{pmatrix},
\end{equation}
with $\Delta(z)$ as defined in \eqref{eq:AsympOfP}. Since $\Delta(z)
\sim z^{N/2} + \cdots + \frac{m}{z}$, the formal monodromy under
counterclockwise rotation is
\begin{equation}\label{eq:MonodromyAtInfinity}
\exp \left(2\pi i \left[ \frac{mR}{\zeta} - 2 m^{(3)} - \bar m R
\zeta \right]\sigma^3 \right).
\end{equation}
We will denote its eigenvalues as $\mu^{\pm 1}$.

When $N$ is odd there is no analog of the mass parameters $m,
m^{(3)}, \bar m$.  Instead, we have a generalization of
\eqref{eq:ISPatInfty},
\begin{align} \label{eq:NoddBC}
\varphi_0 & =\Delta(z) \begin{pmatrix} 0 & (\bar z/z)^{1/4} \\
(z/\bar z)^{1/4} & 0 \end{pmatrix}, \\
 A_0 & = \frac{1}{8} \sigma^3
\left(\frac{dz}{z} - \frac{d\bar z}{\bar z}\right).
\end{align}

\subsection{Examples}

We now illustrate various aspects of our formalism in the cases where $P(z)$ is a polynomial
of degree $N=1, 2, 3, 4$.  Along the way we will encounter
nice ``real-world'' examples of wall-crossing formulae involving finite collections of BPS states.

Because we have an irregular singularity at $z = \infty$, to define the WKB triangulation we will have to
use the modified rules of Section \ref{sec:irregular}.  Applying these rules to the present
case, we will obtain a triangulation of a surface $C'$ which is $\IC\IP^1$ with a disc cut out
around $z = \infty$, with vertices at marked points on that disc, corresponding to the loci where WKB curves
run off to $z = \infty$.
As we have explained in general in Section \ref{sec:WKB-triangulation},
for special values of $\vartheta$ finite WKB curves
will appear, corresponding to the BPS states of our field theory, and
causing $T_\WKB(\vartheta)$ to jump.

\subsubsection{$N=1$}

We begin with $N=1$, so
\begin{equation}
P(z) = z.
\end{equation}
There are no deformations --- normalizable or otherwise --- so $\CB$ is just a single point.  Moreover there is a
corresponding unique solution to the Hitchin equations, so $\CM$ is also just a single point.
This solution can be written explicitly:
\begin{align}
\label{eq:ExactSol}
\varphi & = \begin{pmatrix} 0 &  \vert z \vert^{1/2} e^h \\
\frac{z}{\vert z \vert^{1/2}} e^{-h} & 0 \\ \end{pmatrix},  \\
A & = \left( \frac{1}{8} + \frac{1}{4} \vert z\vert \frac{d}{d\vert
z\vert} h \right) \sigma^3 \left( \frac{dz}{z}- \frac{d\bar z}{\bar
z}
\right).
\end{align}
Here $h(\abs{z})$ is a Painleve III transcendent:  writing $r := \frac{8 R}{3} \vert z\vert^{3/2}$, it obeys
\begin{equation}
\left(\frac{d^2}{dr^2} + \frac{1}{r} \frac{d}{dr}\right)h = \half
\sinh(2h),
\end{equation}
with boundary condition $h(r) \to - \frac{1}{3} \log r + const$ for
$r\to 0$.  It can be shown \cite{McCoy:1976cd} that $h(r) \to
\pi^{-1} K_0(r)$ for $r\to \infty$.  We will return to this solution
in Section \ref{sec:R-To-Infinity} below.

There are $N + 2 = 3$ WKB rays around $z = \infty$, given by
\eqref{eq:PolyRays}, and a single turning point at $z=0$.  The
$\CA$-flatness equation somewhat resembles the Airy equation: in
particular there is a small flat section $s_k$ along each of the
three rays $r_k$. The WKB triangulation $T_\WKB(\vartheta)$ consists
of a single triangle, which rotates in the $z$-plane as
$\vartheta$ varies.  There are no flips and correspondingly no BPS
states.  There are no Fock-Goncharov coordinates,
since the triangulation has no internal edges.

\subsubsection{$N=2$}

Next consider the case $N=2$.  We write
\begin{equation}
P(z) = z^2 + 2m,
\end{equation}
and hence $\Delta(z) = z + \frac{m}{z}$.  The parameter $m$ is
log-normalizable.  The Coulomb branch $\CB$ is a single point and there is no
$U(1)$ gauge field.  The moduli space $\CM$ is also a single point.

The spectral curve $\Sigma$ has genus zero and two punctures lying over $z = \infty$.
In this case $H_1(\Sigma; \IZ)$ is one-dimensional and odd under the exchange of the sheets,
so $\hat \Gamma = H_1(\Sigma; \IZ) \simeq \IZ$.  This one-dimensional lattice is generated
by a single flavor charge, with $m$ the corresponding mass parameter.

In this case we do not know explicit solutions to the Hitchin
equations.  Nevertheless, following the general recipe of the
previous sections, let us examine the WKB triangulation. There are
$N + 2 = 4$ WKB rays, and $N = 2$ turning points.  The generic
behavior of the WKB triangulation is as shown in Figure
\ref{fig:N2plot}. Combinatorially the four boundary edges make up a
square, and the single internal edge gives a triangulation of that
square. \insfig{3.3in}{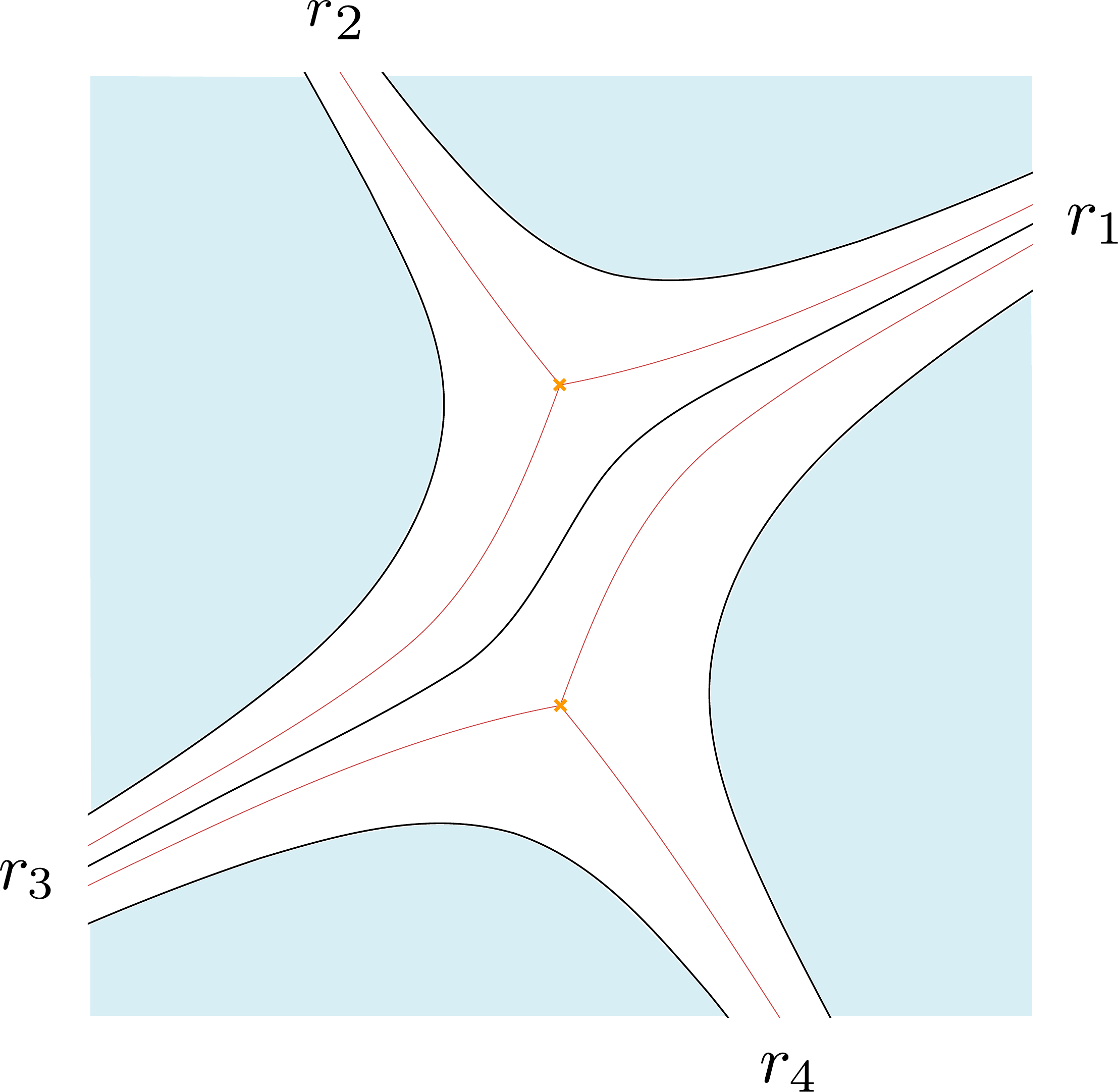}{A generic WKB triangulation for
$N=2$.  Separating WKB curves are shown in red, while the generic WKB curves
chosen for the edges of the triangulation are black.  The shaded region is the union
of the four ``petals'' which we cut out around the irregular singularity at $z = \infty$.}

Since $\hat\Gamma$ is one-dimensional, there is only one independent
Darboux coordinate $\CX_{\gamma}$.  In fact we claim that it is
equal to $\mu$ given in \eqref{eq:MonodromyAtInfinity}.  (Indeed
this is the only reasonable function of $m, m^{(3)}$ which carries
one unit of flavor charge.) To show this we use the asymptotics of
$s_i$. Since two WKB rays are separated by a Stokes ray,   in an
angular sector around $r_2$, $s_1$ and $s_3$ are exponentially
growing and hence (if normalized appropriately) differ by a multiple
of the small solution $s_2$,
\begin{equation}
s_1 = s_3 + a s_2
\end{equation}
for some constant $a$. Similarly, around $r_4$,
$s_1$ and $s_3$ are exponentially large, and hence (after taking
into account the formal monodromy, i.e. the monodromy of the asymptotics)
\begin{equation}
s_1 = \mu s_3 + b s_4
\end{equation}
for some constant $b$.  Since $s_1, \dots, s_4$ are single-valued
these relations hold throughout the plane, and can be
used to simplify the Fock-Goncharov coordinate to
\begin{equation}
\CX_E =  -\frac{(s_1 \wedge s_2) (s_3 \wedge s_4)}{(s_2 \wedge s_3) (s_4
\wedge s_1)} = -\mu^{-1}.
\end{equation}

As $\vartheta$ traverses an arc of length $\pi$, we encounter
one critical value $\vartheta = \vartheta_c$ where a finite WKB
curve appears connecting the two turning points. The WKB
triangulation $T_\WKB(\vartheta)$ experiences a flip at this
$\vartheta$.  See Figure \ref{fig:square-evolution}. The flip
transforms the single Fock-Goncharov coordinate by $\CX^T_{E} =
1/\CX^{T'}_{E}$ (where in both cases $E$ denotes the single
internal edge). On the other hand $\CX^\vartheta_\gamma$ is
unchanged and equal to $-\mu^{-1}$.

\insfig{5.5in}{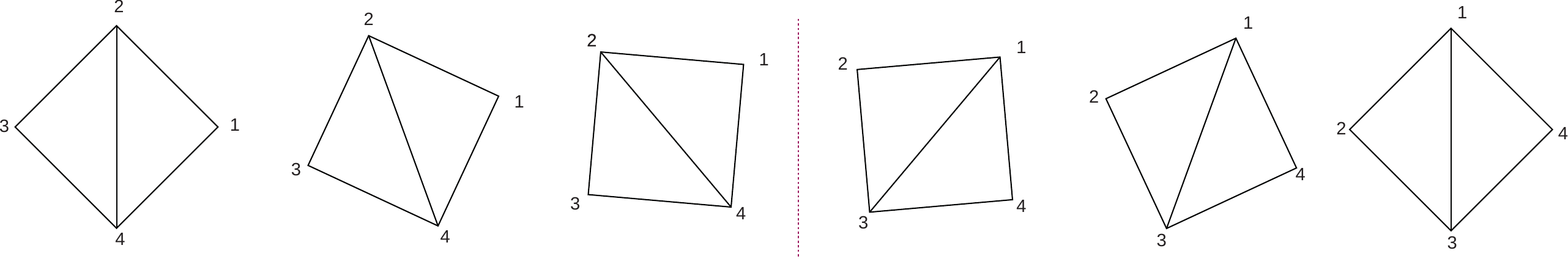}{As the phase $\vartheta$ varies
from $0$ to $\pi$, the WKB triangulation evolves simply, with a
single flip (dotted line). Notice that the initial and final
triangulation are identical, but for a relabeling of the WKB rays.}

What is the meaning of this single flip?
It means that in the scaling limit discussed above we keep a
single BPS particle, of flavor charge $1$.  Based on this
local model, we can make a more general prediction:  near the simplest
singularities in the Coulomb branch where two turning points collide, a single BPS
particle becomes light, corresponding to a BPS string stretched
between the two turning points.

\subsubsection{Intermission: $N=2$ and periodic Taub-NUT space} \label{sec:pnut}

At this point we can make contact with the most fundamental example of a quantum-corrected Coulomb branch
considered in our previous paper \cite{Gaiotto:2008cd}.
This will provide a result which has recently been of use in the
computation of gluon scattering amplitudes \cite{Alday:2009yn}.

The fact that in the $N=2$ example we encountered only a
log-normalizable deformation parameter, and correspondingly $\CM$ is trivial,
might initially be a bit disappointing.  One might have
expected to glean some information about the behavior
of the exact metric on the three-dimensional Coulomb branch near a
singularity of the four-dimensional theory.

Physically, we know what
to expect.  In the original full theory, the flavor symmetry of the
single light BPS particle is actually gauged.  If we simply
ignore the effect of the massive BPS particles, the Coulomb branch
of a $U(1)$ gauge theory on $\IR^3 \times S^1$ coupled to a single
light hypermultiplet is well known as the ``periodic Taub-NUT'' metric
\cite{Ooguri:1996me,Seiberg:1996ns}, and played a central role in the analysis
of \cite{Gaiotto:2008cd}.  It is not a complete hyperk\"ahler metric;
rather, it is well defined up to an arbitrary mass scale $\Lambda$.

In \cite{Gaiotto:2008cd} the periodic Taub-NUT metric was
described in terms of a pair of holomorphic Darboux
coordinates $(\CX_e, \CX_m)$.  $\CX_e$ coincides with the $\CX_\gamma$
discussed here.  The more interesting coordinate is $\CX_m$, which experiences Stokes phenomena,
with Stokes factors $\CK_\gamma$, $\CK_{-\gamma}$ corresponding to the
single massive hypermultiplet and its CPT conjugate.

Can we find a role for $\CX_m$ in the present context?  At
the level of the moduli space of flat connections, it is natural
to consider letting the monodromy parameters $\mu_a$ at the singularities vary.  Indeed,
$\mu_a$ can be interpreted as the complex moment
map for the residual gauge transformation at the $a$-th singularity.
We can let $\mu_a$ vary and at the same time restrict the
gauge transformations to approach the identity at the singularities.
This defines an enlarged symplectic manifold, with two extra complex
coordinates for each singularity \cite{MR1864833}.  In our present
context that means enlarging $\CM$ from complex dimension $0$ to complex
dimension $2$.  As we will now see, $\CX_e$ and $\CX_m$ will be realized as functions on
this extended $\CM$.

Once the gauge transformations at the singularity
have been restricted to the identity, it makes gauge-invariant sense
to pick a specific choice of overall normalization for the small
solutions, for example by prescribing exactly the asymptotic
behavior at the singularity:
\begin{equation} \label{eq:gaugefix1}
s_i(z,\zeta) \exp \left[ \frac{R}{\zeta}\left(\frac{z^2}{2} + m \log
z\right)  -m^{(3)} (\log z - \log \bar z) + R\zeta\left(\frac{\bar
z^2}{2} + \bar m \log \bar z\right)\right] = \begin{pmatrix} 1 \\ 0
\end{pmatrix} + O(1/z)
\end{equation}
or
\begin{equation} \label{eq:gaugefix2}
s_i(z,\zeta) \exp \left[ - \frac{R}{\zeta}\left(\frac{z^2}{2} + m
\log z\right)  + m^{(3)} (\log z - \log \bar z) -
R\zeta\left(\frac{\bar z^2}{2} + \bar m \log \bar z\right) \right] =
\begin{pmatrix} 0 \\ 1 \end{pmatrix} + O(1/z).
\end{equation}
In the example we are now considering, this allows us to make gauge-invariant sense of
individual elements of the Stokes matrices, for example the $a$ in
$s_1 = s_3 + a s_2$.  (Notice that this relation can be consistent with \eqref{eq:gaugefix1}, \eqref{eq:gaugefix2}
only if we make a proper choice of branch cut in $\log z$; we place the cut slightly below $r_3$.)
We think of $a$ as a ``ratio'' (as opposed to ``cross-ratio''),
\begin{equation}
a = \frac{s_1 \wedge s_3}{s_2 \wedge s_3}.
\end{equation}

If the value of $\arg \zeta$ falls in the half plane $\IH_\vartheta$ around a
$\vartheta$ for which there is an edge between $r_1$ and $r_3$,
we can compute the asymptotics of $a$ for small $\zeta$ with
the WKB method. The calculation is straightforward, and gives
our usual exponential form $a \sim c \exp (\pi R \zeta^{-1} Z_{\tilde\gamma})$,
where $Z_{\tilde\gamma}$ is a regularized period of $\pi^{-1} \lambda$,
on a cycle $\tilde \gamma$ which goes from infinity along $r_1$
to infinity along $r_2$, winding around the turning point in
the triangle $123$:
\begin{equation}
Z_{\tilde \gamma} = 2 \int_{z=\sqrt{-2m}}^L \sqrt{z^2 + 2 m}  - L^2-2m \log L = - m \log \frac{m}{-2e} + \CO\left(\frac{1}{L}\right).
\end{equation}
This path intersects $\gamma$ once.  Motivated by these asymptotics we define
$\CX^\vartheta_{\tilde \gamma} = a$ for this value of $\vartheta$.

As we increase the phase $\vartheta$, beyond the critical value $\vartheta_c$ where
the triangulation flips, we should consider instead an $a'$ defined
by $s_4 = s_2 + a' s_1$, or
\begin{equation}
a' = \frac{s_4 \wedge s_2}{s_1 \wedge s_2}.
\end{equation}
The WKB computation of $a'$ is identical to that for $a$, but for a crucial
overall sign, so here the asymptotics suggest that we should define
$\CX^\vartheta_{\tilde \gamma}$ in terms of $1/a'$.
Recall the relations $s_1 = s_3 + a s_2$ and
$s_1 = \mu s_3 + b s_4$.  With our choice of cut for $\log z$,
if we continue these relations all the way to $r_1$, we see
that $b s_4 = (1-\mu) s_1 + a \mu s_2$. Comparing with $s_4=
s_2 + a' s_1$ we see that $b = a \mu$ and $a' = - a^{-1}
(1-\mu^{-1})$.
Then, if ${\rm Re}(im/\zeta)>0$, the asymptotic behavior of
$\CX_{\tilde \gamma}$ remains consistent if we define
$\CX^\vartheta_{\tilde \gamma} = -\frac{1}{a'}$ for this
$\vartheta$.

Now we have defined our coordinates $\CX^\vartheta_\gamma$ and $\CX^\vartheta_{\tilde\gamma}$,
for $\vartheta$ on both sides of the flip associated to the BPS state of charge $\gamma$.
From the explicit formulas above for these coordinates, it quickly follows
that the coordinate transformation across the flip
coincides with $\CK_\gamma$!  Similarly, the coordinate
transformation induced by the BPS state of charge $-\gamma$ turns out to be
$\CK_{-\gamma}$.  With the identification $a^{PT} = -2 i m$ between
the coordinate $a^{PT}$ on the base of periodic Taub-NUT and our
mass parameter, we see that $\CX_\gamma, \CX_{\tilde \gamma}$ have
the same asymptotic behavior as $\CX_e, \CX_m$ (after an appropriate
choice of cutoff) and also transform in the same way as $\vartheta$
crosses the BPS rays.  The uniqueness of the solution of the Riemann-Hilbert problem in
\cite{Gaiotto:2008cd} then guarantees that $\CX_\gamma, \CX_{\tilde
\gamma}$ coincide with $\CX_e, \CX_m$. In particular, it follows
that the formulas of \cite{Gaiotto:2008cd} can be used to compute
$\CX_{\tilde \gamma}$.

We have seen that the Stokes data for the auxiliary flatness equations
associated to this Hitchin system can be computed exactly in terms of the function
$\CX_m(\zeta)$, even though we cannot compute the solution to the
Hitchin equations!

So far we have explained how the functions $\CX_e$ and $\CX_m$ arise in this example,
by considering them as functions on an extension of the moduli space of flat
connections, obtained by introducing some extra parameters associated to the singularity.
But we have not considered this moduli space as a \hk manifold, and hence we have
not found a precise role for the periodic Taub-NUT metric in the context of Hitchin systems.
To do so, we could try letting
the three mass parameters $m, m^{(3)}, \bar{m}$ at the singularity vary, and then
adding one more circle-valued parameter by considering only gauge transformations which reduce to
the identity at the singularity.  However, it is not clear that we can define an \hk metric on the
resulting extended moduli space; the metric and \hk forms diverge
when evaluated on variations of the masses.  One could try to
regularize the divergence by removing a small disk around
each singularity. The metric would depend logarithmically on the
cutoff radius, exactly as in the case of the periodic Taub-NUT metric, and
possibly be incomplete.  As our twistor construction of the \hk metric is completely
local over the base, this potential incompleteness is not an obstacle: the
metric can be computed from the coordinates $\CX_\gamma, \CX_{\tilde \gamma}$
in our standard way, and because these coincide with $\CX_e, \CX_m$, the metric will
coincide with the periodic Taub-NUT metric!

This approach could be extended to more complicated Hitchin systems,
but we will not pursue it further in this paper.

\subsubsection{$N=3$} \label{sec:N3}

We now come to the case $N=3$, where we will first encounter a
wall-crossing formula.  We write
\begin{equation} \label{eq:N3}
P(z) = z^3 - 3 \Lambda^2 z + u,
\end{equation}
where $\Lambda$ is a non-normalizable parameter defining the theory,
and $u$ is a normalizable modulus parameterizing the Coulomb branch $\CB$.
The discriminant of $P(z)$ is $27
((2\Lambda^3)^2-u^2)$, so there are two singular points on $\CB$ at $u = \pm
2 \Lambda^3$, where two zeroes of $P(z)$ collide.  (We take
$\Lambda \neq 0$, so that there is no $u$ for which all three zeroes collide.)

As usual, there is a local system of lattices $\hat\Gamma$ over $\CB$, given by
the odd part of the homology of the family of punctured elliptic curves $\Sigma_u$.
In this case $\hat\Gamma$ has rank $2$.
Altogether $\CB$ strongly resembles the well-known $u$-plane of the $SU(2)$ Seiberg-Witten theory with
$N_f = 0$ (to be considered below in Section \ref{sec:nf0}.)

\insfig{2.7in}{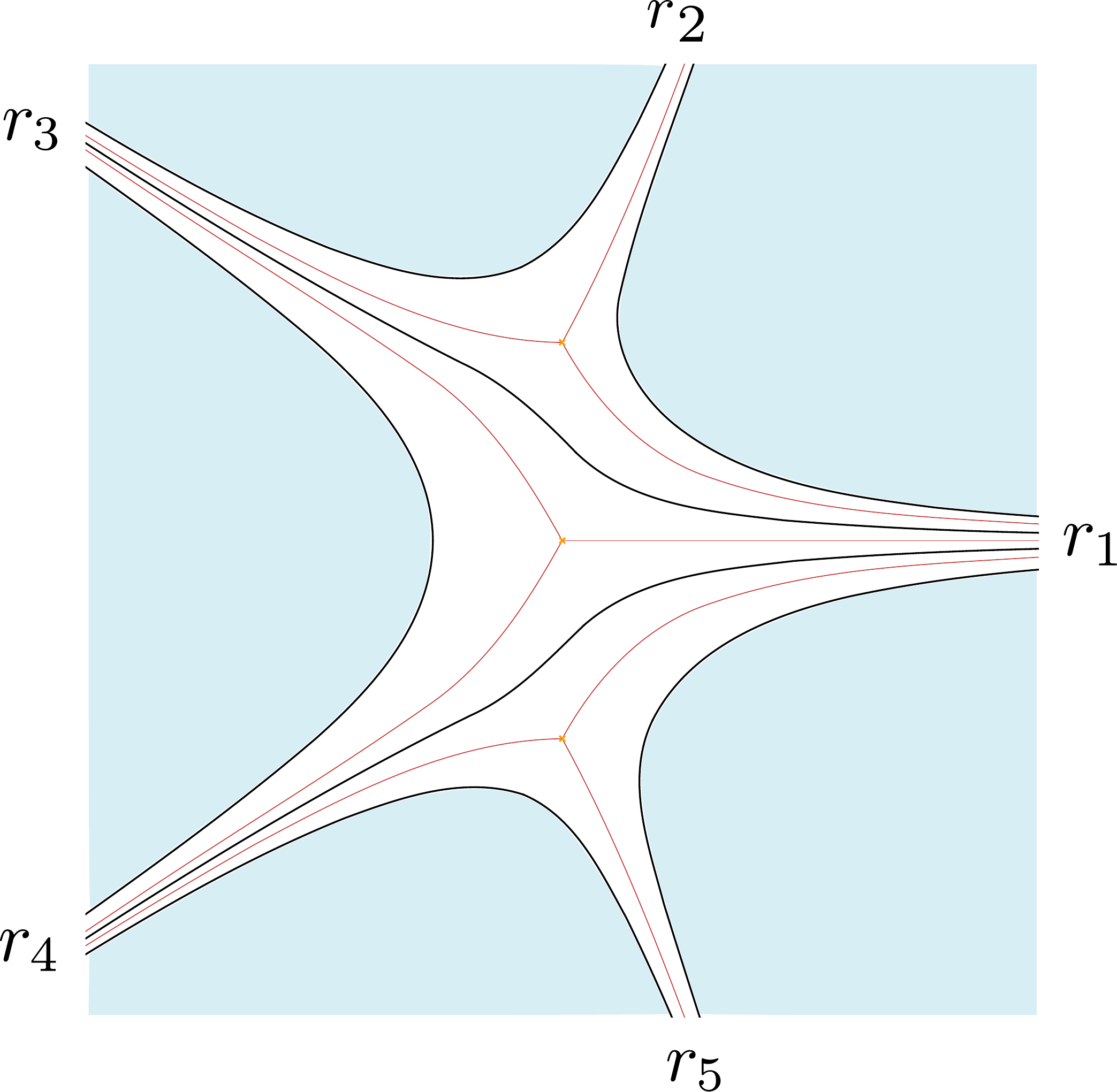}{A sample WKB triangulation in the $N = 3$ example, at generic $\vartheta$.}

Now let us consider the WKB triangulation. There are five WKB rays
(and correspondingly five Stokes rays)  at $z = \infty$ and three
turning points.  A generic $T_\WKB(\vartheta, u)$ is a triangulation
of a pentagon, as depicted in Figure \ref{fig:N3plot}. We can get an
integral basis of $\hat\Gamma_u$ by taking the $\gamma_E^\vartheta$
associated to the two internal edges $E_{1,2}$ of this
triangulation. To be concrete, let us define $\{\gamma_1,
\gamma_2\}$ to be the basis so obtained from $T_\WKB(\vartheta = 0,
u = 0)$ with $\langle \gamma_1, \gamma_2 \rangle = 1$.

As we vary $\vartheta$ from $0$ to $2 \pi$ holding $u$ fixed,
$T_\WKB(\vartheta,u)$ jumps by a flip $4$ times, at
the phases of the periods $\pm Z_{\gamma_1}$ and $\pm Z_{\gamma_2}$. If we
vary $\vartheta$ over a range of $\pi$ then $T_\WKB(\vartheta,u)$ flips twice. These two
flips are generated by two independent finite WKB curves,
representing two BPS particles (the other two flips correspond to
their antiparticles). See the upper strip of Figure
\ref{fig:penta-evolution}. \insfig{5.5in}{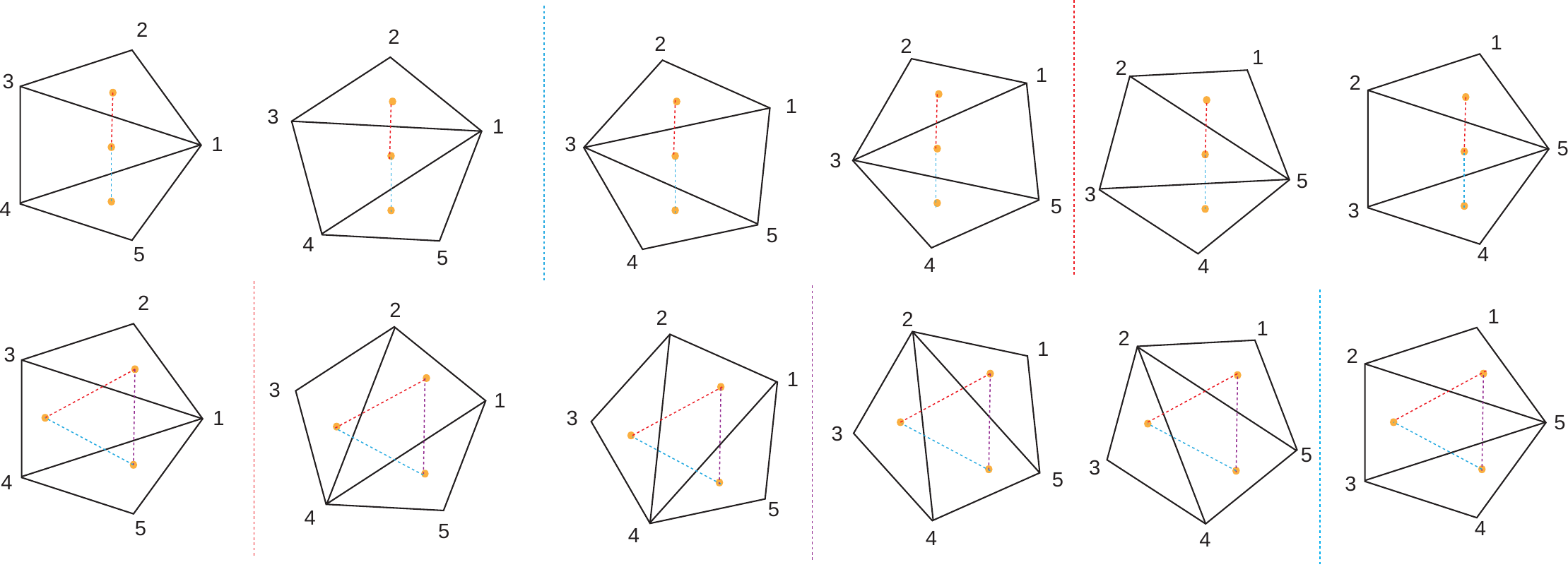}{Upper
strip: at $u=0$, as the phase $\vartheta$ varies from $0$ to $\pi$,
$T_\WKB(\vartheta, u)$ evolves simply, with two flips (dotted
lines).  Lower strip: for $u$ beyond the walls of marginal
stability, $T_\WKB$ evolves in a different manner, involving three
flips (dotted lines).  Notice that $T_\WKB(\vartheta) =
T_\WKB(\vartheta + \pi)$ except for a relabeling of the WKB rays,
which is equivalent to   a pop transformation. } At each of the
singularities in $\CB$, one of these BPS states becomes massless.
(As we remarked above, these singularities arise when a pair of
turning points collides; the massless BPS state then corresponds to
a finite WKB curve connecting this pair of turning points.)

In $\CB$ there is a single closed wall of marginal stability which
divides the $u$-plane into two connected components (again, much as
in the $SU(2)$ Seiberg-Witten theory). So far we have discussed
$u=0$, which lies inside the wall. If we consider some $u$ which
lies outside the wall the behavior of $T_\WKB(\vartheta, u)$ is
different:  there are three flips, induced by three finite WKB
curves, corresponding to three BPS states.  See the lower strip of
Figure \ref{fig:penta-evolution}. (This is most easily seen by
taking $u$ to be large.)

To be more precise, the two singular points $u= \pm 2 \Lambda^3$
divide the wall of marginal stability into two segments. Along one
segment the phases of $Z_{\gamma_{1,2}}$ align.  After crossing this
segment from inside to outside we find three BPS states, of charges
$\gamma_1, \gamma_2, \gamma_1 + \gamma_2$. If we cross the other
segment of the wall, where $Z_{\gamma_1}$ and $Z_{-\gamma_2}$ align,
we find instead BPS states of charges $\gamma_1, \gamma_2,
\gamma_1-\gamma_2$.  To see that these two results are compatible we
recall that $\hat\Gamma$ has the standard Lefschetz monodromy around
the two points $u = \pm 2 \Lambda^3$ where two zeroes of $P(z)$
collide. Generators of the clockwise monodromy around the two
singular points can be taken to be:
\begin{align}\label{eq:LocalSys-1}
M_1 &= (\gamma_1, \gamma_2) \to (\gamma_1, \gamma_2-\gamma_1), \notag \\
M_2 &= (\gamma_1, \gamma_2) \to (\gamma_1+\gamma_2, \gamma_2).
\end{align}

These three finite WKB curves persist in the spectrum as we go to arbitrarily large $\abs{u}$
(indeed there are no other walls of marginal stability where they could disappear).
We infer from this that in a more general theory, near a singularity in the Coulomb branch
where three turning points are coalescing, three light BPS particles
will typically be present, realized as BPS strings joining these three
turning points.

\insfig{4.5in}{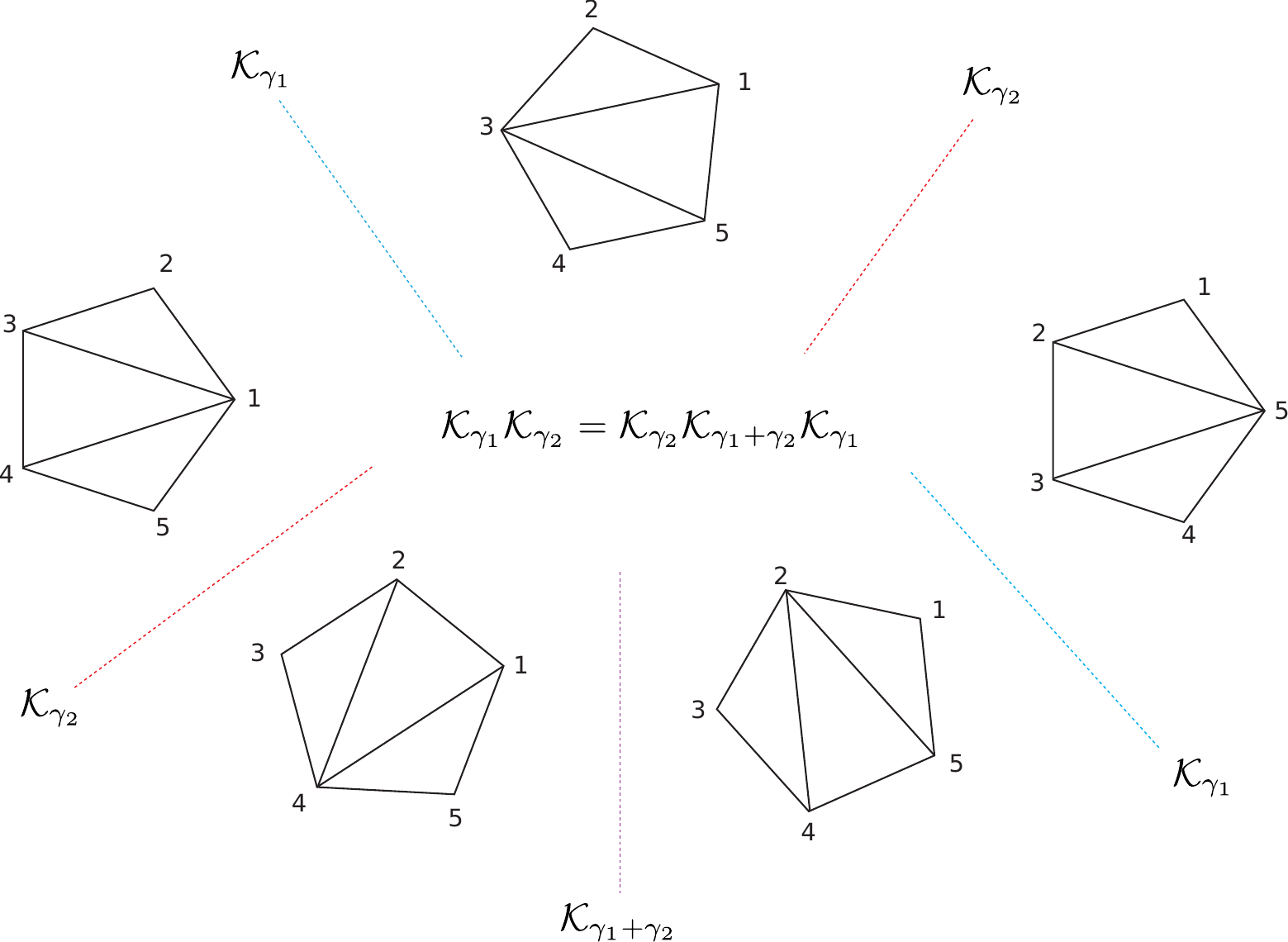}{The two different ways to go from
the triangulation at $\vartheta=0$ to the triangulation at
$\vartheta=\pi$ on the two sides of the wall must correspond to
the same symplectomorphism.  The pentagon identity follows.}

\subsubsection*{The wall-crossing formula}

Now we are ready to see a wall-crossing formula emerge.
Let $u_-$ denote a point inside the wall and $u_+$ a point outside, and fix
some $\vartheta$.
We have two coordinate systems $\CX^{\vartheta, u_-}_\gamma$
and $\CX^{\vartheta+\pi, u_+}_\gamma$ on $\CM$, which are related by
some symplectomorphism.
To evaluate this symplectomorphism, we consider the evolution of the triangulation
$T_\WKB(\vartheta, u)$ along a path from $(\vartheta, u_-)$
to $(\vartheta+\pi, u_+)$.

One possibility is to vary first $\vartheta$ to $\vartheta+\pi$ and
then deform from $u_-$ to $u_+$. As we vary $\vartheta$ the
triangulation undergoes two flips corresponding to the two BPS
states inside the wall.  As we vary $u$, the triangulation does not
jump at all, so long as no $\arg Z_{\gamma}(u)$ crosses $\vartheta$
(which we can always arrange by choosing $u_\pm$ close enough to the
wall and $\vartheta$ generic enough). So the total transformation of
the triangulation involves exactly two flips.

However there is also another possibility:  first deform from $u_-$ to $u_+$ and then
vary $\vartheta$ to $\vartheta+\pi$.  In that case the triangulation undergoes
three flips corresponding to the three BPS states outside the wall.

These two computations must give the same symplectomorphism.  This corresponds to the identity
\begin{equation}\label{eq:simple-pentagon}
\CK_{\gamma_1} \CK_{\gamma_2} = \CK_{\gamma_2} \CK_{\gamma_1+\gamma_2} \CK_{\gamma_1}
\end{equation}
for one section of the wall,
\begin{equation}
\CK_{-\gamma_2} \CK_{\gamma_1} = \CK_{\gamma_1} \CK_{\gamma_1-\gamma_2} \CK_{-\gamma_2}
\end{equation}
for the other.  One can check by direct computation that these relations
are indeed satisfied.  We will refer to either of these basic identities as ``the
pentagon identity.''

\insfig{4.0in}{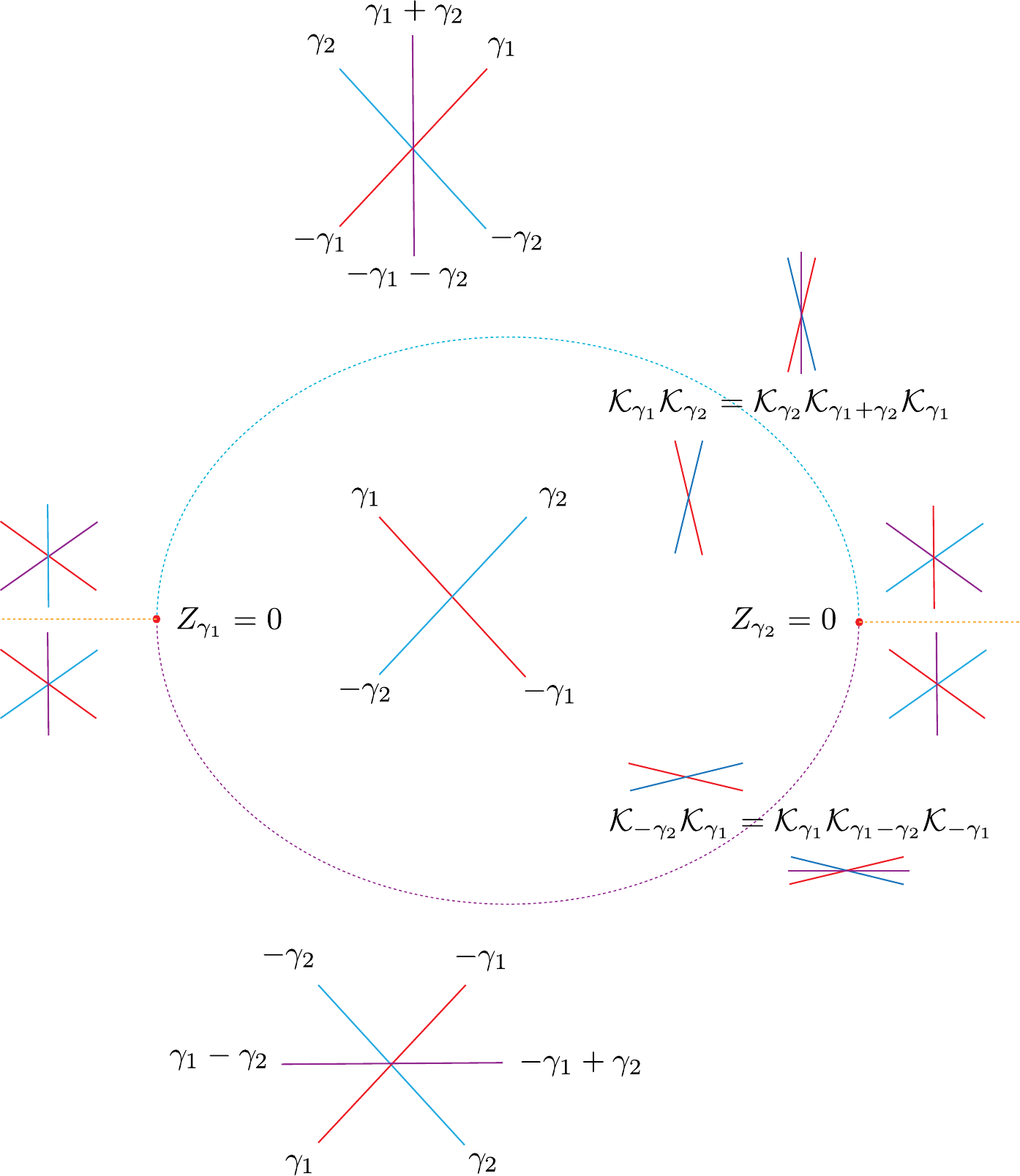}{The Coulomb branch $\CB$ of the $N=3$ theory, with the walls of marginal stability where
$Z_{\pm \gamma_1}$ align with $Z_{\gamma_2}$ marked.  Note that $\hat\Gamma$ has monodromy,
so in order to define the basis $\{\gamma_1, \gamma_2\}$ globally we have to introduce
branch cuts, denoted here in orange.  In each region we show the BPS spectrum and the cyclic
ordering of the BPS rays.  Along the walls we show the wall-crossing formulae.}

All the structures we have been discussing on $\CB$ are illustrated in Figure
\ref{fig:N3uplane}.

\subsubsection{Intermission: Symplectomorphisms and monodromy}

There is an interesting relation between the symplectomorphisms which we encountered above
and the monodromy transformations which arise when we go around a singularity in $\CB$.

Consider the behavior of the BPS rays as $u$ is carried around the simplest type of singularity,
where some $Z_{\gamma_0}(u)$ has a simple zero.  This is the kind
of singularity we were zooming in on in our $N=2$ example above.
The phase of $Z_{\gamma_0}(u)$ rotates by $2 \pi$ as $u$ goes around the singularity.
Hence the BPS ray $\ell_{\gamma_0}$ passes across all the other BPS
rays, followed by the ray $-\ell_{\gamma_0}$.  In other words, as we go around the singularity
we cross a series of walls of marginal stability.

It is easy to check that $\CK_{\gamma_0} \CK_{-\gamma_0}$ is the
transformation
\begin{equation}
   \CX_\gamma \to (-\sigma(\gamma_0))^{\langle \gamma,\gamma_0 \rangle}
   \CX_{\gamma+ \langle \gamma,\gamma_0 \rangle \gamma_0}.
\end{equation}
It follows that as the BPS rays $\ell_{\gamma_0}$ and
$\ell_{-\gamma_0}$ sweep across the spectrum, the associated
Kontsevich-Soibelman transformations implement the expected
monodromy transformations of the charges. (Note that it is
convenient to choose the quadratic refinement $\sigma$ to obey
$\sigma(\gamma_0) = -1$ whenever $\gamma_0$ is the charge of a BPS
hypermultiplet; such a quadratic refinement is invariant under this
monodromy, which thus acts only on the charge label of
$\CX_\gamma$.)

Now how about a singularity where three turning points are
coalescing, like we studied in the $N=3$ case? As we noted above,
near such a singularity three light BPS particles will typically be
present, realized as BPS strings joining the turning points.  The
projections of their charges to the 2-dimensional lattice of gauge
charges relevant for the scaling limit are of the form $\gamma_1$,
$\gamma_1 + \gamma_2$, $\gamma_2$. Naively, as we wind around the
singularity, one might expect that each BPS line will be swept by
the BPS lines of the three light particles, and then the three light
antiparticles.  This is actually incorrect as we can learn by a
simple manipulation of KS transformations. Indeed, we have
\begin{multline}
\CK_{\gamma_2}\CK_{\gamma_1+\gamma_2}\CK_{\gamma_1} \CK_{-\gamma_2}\CK_{-\gamma_1-\gamma_2}\CK_{-\gamma_1}
= \CK_{\gamma_1}\CK_{\gamma_2} \CK_{-\gamma_1}\CK_{-\gamma_2} \\ = (\CK_{\gamma_1} \CK_{-\gamma_1})\CK_{-\gamma_1+\gamma_2}(\CK_{\gamma_2}\CK_{-\gamma_2})
= (\CK_{\gamma_1} \CK_{-\gamma_1})(\CK_{\gamma_2}\CK_{-\gamma_2})\CK_{-\gamma_1}.
\end{multline}
This is not quite the expected monodromy action --- that would have
been implemented just by $(\CK_{\gamma_1}
\CK_{-\gamma_1})(\CK_{\gamma_2}\CK_{-\gamma_2})$. To resolve this
difficulty one should look more closely at the precise dependence of
the small central charges on $u$. For simplicity take $\Lambda\to
0$. Then one   can see from \eqref{eq:N3} that the central charges
of the six light BPS particles behave like the six roots $u^{5/6}$.
It follows that a loop around $u=0$ rotates the central charges only
by $\exp \frac{5\pi i}{3}$. Thus on traveling around this loop a
generic BPS ray undergoes five wall-crossings rather than six, and
the relevant identity is in fact
\begin{align}
\CK_{\gamma_2}\CK_{\gamma_1+\gamma_2}\CK_{\gamma_1} \CK_{-\gamma_2}\CK_{-\gamma_1-\gamma_2}
= (\CK_{\gamma_1} \CK_{-\gamma_1})(\CK_{\gamma_2}\CK_{-\gamma_2}),
\end{align}
which indeed gives the desired monodromy.

\subsubsection{$N=4$} \label{sec:N4}

Finally, let us consider $N=4$. We parameterize
\begin{equation}
P(z) = z^4 + 4\Lambda^2 z^2 + 2 m z + u,
\end{equation}
so that
\begin{equation}
\lambda \sim \pm \left(z^2 + 2\Lambda^2 + \frac{m}{z} + \cdots \right) dz.
\end{equation}
$\Lambda$ is a non-normalizable parameter, $m$ a mass deformation,
and $u$ parameterizes the Coulomb branch.  $\Sigma_u$ is a
twice-punctured elliptic curve (the two punctures lying over $z =
\infty$). The local system $\hat \Gamma$ has fiber $\hat \Gamma_u
\simeq \IZ^3$ with a one-dimensional flavor lattice. The
discriminant is cubic in $u$ and hence there are generically  three
singular points in the $u$-plane, which thus resembles the $u$-plane
of $SU(2)$ theory with $N_f=1$.

\instwofigs{2.75in}{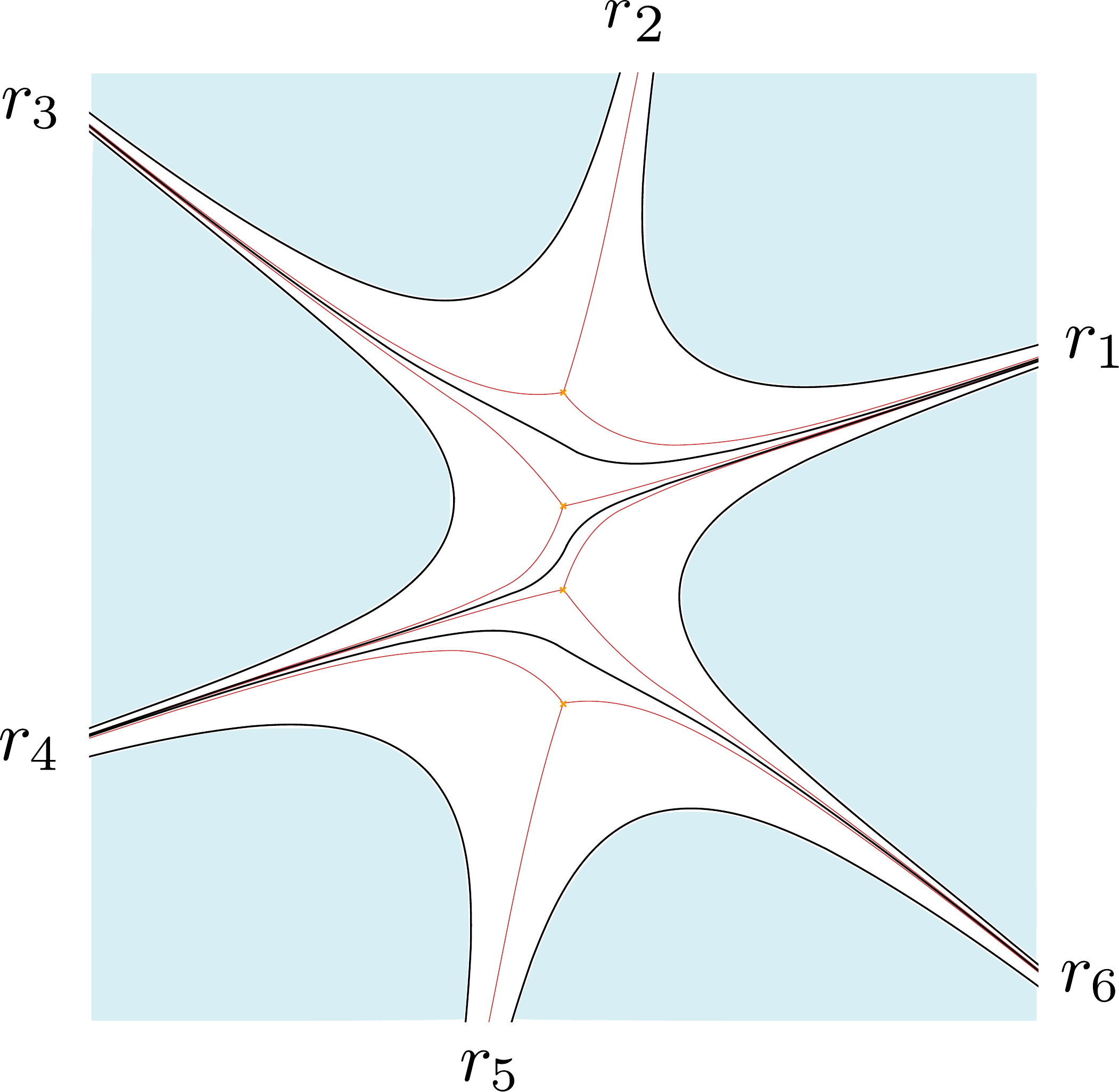}{A generic WKB triangulation for $N=4$
with $m=0$.}{2.75in}{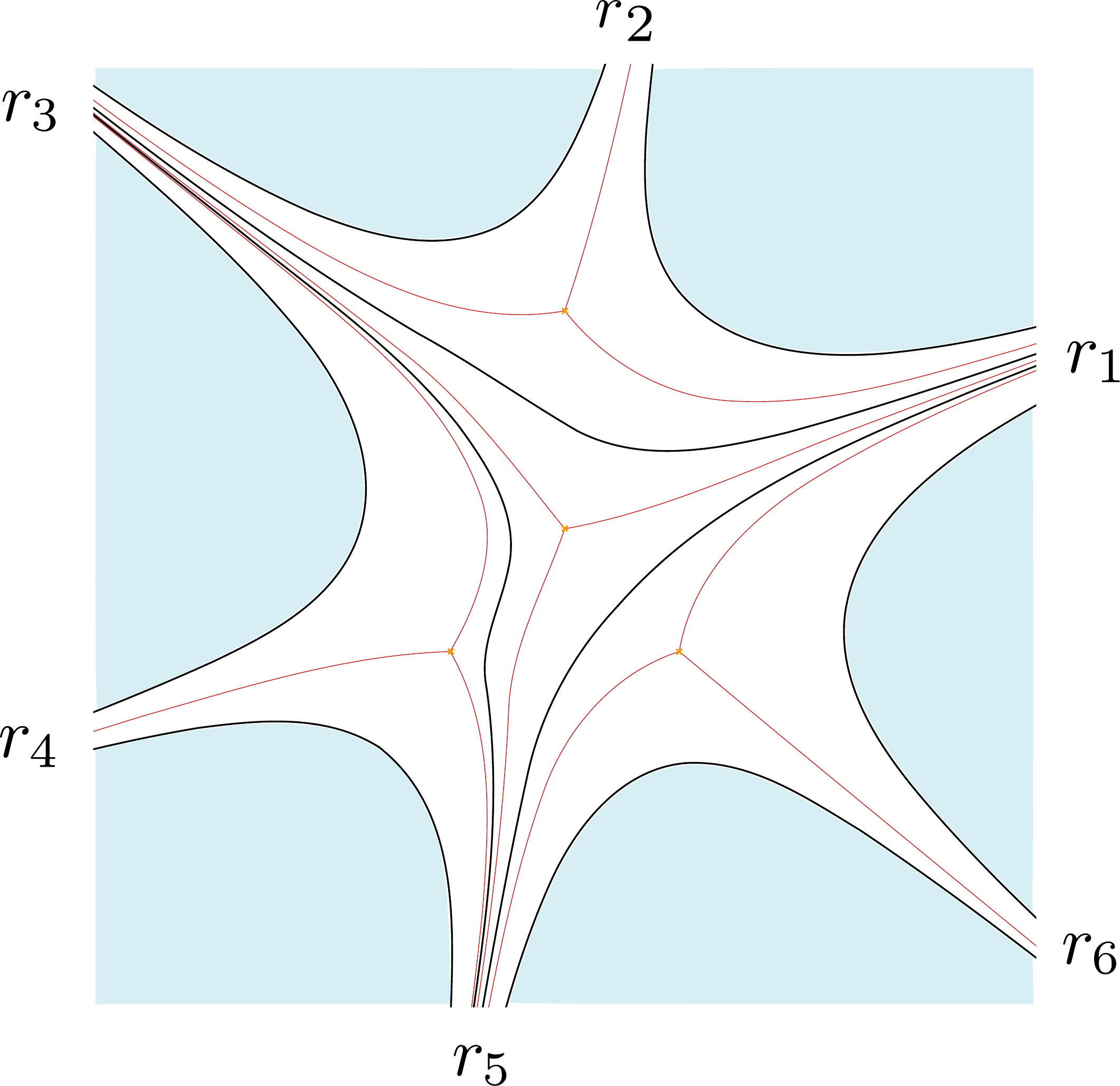}{A generic WKB triangulation for
$N=4$ with large $m$.}

The WKB triangulations now give triangulations of a hexagon, as
in Figures \ref{fig:N4plot}, \ref{fig:N4plotbis}.  All possible
triangulations appear somewhere in parameter space.

Let us now describe the BPS spectrum.

\subsubsection*{The case $m=0$}

It is useful  first to set $m=0$. Then the discriminant degenerates to $256
u(u - 4 \Lambda^4)^2$, so there are two singular points in $\CB$:

\begin{itemize}
\item At $u=0$, two zeroes of $P(z)$ coincide at
$z=0$, while the other two sit at $z = \pm 2 i \Lambda$.  We expect to see a
single BPS particle becoming massless here.
\item At $u=4 \Lambda^4$,
$P(z)$ is a perfect square, and two pairs of zeroes collide, at $z=\pm
i \sqrt{2} \Lambda$.  We expect to see two distinct BPS particles
becoming massless here.  These two particles have the same gauge charges, as the
associated cycles in $\bar\Sigma$ are homologous, but they have
different flavor charges.  (Indeed, the difference between the two corresponding
cycles on $\Sigma$ is a cycle wrapping around $z = \infty$.  If we took some small
$m \neq 0$, $\oint \lambda$ would give the residue $m$ of the simple pole there.)
\end{itemize}

By studying the singularity structure in $\CB$ we have encountered the effects of
three BPS particles.  Now let us explore the BPS spectrum more systematically
by following the $\vartheta$
dependence of $T_\WKB(\vartheta, u)$.  Begin by fixing any $0 < u < 4 \Lambda^4$.
In this case the four turning points are collinear
in the $z$-plane, and we will encounter BPS states associated with strings between
consecutive turning points. Denote the charge of the BPS string joining the
two middle turning points as $\gamma_1$. The charges
of the other two BPS strings will be denoted as $\gamma_2,
\gamma_3$, so that $\gamma_2 + \gamma_3$ is a pure flavor charge
(and in particular, at $m=0$, $Z_{\gamma_2} + Z_{\gamma_3} = 0$.)
We pick our conventions such that
\begin{equation}
\langle \gamma_3, \gamma_1 \rangle = \langle \gamma_1, \gamma_2 \rangle = 1,
\qquad \langle \gamma_2, \gamma_3 \rangle = 0.
\end{equation}
Note that $\CK_{\gamma_2}$ and $\CK_{\gamma_3}$ commute, a fact
which will be used repeatedly below.  There is a wall of marginal
stability in $\CB$, which passes through the two singularities, and
encircles an ellipsoidal region which includes the line $0<u
\Lambda^{-4} <4$. \insfig{5.5in}{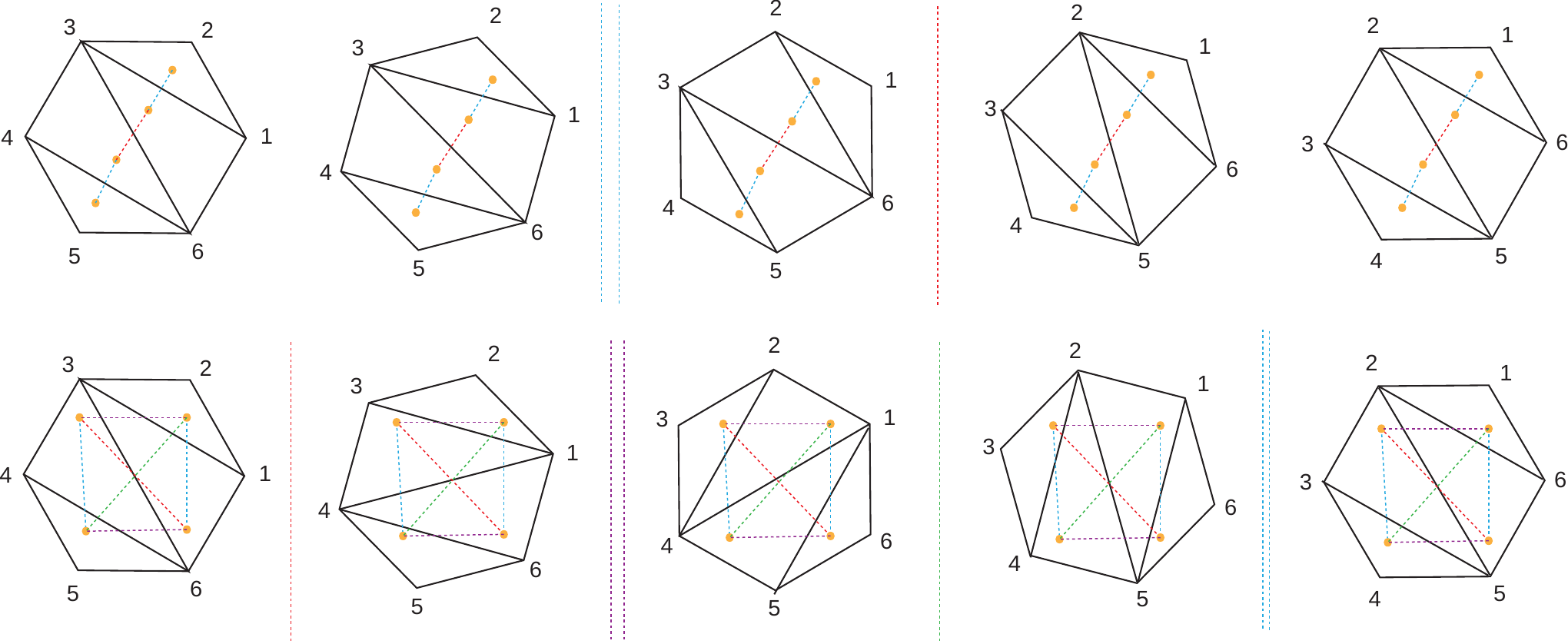}{WKB triangulations
at $m=0$.  Upper strip: at $u=0$, as $\vartheta$ varies from $0$ to
$\pi$, $T_\WKB$ evolves simply, with three flips: A pair of
commuting flips at the first, blue dotted line and a third flip at
the red line.  Lower strip: for $u$ beyond the walls of marginal
stability, $T_\WKB$ evolves in a different way, involving six flips.
Both at the second and at the fourth dotted lines, two flips occur,
corresponding to two BPS particles with the same charges.  Notice
that $T_\WKB(\vartheta)$ and $T_\WKB(\vartheta+\pi)$ are identical,
up to a relabeling of the WKB rays.}  Now by direct computation
(either by hand or using a computer) one can obtain the evolution of
$T_\WKB(\vartheta)$ with $\vartheta$.  In Figure
\ref{fig:exa-evolution} we depict the result, for two $u$, one
inside and one outside the wall of marginal stability. At $u=0$ we
see three flips in $T_\WKB$, corresponding to the three BPS charges
$-\gamma_3$, $\gamma_2$, $\gamma_1$. As we cross the wall of
marginal stability, the phase of the central charges $Z_{\gamma_2}$,
$Z_{-\gamma_3}$ aligns with one of $Z_{\pm \gamma_1}$ (the sign
depending on which segment of the wall we cross). On the other side
of the wall, the flips in $T_\WKB$ happen in a different order, and
we see a total of six BPS hypermultiplets.  There is a BPS string
between each pair of turning points.

Following reasoning analogous to the $N=3$ case above, then, we must have the wall crossing identity:
\begin{equation} \label{eq:k-identity}
\CK_{\gamma_1} \CK_{\gamma_2} \CK_{-\gamma_3}
= \CK_{\gamma_2} \CK_{-\gamma_3} \CK_{\gamma_1+\gamma_2-\gamma_3} \CK_{\gamma_1+\gamma_2}\CK_{\gamma_1-\gamma_3}
\CK_{\gamma_1}
\end{equation}
The ordering of factors in \eqref{eq:k-identity} is easily understood, just using the facts that
$Z_{\gamma_2} = Z_{-\gamma_3}$ and the ordering of the arguments of
$Z_{k_1 \gamma_1 + k_2 \gamma_2}$ as $k_1, k_2$ vary is determined by the ordering of
$k_1 / k_2$.  Again, one can check explicitly that \eqref{eq:k-identity} is
a true identity, but we will give a simpler proof in the next paragraph.

\subsubsection*{The case $m \neq 0$}

For $m \neq 0$ the picture is a bit more complicated.
For small $m$, the particles of charges $\gamma_2$ and $\gamma_3$
become massless at slightly different points in $\CB$.
The wall of marginal stability also splits into several walls.  The
wall-crossing formula \eqref{eq:k-identity} thus ``decomposes'' into three separate
pentagon identities.  One possible decomposition is
\begin{multline}
\CK_{\gamma_1} \CK_{\gamma_2} \CK_{-\gamma_3} =  \CK_{\gamma_2} \CK_{\gamma_1+\gamma_2} \CK_{\gamma_1} \CK_{-\gamma_3}
=  \CK_{\gamma_2} \CK_{\gamma_1+\gamma_2}  \CK_{-\gamma_3}\CK_{\gamma_1-\gamma_3} \CK_{\gamma_1} \\
= \CK_{\gamma_2} \CK_{-\gamma_3} \CK_{\gamma_1+\gamma_2-\gamma_3} \CK_{\gamma_1+\gamma_2}\CK_{\gamma_1-\gamma_3}
\CK_{\gamma_1},
\end{multline}
corresponding to three separate walls of marginal stability, where
$Z_{\gamma_1}$ aligns with $Z_{\gamma_2}$, then $Z_{\gamma_1}$ aligns
with $Z_{-\gamma_3}$, then $Z_{-\gamma_3}$ aligns with
$Z_{\gamma_1 + \gamma_2}$.  A second possibility is
\begin{multline}
\CK_{\gamma_1} \CK_{\gamma_2} \CK_{-\gamma_3} =  \CK_{-\gamma_3} \CK_{\gamma_1-\gamma_3} \CK_{\gamma_1} \CK_{\gamma_2}
=  \CK_{-\gamma_3} \CK_{\gamma_1-\gamma_3}  \CK_{\gamma_2}\CK_{\gamma_1+\gamma_2} \CK_{\gamma_1} \\
= \CK_{\gamma_2} \CK_{-\gamma_3} \CK_{\gamma_1+\gamma_2-\gamma_3} \CK_{\gamma_1+\gamma_2}\CK_{\gamma_1-\gamma_3}
\CK_{\gamma_1}.
\end{multline}
The corresponding sequences of WKB triangulations are shown in Figure
\ref{fig:exa-evolution-mass}. \insfig{5.5in}{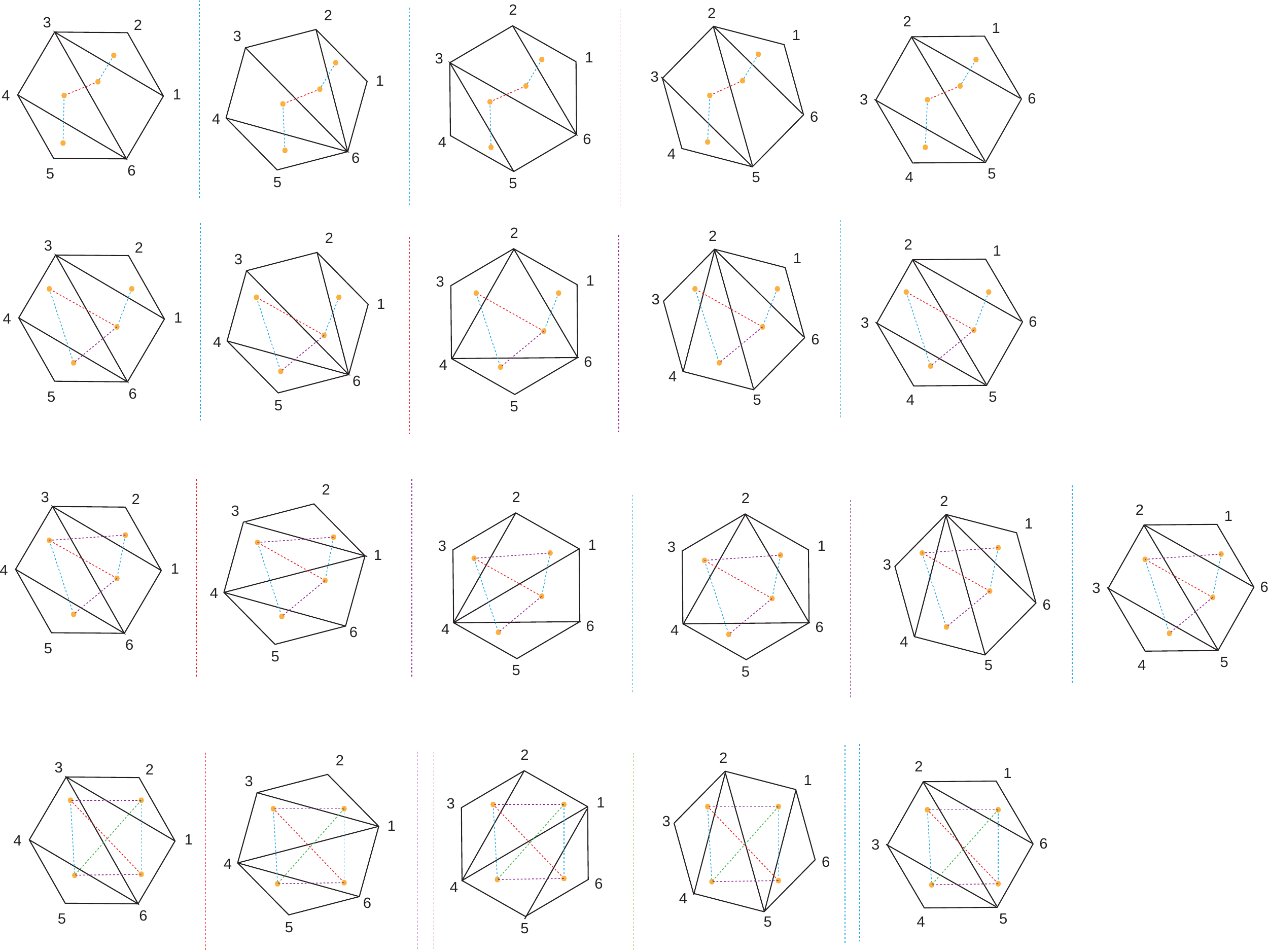}{Upper
strip:  at $u = 0$, $T_\WKB(\vartheta, u)$ evolves simply with $\vartheta$, undergoing
three flips.  Lower strips:  as we vary $u$ toward $\infty$ we cross three walls.
After crossing each wall, we show the new evolution of $T_\WKB(\vartheta, u)$ with $\vartheta$.
In the last strip, to save space, we do not show the
intermediate triangulations between pairs of BPS jumps with the same
gauge charge.}

In Figure \ref{fig:N4uplane-detail-a} we show the structure of the
walls in $\CB$ very close to the singular point where $Z_{\gamma_1} = 0$.
The global structure of the walls is fairly intricate, and depends
on the phase of $m$.

\insfig{3.3in}{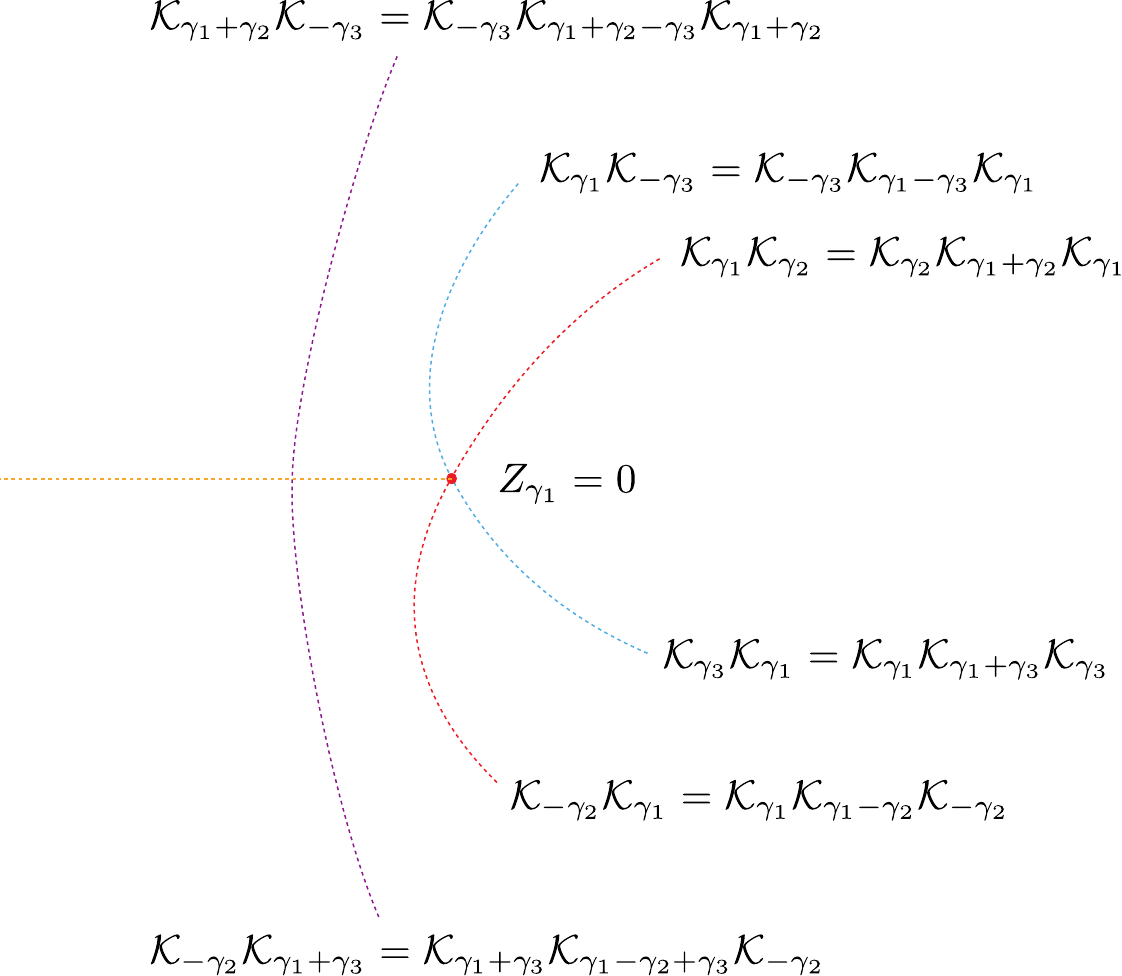}{The walls of marginal stability in the $N=4$ theory,
near a point $u$ with $Z_{\gamma_1}(u)=0$, for a certain choice of small
$m \neq 0$.  (For other phases of $m$, $\gamma_2$ and $-\gamma_3$ may
be exchanged in the figure).  Note that $\hat\Gamma$ has monodromy, hence the need for
the orange cuts.}

Notice that it is possible to go from small $u$ to
large $u$ by a trickier path which passes between the two singularities where
$Z_{\gamma_2} = 0$ and $Z_{\gamma_3} = 0$.  The wall-crossing formula then
arranges itself in a different way, for example as
\begin{multline}
\CK_{\gamma_3} \CK_{\gamma_1} \CK_{\gamma_2} = \CK_{\gamma_1} \CK_{\gamma_1+\gamma_3} \CK_{\gamma_3} \CK_{\gamma_2}
= \CK_{\gamma_1} \CK_{\gamma_2} \CK_{\gamma_1+\gamma_2+\gamma_3}\CK_{\gamma_1+\gamma_3} \CK_{\gamma_3} \\
= \CK_{\gamma_2} \CK_{\gamma_1+\gamma_2}\CK_{\gamma_1}  \CK_{\gamma_1+\gamma_2+\gamma_3}\CK_{\gamma_1+\gamma_3} \CK_{\gamma_3}.
\end{multline}
(Remember that $\CK_{\gamma_3}$ and $\CK_{\gamma_2}$ commute.)
It is an amusing exercise to check the self-consistency of the BPS
spectra generated by crossing from the inner region to the outer
region between different pairs of singular points, and the
monodromies of the charge lattice around the singular points
themselves. The details vary depending on the phase of $m$.

A judicious choice of $m$ can bring other pairs of
singularities, say $Z_{\gamma_1} = 0$ and $Z_{\gamma_2} = 0$, together,
and produce a scaling region in $\CB$ which looks like our $N=3$ example.
In this limit we see a central region with the BPS spectrum of particles which becomes light at the
composite $N=3$ singularity,
with charges $\gamma_2$, $\gamma_1+\gamma_2$, $\gamma_1$, together with the
particle which becomes light at the remaining simple singularity,
with charge $\gamma_3$.  We show the structure of $\CB$ for this
choice of $m$ in Figure \ref{fig:N4uplane-c}.  The corresponding
behavior of $T_\WKB$ is shown in Figure \ref{fig:N4toN3plot}.

\insfig{4.5in}{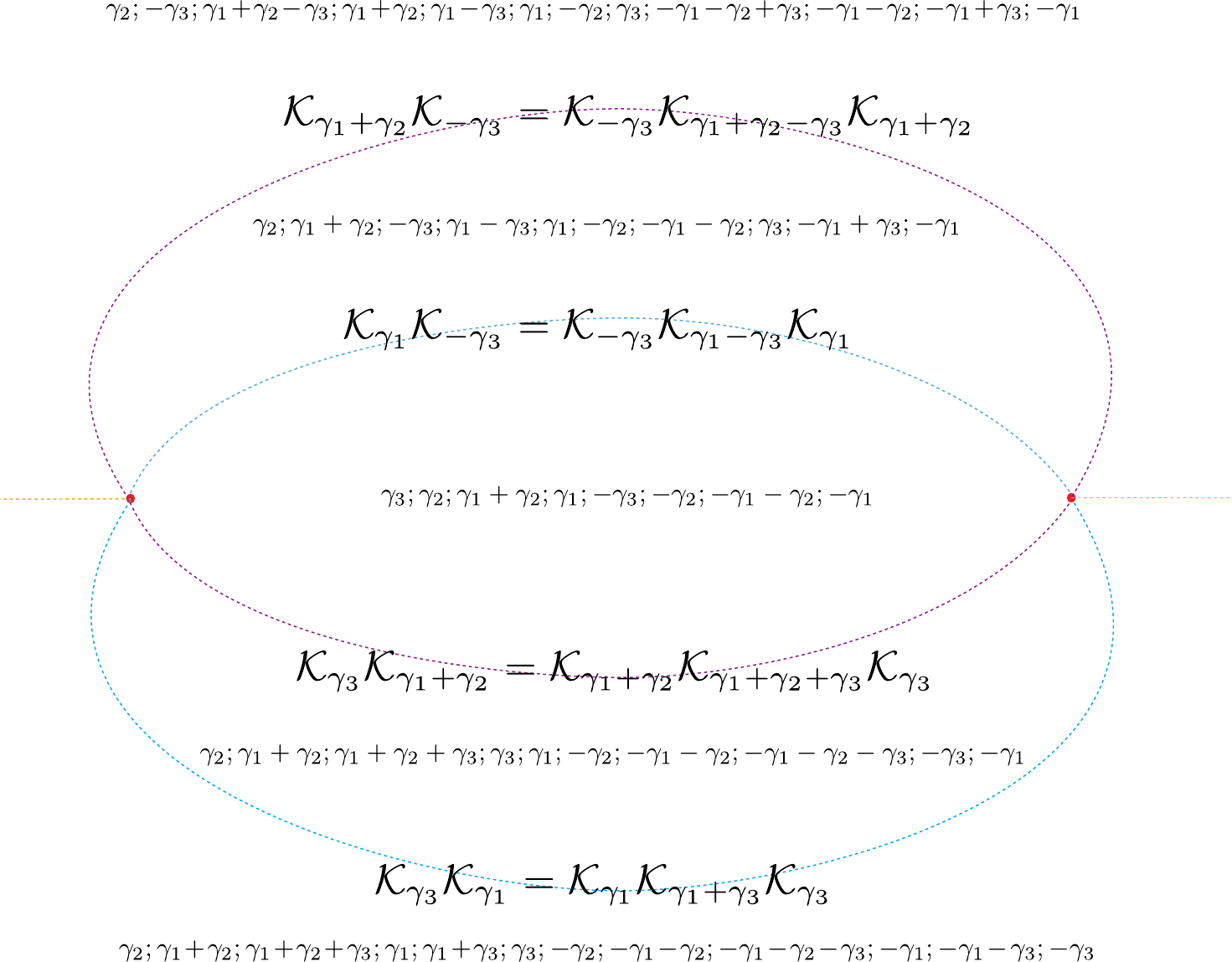}{The walls of marginal stability in the $N=4$ theory, for the special
choice of mass $m$ where the singularities $Z_{\gamma_1}=0$ and
$Z_{\gamma_2}=0$ collide.  Inside each region we list the charges of all BPS
hypermultiplets in the spectrum in that region, in order of their phases (up
to overall cyclic permutation).  On each wall we give the
relevant wall-crossing formula.}

\insfig{3in}{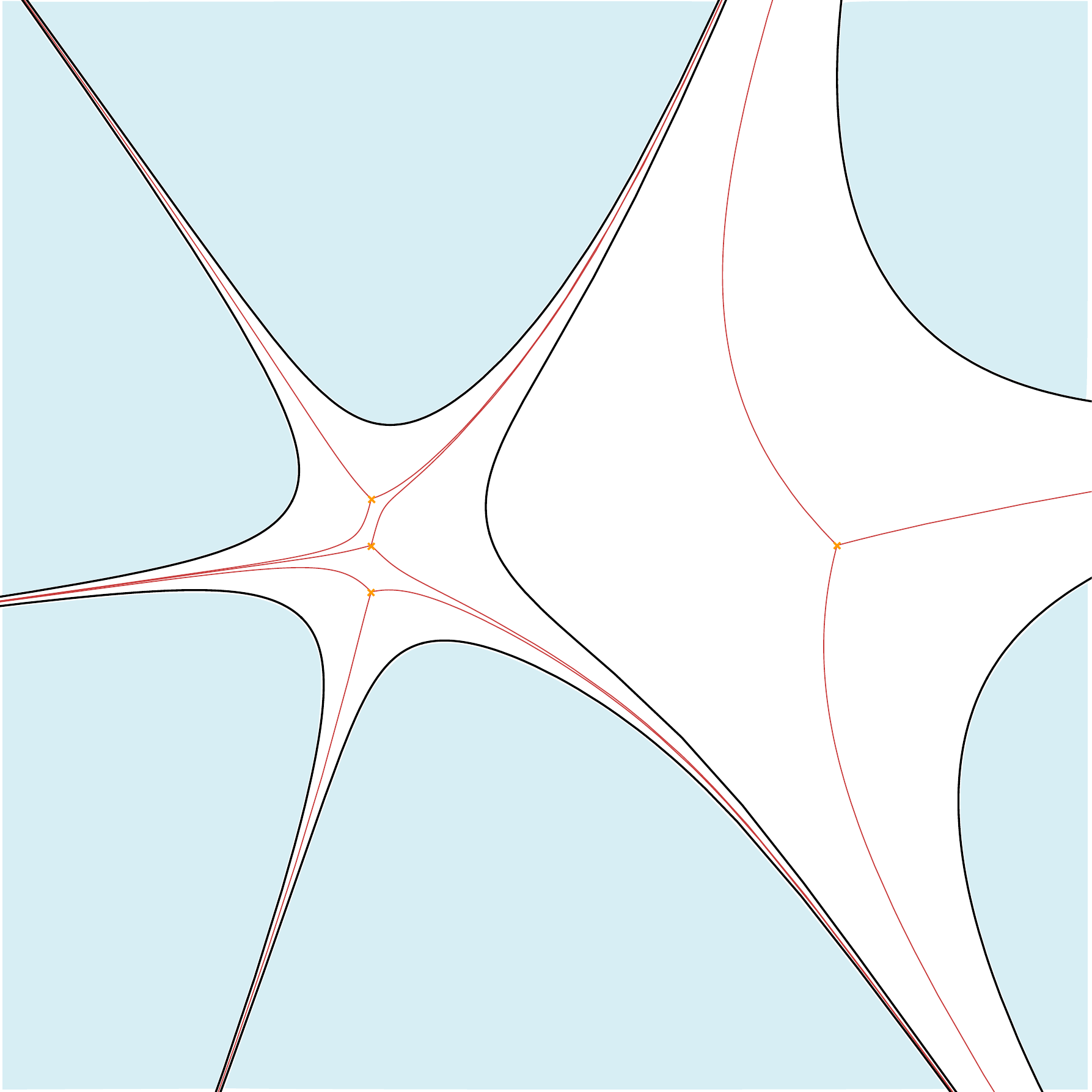}{The triangulation $T_\WKB$ for the
$N=4$ theory, with $m$ and $u$ adjusted so that three turning points are coming together.
On the left, the pentagonal structure of the $N=3$ example emerges inside the $N=4$
triangulation.}

For generic $m$, there is a region at sufficiently large
$\abs{u}$ where the turning points are arranged roughly into a square,
with $6$ BPS states associated to the sides and the diagonals.
From this behavior we learn that in a generic $\CN=2$ theory,
near a singularity in the Coulomb branch where four turning points are
coalescing, we will typically find six light BPS particles,
realized as BPS strings joining any pair of turning points.

\subsection{General $N$ and the associahedron}

Finally, we make some general remarks applicable to all $N$.
The finite WKB curves in these theories were described by Shapere and Vafa in
\cite{Shapere:1999xr}, exactly for the purpose of studying BPS
states.  They showed that:

\begin{itemize}
\item The phase $\vartheta$ at which a finite WKB curve appears
is uniquely determined by the homotopy class of the WKB curve in
the punctured plane $\IC - \{ \CT_i \}$.

\item There is at most one finite WKB curve joining any two turning
points.  Hence there are only a \emph{finite} number of BPS states.

\item  Any two turning points can be joined by a piecewise collection of
finite WKB curves (generically with different $\vartheta$ for
each segment).  Hence there are at least $N-1$ BPS states at any point $u$ on the
Coulomb branch.

\item The minimal number of BPS states, $N-1$, is attained, for example, when all the roots of $P_N(z)$ are
real.  The maximal number, $\half N (N-1)$, is attained, for example, if $P(z) = z^N-1$.

\item The three separating WKB curves emanating from each of the $N$ turning points each asymptote to a different
ray at $z=\infty$.

\item There are no closed WKB curves.
\end{itemize}

The generic WKB curves thus define a triangulation of a convex
polygon with $N+2$ vertices. (We have seen examples of this for $N =
1, 2, 3, 4$ above.) This observation opens up a connection to a rich
branch of mathematics. Triangulations of an $(N+2)$-gon are dual to
trivalent graphs and also correspond to rooted binary trees with
$N+1$ terminal points, or equivalently to ways of parenthesizing a
product of $N+1$ nonassociative variables $x_1 \cdots x_{N+1}$. The
number of such triangulations is the Catalan number $C_N =
\frac{1}{N+1} \binom{2N}{N}$. Moreover, the triangulations of the
$(N+2)$-gon can be considered as the vertices of a simplicial
complex $\CK_{N+1}$, with 1-simplices corresponding to the flips.
This complex  is the 1-skeleton of the $N$-th ``associahedron'' or
Stasheff polytope \cite{MR0158400}. (See for example
\cite{Lee,MR1841954} for recent discussions.) $\CK_3$ is an
interval, and $\CK_4$ is a pentagon. A basic lemma of
\cite{MR0158400} is that the faces of the associahedron are products
of lower-dimensional associahedra.  So in particular the 2-cells are
pentagons (corresponding to the pentagon relation) and squares
(corresponding to the relation that flips on disjoint edges
commute). On the other hand the associahedron itself is simply
connected, and hence the two-skeleton is simply
connected.\footnote{This statement is also closely related to the
MacLane coherence theorem in category theory \cite{MR0330253}.} In
our context this implies that all the wall-crossing formulae that
arise in these examples are consequences of the pentagon relation.
We have seen this fact in the $N=3, 4$ examples above.

The appearance of the associahedron in this class of examples raises
the question of whether the theory of $A_\infty$ algebras has any
interesting role to play here. Another natural question is whether
there is a simple algorithm for computing the spectrum of BPS states
given only the polynomial $P(z)$. The computation of the spectrum
generator in Section \ref{sec:UniversalStokesMatrix} answers this
second question in the affirmative.

\section{$SU(2)$ gauge theory}\label{section:SU2Examples}

The examples we have just studied in Section
\ref{section:Superconformal} only had BPS hypermultiplets, so the
WKB triangulations $T_\WKB(\vartheta)$ only exhibited simple flips
as $\vartheta$ varied.  Next we would like to consider some examples
with BPS vectormultiplets and the corresponding \juggle
transformations.  We will look at a well-known set of theories:
$SU(2)$ gauge theories with $N_f=0, 1, 2, 3, 4$. Some aspects of
their spectra are well understood.  See in particular
\cite{Bilal:1996sk,Ferrari:1996sv,Ferrari:1997gu} for explicit
results on the BPS spectrum in various parameter ranges and
\cite{Banks:1996nj,Douglas:1996js,Gaberdiel:1998mv,Mikhailov:1998bx,DeWolfe:1998bi}
  for an approach using $F$-theory.  However, a complete
description of their BPS spectra is not available. In this
section we will make progress towards such a complete
description, giving the spectra in many regions of parameter
space. In fact, our methods could be used to obtain the
spectrum at any point in moduli space.

\subsection{$N_f=0$} \label{sec:nf0}

This is the canonical example of Seiberg-Witten theory. It can be
constructed as in Section \ref{sec:physics} from two D4-branes
stretched between two NS5-branes (see equation
\eqref{eq:Standard-SU2-curve} et. seq.)  The quadratic differential
is simply
\begin{equation}
\lambda^2 = \left( \frac{\Lambda^2}{z^3} + \frac{2u}{z^2} + \frac{\Lambda^2}{z} \right) dz^2.
\end{equation}
The corresponding Hitchin system has the mildest possible irregular
singularities at $z=0, \infty$, with a single WKB ray emerging from
each (corresponding to $L=1$ in the notation of Section
\ref{sec:irregular}). There are two turning points, $z_{tp}^\pm = -
\frac{u}{\Lambda^2} \pm \sqrt{(\frac{u}{\Lambda^2} )^2-1}$, which
collide at $z_{tp}^\pm = \mp 1$ when  $u = \pm \Lambda^2$. These two
values of $u$ are singularities in $\CB$. At each of the
singularities a single BPS particle becomes massless. As $u$ varies
along the interval $-\Lambda^2 < u < \Lambda^2$ the two turning
points separate, move in opposite directions around $z=0$, and then
rejoin. The BPS spectrum anywhere on this interval, for example at
$u=0$, consists of two BPS hypermultiplets.  They are described by
strings joining the two turning points but passing on opposite sides
of the singularity at $z = 0$. They share both endpoints, hence
their charges $\gamma_{1,2}$ satisfy $\langle \gamma_1, \gamma_2
\rangle = 2$. $\gamma_{1,2}$ generate the charge lattice $\hat
\Gamma$. A typical WKB triangulation $T_\WKB(\vartheta,u)$ for $u$
in this interval, and its transformations as $\vartheta$ varies, are
shown in Figures \ref{fig:Nf0plot}, \ref{fig:Nf0evolution-a}.
\insfig{4.2in}{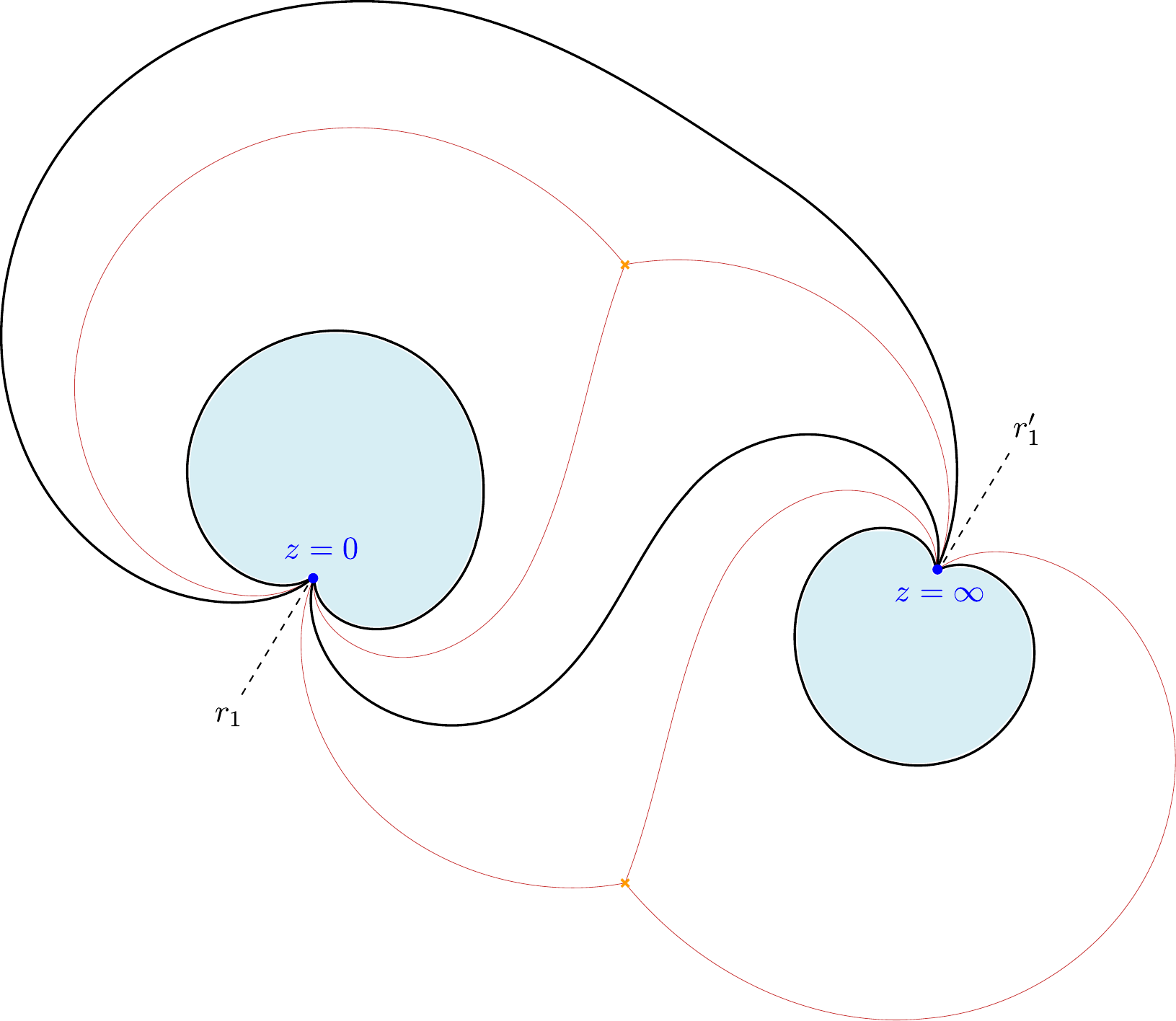}{A generic $T_\WKB$ for the $SU(2)$ theory
with $N_f=0$.  For clarity $z=\infty$ has been mapped to finite
distance.} \insfig{6in}{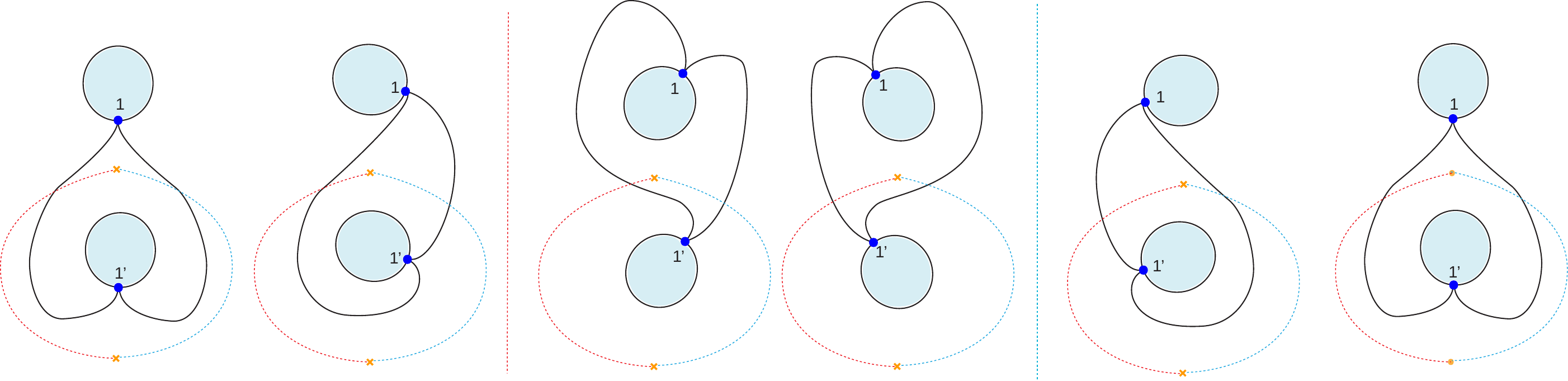}{The evolution of
$T_\WKB(\vartheta, u)$ for the $SU(2)$ theory with $N_f=0$, for $u$
in the strong coupling region around $u=0$. For clarity $z=0$ and
$z=\infty$ have been mapped to finite distance. There are two flips,
corresponding to two BPS states $\gamma_1$, $\gamma_2$ with $\langle
\gamma_1, \gamma_2\rangle=2$. One finite WKB curve is the dotted red
and the other is the dotted blue curve. The red (blue) dotted
vertical line indicates the flips due to the existence of the red
(blue) BPS state.}

There is a single ellipsoidal wall of marginal stability in $\CB$, passing
through the two singular points.  The region inside this wall (which we have just
been discussing) is also called the ``strong coupling'' region, while the region
outside is the weak coupling region.

Note that all of the WKB triangulations in this theory are just
annuli traversed by two internal edges --- exactly the prerequisite
situation for our discussion of limit triangulations in Section
\ref{sec:vector-jump}. When $u$ is in the strong coupling region the
evolution with $\vartheta$ is simple, and illustrated in Figure
\ref{fig:Nf0evolution-a}. On the other hand, in the weak coupling
region we   encounter a new feature, namely a BPS vectormultiplet.
For an appropriate $\vartheta_c$ a one-parameter family of closed
WKB curves appears, corresponding to a single vectormultiplet of
charge $\gamma_1+\gamma_2$,   as shown in Figure
\ref{fig:Nf0plot-limit}. Thus, if  $u$ is in the weak coupling
region, as the phase of $\vartheta$ varies, $T_\WKB(\vartheta,u)$
undergoes infinitely many elementary flips, corresponding to a
spectrum of BPS hypermultiplets with charges of the form $(n+1)
\gamma_1 + n \gamma_2$, at BPS rays which accumulate at the phase of
$Z_{\gamma_1 + \gamma_2}$.  At the critical phase for the
vectormultiplet   $T_\WKB$ transforms by a \juggle.  This is
followed by another infinite sequence of flips, corresponding to a
spectrum of BPS hypermultiplets, with charges of the form $n
\gamma_1 + (n+1) \gamma_2$. So altogether we encounter the standard
weak coupling BPS spectrum of the theory: a W-boson of electric
charge $\gamma_1+\gamma_2$ and an infinite tower of dyons.

\insfig{4.5in}{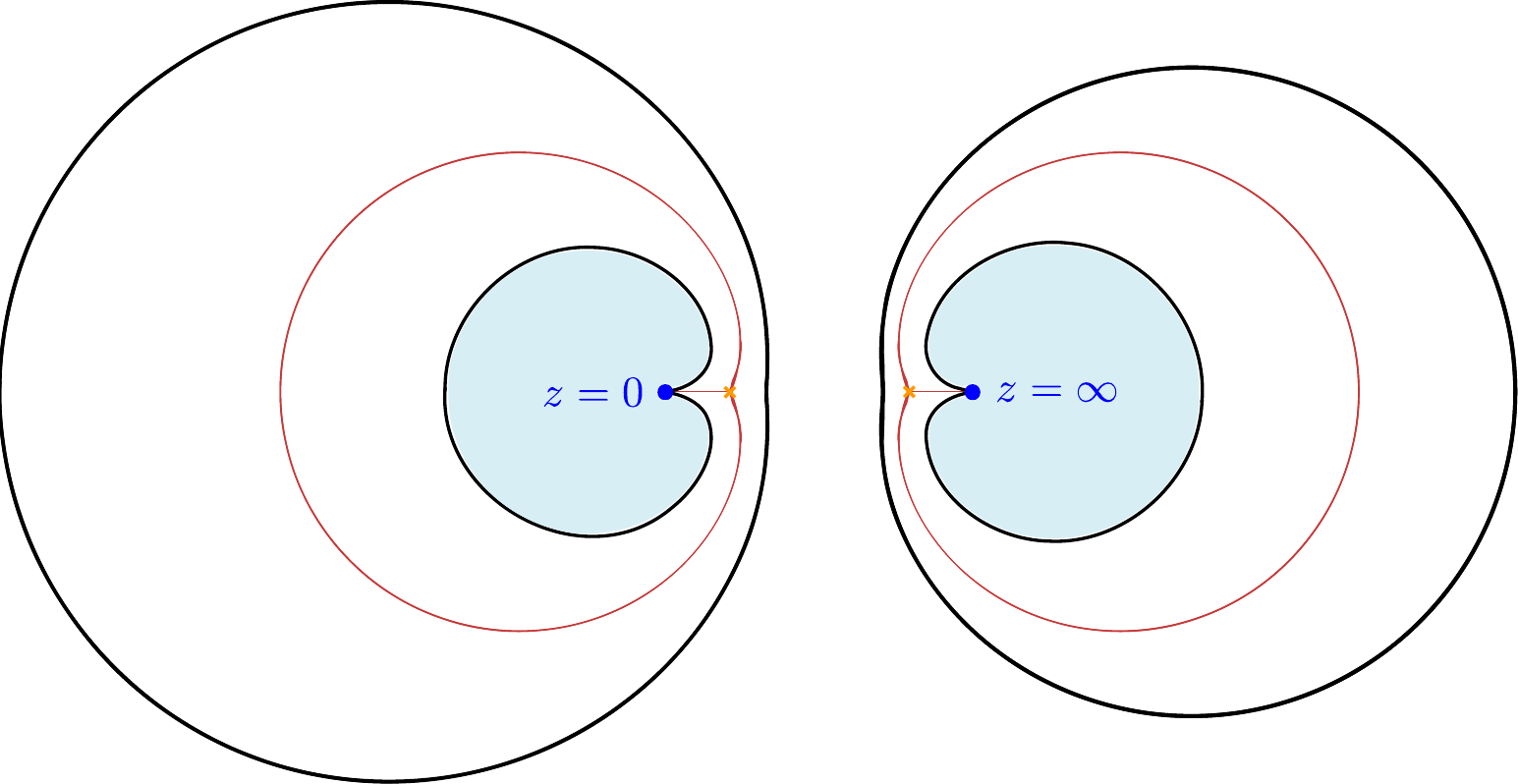}{The limit WKB
triangulation for the $SU(2)$ theory with $N_f=0$.  For clarity $z=\infty$
has been mapped to finite distance, and a few extra closed WKB curves are shown.}

Comparing the spectrum on the two sides of the wall, we encounter the important wall-crossing formula
for charges with $\langle \gamma_1, \gamma_2\rangle=2$:
\begin{equation} \label{eq:wcf-prod2}
\CK_{\gamma_1} \CK_{\gamma_2}
= \CK_{\gamma_2} \CK_{\gamma_1+2 \gamma_2}\CK_{2\gamma_1+3\gamma_2} \cdots  \CK_{\gamma_1+\gamma_2}^{-2} \cdots \CK_{3\gamma_1+2\gamma_2}\CK_{2\gamma_1+\gamma_2}\CK_{\gamma_1}.
\end{equation}

More precisely, this is the wall-crossing formula relevant for one of the two
sections of the wall.  For the other section, the relevant formula is
\begin{equation}
\CK_{\gamma_2}\CK_{-\gamma_1}
= \CK_{-\gamma_1} \CK_{-2\gamma_1+ \gamma_2}\CK_{-3\gamma_1+2\gamma_2} \cdots  \CK_{-\gamma_1+\gamma_2}^{-2} \cdots \CK_{-2\gamma_1+3\gamma_2}\CK_{-\gamma_1+2\gamma_2}\CK_{\gamma_2}
\end{equation}
Again, the two spectra appearing on the right side are compatible with one another, once we
take account of the expected monodromy around the singularities:
\begin{align}
M_1 = (\gamma_1, \gamma_2)& \to (\gamma_1, \gamma_2-2 \gamma_1), \\
M_2 = (\gamma_1, \gamma_2)& \to (\gamma_1+2 \gamma_2, \gamma_2).
\end{align}

\subsection{$N_f=1$} \label{sec:nf1}

If we add a single flavor to the $SU(2)$ gauge theory, the quadratic
differential is modified to
\begin{equation}
\lambda^2 = \left( \frac{\Lambda^2}{z^3} + \frac{3u}{z^2}+ \frac{2
\Lambda m}{z} + \Lambda^2 \right) (dz)^2.
\end{equation}
Again, this may be easily derived using the methods of Section
\ref{sec:physics}. The expansion around $z=\infty$ now shows an
irregular singularity with two WKB rays. The mass parameter $m$
coincides with the residue at this singularity:
\begin{equation}
\lambda \sim \left( \Lambda + \frac{m}{z} + \cdots \right) dz.
\end{equation}

There are three turning points. If we set $m = 0$, there are
three singularities in $\CB$, at $u^3 = \frac{1}{4}
\Lambda^6$.  At each singularity, a single BPS state becomes
massless. In a ``strong coupling'' region around $u=0$, meeting these three singularities,
the BPS spectrum consists exactly of these three BPS states.  They
are realized as strings joining consecutive pairs of turning points
clockwise around $z=0$. Their charges $\gamma_{1,2,3}$ are a basis
of the charge lattice $\hat \Gamma$, satisfying $\langle
\gamma_1, \gamma_2\rangle=\langle \gamma_2, \gamma_3\rangle=\langle
\gamma_3, \gamma_1\rangle=1$.  Furthermore, $\sum \gamma_i$
can be wrapped around $z=0$, and thus corresponds to a
pure flavor charge.
 \insfig{4.5in}{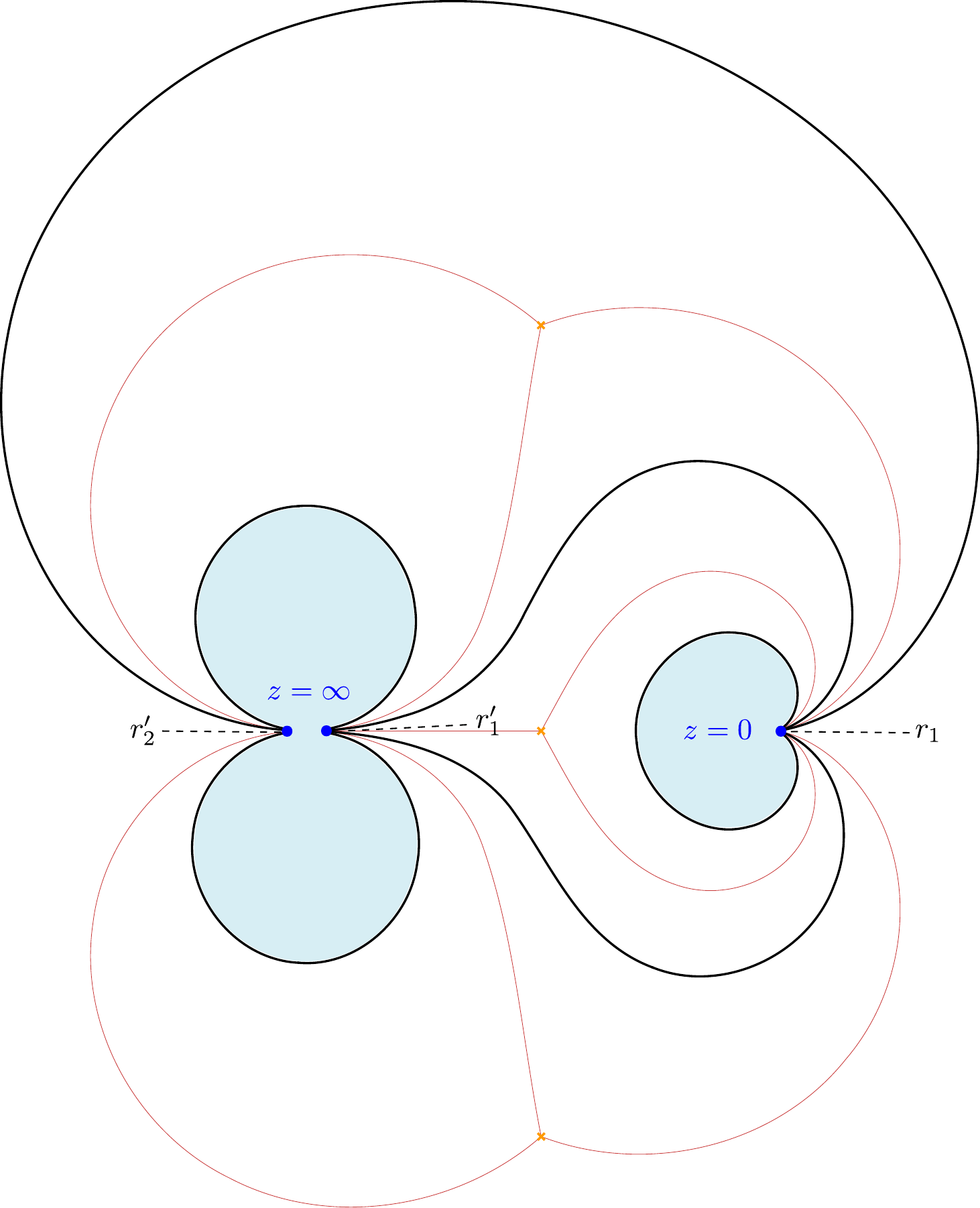}{$T_\WKB$ for the $SU(2)$ theory with $N_f=1$.
For clarity $z=0$ and $z=\infty$ have been mapped to finite distance.}
\insfig{5.5in}{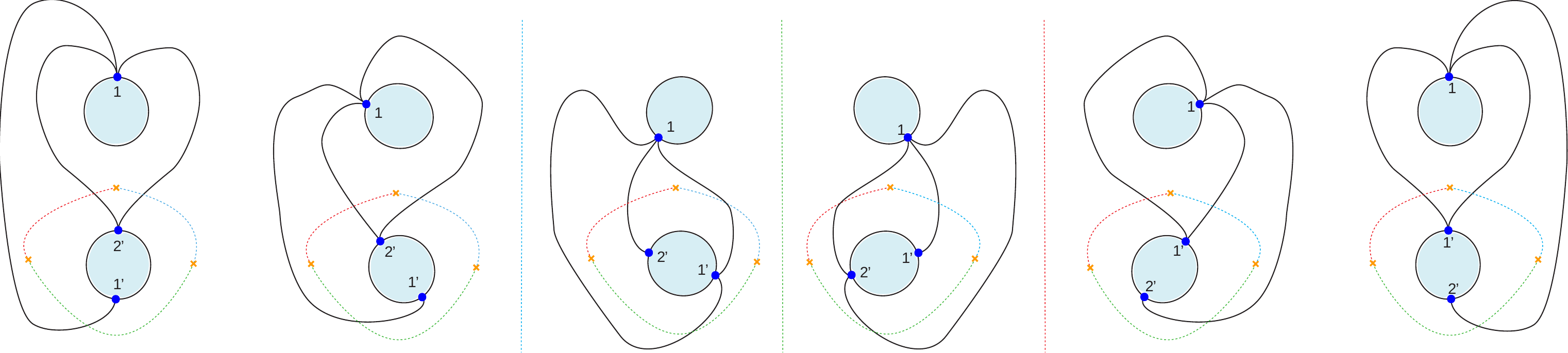}{The evolution of $T_\WKB(\vartheta, u)$
with $\vartheta$,
in the $SU(2)$ theory with $N_f=1$, for $u$ in the strong coupling region
around $u=0$. For clarity $z=\infty$ has been mapped to
finite distance. There are three flips, corresponding to the three
BPS states with charges $\gamma_1$, $\gamma_2$, $-\gamma_3$.}

At $m=0$, $Z_{-\gamma_3} = Z_{\gamma_1 + \gamma_2}$, and there is a
single wall of marginal stability, where the central charges of the
three BPS states align.  This wall passes through the three
singularities.

We now reverse the logic that we have been using in examining these
examples. Thus far, we have been deducing the BPS spectra from the $\vartheta$-dependence
of $T_\WKB$, and then writing down the identities between products of
symplectomorphisms which arise when we cross a wall.  We will
now instead use these identities to determine what the BPS spectra must be.

Consider for instance a section
of the wall where $Z_{\gamma_1}$, $Z_{\gamma_2}$, $Z_{-\gamma_3}$ are
aligned. In the strong coupling region, the triangulation
undergoes three flips as $\vartheta$ crosses the corresponding phases, which
combine to give the symplectomorphism $\CK_{\gamma_1} \CK_{-\gamma_3}
\CK_{\gamma_2}$.  On the other side, the same symplectomorphism has
to be decomposed into factors for which the phases are ordered in the opposite direction.

To find such a decomposition we can play around with the identities we already know.
There is no problem bringing $\CK_{\gamma_1}$ to the right, by two applications
of the pentagon identity
\begin{equation}
\CK_{\gamma_1} \CK_{-\gamma_3} \CK_{\gamma_2} = \CK_{-\gamma_3} \CK_{\gamma_1-\gamma_3}
\CK_{\gamma_2} \CK_{\gamma_1+\gamma_2}\CK_{\gamma_1}.
\end{equation}
However, the factors are still not properly ordered: we need to bring
$\CK_{\gamma_2}$ to the left, across $\CK_{-\gamma_3}
\CK_{\gamma_1-\gamma_3}$.  We have $\langle \gamma_1-\gamma_3, \gamma_2 \rangle=2$,
so we need to use the identity \eqref{eq:wcf-prod2}
we encountered in the $N_f=0$ theory (with a change of basis), giving
\begin{multline}
\CK_{-\gamma_3} \CK_{\gamma_1-\gamma_3}\CK_{\gamma_2} \CK_{\gamma_1+\gamma_2}\CK_{\gamma_1}
= \\ \CK_{-\gamma_3} \CK_{\gamma_2} \CK_{2 \gamma_2 +\gamma_1-\gamma_3} \CK_{3 \gamma_2 +2\gamma_1-2\gamma_3}\cdots
\CK_{\gamma_1-\gamma_3+ \gamma_2}^{-2}\cdots \CK_{2\gamma_2 +3\gamma_1-3\gamma_3}\CK_{\gamma_2 +2\gamma_1-2\gamma_3}\CK_{\gamma_1-\gamma_3} \CK_{\gamma_1+\gamma_2}\CK_{\gamma_1}.
\end{multline}
We are almost done.  The factors $\CK_{-\gamma_3}$ and
$\CK_{\gamma_1+\gamma_2}$ are associated to charges with $Z_{-\gamma_3} = Z_{\gamma_1 + \gamma_2}$,
so in the properly reordered product they should be adjacent to the vectormultiplet of
charge $\gamma_1 - \gamma_3 + \gamma_2$.  To get them there we need to carry each through an
infinite set of other factors; at each step we use the pentagon identity, and obtain the final result
\begin{multline}
\CK_{\gamma_1} \CK_{-\gamma_3} \CK_{\gamma_2} = \\
\CK_{\gamma_2}\CK_{\gamma_2-\gamma_3} \CK_{2 \gamma_2 +\gamma_1-\gamma_3}\CK_{2 \gamma_2 +\gamma_1-2\gamma_3} \CK_{3 \gamma_2 +2\gamma_1-2\gamma_3}\CK_{3 \gamma_2 +2\gamma_1-3\gamma_3}\cdots
\CK_{-\gamma_3} \CK_{\gamma_1-\gamma_3+ \gamma_2}^{-2} \CK_{\gamma_1+\gamma_2}\cdots \\ \cdots \CK_{2\gamma_2 +3\gamma_1-3\gamma_3}\CK_{2\gamma_2 +3\gamma_1-2\gamma_3}
\CK_{2\gamma_2 +3\gamma_1-2\gamma_3}\CK_{\gamma_2 +2\gamma_1-2\gamma_3}\CK_{2\gamma_1+\gamma_2-\gamma_3}\CK_{\gamma_1-\gamma_3}\CK_{\gamma_1}.
\end{multline}
This indeed corresponds to the BPS spectrum in the weak coupling
region outside the wall.  Notice that the two ``electric'' hypermultiplets of charges $-\gamma_3$ and
$\gamma_1 + \gamma_2$ are the basic matter particles in the Lagrangian
of the $SU(2)$ theory with $N_f=1$, and they have the same gauge
charge, half that of the W boson.

As we vary $\vartheta$ in the weak coupling region, $T_\WKB$
undergoes an infinite sequence of flips and then a transformation
similar to the ``\juggle'' we described in Section
\ref{sec:vector-jump} and encountered in the $N_f = 0$ theory.
However, it is not quite the same; in particular, in the limiting
triangulation one of the boundaries of the annulus has \ti{two}
vertices on it (coming from the two WKB rays entering the
singularity at $z=0$). See Figure \ref{fig:Nf1plot-limit}. This is a
non-generic situation, which occurs here because we chose the
special value $m=0$. It is related to the fact that the coordinate
transformation across the critical $\vartheta$ must be
$\CK_{-\gamma_3} \CK_{\gamma_1-\gamma_3+ \gamma_2}^{-2}
\CK_{\gamma_1+\gamma_2}$ corresponding to a vectormultiplet and two
hypermultiplets, in contrast to the usual situation where we have
only the vectormultiplet at the critical $\vartheta$.

\insfig{5in}{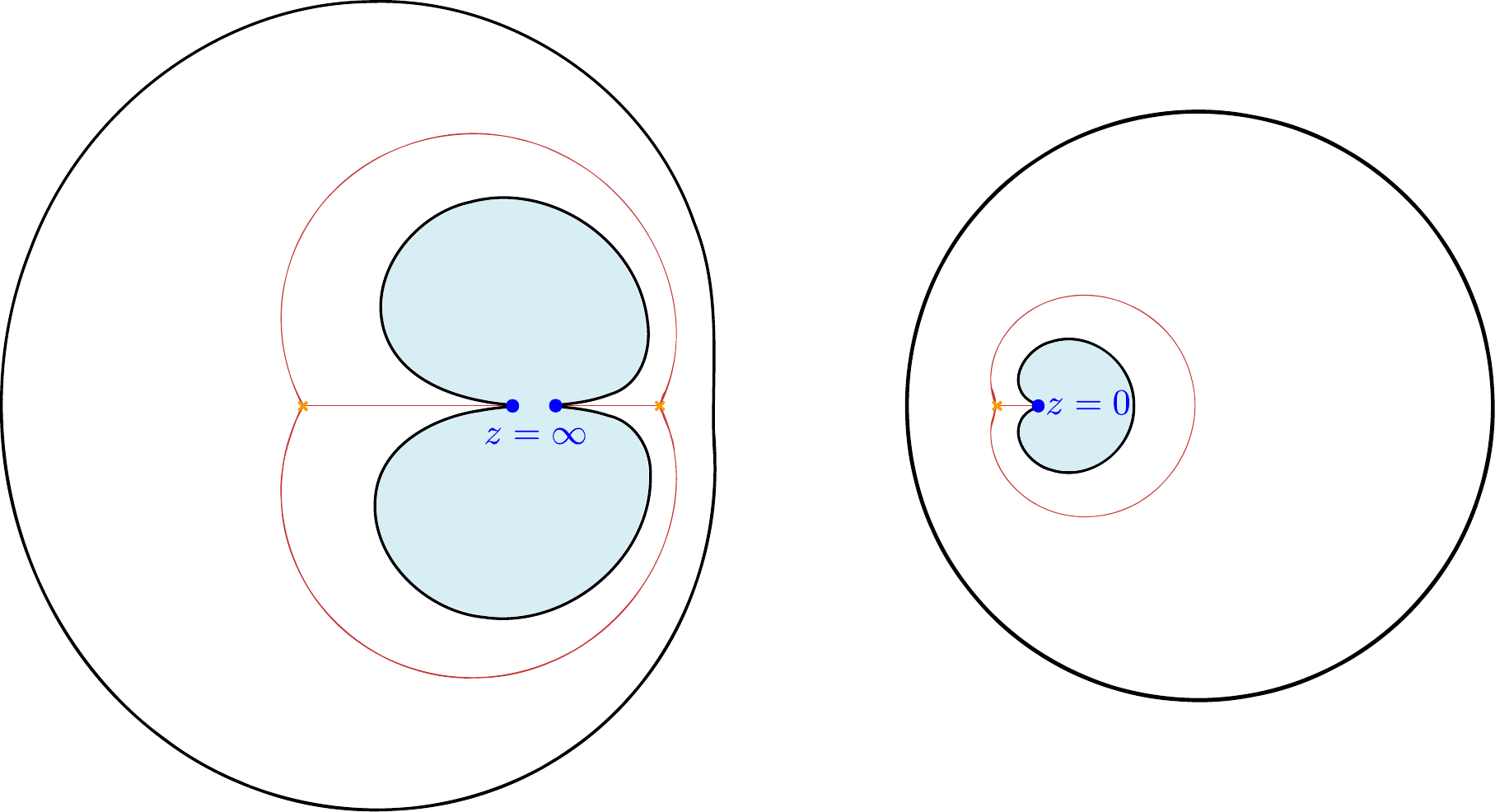}{The limit $T_\WKB$ for the $SU(2)$
theory with $N_f=1$ and $m=0$.  For clarity $z=0$ and $z=\infty$
have been mapped to finite distance, and an extra closed WKB curve is
shown.}

\insfig{6in}{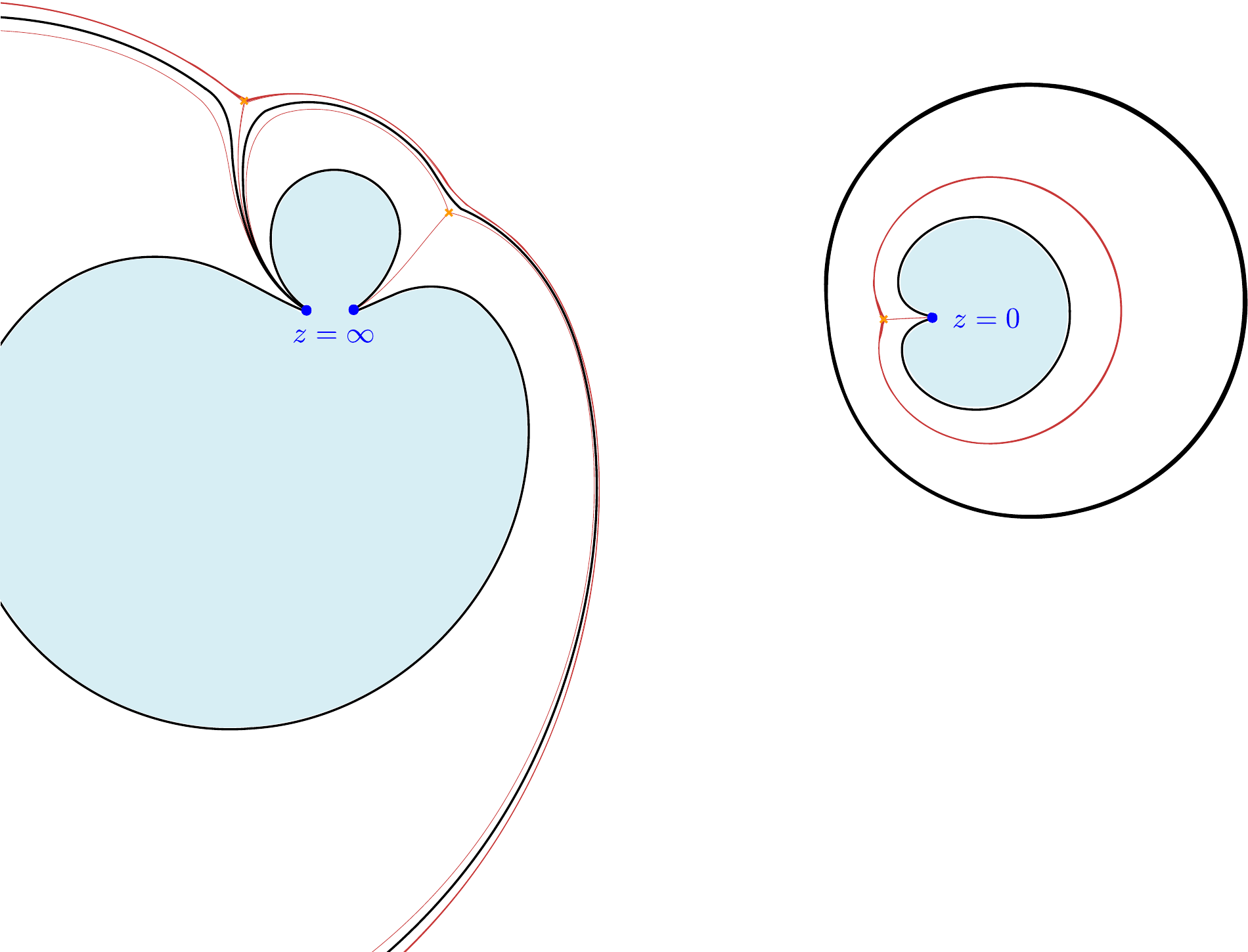}{The limit $T_\WKB$ for the $SU(2)$
theory with $N_f=1$ and $m \neq 0$.  For clarity $z=0$ and
$z=\infty$ have been mapped to finite distance, and an extra closed
WKB curve is shown.}

As soon as we perturb $m$ slightly away from zero, we reach a more
conventional setup. As we move from the strong coupling region
towards large $u$, the sequence of elementary operations we
described, or some close analog, happens at a sequence of walls of
marginal stability. At large but finite $u$, the central charges of
the hypermultiplets associated with $\CK_{-\gamma_3}$ and
$\CK_{\gamma_1+\gamma_2}$ are not exactly aligned with the central
charge associated to $\CK_{\gamma_1-\gamma_3+ \gamma_2}^{-2}$. Hence
in rearranging the product the two factors $\CK_{-\gamma_3}$ and
$\CK_{\gamma_1+\gamma_2}$ should only be moved a finite number of
steps.  The ``core'' of the infinite product around
$\CK_{\gamma_1-\gamma_3+ \gamma_2}^{-2}$ then corresponds to the
standard sequence of flips and \juggles as in Section
\ref{sec:vector-jump}.  The limit WKB triangulation is shown in
Figure \ref{fig:Nf1plot-limit-pert}; now both boundaries of the
annulus have only one vertex.

It is amusing to bring $m$ close to the value
$\frac{3}{2} \Lambda$, where two singularities in $\CB$
coalesce.  We expect to see the $N=3$ theory from Section \ref{sec:N3}
emerge in a scaling region near the two singularities.  Indeed, it is instructive to look at the
shape of the WKB foliation (see Figure \ref{fig:Nf1plotN3}) in the
scaling region, and see the pentagon appear near the region where
the three turning points are converging.
\insfig{4.5in}{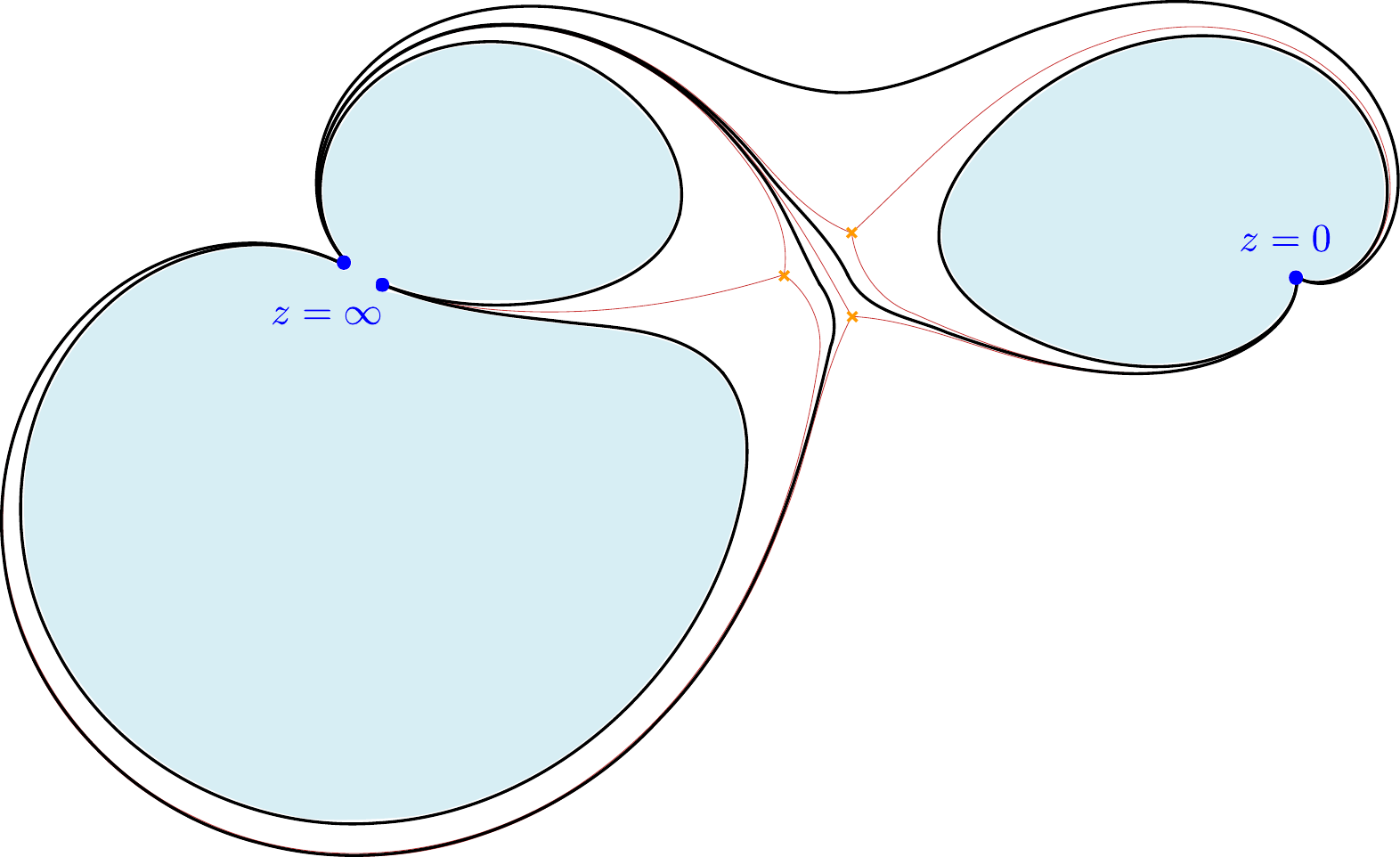}{The WKB triangulation for the $SU(2)$ theory with $N_f=1$,
with $m$ adjusted near the scaling limit in which the $N=3$ theory
from Section \protect\ref{sec:N3} emerges.}

\subsection{$N_f=2$, first realization}

Using the brane setup of Section \ref{sec:physics},
there are two ways to add a second flavor to the $SU(2)$ gauge
theory.  Correspondingly, there are two distinct Hitchin systems, for which the metric
on $\CM$ is expected to be the same.  In this subsection we
deal with the symmetric possibility, where the quadratic differential is
\begin{equation}
\lambda^2 = \left( \frac{\Lambda^2}{z^4} + \frac{2 \Lambda m_1}{z^3} + \frac{4u}{z^2} +  \frac{2 \Lambda m_2}{z}+ \Lambda^2 \right) dz^2
\end{equation}
The expansion around either $z = 0$ or $z = \infty$ then shows an
irregular singularity with two WKB rays. The two mass parameters
$m_{1,2}$ are the residues of the two poles.

There are four turning points, and four singularities in $\CB$.
The reader will probably guess that, once again, there will be a strong
coupling region in $\CB$ where the BPS spectrum consists only of the four
BPS particles which are massless at the four singularities.

If we set $m_{1,2} = 0$, a surprising simplification occurs. The
Seiberg-Witten differential for the $N_f = 2$ theory, and in fact
the whole Hitchin system, can be mapped to those of the $N_f = 0$
theory, by the simple coordinate transformation $z^2 \to \hat z$.
This allows us to borrow much of the discussion from the $N_f = 0$
case. Each turning point of the $N_f = 0$ theory maps to two turning
points of the $N_f = 2$ theory.  Each BPS hypermultiplet of the
$N_f=0$ theory maps to a pair of BPS hypermultiplets in the $N_f=2$
theory. See Figure \ref{fig:Nf2plot-a}.  This pair of
hypermultiplets have the same gauge charges, but their flavor
charges differ:  the difference of the two corresponding paths can
be deformed to a sum of paths around $0$ and infinity.  We call the
charges of one pair $\gamma_1^{1,2}$, and the other pair
$\gamma_2^{1,2}$, with $\langle \gamma_1^i, \gamma_2^j \rangle = 1$.
In addition $\langle \gamma_1^i, \gamma_1^j\rangle =\langle
\gamma_2^i, \gamma_2^j\rangle =0$.  If we turn on masses then
$Z_{\gamma_1^1 - \gamma_1^2} = m_1+m_2$, while $Z_{\gamma_2^1 -
\gamma_2^2} = m_1 - m_2$.

\insfig{3.2in}{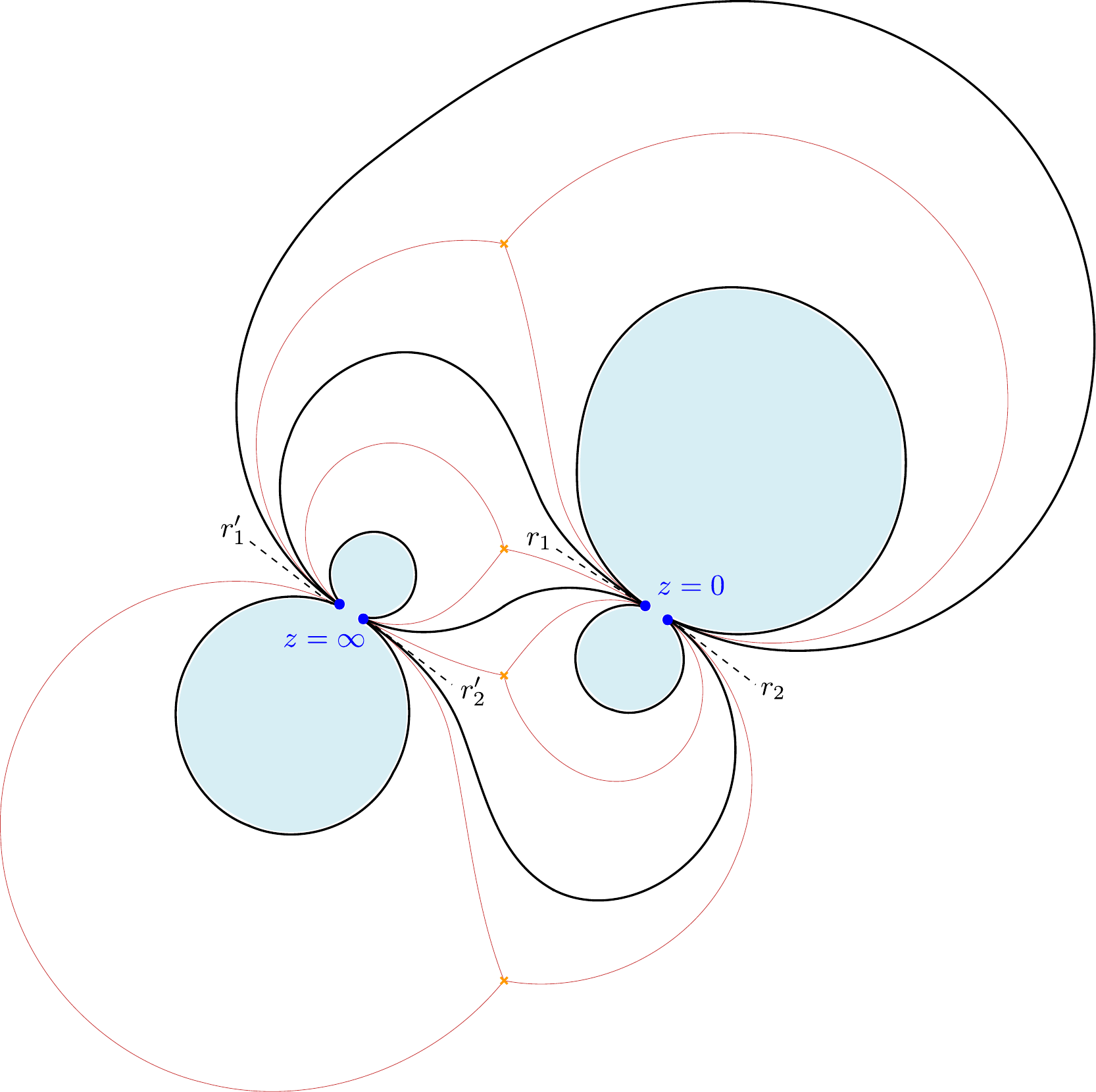}{A typical WKB triangulation for the first realization
of the $SU(2)$ theory with $N_f=2$, $m_1 = m_2 = 0$.  For clarity $z=0$ and $z=\infty$
have been mapped to finite distance.}
\insfig{3.2in}{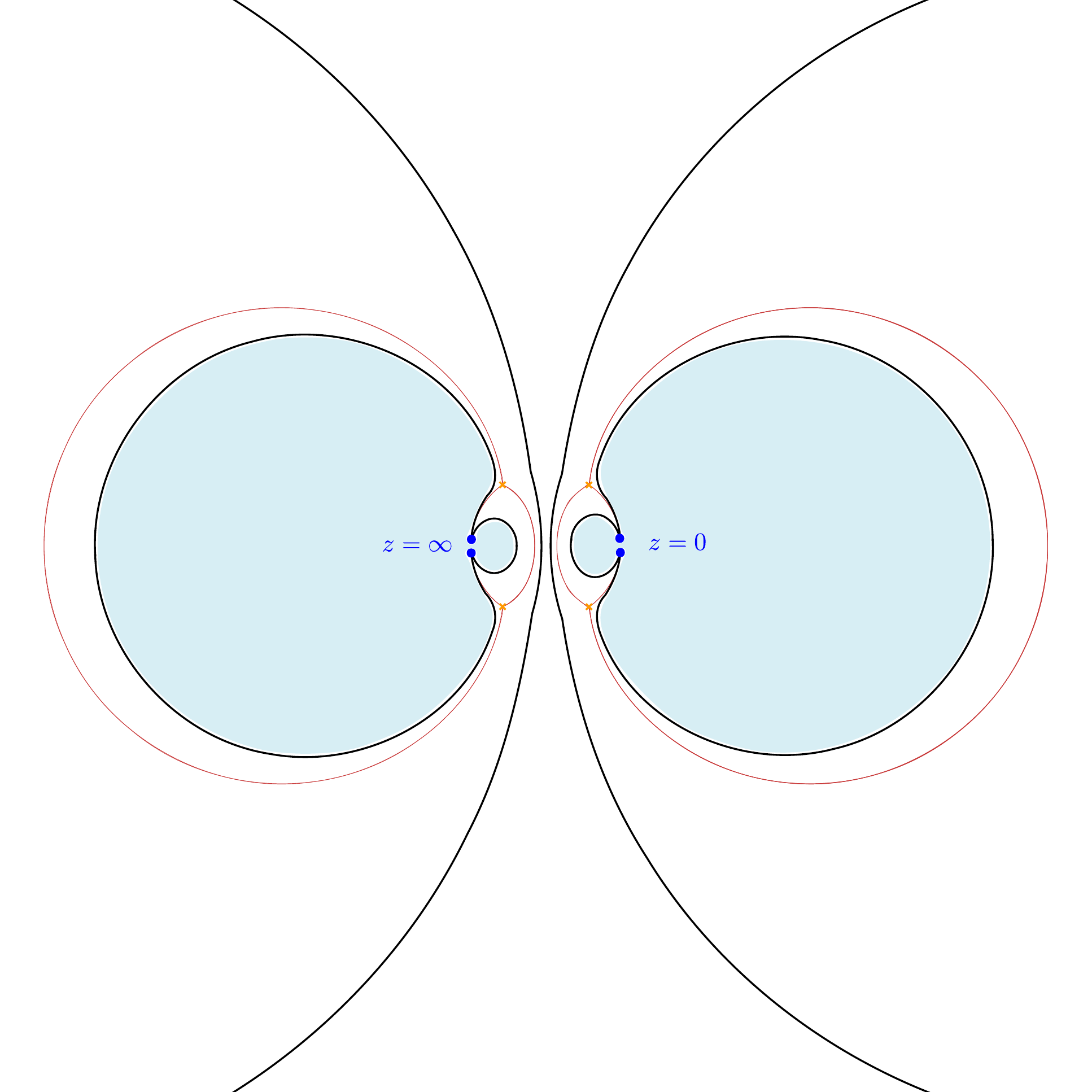}{The limit WKB triangulation $T_\WKB(\vartheta_c)$ where a vectormultiplet appears,
in the first realization of the $SU(2)$ theory with $N_f=2$, $m_1 = m_2 = 0$.  For clarity
$z=\infty$ has been mapped to finite distance, and we show two closed
WKB curves among the 1-parameter family representing the vectormultiplet.}

Following this reasoning we can guess the shape of the relevant wall-crossing formula ---
it should be a kind of doubling of the formula \eqref{eq:wcf-prod2} for $N_f = 0$, of the form
\begin{equation}
\CK_{\gamma^1_1} \CK_{\gamma^2_1} \CK_{\gamma^1_2} \CK_{\gamma^2_2}
= \CK_{\gamma_2}\CK_{\gamma_2} \CK_{\gamma_1+2 \gamma_2}\CK_{\gamma_1+2 \gamma_2} \cdots  ??? \cdots \CK_{2\gamma_1+\gamma_2}\CK_{2\gamma_1+\gamma_2}\CK_{\gamma_1}\CK_{\gamma_1}.
\end{equation}
On the right side it is not obvious \emph{a priori} which of the
cycles $\gamma_{1,2}^i$ each $\gamma_{1,2}$ represents, though it
may be determined from inspection of the flips in the WKB
triangulation. Also, it is not obvious what should replace the
\juggle transformation, as we have two vertices on each of the
circles surrounding the annular region --- see Figure
\ref{fig:Nf2plot-a-limit}.  We can of course simply play around with
the wall-crossing formulae we already know to fully determine the
right hand side.  We begin by bringing $\CK_{\gamma_1^2}$ to the
right, obtaining
\begin{align}
\CK_{\gamma_1^1}\CK_{\gamma_1^2} \CK_{\gamma_2^1}\CK_{\gamma_2^2} =
\CK_{\gamma_1^1} \CK_{\gamma_2^1}\CK_{\gamma_2^1+\gamma_1^2}\CK_{\gamma_2^2}\CK_{\gamma_2^2+\gamma_1^2} \CK_{\gamma_1^2},
\end{align}
and then do the same for $\CK_{\gamma_1^2}$, giving
\begin{multline}
\CK_{\gamma_1^1} \CK_{\gamma_2^1}\CK_{\gamma_2^1+\gamma_1^2}\CK_{\gamma_2^2}\CK_{\gamma_2^2+\gamma_1^2} \CK_{\gamma_1^2}
= \\ \CK_{\gamma_2^1} \CK_{\gamma_2^1+\gamma_1^1} \CK_{\gamma_2^1+\gamma_1^2} \CK_{\gamma_2^1+\gamma_1^2+\gamma_1^1} \CK_{\gamma_2^2} \CK_{\gamma_2^2+\gamma_1^1}
\CK_{\gamma_2^2+\gamma_1^2}\CK_{\gamma_2^2+\gamma_1^2+\gamma_1^1}  \CK_{\gamma_1^2}  \CK_{\gamma_1^1}.
\end{multline}
We still need to bring $\CK_{\gamma_2^2}$ to the left, but it will cross $\CK_{\gamma_2^1+\gamma_1^2+\gamma_1^1}$, leading to an
infinite product like those we encountered in the $N_f = 0$ example:
\begin{multline}
\CK_{\gamma_2^1} \CK_{\gamma_2^1+\gamma_1^1} \CK_{\gamma_2^1+\gamma_1^2} \CK_{\gamma_2^1+\gamma_1^2+\gamma_1^1} \CK_{\gamma_2^2} \CK_{\gamma_2^2+\gamma_1^1}
\CK_{\gamma_2^2+\gamma_1^2}\CK_{\gamma_2^2+\gamma_1^2+\gamma_1^1}  \CK_{\gamma_1^2}  \CK_{\gamma_1^1}
= \\ \CK_{\gamma_2^1} \CK_{\gamma_2^1+\gamma_1^1} \CK_{\gamma_2^1+\gamma_1^2} \CK_{\gamma_2^2}\CK_{\gamma_2^1+2\gamma_2^2+\gamma_1^2+\gamma_1^1}\cdots
\CK_{\gamma_2^1+\gamma_2^2+\gamma_1^2+\gamma_1^1}^{-2}\cdots \\ \cdots
 \CK_{2\gamma_2^1+\gamma_2^2+2\gamma_1^2+2\gamma_1^1}\CK_{\gamma_2^1+\gamma_1^2+\gamma_1^1} \CK_{\gamma_2^2+\gamma_1^1}
\CK_{\gamma_2^2+\gamma_1^2}\CK_{\gamma_2^2+\gamma_1^2+\gamma_1^1}  \CK_{\gamma_1^2}  \CK_{\gamma_1^1}.
\end{multline}
The factors
$\CK_{\gamma_2^1+\gamma_1^1}\CK_{\gamma_2^1+\gamma_1^2}$ and
$\CK_{\gamma_2^2+\gamma_1^2}\CK_{\gamma_2^2+\gamma_1^1}$
correspond to states with the same phase as the vector
multiplet $\CK_{\gamma_2^1+\gamma_2^2+\gamma_1^2+\gamma_1^1}^{-2}$.
Hence they need to be brought across the infinite product, through a
sequence of pentagon identities, giving the final result
\begin{multline}
\CK_{\gamma_1^1}\CK_{\gamma_1^2} \CK_{\gamma_2^1}\CK_{\gamma_2^2}
= \\ \CK_{\gamma_2^1}  \CK_{\gamma_2^2}\CK_{\gamma_1^1+\gamma_2^2+\gamma_2^1} \CK_{\gamma_1^2+\gamma_2^2+\gamma_2^1}\CK_{\gamma_1^2+\gamma_1^1+\gamma_2^2+2\gamma_2^1}
\CK_{\gamma_1^2+\gamma_1^1+\gamma_2^1+2\gamma_2^2}\cdots \\ \cdots\CK_{\gamma_2^1+\gamma_1^1} \CK_{\gamma_2^1+\gamma_1^2}
\CK_{\gamma_2^1+\gamma_2^2+\gamma_1^2+\gamma_1^1}^{-2}\CK_{\gamma_2^2+\gamma_1^1}
\CK_{\gamma_2^2+\gamma_1^2} \cdots \\ \cdots \CK_{\gamma_1^2+2\gamma_1^1+\gamma_2^2+\gamma_2^1}
\CK_{2\gamma_1^2+\gamma_1^1+\gamma_2^2+\gamma_2^1}\CK_{\gamma_1^2+\gamma_1^1+\gamma_2^1} \CK_{\gamma_1^2+\gamma_1^1+\gamma_2^2}  \CK_{\gamma_1^2}  \CK_{\gamma_1^1}.
\end{multline}
As in the $N_f = 1$ example, if we turn on $m_{1,2} \neq 0$, the single wall will fragment into
several walls.  By playing around with these
parameters one can produce some entertaining results. For example, if we
set $m_1 = m_2 = 2 \Lambda$, we get a neat example,
\begin{equation}
\lambda^2 = \left( \frac{\Lambda^2(z+1)^4}{z^4} + \frac{16 \tilde u}{z^2} \right) dz^2,
\end{equation}
where three out of four singular points in $\CB$ coalesce at
$\tilde u = 0$, and we can recover the $N=4$ Argyres-Douglas scaling limit.  One of
the four particles in the strong coupling region becomes very
massive, and the other three coincide with the basic spectrum of the
$N=4$ Argyres-Douglas theory.

\subsection{Intermission: non-abelian flavor symmetries and WKB
triangulations}

We noticed in Section \ref{sec:flavor} how the WKB flow lines behave when
the mass parameter at a regular singularity $\CP$ goes to zero, and a
non-abelian $SU(2)$ flavor symmetry is restored.  It is useful to
consider the behavior of the whole WKB foliation in that limit.  As
$m \to 0$, a turning point $\CT$ moves toward $\CP$.  At generic $\vartheta$,
one of the three separating WKB emerging from $\CT$ flows toward $\CP$.  The
other two wrap around $\CP$ in opposite directions, and then flow
away while remaining very close to each other, ending up at some other
singularity.  ($T_\WKB$ is thus degenerate, with a single
edge coming out of $\CP$.)
As we vary $\vartheta$, we will occasionally
meet pairs of BPS states with very close phases, when these
two separatrices become finite WKB curves landing on another turning point
$\CT'$.  The charges of these BPS states differ by
(twice) a cycle wrapping around the singularity.  In other words, these finite
WKB curves represent a doublet of BPS particles with charges $\gamma \pm
\gamma_+$, where $\gamma_+$ is the flavor charge associated to the singularity $\CP$.

For $\vartheta$ in the narrow window between the phases of $Z_{\gamma \pm \gamma_+}$, these
two separatrices pass on opposite sides of the turning point $\CT'$, and go to different
singularities.  The corresponding transformations of the WKB
triangulation are straightforward.  The first BPS state induces a flip
of the non-degenerate edge of the degenerate triangle associated to
$\CP$. Then two edges end on $\CP$.  The second BPS state induces a flip
of the old edge ending on $\CP$, so we get again a degenerate
triangle. \insfig{4.5in}{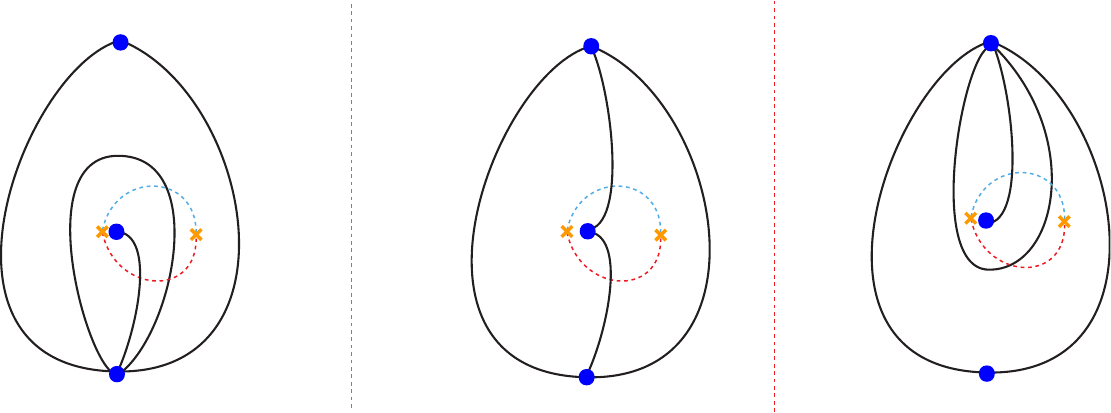}{The two flips in $T_\WKB$ associated
to a $SU(2)$ flavor doublet of BPS hypermultiplets, for small $SU(2)$ breaking
mass.}

\subsection{$N_f=2$, second realization}

Next we consider with the asymmetric realization of the
$N_f=2$ theory, where the irregular singularity at infinity is the
same as the one in the $N_f=0$ theory, but the irregular singularity at $z=0$ is
replaced by two regular singularities:
\begin{equation}
\lambda^2 = \frac{P_3(z)}{z^2 (z-1)^2} = \left( \frac{m_+^2}{z^2}+\frac{m_-^2}{(z-1)^2} + \frac{\Lambda^2 +u}{2 z} + \frac{\Lambda^2 -u}{2 (z-1)} \right) dz^2
\end{equation}
Here the parameters $u, \Lambda$ do not necessarily coincide with the
ones in the previous subsection.  The expansion around
$z=\infty$ shows an irregular singularity with a single WKB ray.
There are three turning points.  There are four singular points in $\CB$.

For very small masses $m_\pm$, the typical $T_\WKB$ here has
two degenerate triangles attached to the singularity at $z =
\infty$.  The strong-coupling spectrum just includes two pairs
of BPS states, one for each singularity, as shown in Figure
\ref{fig:Nf2evolution-a}. \insfig{5.5in}{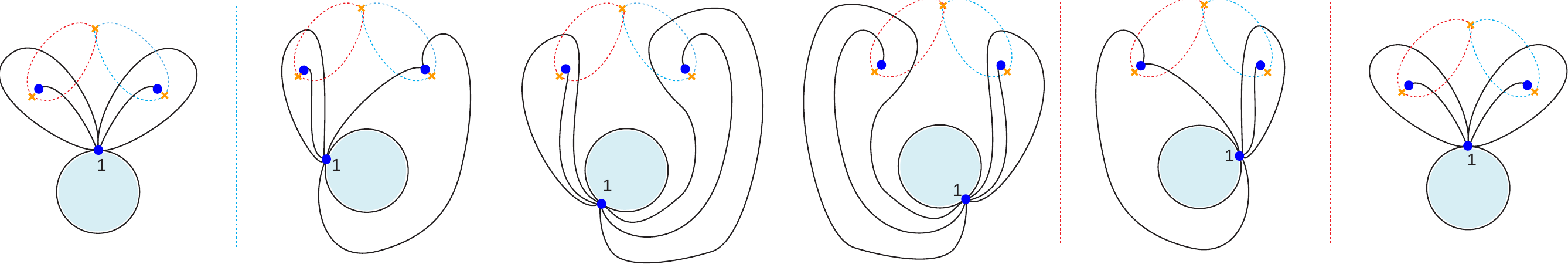}{The
evolution of the WKB triangulation for the second realization
of the $N_f=2$ theory, in the strong coupling region around
$u=0$, for small $m_\pm$.  For clarity $z = \infty$ has been
mapped to finite distance.  There are two pairs of flips,
corresponding to two BPS states with charges $\gamma^i_1$,
$\gamma^i_2$, with $\langle \gamma^i_1,\gamma^j_2\rangle=1$.
Notice that the two middle pictures differ by a pop, which does
not correspond to a BPS state.} Comparison with the previous
section (or the analysis of Section \ref{sec:physics}) shows
that $m_\pm = m_1 \pm m_2$. The weak-coupling spectrum
manifests itself as a sequence of pairs of BPS states
associated to either the singularity at $z=0$ or at $z=1$ (as
described in the last section), which produces tightly wound
degenerate triangles. For general $m_\pm$, less degenerate
triangulations may occur, as in Figure \ref{fig:Nf2plot-b}.
\insfig{2.5in}{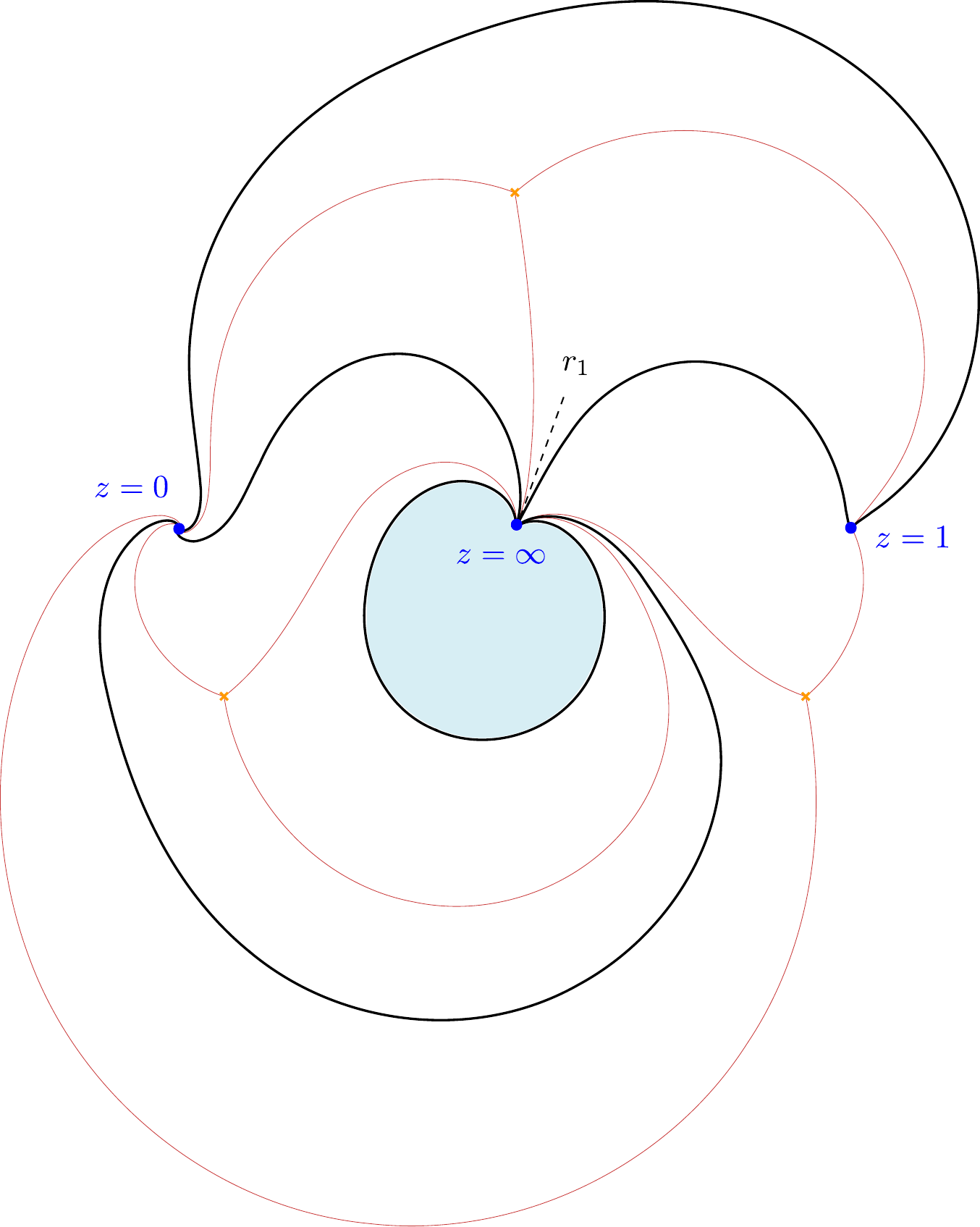}{A typical WKB triangulation for the
$SU(2)$ theory with $N_f=2$, second realization, with generic
masses. For clarity $z=\infty$ has been mapped to finite
distance.}

It is not at all obvious that the two \hk\ metrics corresponding
to the two realizations of the $N_f=2$ theory are
the same. Indeed, this is an example of the surprising isomorphism of
Section \ref{sec:surprise}. A proof that the two Hitchin systems for
$N_f=2$ have the same \hk metric goes as follows: the two systems
have the same spectral curves (although they are embedded
differently in $T^* \IC \IP^1$) and the same central charge
functions.  Now suppose we show (say, by explicit examination of
$T_\WKB$) that the spectra of BPS states in
the two systems coincide at some strong coupling point. Then the
wall-crossing formula ensures that the spectra agree everywhere on
$\CB$, and moreover the functions $\CX^\vartheta_\gamma$ must also
agree, as they are solutions of the same Riemann-Hilbert problem,
with the same asymptotics. Hence the twistor spaces and the \hk
metrics for the two Hitchin systems must coincide.

\subsection{$N_f=3$}

The $SU(2)$ gauge theory with $N_f=3$ is associated to a $SU(2)$
Hitchin system which is rather similar to the second realization of
the $N_f=2$ theory.  The quadratic differential is
\begin{equation}
\lambda^2 = \frac{P_4(z)}{z^2 (z-1)^2} = \left(
\frac{m_+^2}{z^2}+\frac{m_-^2}{(z-1)^2} + \frac{2\Lambda m +u}{2 z}
+ \frac{2 \Lambda m -u}{2 (z-1) } + \Lambda^2\right) dz^2
\end{equation}
The expansion around $z=\infty$ shows an irregular singularity with
two WKB rays. See Figure \ref{fig:Nf3plot}.

There are four turning points in the $z$-plane, and five singularities in $\CB$.
The three mass parameters $m$, $m_\pm$ appear on a different footing.
(To treat them symmetrically we would have to realize the theory in
terms of an $SU(3)$ Hitchin system, which goes beyond the scope of this paper.)

\insfig{2.5in}{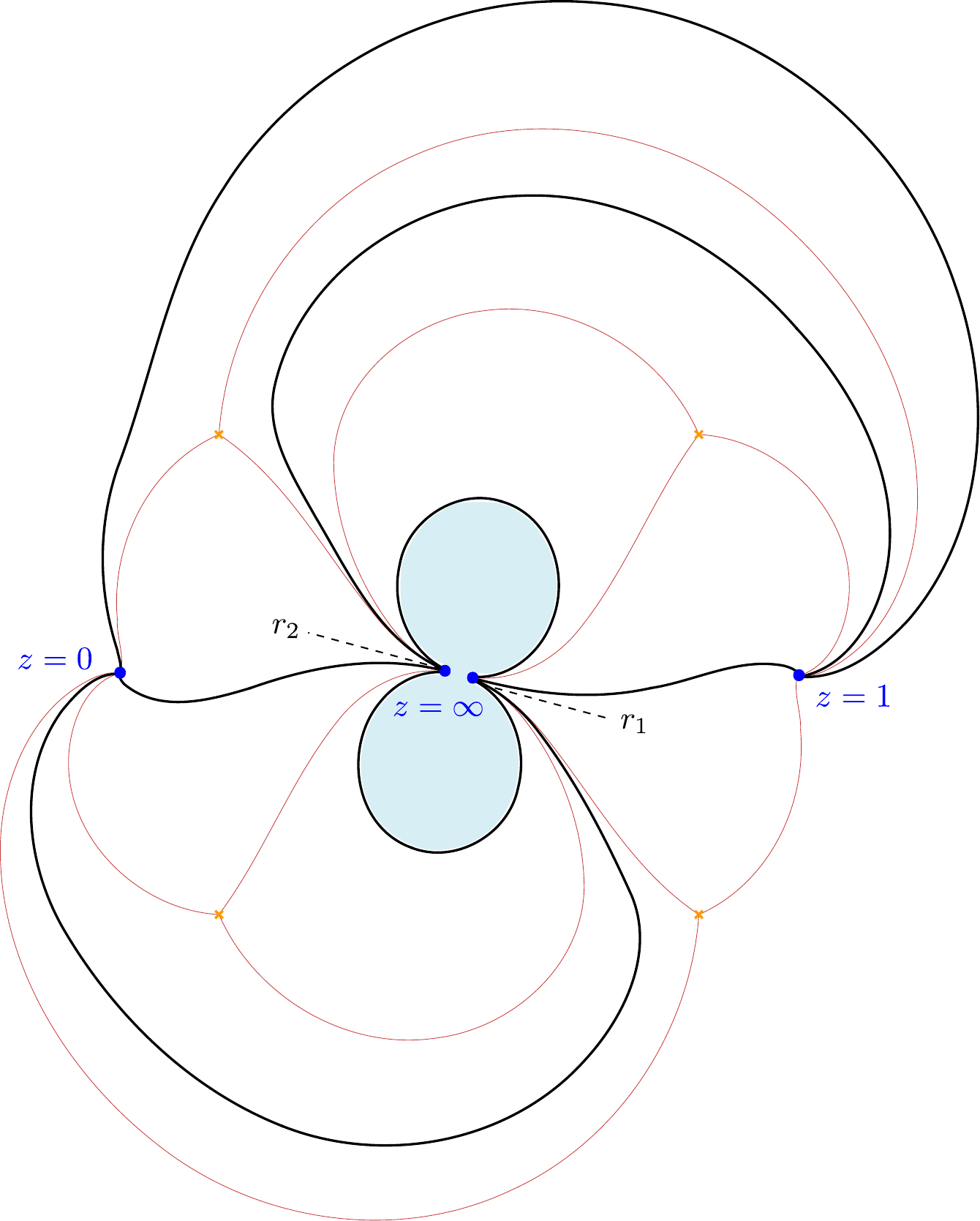}{A typical WKB
triangulation for the $SU(2)$ theory with $N_f=3$.  For clarity $z=\infty$
has been mapped to finite distance.}

For very small $m$ and $m_\pm$, four of the singularities in
$\CB$ come together near $u=0$. The BPS particles which become
massless at these four singularities have the same gauge
charges, but different flavor charges. See Figure
\ref{fig:Nf3evolution-a}. \insfig{5.5in}{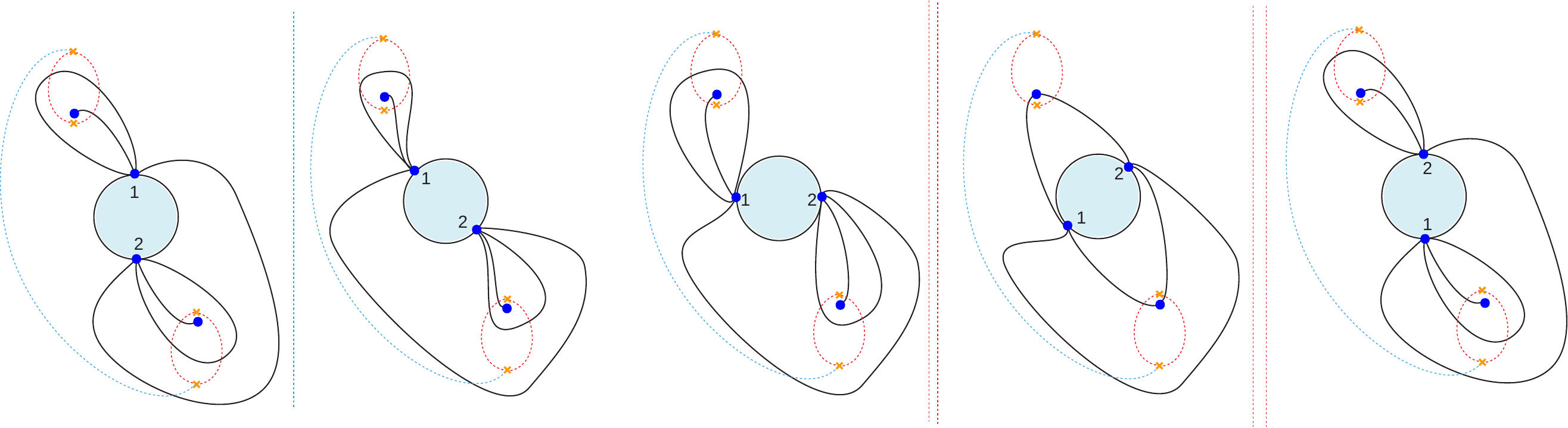}{The
evolution of $T_\WKB(\vartheta)$ in the $N_f=3$ theory, in the
strong coupling region of $\CB$ around $u=0$, for small $m$ and
$m_\pm$. For clarity $z=\infty$ has been mapped to finite
distance.  As $\vartheta$ varies there are five flips,
corresponding to BPS states with charges $\gamma^i_1$,
$\gamma_2$ with $\langle \gamma^i_1,\gamma_2\rangle=1$. The
second and third pictures differ by a pop, which does not
correspond to a BPS state.} We know that for small enough
$m_\pm$ a turning point will be close to each of the regular
singularities, and BPS states will typically appear in doublets
of the corresponding $SU(2)_\pm$ subgroups of the $SU(4)$
flavor symmetry. $m$ is the mass parameter for a remaining
$U(1)\in SU(4)$. The four particles actually sit in a $\bf{4}$
of $SU(4)$ \cite{Seiberg:1994aj}. Indeed the $\bf{4}$
decomposes as $(\bf{2}, \bf{1})_{+1} \oplus (\bf{1},
\bf{2})_{-1}$ under $SU(2) \times SU(2) \times U(1)$. We see
from Figure \ref{fig:Nf3evolution-a} that the two doublets
indeed have charges which differ by the flavor charge at
infinity.

There is a second singularity at $u\sim \frac{\Lambda^2}{4}$, where
two turning points coalesce. All in all, there is a strong coupling
region containing the interval $0 < u < \frac{\Lambda^2}{4}$, with a
simple BPS spectrum: the quadruplet of particles of charge
$\gamma_1^i$ (all with the same gauge charge) and a single
$\gamma_2$ with $\langle \gamma_1^i,\gamma_2\rangle=1$. In addition
$\langle \gamma_1^i, \gamma_1^j \rangle =0$ for all $i,j$. The
wall-crossing formula then predicts the correct known spectrum in
the weak coupling region, as we can see by a simple but tedious
sequence of the identities we have used before. To write the final
result let $\tilde \gamma =
2\gamma_2+\gamma_1^1+\gamma_1^2+\gamma_1^3+\gamma_1^4$.  Then:
\begin{multline}
\left( \prod_{i=1}^4 \CK_{\gamma_1^i}\right) \CK_{\gamma_2} =
\CK_{\gamma_2} \left( \prod \CK_{\gamma_2+\gamma_1^i}\right) \CK_{\gamma_2+\tilde \gamma}\left( \prod \CK_{\tilde \gamma-\gamma_1^i}\right) \cdots \\
\cdots \CK_{2\gamma_2+\gamma_1^1+\gamma_1^2+\gamma_1^3+\gamma_1^4}^{-2} \left(\prod_{i<j}\CK_{\gamma_2+\gamma_1^i+\gamma_1^j}\right) \cdots \\
\cdots \left( \prod \CK_{\tilde \gamma-\gamma_2-\gamma_1^i}\right)\CK_{\tilde \gamma-\gamma_2}\left( \prod_{i=1}^4 \CK_{\gamma_1^i}\right).
\end{multline}

\subsection{$N_f=4$: The superconformal
case}\label{subsec:SuperConformalCase}

Finally, we have learned enough to discuss the $N_f=4$ theory properly.  In this subsection we
will only consider the ``balanced'' realization of the theory, with four regular
singular points and four turning points on $\IC\IP^1$.
The Seiberg-Witten curve looks like
\begin{equation}
\lambda^2 = \frac{P_4(z)}{D_4(z)^2}(dz)^2= \frac{P^0_4(z)+ u
D_4(z)}{D_4(z)^2}(dz)^2.
\end{equation}
In the second equality we introduced a basepoint and defined a
normalizable $u$-parameter.  In addition to $u$, the theory has four
mass parameters, and an exactly marginal gauge coupling. There are
$6$ singularities in the $u$ plane $\CB$.

If all the masses vanish (or if $u$ is much
larger than the masses) the WKB flows can become very intricate. In
this regime the WKB flows determined by $\lambda$ are well approximated by those for
\begin{equation} \lambda_0=\frac{dz}{\sqrt{D_4(z)}} = d\upsilon,
\end{equation}
where $\upsilon$ is a uniformizing coordinate on the elliptic curve
$y^2 = D_4(z)$; so these flows are just straight lines on a torus,
of inclination determined by $\vartheta$.  For any rational slope,
we get closed WKB curves. In particular, there is an infinite
spectrum of $W$-bosons forming an $SL(2,\IZ)$ duality orbit.

At the same $\vartheta$ where the W-bosons appear, eight
hypermultiplets should also appear, in representations ${\bf 8}_v$, ${\bf 8}_s$
or ${\bf 8}_c$ of the $Spin(8)$ flavor symmetry group.  We can understand
them as follows:  since we are considering very small masses,
each turning point is very close to
one of the four regular singular points.  BPS hypermultiplets joining two
turning points thus arise in doublets of the two $SU(2)$ flavor subgroups corresponding to
the two endpoints.  If we denote the four singular points as
$a, b, c, d$, and the corresponding subgroups of $Spin(8)$ as
$SU(2)_a$, $SU(2)_b$, $SU(2)_c$, $SU(2)_d$, then we have the
decompositions
\begin{align}
{\bf 8}_v = ({\bf 2}_a, {\bf 2}_b, {\bf 1}_c, {\bf 1}_d) \oplus ({\bf 1}_a, {\bf 1}_b, {\bf 2}_c, {\bf 2}_d), \notag \\
{\bf 8}_s = ({\bf 2}_a, {\bf 1}_b, {\bf 2}_c, {\bf 1}_d) \oplus ({\bf 1}_a, {\bf 2}_b, {\bf 1}_c, {\bf 2}_d), \notag \\
{\bf 8}_c = ({\bf 2}_a, {\bf 1}_b, {\bf 1}_c, {\bf 2}_d) \oplus ({\bf 1}_a, {\bf 2}_b, {\bf 2}_c, {\bf 1}_d).
\end{align}
Each closed WKB curve divides the four singular points into two pairs, either $(ab,cd)$, $(ac,bd)$, or
$(ad,bc)$.  The corresponding pair of turning points, say $(ab,cd)$,
supports the 8 BPS hypermultiplets in, say, ${\bf 8}_v$.

As soon as we turn on finite mass parameters, the spectrum
simplifies considerably.  Each singularity becomes an attractor,
preventing most of the closed WKB curves from appearing.  Indeed, it
is entertaining to introduce a different auxiliary torus by writing
\begin{equation}
\lambda = \frac{dz}{\sqrt{P_4(z)}}\frac{P_4(z)}{D_4(z)}= d\upsilon
\frac{P_4(z)}{D_4(z)},
\end{equation}
where now  $d\upsilon = dz/\sqrt{P_4(z)} $ is a uniformizing
parameter for the torus  $y^2 = P_4(z)$. The WKB curves lift to flow
lines for the meromorphic differential $\lambda$ on this torus. Each
zero of $P_4$ becomes a double zero on the torus, and each zero of
$D_4$ lifts to a pair of poles. If the mass parameters are small,
the pair of poles will be very close to the double zero. A local
picture of the flow shows a localized disturbance, with a
well-defined ``absorption cross-section'' for the flows. We can even
compute that cross section. We integrate the local form of
$\lambda$,
\begin{equation}
w = \int \lambda \sim \int_{\upsilon=0}^{\upsilon} \frac{m
(\upsilon')^2}{1-(\upsilon')^2}d\upsilon' = - m \upsilon +
\frac{m}{2}\log \frac{1+\upsilon}{1-\upsilon}.
\end{equation}
Since $- m \upsilon + \frac{m}{2}\log
\frac{1+\upsilon}{1-\upsilon}\sim \frac{m}{3}\upsilon^3 + \cdots$
there are six flows of ${\rm Im}(e^{-i \vartheta} w) =0$ emerging
from $\upsilon=0$. The flows which asymptote to parallel lines have
a spacing  $ \pi \mathrm{Re}\left( e^{-i \vartheta}m \right)$. See
Figure \ref{fig:perturbed-flow}.  \insfig{2.5in}{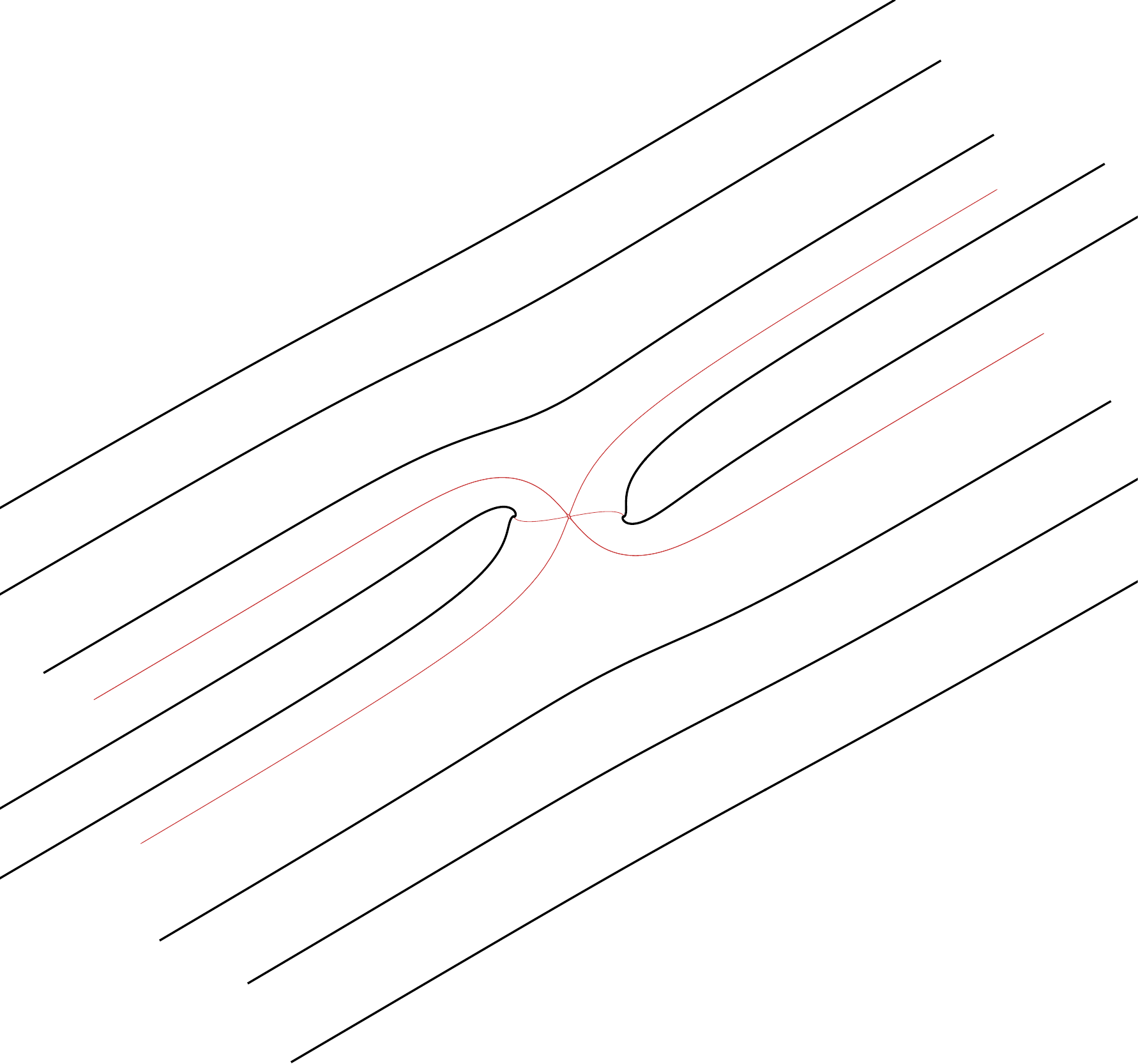}{The
local flow on the torus near a group of zeroes and poles of
$\lambda$.}

However small the mass parameter, the singularity will surely end up
attracting any WKB curve which winds too many times around the torus,
unless $\vartheta$ is very close to the phase of the mass parameter.
So unless the phases of all mass parameters are aligned, it appears
that only a finite number of $SL(2,\IZ)$ images of the W-boson will
survive at finite $u$.  Adjusting the mass parameters appropriately,
it should even be possible to find a ``strong coupling region''
where no vectormultiplet BPS states survive. That would clearly be
a good point from which to start our analysis.

With that in mind, let's consider the following symmetric choice of
parameters:
\begin{align}
\lambda^2 &= \frac{z^4-u(z^4-1)}{(z^4-1)^2}= \notag \\ &= \frac{1}{16 (z-1)^2}-\frac{1}{16
   (z-i)^2}-\frac{1}{16 (z+i)^2}+\frac{1}{16 (z+1)^2}+\frac{1-4 u}{16 (z-1)}-\frac{i (4 u-1)}{16 (z-i)}+\frac{i (4 u-1)}{16
   (z+i)}+\frac{4 u-1}{16 (z+1)}.
\end{align}
We have adjusted the gauge coupling to $\tau=i$, which gives a useful
discrete ${\mathbb Z}_4$ symmetry to the problem.  The singularities are
$z_a = 1$, $z_b = i$, $z_c = -1$, $z_d = -i$. The mass parameters
turn out to be $m_a = \frac{1}{4} z_a$.  Their sign is arbitrary:  we
used a convention such that $\sum m_a=0$ for this setup. An
example of a WKB triangulation is shown in Figure \ref{fig:Nf4plot}.
\insfig{3.0in}{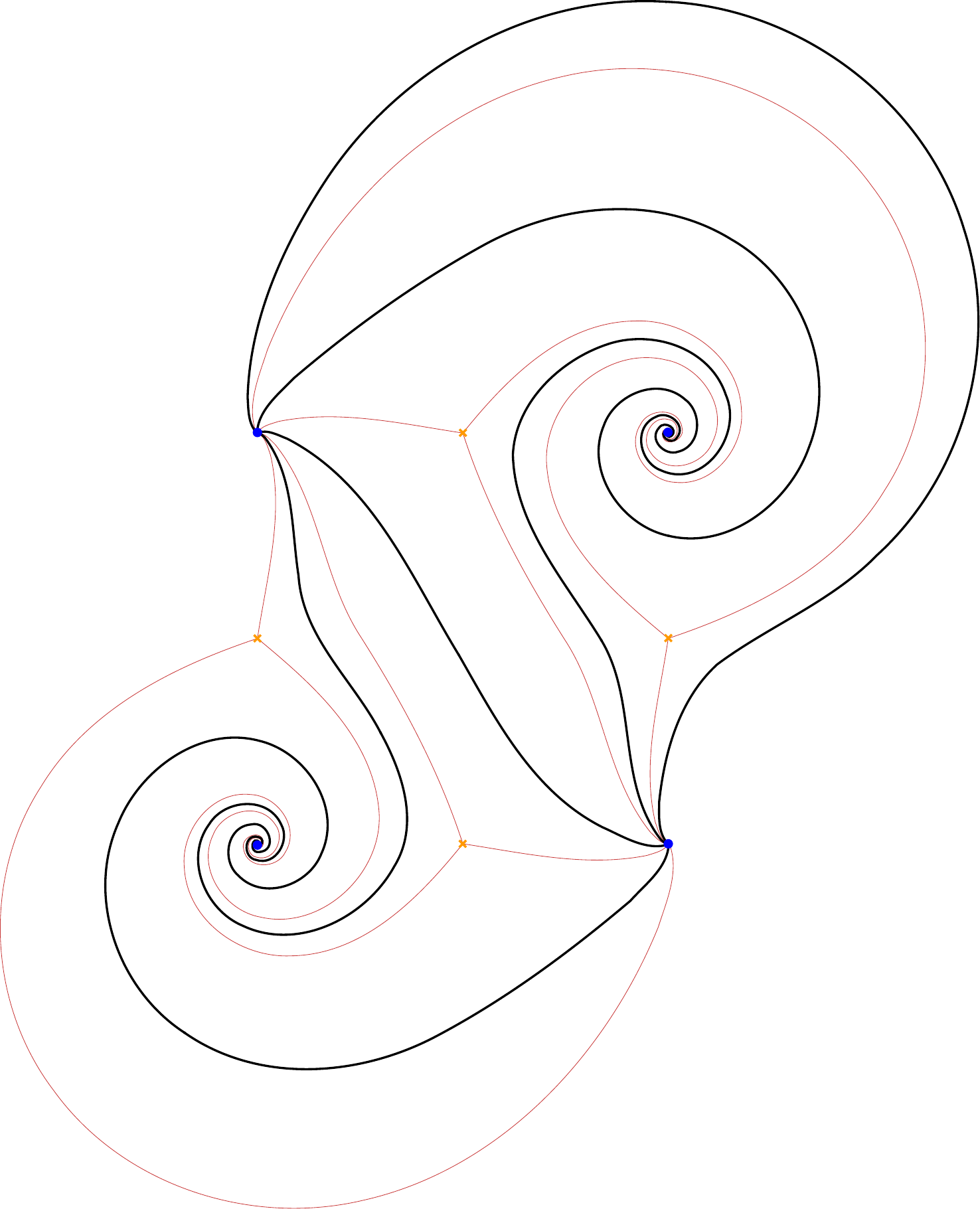}{A typical WKB triangulation for the $SU(2)$ theory with $N_f=4$,
at the special ${\mathbb Z}_4$ symmetric point.}

The ${\mathbb Z}_4$ symmetry ensures that the six singularities in the $u$
plane coalesce in two groups of three, giving two singular
points of the $N=4$ type at $u=0, 1$.  As $u$ approaches these two points,
the turning points coalesce at $z=0$ or $z=\infty$ respectively. We know that near each
$N=4$ singularity there should be $6$ light BPS particles, joining
the turning points in all possible ways (passing either near $z=0$
or near $z=\infty$). It is easy to see that on the segment $0<u<1$
these two groups of $6$ particles exhaust the BPS spectrum! This can
be checked, say, at $u=1/2$, and remains true on the whole segment
because the phase of the central charges are actually constant
there, so no wall-crossing may happen.

We expect to see an ellipsoidal ``strong coupling'' region where the
spectrum consists of those $12$ particles. We will need a convenient
labeling of the charges for these BPS particles. First of all, we
can forget the flavor charges and project to a
convenient $\IZ^2$ lattice of gauge charges, inspired by the labeling
in the $N=4$ example.  We will meet only charges
\begin{equation}(0,1), (1,1), (1,0), (1,-1). \end{equation}
Indeed, the turning points form a square; BPS strings joining
opposite sides have the same gauge charge $(0,1)$ or $(1,0)$, while
the diagonals have gauge charge $(1,1)$ or $(1,-1)$.
\insfig{7in}{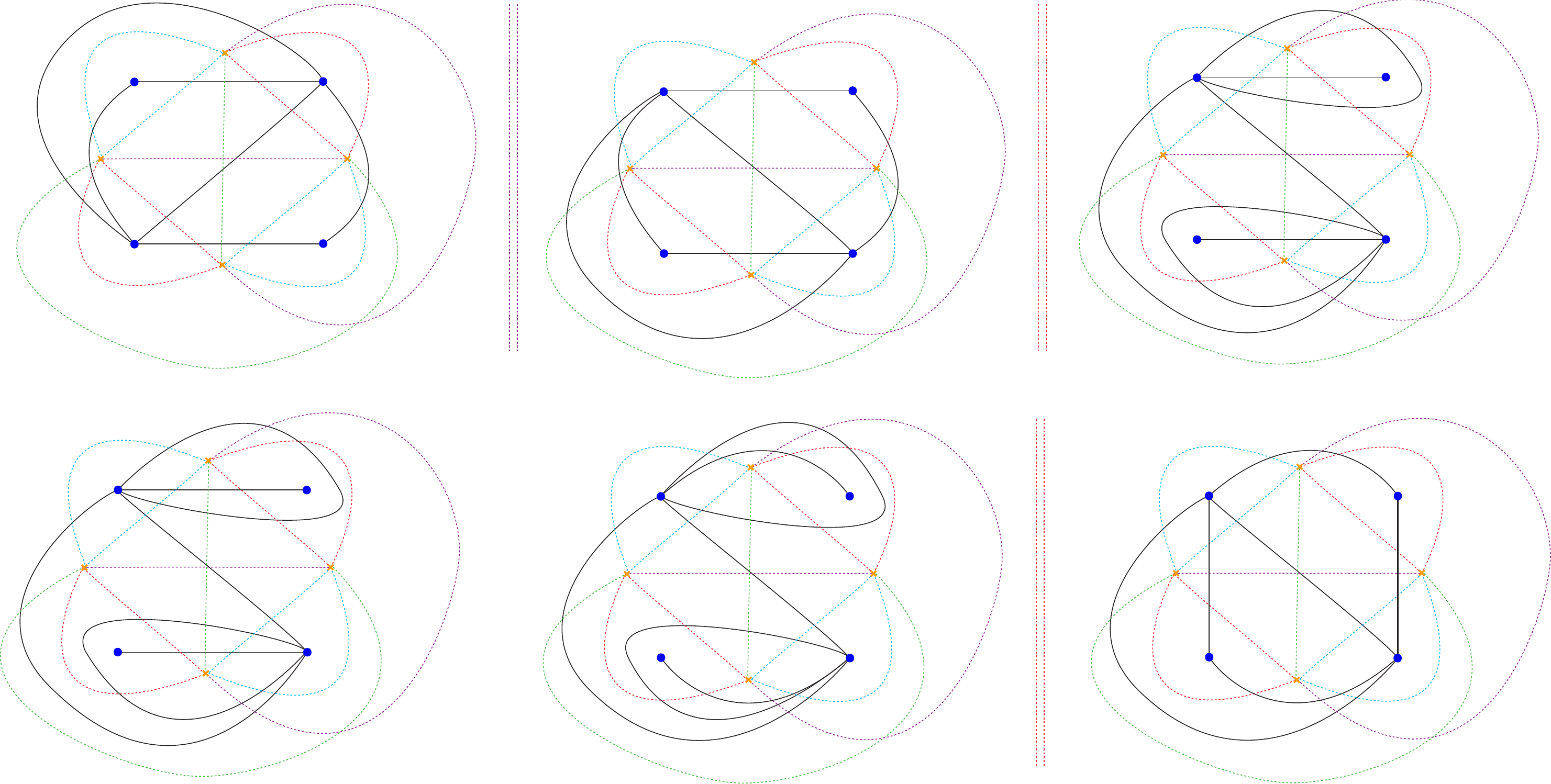}{The evolution of the WKB triangulation
for the $SU(2)$ theory with $N_f = 4$, for the parameters specified in the text.
We only show half of the overall evolution, i.e.  six out of twelve
flips. Because of the $\IZ_4$ symmetry of the problem, the second
half of the sequence is identical to the first half, with pictures
rotated by $90$ degrees. We split the sequence further in two rows,
the last triangulation in the first row coincides with the first in
the second row. We see first a flip of the two diagonal edges,
corresponding to the purple BPS states, then a flip of two side
edges corresponding to two of the four red BPS states, then in the
next row a pop of the northeast singularity and of the southwest
one, followed by flips corresponding to the other two red BPS
states. The other half of the evolution involves another flip of the
diagonal edges due to the green BPS states, flips corresponding to
two blue BPS states, a pop of the northwest and of the southeast
singularities, and another flip from the other two blue BPS states.}

In Figure \ref{fig:Nf4evolution-a} we show the evolution of $T_\WKB$,
from which we can read off the BPS particles at $u = \half$.
There are four particles of gauge charge $(0,1)$, four of
charge $(1,0)$, two of charge $(1,1)$ and two of charge $(1,-1)$,
with various combinations of flavor charges.  Let us ignore the
flavor charge information for the moment.  The triangulations
undergo a simple sequence of flips, which involves the required
appearance of degenerate triangles for each regular singularity, at
$\vartheta=0,\pi$ or $\vartheta=\pm \pi/2$.

Reading from the figure, the corresponding sequence of KS
transformations is (suppressing flavor information, but showing a
$\pm$ sign distinguishing particles which become light at $u=0$ vs.
$u=1$):
\begin{equation}\label{eq:12ParticlesK}
\CK_{-1,1;+}\CK_{1,-1;-}\CK_{0,1;+}^2 \CK_{0,-1;-}^2 \CK_{1,1;+}\CK_{-1,-1;-}\CK_{1,0;+}^2 \CK_{-1,0;-}^2.
\end{equation}

If we move to a region of large $u$, the phases of the charges,
say, $(1,0;+)$ and $(1,0;-)$ will approach each other, as the flavor
information becomes irrelevant.  This will require a complete
overhaul of the product of symplectomorphisms.  A first step is the exchange of consecutive
pairs of ``side'' and ``diagonal'' BPS states (now a $*$ indicates a
mixed flavor charge, with both $+$ and $-$ contributions):
\begin{multline}
(\CK_{1,0;-}^2 \CK_{-1,1;+})(\CK_{1,-1;-}\CK_{0,1;+}^2)( \CK_{0,-1;-}^2\CK_{1,1;+})(\CK_{-1,-1;-}\CK_{1,0;+}^2) = \\
(\CK_{-1,1;+}\CK_{0,1;*}^2\CK_{1,1;*}\CK_{1,0;-}^2 )(\CK_{0,1;+}^2\CK_{1,1;*}\CK_{1,0;*}^2\CK_{1,-1;-}) \\ (\CK_{1,1;+}
\CK_{1,0;*}^2\CK_{1,-1;*} \CK_{0,-1;-}^2)(\CK_{1,0;+}^2\CK_{1,-1;*}\CK_{0,-1;*}^2\CK_{-1,-1;-}).
\end{multline}
We have applied the pentagon twice to  each of the four pairs of
transformations on the left.  We already see something interesting:
now there are $8$ particles of gauge charge $(1,0)$, $8$ particles
of gauge charge $(0,1)$. Some of the octets of hypermultiplets of
the very large $u$ region are making their appearance. On the other
hand we still see only $4$ particles of charge $(1,1)$.

The next steps towards the large $u$ region generate vector
multiplets: we recognize in pairs like $\CK_{1,0;-}^2 \CK_{0,1;+}^2$
the left hand side of the $N_f=2$ wall-crossing formula, and in
pairs like $\CK_{1,-1;-}\CK_{1,1;+}$ the left hand side of the
$N_f=0$ wall crossing formula.  The latter will give rise to a vectormultiplet
of gauge charge $(2,0)$, the former to a vectormultiplet
of gauge charge $(2,2)$, accompanied by $4$ hypermultiplets of
charge $(1,1)$. These, together with the other $4$ hypers already
present at the previous step, form an octet.

From here on, the sequence of wall-crossings becomes increasingly
intricate. As $u$ grows, the phases of each member of the
various octets move toward the phase of the
corresponding W-boson. In the process one meets a
recursive structure of wall-crossing formulae from the $N_f=0$ or $N_f=2$
theories; applying them repeatedly here gives rise to the infinitely dense
structure of $SL(2,\IZ)$ S-dual images of the W-bosons.

It is interesting to recover the flavor structure from our considerations here.
Our special choice of masses breaks all $SU(2)_{a,b,c,d}$ subgroups to their
Cartan subgroups, but actually preserves some non-manifest part of
$SO(8)$.  In the three possible triality frames, the $SO(8)$ Cartan
generator has eigenvalues $(m_a \pm m_c, m_b \pm m_d)$, $(m_a \pm
m_b, m_c \pm m_d)$, $(m_a \pm m_d, m_c \pm m_b)$.  In the first
frame, two eigenvalues are actually zero:  a hidden
$SO(4) \sim SU(2)_L \times SU(2)_R$ is unbroken. This is essentially the reason for the
reappearance here of the wall-crossing formulae from the $N_f=2$ example. ${\bf 8}_v$ contains a
vector of this $SO(4)$ and four singlets.  ${\bf 8}_s$ and ${\bf 8}_c$ each contain
two doublets ${\bf 2}_L$ and two doublets ${\bf 2}_R$.

The wall-crossing we described is consistent with the idea that
the original four particles of charge $(1,0)$ in the strong coupling region
should be viewed as part of ${\bf 8}_s$; more precisely $(1,0;+)$ can be taken to be a doublet of $SU(2)_L$
and $(1,0;-)$ a doublet of $SU(2)_R$.
Similarly, the four particles of charge $(0,1)$ should be doublets of each kind in $8_c$: $(0,1;+)$ can
be taken to be a doublet of $SU(2)_L$ and $(0,1;-)$ a doublet of $SU(2)_R$.
The particles of charges $(1,1)$ (and similarly $(1,-1)$) should be all singlets of $8_v$.
Indeed it is easy to see that the first type of wall-crossing,
\begin{equation}
\CK_{1,0;-}^2 \CK_{-1,1;+}=\CK_{-1,1;+}\CK_{0,1;*}^2\CK_{1,1;*}\CK_{1,0;-}^2,
\end{equation}
produce the missing doublets in ${\bf 8}_s$ and ${\bf 8}_c$, and the missing singlets of ${\bf 8}_v$. The $N_f=2$
type wall-crossing of $SU(2)_L$ doublets and $SU(2)_R$ doublets like $\CK_{1,0;-}^2 \CK_{0,1;+}^2$ produces
the missing vectors in ${\bf 8}_v$.

A simple way to describe the full charge lattice is the following.
The six charges which become light at $u\to 0$ (which we have
labeled $+$) satisfy the same linear relations as we found in the
general $N=4$ analysis, and in particular span a three-dimensional
lattice. The same is true of the six charges labeled $-$. The full
charge lattice is the orthogonal sum of these two three-dimensional
lattices.

In principle, one could determine the spectrum anywhere in parameter space
from the strong coupling spectrum and the wall-crossing formula. In the next section we will
propose a better method.

\section{The spectrum generator}
\label{sec:UniversalStokesMatrix}

In this section, we give an algebraic method of determining the BPS spectrum of any theory
related to an $SU(2)$ Hitchin system.

Consider the evolution of the WKB triangulation $T_\WKB(\vartheta,
\lambda^2)$ as $\vartheta$ varies continously through an arc of
length $\pi$. As we saw in Section \ref{subsec:Jumps-WKB-triang},
generically there are three species of transformations of $T_\WKB$
which might occur at critical values $\vartheta = \vartheta_c$. Two
of the three are flips and \juggles, corresponding respectively to
the BPS hypermultiplets and vectormultiplets we want to detect, with
$\arg -Z = \vartheta_c$.  As $\vartheta$ varies we encounter exactly
half of the BPS particles of the theory (given any BPS particle we
encounter either the particle or its antiparticle). Each corresponds
to a symplectomorphism acting on the coordinates
$\CX^\vartheta_\gamma$. The third possible transformation is a pop,
which does not correspond to any of the BPS particles in which we
are interested.  Fortunately, these pops always occur within
degenerate triangles. Thanks to the analysis of Section
\ref{sec:no-flavor-jump} these transformations actually give the
trivial symplectomorphism, and hence can be ignored.

Composing all these symplectomorphisms in order, we arrive at a transformation $\bf S$ relating $\CX^\vartheta$ and
$\CX^{\vartheta+\pi}$:
\begin{equation} \label{eq:stokes-decomp}
{\bf S} = \prod_{\gamma: \vartheta < \arg -Z_{\gamma}(u) < \vartheta+\pi}^\ccwarrow \CK_\gamma^{\Omega(\gamma;u)}.
\end{equation}
Here the product from left to right is taken in increasing order of
$\arg -Z_\gamma$. This $\bf S$ is a Stokes matrix for the
Riemann-Hilbert problem of \cite{Gaiotto:2008cd}. We will call it
the \emph{spectrum generator}.

Our trick is to notice that the technology of the previous sections
provides a neat way of computing $\bf S$ \ti{without} following the
continuous evolution of the triangulation. Once $\bf S$ is known,
the decomposition \eqref{eq:stokes-decomp} uniquely determines the
$\Omega(\gamma;u)$.

How will we compute $\bf S$?  Notice that the WKB foliations with
phases $\vartheta$ and $\vartheta+\pi$ are identical.  It follows
that the undecorated WKB triangulations are also the same.  What
about the decorations? In the case of regular singularities, if we
change $\vartheta$ to $\vartheta + \pi$, it follows from Section
\ref{sec:small-flat} that the decoration switches from one monodromy
eigenvector to the other --- this is the transformation we called a
pop. Similarly, it follows from the local analysis in Section
\ref{sec:irregular}  that the same conclusion holds for irregular
singularities. So the decorated triangulation
$T_\WKB(\vartheta+\pi)$ differs from $T_\WKB(\vartheta)$ by a pop
transformation at every vertex --- what we will refer to as an
\emph{omnipop.}

So to determine $\bf S$ our job is to work out the explicit
transformation corresponding to an omnipop.

\subsection{Deriving the transformation under an
omnipop}\label{subsec:SuperPop}

Given a vertex $P$, the decoration of $T_\WKB(\vartheta)$ provides a
distinguished section (up to
scale) near $P$. In the case of a regular singularity, $P$ is the
singular point, and for an irregular singular point, $P$ is one of the
distinguished points on the small circle surrounding the
singularity.  Choose a scale and let $s_P$ be a distinguished
section, while $\tilde s_P$ is the new distinguished section resulting
from a pop.  Similarly, we write $\CX_{P_1 P_2}$ and $\tilde
\CX_{P_1 P_2}$ for the original and new Fock-Goncharov coordinates
at the edge $E_{P_1 P_2}$ joining vertices $P_1, P_2$ computed using
the sections $s_{P_i}$ and $\tilde s_{P_i}$, respectively.

It follows from the definitions in Section \ref{sec:homology} that,
for all $E$, $\gamma^{\vartheta+\pi}_E = - \gamma^{\vartheta}_E$.
Combining this with the definitions
\begin{equation}
\CX_{\gamma^{\vartheta+\pi}_E}^{ \vartheta+\pi } =
\CX^{T_\WKB(\vartheta+\pi, \lambda^2)}_E, \qquad
\CX_{\gamma^{\vartheta }_E}^{ \vartheta } = \CX^{T_\WKB(\vartheta,
\lambda^2)}_E,
\end{equation}
we have
\begin{equation}\label{eq:X-to-tildeX}
\CX^{\vartheta+\pi}_\gamma = \CX^{\vartheta}_\gamma \cdot
\left(\CX_E^{T_\WKB(\vartheta,\lambda^2)} \tilde
\CX_E^{T_\WKB(\vartheta,\lambda^2)} \right)^{-1}
\end{equation}
for $\gamma = \gamma^{\vartheta}_E$.
 Therefore, we seek a
simple formula for $\CX_E^T \tilde \CX_E^T$ for a fixed decorated
triangulation $T$.  To ease the notation we will drop the
superscript $T$ in the computations that follow, but bear in mind
that we are working within a fixed decorated triangulation.

For any edge $PQ$ it will be convenient to define $A_{PQ}$ by
\begin{equation} \label{eq:rewrite-A}
A_{PQ} := - \frac{(s_Q \wedge \tilde s_P)(s_P \wedge \tilde
s_Q)}{(s_P\wedge \tilde s_P)(s_Q \wedge \tilde s_Q)}.
\end{equation}
Note that $A_{PQ}=A_{QP}$.
This is a useful definition because one can easily show that\footnote{In
order to obtain \eqref{eq:def-A} from
\eqref{eq:rewrite-A} we use  the fact that in a two-dimensional
vector space, for any three vectors $v_1, v_2, v_3$, we have
\begin{equation}\label{eq:3-termIdent}
(v_1\wedge v_2) v_3 + (v_3\wedge v_1) v_2 + (v_2 \wedge v_3) v_1 = 0,
\end{equation}
or equivalently, for any four vectors $v_1, v_2, v_3, v_4$, we have
\begin{equation}
(v_1\wedge v_2)(v_3 \wedge v_4)  + (v_3 \wedge v_1)(v_2\wedge v_4)+
(v_2\wedge v_3)(v_1\wedge v_4) = 0.
\end{equation}
We will use this relation repeatedly in what follows.}
\begin{equation} \label{eq:def-A}
1 + A_{PQ} =  \frac{(s_P \wedge s_Q)(\tilde s_P \wedge \tilde
s_Q)}{(\tilde s_P \wedge s_P)(\tilde s_Q \wedge s_Q)},
\end{equation}
and therefore, if $E$ is the edge $ac$ in the quadrilateral $Q_E$
with vertices $abcd$ in counterclockwise order,
\begin{equation}\label{eq:X-times-tildeX}
\tilde \CX_E \CX_E = \frac{ (1+ A_{ab}) (1+ A_{cd})
}{(1+A_{bc})(1+A_{da})}.
\end{equation}
Therefore we can solve our problem if we can express $A_{ab}$ in
terms of the coordinates $\CX^T_E$.

\insfig{3in}{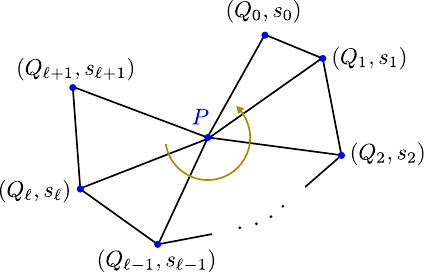}{Partial star-shaped neighborhood of a vertex
$P$, used in the computation of $\Sigma(P;Q_{\ell+1}\to Q_0)$.
The brown arrow intersects the edges
which occur in \protect\eqref{eq:def-sigma}. The final edge intersected
by the brown arrow only occurs in the final summand. The result \protect\eqref{eq:key-star-lemma} is
expressed in terms of the decorations at $P$ and $Q_\ell$, as well as at
the bounding edges $Q_0,Q_{\ell+1}$, whose edges do not occur in the
expression for $\Sigma(P;Q_{\ell+1}\to Q_0)$.}

We begin with a key lemma. Consider a partial star-shaped
neighborhood of a decorated point $(P, s_P)$  with decorated vertices
$(Q_0, s_0), (Q_1, s_1), \dots, (Q_{\ell+1}, s_{\ell+1})$ in
\emph{clockwise} order, as in Figure \ref{fig:star-lemma}. Denote
the edge coordinates $\CX_{P,Q_j}$ simply by $\CX_{P,j}$.
We claim that the quantity
\begin{equation} \label{eq:def-sigma}
\Sigma(P;  Q_{\ell+1}\to Q_0):= 1 + \CX_{P,\ell} +
\CX_{P,\ell}\CX_{P,\ell-1} + \cdots + (\CX_{P,\ell}\cdots \CX_{P,1})
\end{equation}
can be written simply in terms of the sections $s_P,s_0,s_\ell,
s_{\ell+1}$, which have been continued in a single-valued fashion in
the simply connected region formed by the triangles. Indeed we have
\begin{equation}\label{eq:key-star-lemma}
\Sigma(P;  Q_{\ell+1}\to Q_0) = \frac{(s_0 \wedge s_{\ell+1})(s_P
\wedge s_\ell)}{(s_{\ell+1}\wedge s_\ell)(s_0\wedge s_P)}.
\end{equation}
To prove this, first check it for $\ell=1$, using the identity
(\ref{eq:3-termIdent}). Then, for the inductive step note that if we
add a further decorated vertex $(Q_{\ell+2},s_{\ell+2})$ then
\begin{equation}
\begin{split}
\Sigma(P;  Q_{\ell+2}\to Q_0)& = 1 + \Sigma(P;  Q_{\ell+1}\to
Q_0)\CX_{P,\ell+1} \\
& = 1  -\frac{(s_0 \wedge s_{\ell+1})(s_P \wedge
s_\ell)}{(s_{\ell+1}\wedge s_\ell)(s_0\wedge s_P)}\frac{(s_P\wedge
s_{\ell+2})(s_{\ell+1}\wedge s_{\ell})}{(s_{\ell+2}\wedge
s_{\ell+1})(s_\ell \wedge s_P)} \\
& = 1 + \frac{(s_P\wedge s_{\ell+2})(s_0 \wedge
s_{\ell+1})}{(s_{\ell+2}\wedge
s_{\ell+1})(s_0 \wedge s_P)} \\
& = \frac{ (s_0\wedge s_P)(s_{\ell+2}\wedge s_{\ell+1})+ (s_0 \wedge
s_{\ell+1})(s_P\wedge s_{\ell+2})}{(s_{\ell+2}\wedge
s_{\ell+1})(s_0 \wedge s_P)} \\
& =\frac{(s_0 \wedge s_{\ell+2})(s_P \wedge
s_{\ell+1})}{(s_{\ell+2}\wedge
s_{\ell+1})(s_0\wedge s_P)}\\
\end{split}
\end{equation}
where in the last line we have again used equation
(\ref{eq:3-termIdent}).

\insfig{5in}{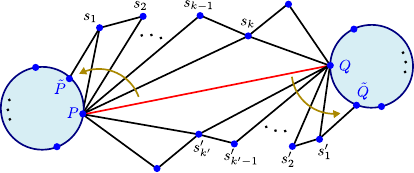}{Configuration of edges involved in computing
$A_{PQ}$ when $P, Q$ are both irregular singular points. The
elementary pop transformations at $P$ and $Q$ associate the decorations
at $\tilde P, \tilde Q$ to the vertices $P, Q$ respectively.}

Now, to compute $A_{PQ}$ we need to distinguish the cases where
$P,Q$ are regular or irregular singular points. Let us consider
first the case where both $P,Q$ are irregular singular points. Then
we have the situation illustrated in Figure \ref{fig:two-ISP}.  Note
that we have denoted $\tilde P$ as the distinguished point from the
WKB ray one step counterclockwise from $P$, so that for an
elementary pop, $\tilde s_P = s_{\tilde P}$, and similarly for $Q$.
Here we parallel-transport $s_{\tilde P}$ along the edge going from
$P$ to $\tilde P$.  Applying our lemma we have
\begin{equation}
\Sigma(P; Q \to \tilde P) = \frac{ (\tilde s_P\wedge s_Q)(s_P \wedge
s_k)}{(s_Q \wedge s_k)(\tilde s_P \wedge s_P)},
\end{equation}
\begin{equation}
\Sigma(Q; P \to \tilde Q) =\frac{ (\tilde s_Q\wedge s_P)(s_Q \wedge
s'_{k'})}{(s_P \wedge s'_{k'})(\tilde s_Q \wedge s_Q)},
\end{equation}
and by definition
\begin{equation}
\CX_{PQ} = - \frac{ (s_P\wedge s'_{k'})(s_Q \wedge
s_k)}{(s'_{k'}\wedge s_Q)(s_k \wedge s_P)}.
\end{equation}
Taking the product of these three expressions and cancelling factors
we find:
\begin{equation}\label{eq:two-ISP}
\CX_{PQ} \Sigma(P; Q \to \tilde P)\Sigma(Q; P \to \tilde Q)= -
\frac{(\tilde s_P \wedge s_Q)(\tilde s_Q \wedge s_P)}{(\tilde s_P
\wedge s_P)(\tilde s_Q \wedge s_Q)} = A_{PQ}.
\end{equation}
This gives us the desired expression for $A_{PQ}$ in terms of the
coordinates $\CX_E^T$. This expression even works if $P,Q$ are
consecutive points on the same boundary circle, since we defined
$\CX_{PQ}=0$ in that case.

\insfig{2.2in}{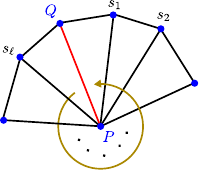}{Configuration of edges involved in
the computation of $\Sigma(P;Q\to Q)$, for $P$ a regular singular point.
Here one must take into account the
monodromy around $P$, since in the inductive proof of \protect\eqref{eq:key-star-lemma}
it is assumed that the $s_i$ are single-valued
in the region determined by the triangles. Thus, in the final
inductive step one must move a cut.}

Now let us consider the case where $P$ is a regular singular point
as shown in Figure \ref{fig:star-RSP}.  A new point arises in
computing $\Sigma(P;Q\to Q)$:  we must take into account
the monodromy around the regular singular point. If $M_P$ is the
clockwise monodromy operator around $P$, we use expression
(\ref{eq:key-star-lemma}) above with $s_{\ell+1}=s_Q$ and $s_0 =
M^{-1} s_{Q}$:
\begin{equation}
\Sigma(P; Q\to Q) = \frac{ (s_P \wedge s_\ell)(M^{-1} s_Q \wedge
s_Q)}{(s_Q \wedge s_\ell)(M^{-1} s_Q \wedge s_P)}.
\end{equation}
This can be put in a more useful form by expanding $s_Q$ in the basis
$s_P, \tilde s_P$, where $M s_P = \mu_P s_P$ and $M \tilde s_P =
\mu_P^{-1} \tilde s_P$. Then we can expand $s_Q = c_1 s_P + c_2
\tilde s_P$ and compute
\begin{equation}
M^{-1} s_Q \wedge s_Q = c_1 c_2 (\mu_P^{-1} - \mu_P) s_P \wedge
\tilde s_P,
\end{equation}
\begin{equation}
M^{-1} s_Q \wedge s_P =  c_2 \mu_P \tilde s_P \wedge s_P.
\end{equation}
Thus,  using $c_1 = (s_Q \wedge \tilde s_P)/(s_P \wedge \tilde s_P)$
we get
\begin{equation}\label{eq:SimpleRSP}
\Sigma(P; Q \to Q) =   (1- \mu_P^{-2}) \frac{ (s_P \wedge
s_\ell)(s_Q \wedge \tilde s_P)}{(s_Q \wedge s_\ell)(s_P \wedge
\tilde s_P)}.
\end{equation}

\insfig{3in}{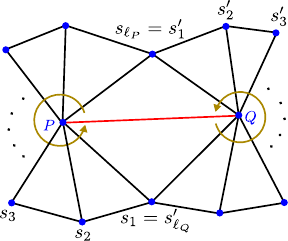}{Configuration of edges involved in computing $A_{PQ}$
when $P, Q$ are both regular singular points.}

Now, consider an edge between two regular singular points $P$ and
$Q$ as in Figure \ref{fig:two-RSP}.  Using the result
(\ref{eq:SimpleRSP}) for both $P$ and $Q$ we obtain
\begin{equation}\label{eq:two-RSP}
 \CX_{PQ}\Sigma(P; Q\to Q) \Sigma(Q; P \to P) = (1-\mu_P^{-2})(1-\mu_Q^{-2})
A_{PQ}
\end{equation}
from which we obtain $A_{PQ}$ in terms of the $\CX_E^T$.

Finally, suppose $P$ is an irregular singular point and $Q$ is a
regular singular point. Then similarly to the above we have
\begin{equation}\label{eq:ISP-RSP}
A_{PQ} =  (1-\mu_Q^{-2})^{-1} \CX_{PQ} \Sigma(P; Q \to \tilde P)
\Sigma(Q;P \to P).
\end{equation}

Taken together,
(\ref{eq:X-to-tildeX}), (\ref{eq:X-times-tildeX}), (\ref{eq:two-ISP}),
(\ref{eq:two-RSP}), (\ref{eq:ISP-RSP}) completely solve the problem
of determining the spectrum generator ${\bf S}$ as
an elementary symplectic transformation.  In the following sections
we will illustrate the kinds of formulae which appear in several
examples.

Let us conclude with two remarks:

\begin{itemize}
\item The reader might be disturbed by the minus sign appearing
in the final expression for $A_{ab}$ in the case of a regular
singular point.  After all, any two triangulations should be connected
by a sequence of flips, and the corresponding coordinate transformations
only involve positive signs.  Fortunately it is
easy to rearrange the final formula (\ref{eq:X-times-tildeX}) for the transformation of a
cross-ratio in such a way that only positive signs appear. Indeed, in (\ref{eq:X-times-tildeX})
each of the labels $abcd$ appears once in the numerator and once in the
denominator.  Multiplying both numerator and denominator
by $(1-\mu_a^{2})(1 -\mu_b^{2})(1 -\mu_c^{2})(1 -\mu_d^{2})$,
the factors in (\ref{eq:X-times-tildeX}) can be combined into
four blocks of the form
\begin{multline}
(1-\mu_a^{2})(1 -\mu_b^{2})(1+A_{ab}) =  \\
1-\mu_a^{2}
-\mu_b^{2}+\mu_a^{2} \mu_b^{2}+\CX_{a,b} \left( \sum_{i=0}^{k-1}
\prod_{j=1}^i \CX_{a,j} \right) \left( \sum_{i'=0}^{k'-1}
\prod_{j'=1}^{i'} \CX_{b,-j'} \right),
\end{multline}
where there are $k$ vertices in the star region around $a$ and $k'$
vertices in the star region around $b$.  In this sum, the term with
$i=k-1$, $i'=0$ equals $\mu_a^{2}$, and the term with $i=0$,
$i'=k'-1$ equals $\mu_b^{2}$.  These two terms then cancel the
terms with negative signs. This cancellation is needed e.g. to show
that the spectrum generators ${\bf S}$ of the two realizations of the
$SU(2)$ theory with $N_f=2$ coincide, as one realization involves
regular singularities (and hence negative signs can appear) while
the other does not.

\item In a region where
$\abs{\mu_P} < 1$, the expression \eqref{eq:def-sigma} defining $\Sigma(P;Q\to Q)$ can be rewritten in a suggestive form,
\begin{equation}
\frac{1}{(1 - \mu_P^{2})}\Sigma(P;Q\to Q) =1+ \sum_{i=1}^{\infty}
\prod_{j=1}^i \CX_{P,j},
\end{equation}
by expanding the denominator in a geometric series and remembering that $\prod_{j=1}^k
\CX_{P,j}=\mu_P^{2}$.
(In this formula it is better to label the vertices in
counterclockwise order.)

\end{itemize}

\subsection{$N=3$}

We first consider the $N=3$ example discussed in Section \ref{sec:N3}. For
notational simplicity we remove the $\vartheta$, and indicate
$\CX^{\vartheta+\pi}$ as $\tilde \CX$. We use the basis $\gamma_1$,
$\gamma_2$ of $\hat\Gamma$ corresponding to edges $13$, $14$ in Figure
\ref{fig:N3-universal} with the orientation at $\vartheta$ (as opposed to $\vartheta+\pi$),
in other words, $\gamma_1 := \gamma_{13}^\vartheta$ and similarly for $\gamma_2$.
Then from equation (\ref{eq:X-to-tildeX}) and
(\ref{eq:X-times-tildeX}) we have
\begin{equation}
\tilde \CX_{\gamma_1} = \CX_{\gamma_1} \frac{ (1+
A_{23})(1+A_{41})}{(1+A_{12})(1+A_{34})}.
\end{equation}
Now $A_{12}= A_{23}=A_{34}=0$, while equation (\ref{eq:two-ISP})
simply gives $A_{41}=\CX_{\gamma_2}$. Similarly,
\begin{equation}
\tilde \CX_{\gamma_2} = \CX_{\gamma_2} \frac{ (1+
A_{34})(1+A_{51})}{(1+A_{13})(1+A_{45})}
\end{equation}
and $A_{34}=A_{45}=A_{51}=0$, but, by equation (\ref{eq:two-ISP})
\begin{equation}
A_{13} = \CX_{\gamma_1}(1+\CX_{\gamma_2}).
\end{equation}
So, altogether we obtain the transformation ${\bf S}$:
\begin{align}
\tilde \CX_{\gamma_1} &= \CX_{\gamma_1} (1+\CX_{\gamma_2}), \label{eq:n31} \\
\tilde \CX_{\gamma_2} &= \CX_{\gamma_2} (1+\CX_{\gamma_1}+\CX_{\gamma_1}\CX_{\gamma_2})^{-1}. \label{eq:n32}
\end{align}

\insfig{4.2in}{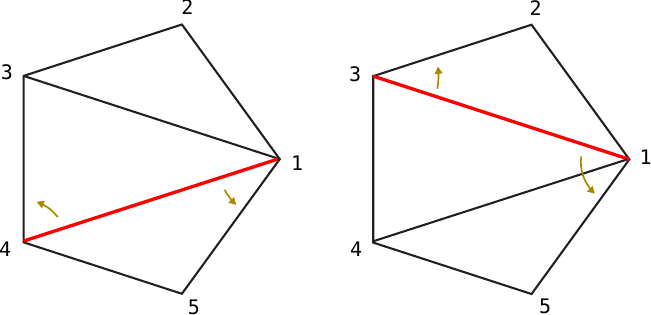}{Diagram for computing the factors $A_{ab}$ appearing in the
spectrum generator of the $N=3$ theory.  On the left is the example $ab=14$, where we simply get
$A_{14} = \CX_{\gamma_2}$ (since the brown arrows do not meet any edges).  On the right is
$ab=13$ where we get $A_{13} = \CX_{\gamma_1}(1+\CX_{\gamma_2})$.}

Above we promised that given ${\bf S}$ one can recover the BPS
spectrum.  In this simple example that means recovering the
decomposition ${\bf S} = \CK_{\gamma_1} \CK_{\gamma_2}$ starting
from \eqref{eq:n31}, \eqref{eq:n32}.  The general algorithm for
doing this was sketched in Section \ref{subsec:wallcrossing}; let us
see how it would work here. We are seeking a decomposition of the
form
\begin{equation}
{\bf S} = \prod_{m,n \ge 0} \CK_{m \gamma_1 + n \gamma_2}^{\Omega(m \gamma_1 + n \gamma_2;u)}
\end{equation}
where the product is taken in increasing order of $m/n$, from $0$ to $\infty$.
First specialize to $\CX_{m \gamma_1 + n \gamma_2} = 0$ for all $m+n \ge 2$.  After this specialization
all $\CK_{m \gamma_1 + n \gamma_2}$ for $m+n \ge 2$ become identity operators, so the decomposition reduces
to
\begin{equation}
{\bf S} = \CK_{\gamma_1}^{\Omega(\gamma_1;u)} \CK_{\gamma_2}^{\Omega(\gamma_2;u)}.
\end{equation}
The action of this operator is easily computed as
\begin{align}
\tilde \CX_{\gamma_1} &= \CX_{\gamma_1} (1+\CX_{\gamma_2})^{\Omega(\gamma_2;u)}, \\
\tilde \CX_{\gamma_2} &= \CX_{\gamma_2} (1+\CX_{\gamma_1})^{-\Omega(\gamma_1;u)}.
\end{align}
Comparing this with the known action of ${\bf S}$ by \eqref{eq:n31}, \eqref{eq:n32}
and recalling that we have set $\CX_{\gamma_1 + \gamma_2} = 0$,
we see that the two are consistent only if $\Omega(\gamma_1;u) = 1$ and $\Omega(\gamma_2;u) = 1$.
We can now continue to the next order by specializing to $\CX_{m \gamma_1 + n \gamma_2} = 0$ for all $m+n \ge 3$;
we would then have three new unknown $\Omega$ appearing,
\begin{equation}
{\bf S} = \CK_{\gamma_1} \CK_{2 \gamma_1}^{\Omega(2 \gamma_1;u)} \CK_{\gamma_1 + \gamma_2}^{\Omega(\gamma_1+\gamma_2;u)} \CK_{2 \gamma_2}^{\Omega(2 \gamma_2;u)} \CK_{\gamma_2}.
\end{equation}
Computing the action of this operator we find that it is consistent with \eqref{eq:n31}, \eqref{eq:n32}
only if all of these three unknown $\Omega$ actually vanish.  Similar computations to higher orders give the
same result --- all of the new $\Omega$ which appear at each order turn out to vanish.
This can be carried out as long as one has
patience or computer time, but eventually one might be inspired to conjecture
that in fact all of the higher $\Omega$ vanish, i.e. that one has an exact identity ${\bf S} = \CK_{\gamma_1} \CK_{\gamma_2}$.  Once conjectured this identity is of course easy to verify.

Note that had we sought a decomposition in the order of decreasing $m/n$, by the same algorithm we would have
arrived at the formula ${\bf S} = \CK_{\gamma_2} \CK_{\gamma_1 + \gamma_2} \CK_{\gamma_1}$.  The two
different decompositions of course correspond to the two sides of the wall of marginal stability.

\subsection{$N=4$}

Next we consider the $N=4$ example from Section \ref{sec:N4}. Let us focus
on the triangulation of Figure \ref{fig:N4-universal} (a) and
identify
\begin{equation}\label{eq:N=4match}
\CX_{\gamma_1} = \CX_{14}, \qquad \CX_{\gamma_2} = \CX_{15}, \qquad
\CX_{\gamma_3} = \CX_{13}.
\end{equation}
Then (\ref{eq:X-to-tildeX}) becomes
\begin{align}
\tilde \CX_{\gamma_1} & = \CX_{\gamma_1} (1+
A_{51})(1+A_{13})^{-1},\\
\tilde\CX_{\gamma_2} & = \CX_{\gamma_2} (1+A_{14})^{-1}, \\
\tilde \CX_{\gamma_3} & = \CX_{\gamma_3} (1+ A_{14}).
\end{align}
Evaluating $A_{13}, A_{15}, A_{14}$ using equation (\ref{eq:two-ISP})
and the identifications (\ref{eq:N=4match}) we obtain:
\begin{align}
\tilde \CX_{\gamma_1} &= \CX_{\gamma_1} (1+\CX_{\gamma_2})(1+\CX_{\gamma_3} + \CX_{\gamma_3}\CX_{\gamma_1}+\CX_{\gamma_3}\CX_{\gamma_1}\CX_{\gamma_2})^{-1}, \label{eq:n41} \\
\tilde \CX_{\gamma_2} &=  \CX_{\gamma_2}(1+\CX_{\gamma_1}+\CX_{\gamma_1}\CX_{\gamma_2})^{-1}, \label{eq:n42} \\
\tilde \CX_{\gamma_3} &=  \CX_{\gamma_3}(1+\CX_{\gamma_1}+\CX_{\gamma_1}\CX_{\gamma_2}). \label{eq:n43}
\end{align}
This transformation indeed has the KS decomposition ${\bf S} = \CK_{\gamma_3}
\CK_{\gamma_1} \CK_{\gamma_2}$ for an appropriate choice of $(u,
\vartheta)$, agreeing with our expectations. Note that
$\CX_{\gamma_1}$ comes from a quadrilateral with
two internal edges, which correspond to the two multiplicative
factors in \eqref{eq:n41}.

Similarly, making the identifications
\begin{equation}\label{eq:N=4match-b}
\CX_{\gamma_1} = \CX_{63}, \qquad \CX_{\gamma_2} = \CX_{13}, \qquad
\CX_{-\gamma_3} = \CX_{46},
\end{equation}
appropriate to the triangulation in Figure \ref{fig:N4-universal} (b),
we find that
\begin{align}
\tilde \CX_{\gamma_1} &= \CX_{\gamma_1} (1+\CX_{\gamma_2})(1+\CX_{-\gamma_3}), \label{eq:n44} \\
\tilde \CX_{\gamma_2} &=  \CX_{\gamma_2}\left[1+\CX_{\gamma_1} (1+\CX_{\gamma_2})(1+\CX_{-\gamma_3})\right], \label{eq:n45} \\
\tilde \CX_{-\gamma_3} &=  \CX_{-\gamma_3}\left[1+\CX_{\gamma_1} (1+\CX_{\gamma_2})(1+\CX_{-\gamma_3})\right]. \label{eq:n46}
\end{align}
In this case $\CX_{\gamma_{2,3}}$ come from quadrilaterals which
have the same single internal edge.   Moving counterclockwise from
that edge at either of its ends, we meet an internal edge, which
contributes to the multiplicative factor in \eqref{eq:n45},
\eqref{eq:n46}. $\CX_{\gamma_1}$ comes from a quadrilateral with two
internal edges, corresponding to the two factors in \eqref{eq:n44}.
The transformation \eqref{eq:n44}-\eqref{eq:n46} has the
decomposition (for an appropriate value of $(u,\vartheta)$)
${\bf S} = \CK_{\gamma_1} \CK_{\gamma_2}\CK_{-\gamma_3}$. Case (d) is similar
and will be left to the reader.

Finally, consider the third type of triangulation of the hexagon,
with a triangle of internal edges, as in Figure \ref{fig:N4-universal}
(c):
\begin{align}
\tilde \CX_{-\gamma_1} &= \CX_{-\gamma_1} (1+\CX_{\gamma_1+\gamma_2}+\CX_{\gamma_1+\gamma_2}\CX_{-\gamma_1})^{-1}(1+\CX_{\gamma_3}+\CX_{\gamma_3}\CX_{\gamma_1+\gamma_2}), \\
\tilde \CX_{\gamma_1+\gamma_2} &=  \CX_{\gamma_1+\gamma_2}(1+\CX_{\gamma_3}+\CX_{\gamma_3}\CX_{\gamma_1+\gamma_2})^{-1}(1+\CX_{-\gamma_1}+\CX_{-\gamma_1}\CX_{\gamma_3}),\\
\tilde \CX_{\gamma_3} &=  \CX_{\gamma_3}(1+\CX_{-\gamma_1}+\CX_{-\gamma_1}\CX_{\gamma_3})^{-1}(1+\CX_{\gamma_1+\gamma_2}+\CX_{\gamma_1+\gamma_2}\CX_{-\gamma_1}).
\end{align}
The two factors in each transformation correspond to the fact that
each quadrilateral has two internal edges. Here we find ${\bf S} =
\CK_{-\gamma_1} \CK_{\gamma_3}\CK_{\gamma_2}\CK_{\gamma_1+\gamma_2}$
(after some surprising simplifications).

\insfig{5.5in}{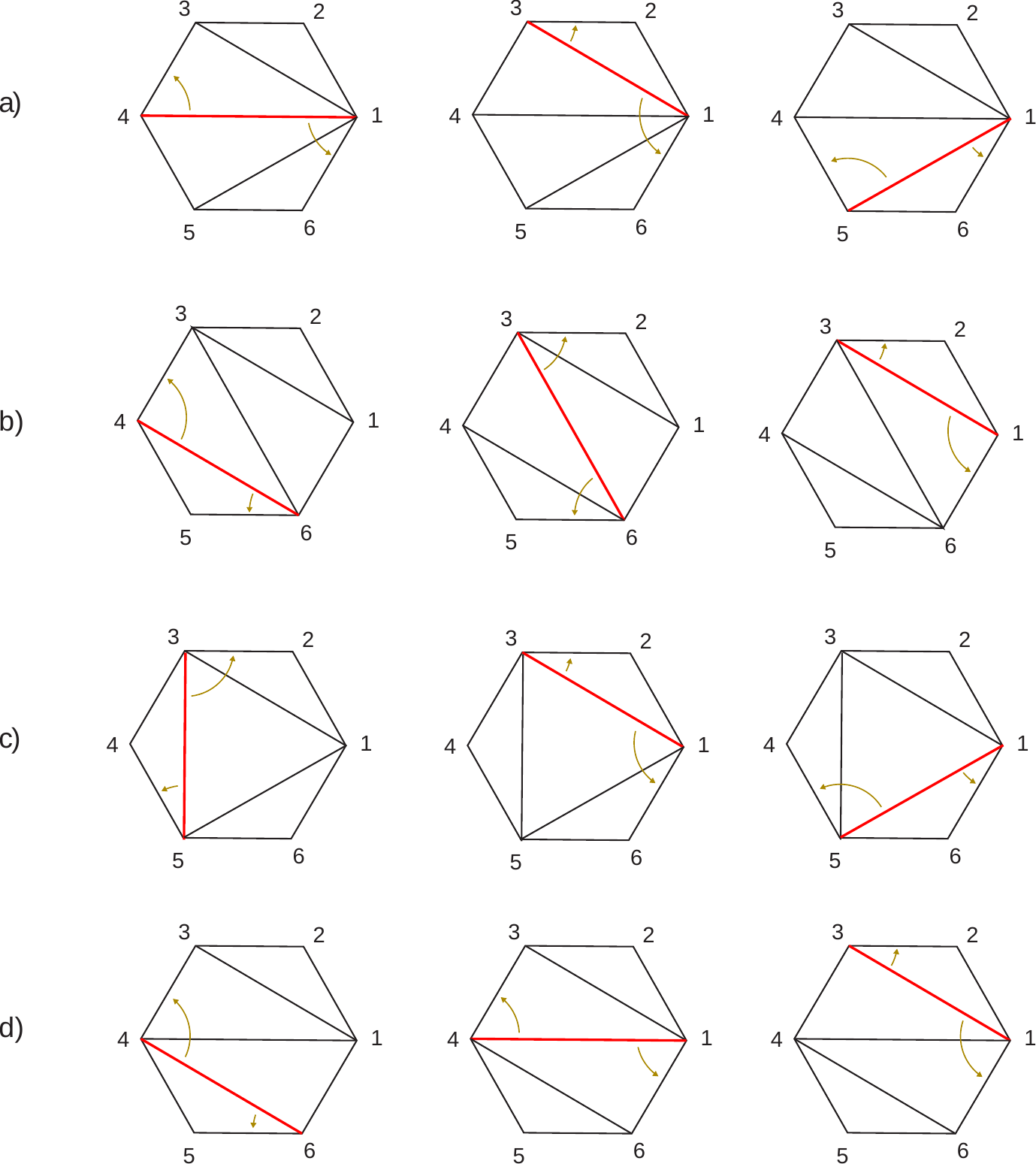}{Diagrams for computing the factors $A_{ab}$
appearing in the spectrum generator
in the $N=4$ example.  The four different triangulations we show
occur as $T_\WKB(u, \vartheta)$ for different values
of $(u,\vartheta)$.  In a), $ab=14$ defines $\gamma_1$, $ab=13$
defines $\gamma_3$ and $ab=15$ defines $\gamma_2$. In b),
$ab=63$ defines $\gamma_1$, $ab=13$ defines $\gamma_2$ and
$ab=46$ defines $-\gamma_3$.  In c), $ab=35$ defines $-\gamma_1$,
$ab=31$ defines $\gamma_3$  and $ab=15$ defines $\gamma_1 +
\gamma_2$.  In d), $ab=46$, $ab=41$ and $ab=31$.}

\subsection{$SU(2)$, $N_f=0$}

Now we turn to the $SU(2)$ theory with $N_f=0$.  In Section \ref{sec:nf0} we explained that
in the strong coupling region we have ${\bf S} = \CK_{\gamma_1} \CK_{\gamma_2}$, where $\langle \gamma_1,\gamma_2 \rangle=2$.
This transformation acts by
\begin{align}
\tilde \CX_{\gamma_1} &= \CX_{\gamma_1} (1+\CX_{\gamma_2})^2, \\
\tilde \CX_{\gamma_2} &= \CX_{\gamma_2} \left[1+\CX_{\gamma_1}(1+\CX_{\gamma_2})^2\right]^{-2}.
\end{align}
This transformation can be obtained from $T_\WKB$ using the rules we have described.
In particular, the fact that each quadrilateral in $T_\WKB$ has two coinciding internal edges leads to the
overall powers of $2$ in the multiplicative factors.  See Figure \ref{fig:Nf0-universal}.
\insfig{3.5in}{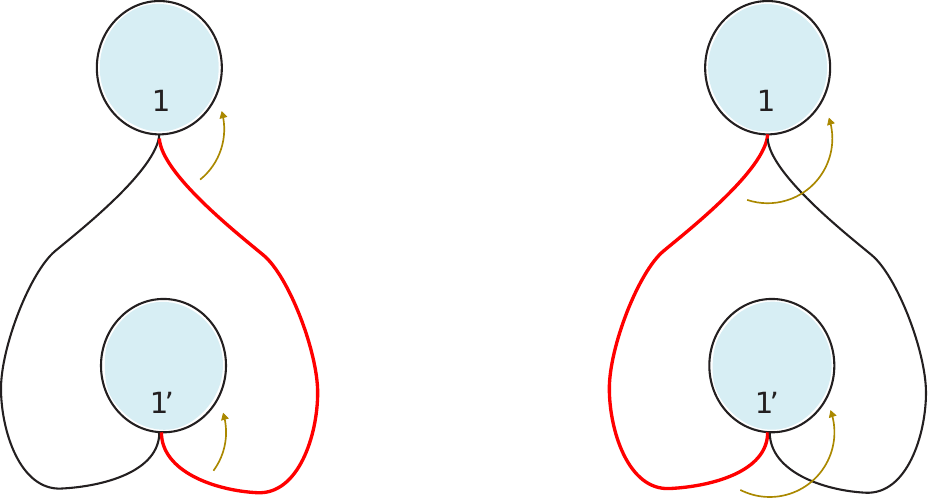}{Diagram for computing the factors $A_{ab}$ appearing in the
spectrum generator of the $SU(2)$ theory with $N_f=0$.}

\subsection{$SU(2)$, $N_f=1$}

Next consider the $SU(2)$ theory with $N_f = 1$, which we described in Section \ref{sec:nf1}.
Here at strong coupling we found that ${\bf S} = \CK_{\gamma_1} \CK_{-\gamma_3}\CK_{\gamma_2}$, with $\langle
\gamma_1,\gamma_2\rangle=\langle \gamma_2,\gamma_3\rangle=\langle
\gamma_3,\gamma_1\rangle=1$.  This transformation acts by
\begin{align}
\tilde \CX_{\gamma_1} &= \CX_{\gamma_1} (1+\CX_{\gamma_2}) (1+ \CX_{-\gamma_3}+ \CX_{-\gamma_3}\CX_{\gamma_2}),\\
\tilde \CX_{\gamma_2} &= \CX_{\gamma_2} (1+ \CX_{-\gamma_3}+ \CX_{-\gamma_3}\CX_{\gamma_2})^{-1} \left[1+   \CX_{\gamma_1} (1+\CX_{\gamma_2}) (1+ \CX_{-\gamma_3}+ \CX_{-\gamma_3}\CX_{\gamma_2})\right]^{-1}, \\
\tilde \CX_{-\gamma_3} &= \CX_{-\gamma_3}(1+\CX_{\gamma_2}) \left[1+   \CX_{\gamma_1} (1+\CX_{\gamma_2}) (1+ \CX_{-\gamma_3}+ \CX_{-\gamma_3}\CX_{\gamma_2})\right]^{-1}.
\end{align}
Once again, it is straightforward to obtain these factors from the triangulation using our rules.  See Figure \ref{fig:Nf1-universal}.
\insfig{4.1in}{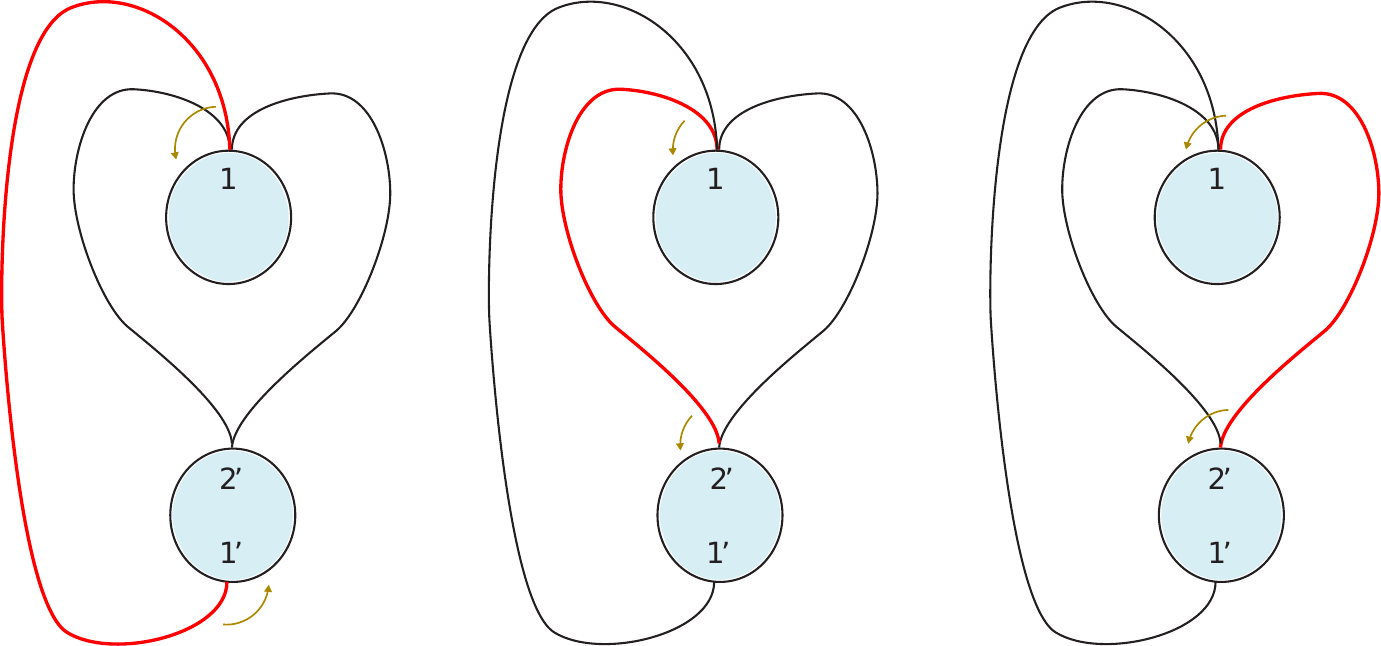}{Diagram for computing the factors $A_{ab}$ appearing in the spectrum
generator of the $N_f=1$ theory.}

\subsection{$SU(2)$, $N_f=4$}

As our final example, we describe the spectrum
generator ${\bf S}$ for the $N_f=4$ theory.  We will consider a simple situation
where $T_\WKB$ is the graph of a tetrahedron:  we have
six edges, labeled by all pairs $ij$, $i<j$, $i,j=1, \dots, 4$,
as in Figure \ref{fig:Nf4-universal}.
\insfig{2in}{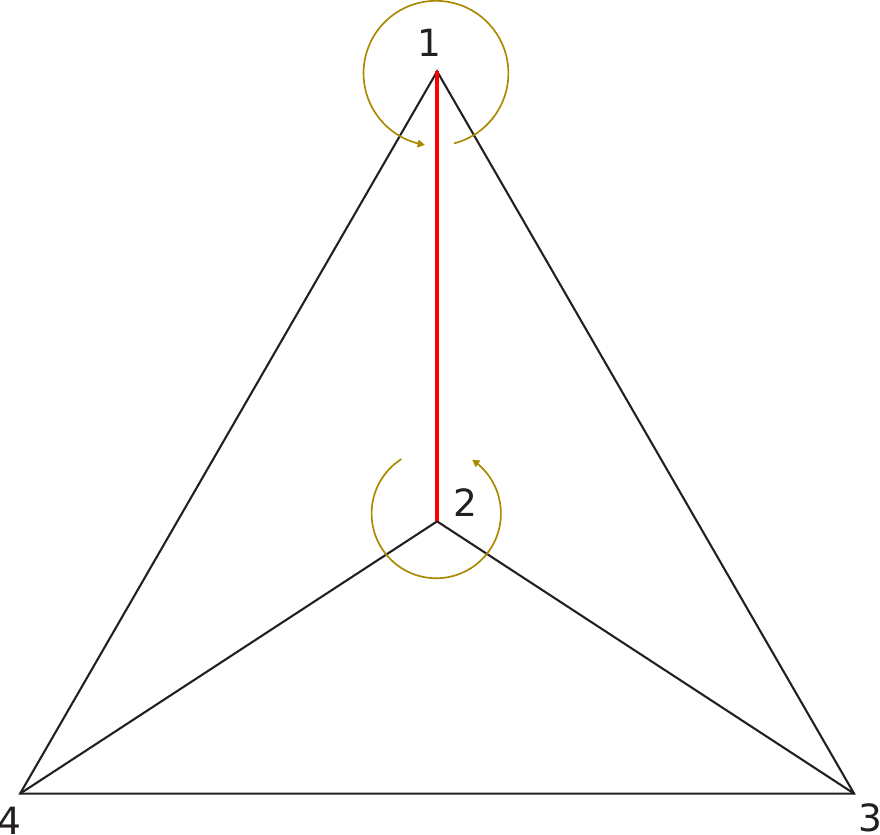}{Diagram for computing the factor $A_{12}$ appearing in the
spectrum generator of the $SU(2)$ theory with $N_f = 4$.}
Applying our rules we have
\begin{equation} \label{eq:nf4-u}
\tilde \CX_{12} = \CX_{12} \frac{ (1+ A_{13}) (1+
A_{24})}{(1+A_{23})(1+A_{14})},
\end{equation}
and 5 more transformations related to this by using tetrahedral
symmetry.  We can simplify the factors appearing in \eqref{eq:nf4-u} slightly by multiplying
numerator and denominator by
$(1-\mu_1^{2})(1-\mu_2^{2})(1-\mu_3^{2})(1-\mu_4^{2})$. Then,
for example, using $\mu_2^{2} = \CX_{12} \CX_{24} \CX_{23}$ and
$\mu_3^{2} = \CX_{34} \CX_{23}\CX_{13}$, we obtain
\begin{multline}
(1-\mu_2^{2})(1-\mu_3^{2})(1+ A_{23}) =
1 + \CX_{23} + \CX_{12}\CX_{23} + \CX_{23}\CX_{34} + \\
+\CX_{12} \CX_{13}\CX_{23}\CX_{34} + \CX_{12} \CX_{23} \CX_{24}
\CX_{34} + \CX_{12}\CX_{13} \CX_{23}\CX_{24} \CX_{34} +
\CX_{12}\CX_{13} \CX_{23}^2 \CX_{24}\CX_{34}.
\end{multline}

Based on the description of the spectrum which we gave in Section
\ref{subsec:SuperConformalCase}, we expect that the above
transformation can be decomposed as
\begin{equation}\label{eq:12ParticlesKii}
\CK_{1,-1;-}\CK_{0,1;+}^2 \CK_{0,-1;-}^2 \CK_{1,1;+}\CK_{-1,-1;-}\CK_{1,0;+}^2 \CK_{-1,0;-}^2\CK_{1,-1;+},
\end{equation}
but we have not explicitly checked this.

\section{Categorical matters} \label{sec:category}

The mathematically oriented reader may wonder how closely the constructions in this paper
can be related to the approach to Donaldson-Thomas invariants
employed in \cite{ks1}.  In particular, in that approach the starting point
is an appropriate category and a family of stability conditions thereon.  How are these data
realized in our examples?  In this section we make a few observations
which might help point the way.  We are very rough and avoid several important issues,
in particular the role of $\IZ$-gradings in the story.

Let us begin with the case $K=2$ (on which we have been
concentrating for the last few sections). By analogy to the Fukaya
category of a Calabi-Yau threefold, we can then set up the basic
story as follows. We have a Riemann surface $C$ and a space $\CB$ of
meromorphic quadratic differentials on $C$. Let $\CB' \subset \CB$
be the regular locus, consisting of quadratic differentials with
only \ti{simple} zeroes. For any point $u \in \CB'$ we have a
corresponding double covering $\Sigma_u \to C$ equipped with a 1-form $\lambda$.
In the usual discussion of stability conditions on the Fukaya category one
considers a family of Calabi-Yau manifolds which are symplectically
isomorphic and so can locally be identified with one fixed
symplectic manifold. Similarly here, the various coverings
$\Sigma_u$, when considered just as topological branched covers of
$C$, can be identified with a single fixed $\Sigma$.  Then:

\begin{itemize}
\item The objects of our category should be non-intersecting collections of oriented closed paths $\gamma$ on $\Sigma$ (or perhaps complexes
of such collections of paths), which are anti-invariant under the exchange of the two sheets.

\item The space of morphisms between two paths $\gamma_1$, $\gamma_2$ should have a basis element for each point of $\gamma_1 \cap \gamma_2$.

\item The K-theory class of a path $\gamma$ should be its homology class.

\item The space of stability conditions should be the universal cover $\tilde{\CB}$ of $\CB'$.
\end{itemize}

Given a point of $\tilde{\CB}$
we can define the \ti{phase function} on a path $\gamma$ to be the
phase of $\lambda \cdot \partial_t$ where $\partial_t$ is the positively oriented tangent vector.
The stable objects in which we are ultimately interested are paths which have constant phase.
This is an analogue of the special Lagrangian condition which one imposes to
define stable objects in the Fukaya category.
The central charge function associated to a stability condition is the one we have been using throughout this paper,
$Z_\gamma = \frac{1}{\pi} \oint_\gamma \lambda$.

If we have a pair of intersecting paths $\gamma_1$, $\gamma_2$, such that their phases near the intersection point obey
$\vartheta_2 < \vartheta_1 < \vartheta_2 + \pi$, we can define a new
path $\gamma_1 \# \gamma_2$ by smoothing the intersection.
We expect that a suitably general $\gamma$ admits such a decomposition,
\begin{equation}
\gamma = \gamma_1 \# \cdots \# \gamma_n,
\end{equation}
where the individual $\gamma_i$ are stable, and
the phases of the central charges of the constituents $\gamma_i$
are monotonically decreasing.
The existence and
uniqueness of such decompositions is an essential prerequisite for
the approach of \cite{ks1} to the wall-crossing formula.

One approach to obtaining such decompositions has been described in
\cite{thomas-yau} using a variant of mean curvature flow. The
technology of this paper suggests a possible alternative.  Namely,
as we described in Section \ref{subsec:GlobalBehaviorWKB}, for any
phase $\vartheta$ we have a corresponding decomposition of $C$ into
cells, bounded by the separating WKB curves.  As we rotate
$\vartheta$ clockwise through generic values, these cells vary
continuously, and we can likewise deform $\gamma$ continuously so
that its incidence relations with the separating WKB curves are
unchanged. At some critical $\vartheta = \vartheta_c$ where a BPS
hypermultiplet appears, the topology of the cell decomposition
changes: one of the cells collapses.  This collapsing cell can
``trap'' a segment of $\gamma$, as indicated in Figure
\ref{fig:collapsing-cell}. \insfig{5.6in}{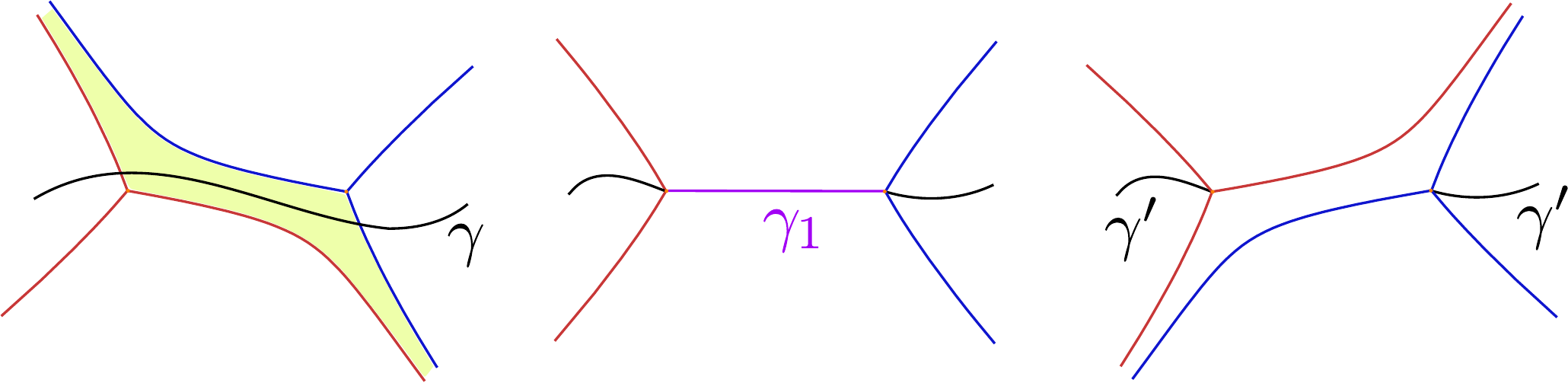}{As
$\vartheta$ varies one of the cells defined by the WKB foliation
(shaded) collapses, trapping a segment of the projection of $\gamma$
to $C$.  This gives a natural splitting of $\gamma$ into $\gamma_1$
and $\gamma'$.} This gives a natural splitting of $\gamma$ into
$\gamma_1$ and $\gamma'$, where $\gamma_1$ is the trapped segment
(which is a stable object of phase $\vartheta_c$) and $\gamma'$ is
the rest (perhaps disconnected). Continuing in this way $\gamma'$
will be further decomposed into objects $\gamma_i$.  Note that the
$\gamma_i$ naturally come out with their phases ordered, since we
are rotating $\vartheta$ in a definite direction! In this way we can
determine geometrically a ``Harder-Narasimhan filtration'' of an
unstable object.

As we remarked in Section \ref{sec:iib}, one actually expects that the theories
we are considering can be realized directly via Type IIB string theory on
an appropriate non-compact Calabi-Yau threefold.  Presumably the category
we are describing here should be identified with an appropriate version of
the Fukaya category of that threefold.

Finally we note that all of these considerations should have a
natural extension to the case $K > 2$ where we have to consider not
only BPS strings but also BPS string webs.  The crucial question, of
course, is to determine the structure which generalizes the
triangulations $T_\WKB(\vartheta,\lambda)$, controls the WKB
approximation, and encodes the spectrum generator for the theory.
Since the first preprint version of this paper appeared, we have made some progress in this direction,
to appear in \cite{gmn5-to-appear}.

\section{$R\to \infty$ limit}\label{sec:R-To-Infinity}

As we reviewed in Section \ref{sec:review}, the expectation from \cite{Gaiotto:2008cd}
is that as $R \to \infty$ one should have
\begin{equation}\label{eq:RtoInfty2}
\CX_{\gamma}^\vartheta \sim  \exp \left( \pi R\zeta^{-1} Z_\gamma + i \theta_\gamma +
\pi R \zeta \bar Z_\gamma \right) \left( 1 + \CO(e^{-const. R}) \right)
\end{equation}
so long as $\zeta$ lies in the half-plane $\IH_\vartheta$
centered on $e^{i \vartheta}$.
We would like to check that \eqref{eq:RtoInfty2} indeed holds for the
$\CX_\gamma^\vartheta$ we have defined.

We defined $\CX_\gamma^\vartheta$ as
functions on the Hitchin moduli space $\CM$, so in order even to
make sense of the equation \eqref{eq:RtoInfty2}, we must make some
identification between the moduli spaces $\CM$ for different $R$.
For this we use the fact that $\CM$ can always be canonically
identified with the ($R$-independent) moduli space of Higgs bundles,
as remarked in Section \ref{sec:higgs}.  On this moduli space,
moreover, there is a natural definition of the angular coordinates
$\theta_\gamma$ as we will see below.

Having made sense of \eqref{eq:RtoInfty2}, our strategy for proving
it is to give an explicit but \ti{approximate} description of the
solution of Hitchin's equations corresponding to any fixed $(u,
\theta)$.  This approximate solution is very close to the true
solution in the limit $R \to \infty$; indeed, by studying linearized
perturbations we will argue that it differs from the exact solution
$(A,\varphi)$ only by corrections which are exponentially suppressed
as $R\to \infty$.   On the other hand, the approximate solution is
exactly diagonal along the edges of the WKB triangulation, and hence
forms a very convenient starting point for a WKB analysis at large
$R$. This WKB analysis then gives the desired \eqref{eq:RtoInfty2}.

In this section we will encounter a bit more of the fine structure of $\CM$ than we have seen
in other parts of this paper --- we have to pay a bit of attention to issues such as finite coverings
depending on the precise choice of gauge group.  The simplest version of the story arises
if we take the gauge group to be $PSU(2)$ rather than $SU(2)$, so for simplicity, we
restrict to that case here.  Related issues are discussed in the end of Appendix \ref{app:MonodromyX}.

\subsection{A parameterization of Higgs bundles} \label{sec:higgs-param}

As mentioned above, to identify the moduli spaces $\CM$ at different values of $R$,
it is convenient to use the Higgs bundle picture which is manifestly $R$-independent.
We begin by giving a more explicit description of the gauge equivalence classes of Higgs
bundles.

So suppose we have a Higgs bundle $(V, \varphi, \bar\partial)$.
As usual we write $\Tr \varphi^2 = 2 \lambda^2$ and
assume that $\lambda^2$ has only simple zeroes.
$\lambda^2$ determines a point on the base $\CB$ of the Hitchin fibration.
Now we want to parameterize the torus fiber.

Choose any spin structure $K^\half$ on $C$,
and identify $V$ with $K^{-\half} \oplus K^{\half}$.  (The specific choice of spin structure here
is irrelevant since different choices give the same $PSL(2,\IC)$ bundle.)
Then up to gauge equivalence we may take $\varphi \in \End(K^{-\half} \oplus K^{\half}) \otimes K$ of the form
\begin{equation} \label{eq:phi-simple}
\varphi = \begin{pmatrix} 0 & 1 \\ \lambda^2 & 0 \end{pmatrix}.
\end{equation}
Having done so, $\bar\partial \varphi = 0$ implies that
$\bar\partial$ differs from the standard holomorphic structure $\bar\partial_0$ on
$K^{-\half} \oplus K^{\half}$ only by a matrix commuting with $\varphi$.  A convenient way of writing this is
\begin{equation} \label{eq:bp-simple}
\bar \partial = \bar \partial_0 + \begin{pmatrix} 0 & \lambda^{-1} \\ \lambda & 0 \end{pmatrix} a^{0,1},
\end{equation}
where $a^{0,1}$ is a $(0,1)$-form on the spectral curve $\Sigma$,
which is odd under the exchange of the two
sheets (so that $\bar\partial$ is well defined).

Not all $a^{0,1}$ give gauge inequivalent Higgs bundles:  we should
divide out by gauge transformations given by any section $g_c$ of
$\End(K^{-\half} \oplus K^{\half})$ which commutes with
\eqref{eq:phi-simple}.  Such a $g_c$ can be written as
\begin{equation}
g_c = \exp \left[i \begin{pmatrix} 0 & \lambda^{-1} \\ \lambda & 0 \end{pmatrix}  f \right]
\end{equation}
with $f$ a complex-valued function on $\Sigma$, odd under
exchanging sheets.  While $g_c$ must be single-valued, $f$ need not be:
upon going around a cycle $\gamma \in H_1(\Sigma; \IZ)$ it may
shift by
\begin{equation} \label{eq:f-multi}
f \to f + 2 \pi n_\gamma
\end{equation}
for some $n_\gamma \in \IZ$.\footnote{There is a delicate point here.  We actually consider Higgs bundles with
structure group $PSL(2,\IC)$, so $g_c$ need only be single-valued in that group.  One might then have expected that $n_\gamma$
should be allowed to be half-integer.  Actually, requiring that $g_c$ is well defined near the zeroes of $\lambda$ turns
out to imply that $n_\gamma$ is an integer.}
Such a gauge transformation shifts
\begin{equation} \label{eq:gtf}
a^{0,1} \to a^{0,1} + i \bar \partial f.
\end{equation}
Using this freedom we can arrange that
\begin{equation} \label{eq:a-flatness}
d a^{0,1} - \overline{d a^{0,1}} = 0,
\end{equation}
in other words,
$a^{0,1}$ is actually the $(0,1)$ part
of an imaginary \emph{flat} 1-form
\begin{equation}
\tilde a = a^{0,1} - \overline{a^{0,1}}.
\end{equation}
Even after fixing \eqref{eq:a-flatness} there is still some
gauge freedom left:  if $f$ is real then the gauge
transformation \eqref{eq:gtf} preserves \eqref{eq:a-flatness},
and transforms
\begin{equation}
\tilde a \to \tilde a + i d f.
\end{equation}
Hence the fiber of the Hitchin fibration is parameterized by the imaginary flat odd 1-forms $\tilde a$
modulo this equivalence.
More explicitly, we can give coordinates on the fiber as
\begin{equation}
\theta_\gamma := i \oint_\gamma \tilde a
\end{equation}
for $\gamma \in \hat\Gamma$.  A multivalued gauge
transformation as in \eqref{eq:f-multi} shifts $\theta_\gamma
\to 2 \pi n_\gamma + \theta_\gamma$, so $\theta_\gamma \in \IR
/ 2 \pi \IZ$.

So we have obtained the angular coordinates $\theta_\gamma$ on the fiber
of the Hitchin fibration.  We can also think of them in
terms of $\pb$-operators on a topologically trivial complex line
bundle modulo complex abelian gauge transformations. If we were to
consider all $\pb$-operators then we would get $\CA^{0,1}/\CG_c$,
which is just the  Jacobian of the curve $\Sigma$. Since $a^{0,1}$
is odd under deck transformations, we in fact get the Prym
subvariety.

\subsection{Reformulating Hitchin's equations}

As we described in Section \ref{sec:higgs}, to each of the Higgs
bundles we have just described, there is a corresponding solution
of Hitchin's equations \cite{MR89a:32021,MR965220,MR1040197,wnh}.
Although the Higgs bundle does not depend on $R$, its corresponding
solution $(A, \varphi)$ certainly does, and we are interested in studying its
behavior in the $R \to \infty$ limit.

The passage from the Higgs bundle $(V, \varphi, \bar\partial)$
to the desired $(A, \varphi)$ proceeds as follows.
First, the operator $\bar\partial$ is identified with the $(0,1)$ part of
$d + A$.  Next, we fix a Hermitian metric on $V$, and then
define the $(1,0)$ part of $d + A$ to be minus the adjoint of $\bar\partial$ in this metric, so that
the full connection is unitary.  We can then ask whether this connection, together
with $\varphi$, gives a solution of Hitchin's equations or not.
(They obviously solve $\bar\partial_A \varphi = 0$, so the real question is whether
$F + R^2 [\varphi, \bar\varphi] = 0$.)
For some choice of metric on $V$, called the \ti{harmonic metric}, this
will indeed be the case.

A convenient way of specifying the harmonic metric is to give the change-of-basis matrix $B$
between the basis we used in Section \ref{sec:higgs-param} and a unitary basis.
It will be convenient to choose a patch
$U \subset C$, with local coordinate $z$, and a
trivialization $(dz)^\half$ of $K^\half$ over $U$.  Then define $p(z)$ by
\begin{equation}
\lambda^2 = p(z) (dz)^2
\end{equation}
and a (multivalued) function $\eta$ by
\begin{equation}
\eta := \left( \frac{p}{\bar p} \right)^{1/8} = \frac{p^{1/4}}{\vert p\vert^{1/4}}.
\end{equation}
Given a metric there is some freedom in the choice of unitary basis:
we fix that freedom by requiring that in this basis $\varphi$ should
be purely off-diagonal and its upper right entry real. Then the most
general possible change-of-basis matrix takes the form
\begin{equation} \label{eq:M}
B = \begin{pmatrix} \abs{p}^{\qtr} e^{h/2} & 0 \\ 0 &
\abs{p}^{-\qtr} e^{-h/2} \end{pmatrix} \exp[  \varphi f_c/p^{1/2} ]
\end{equation}
where $h$ is a real-valued function on $U$.
In the unitary basis one then gets
\begin{equation}\label{eq:RealSect}
\varphi   =  \begin{pmatrix} 0 & \vert p\vert^{1/2} e^h  \\
\frac{p}{\vert p\vert^{1/2}} e^{-h} & 0 \\ \end{pmatrix}
\end{equation}
and
\begin{equation}\label{eq:SpecGaugeField}
A = a_{\bar z} d\bar z \begin{pmatrix} 0 & \bar\eta^2 e^h  \\
\eta^2 e^{-h} & 0 \\ \end{pmatrix} +
a_{ z} d  z \begin{pmatrix} 0 &  \bar\eta^2 e^{-h}  \\
 \eta^2 e^{h} & 0 \\ \end{pmatrix} + b \sigma^3,
\end{equation}
where $a^{0,1} = a_{\bar z} d \bar z$, we defined $a_z: = - (a_{\bar
z})^*$, and
\begin{equation}\label{eq:bee}
b = (\p-\pb) \log \left( \vert p \vert^{\qtr} e^{h/2} \right)  =
\frac{d\eta}{\eta} + \half (\p-\pb) h.
\end{equation}
Substitution into Hitchin's equations
then requires that $h$ and $f_c$ satisfy
\begin{equation}\label{eq:DeformedpSG}
\p_z \pb_{\bar z} h - (\vert p \vert R^2 + \vert a_{\bar z} + \p_{\bar z } f_c \vert^2)(e^{2h} - e^{-2h}) = 0,
\end{equation}
\begin{equation}\label{eq:DefFlat}
 e^{-h}\p_z \left( e^{2h} (\p_{\zb } f_c + a_{\bar z} )\right) + e^h\pb_{\bar z}
 \left(e^{-2h} (\p_z \bar f_c + (a_{\bar z})^* )\right) = 0.
\end{equation}
(Note that \eqref{eq:DefFlat} is not
equivalent to its complex conjugate.)
As in the previous section,
we partially fix the choice of $a_{\bar z}$ by taking the corresponding imaginary
one-form $\tilde a$ to be flat. By a gauge transformation of $\tilde a$ we can
moreover take $f_c$ to be real.\footnote{Below we will
further partially fix $\tilde a$ by taking it to vanish in some
neighborhood of the turning points. In this case we cannot
necessarily simultaneously take $f_c$ to be real in this
neighborhood. However, we can continue to take it to be real  in
some neighborhood away from the turning points.}

\subsection{The trivial solution} \label{sec:trivial}

One obvious choice would be to take $h=0$.
In that case  \eqref{eq:DeformedpSG} is obviously solved, while \eqref{eq:DefFlat}
becomes
\begin{equation}\label{eq:flatreduced}
2 \p_z \p_{\zb } f_c + \p_z a_{\bar z} + \pb_{\bar z}
a_{\bar z}^* = 0,
\end{equation}
which is also solved if we pick $f_c = 0$, since we have chosen $\tilde{a}$ flat.

The only trouble with this solution is that it does not lead to a regular solution of Hitchin's equations.
Indeed, with $h = f_c = 0$, our change-of-basis matrix $B$ given in \eqref{eq:M}
becomes singular at the turning points, where $p = 0$.  Requiring that $B$ is regular imposes a
boundary condition which will force $h$ to have a singularity at each turning point.

In the next subsection we turn to the discussion of this honest regular solution of Hitchin's equations.
 We will argue that, if we fix a point $\CP$ which is not a turning point, then, for sufficiently large
 $R$,   $h(\CP)$ and $f_c(\CP)$ decay exponentially fast as $R \to
 \infty$.    It follows that the ``trivial'' solution we considered here
is actually exponentially close to the exact one.

\subsection{The regular solution}

For simplicity, we will
first analyze the case $\theta_\gamma=0$.  In this case we can take
$\tilde a = 0$, and clearly then $f_c = 0$ will solve \eqref{eq:DefFlat}.
This leads to an important simplification in \eqref{eq:DeformedpSG}:  it
becomes simply the equation of motion of the sinh-Gordon theory for a
scalar field $h$ on a surface with metric $ds^2 = \vert p(z) \vert \vert dz \vert^2$,
\begin{equation}\label{eq:SG}
 \p_z \pb_{\bar z} h - R^2 \vert p \vert (e^{2h} - e^{-2h}) = 0.
 \end{equation}
Multiplying \eqref{eq:SG} by $h$ and integrating shows that there are no
nonsingular solutions with $h \neq 0$ on any compact Riemann surface.
This is just as well, since the only reason we want a solution with
$h \neq 0$ is to deal with our nontrivial boundary conditions, which we now describe.

First, as remarked above, near each zero $z_a$ of $p$, the regularity
of $B$ requires
\begin{equation}
h \sim \log \vert z-z_a \vert^{-1/2}.
\end{equation}
We should also discuss the behavior near the singular points $\CP_i \in C$.
Our boundary conditions on Hitchin's equations require that $h \to 0$ there.
It then follows from \eqref{eq:bee} that $A \to \frac{d\eta}{\eta}\sigma^3$.
In particular, the boundary condition on $A$ at $\CP_i$ can only be satisfied if
$m_i^{(3)}=0$.  (This is what we expected:  we have fixed all $\theta_\gamma=0$, and $m_i^{(3)} = \theta_\gamma$
for $\gamma = C_i$.)

We now argue using perturbation theory that solutions
satisfying our boundary conditions exist.\footnote{Physically, it is
obvious that they exist. These solutions are the semiclassical field
configuration in the sinh-Gordon model in the presence of vertex
operator sources at the points $z=z_a$.}  Let $h_{\rm ax}$ be an
approximate solution. The perturbative equation for the true
solution $h= h_{\rm ax} + \delta h$ is
\begin{equation}
L(h_{\rm ax}) \delta h = E_0(h_{\rm ax}) + \CI(\delta h),
\end{equation}
where $L(h_{\rm ax})$ is given by
\begin{equation}\label{eq:L-Operator}
L(h_{\rm ax}) := \p_z \pb_{\bar z} - 4 R^2 |p| \cosh(2h_{\rm ax}),
\end{equation}
the source term to begin the perturbation expansion is
\begin{equation}
E_0(h_{\rm ax}) = - ( \p \pb h_{\rm ax} - 2 R^2\vert p \vert
\sinh(2h_{\rm ax}) ),
\end{equation}
and $\CI$ are the interaction terms:
\begin{equation}
\CI = 2 R^2 \vert p \vert \biggl( \sinh(2h_{\rm ax}) (\cosh(2\delta
h ) -1) + \cosh(2h_{\rm ax}) ( \sinh(2\delta h) - 2 \delta h)
\biggr).
\end{equation}
Let $G(x,y)$ be the Green's function for the operator $L(h_{\rm
ax})$. Think of this as the operator in Euclidean space for a massive
scalar field in two dimensions, with mass-squared $4R^2 \cosh(h_{\rm ax}) > 4 R^2$,
on a surface with metric $ds^2 = \vert p(z)\vert \vert dz\vert^2 $.
(Near the zeroes of $p$ the Green's function reduces to
that of a scalar field of mass-squared $const. R^2$.)
Therefore, away from turning points, for $x, y$ at fixed separation
in the metric $\vert p(z)\vert \vert
dz\vert^2$, and as $R\to \infty$, we have
\begin{equation}
\vert G(x,y)\vert \sim e^{- 2R d(x,y)}
\end{equation}
\emph{or even smaller} (because the mass can get bigger), where
$d(x,y)$ is the geodesic distance from $x$ to $y$.  On the other
hand, as $d(x,y)\to 0$ the mass becomes irrelevant to the
short-distance behavior and $G(x,y) \sim  - \frac{2}{\pi} \log \vert
z(x) - z(y) \vert $.  We have
\begin{equation}\label{eq:FirstOrderh}
\delta h(x) = \int G(x,y) E_0(h_{\rm ax}(y)) d^2 y
\end{equation}
Now we construct a suitable $h_{\rm ax}$.
By a change of
variable such that $dw = 2 R \sqrt{p(z)}dz$ we can bring the
equation to the form
\begin{equation}\label{eq:2DSG}
\p_w \p_{\bar w} h = \half \sinh (2h).
\end{equation}
In view of our boundary
conditions it is natural to search for a solution which is radially
symmetric in the $w$-coordinate around $z_a$. Thus, working in some
neighborhood  $D_a = \{ z: |z-z_a| = \rho_a \}$  we take $w =
\int_{z_a}^z 2 R \sqrt{p(z)}dz$. Of course, $w$ is only locally
well-defined, but in a suitable neighborhood $D_a$ of $z_a$ it will
be undefined only up to a sign so that $\vert w \vert$ is
well-defined. It therefore makes sense to search for radially
symmetric solutions $h$ which are only functions of $\vert w \vert$
in such a neighborhood. Letting $x= 2\vert w\vert$,    the equation
for $h = \tilde h(x)$ reduces to
\begin{equation}
\left(\frac{d^2}{dx^2} + \frac{1}{x} \frac{d}{dx}\right)\tilde h =
\half \sinh(2\tilde h)
\end{equation}
This is the Painlev\'e III equation and has been well-studied in
connection with correlation operators of spin and disorder operators
in the massive Ising model. (See, for example \cite{Fateev:1998xb}.)
It is known that if
\begin{equation}\label{eq:conn1}
\tilde h(x) = 2 \sigma \log(8/x) - \log(\gamma(\half -\sigma)) +
\CO(x^{2\pm 4 \sigma})
\end{equation}
for $x\to 0$ then
\begin{equation}\label{eq:conn2}
\tilde h(x) \to \frac{2\sin(\pi \sigma)}{\pi} K_0(x)
\end{equation}
for $x\to \infty$ \cite{McCoy:1976cd}. For our boundary conditions
we take $\sigma = 1/6$. Let us denote the resulting solution,
defined in $D_a$ as $h_P^{(a)}$. Note that at any fixed value of
$z\in D_a$, for $R \to \infty$ we have $h_P^{(a)}\sim \pi^{-1}
K_0(\frac{8R}{3} \vert z\vert^{3/2} (1+ \cdots ) ) $ and this is
exponentially small.

Now let   $C- \amalg_a D_a = \CF$ be the complement of the regions
near the turning points. We will refer to this as the ``fatgraph
region'' because we regard it as a thickened version of a WKB
triangulation. Choose
\begin{equation}
h_{\rm ax} = \begin{cases} h_P^{(a)} &  {\rm in} \quad  D_a \\  0  &
{\rm in} \quad \CF
\\ \end{cases}
\end{equation}
This choice is convenient since the integral in
(\ref{eq:FirstOrderh}) only receives support from the boundary
$\amalg \p D_a$ of the fatgraph region.

The contribution of the boundary of a disk $D_a$ to $\delta h(x)$
can be written as:
\begin{equation}
- \frac{1}{4} \int_{0}^{2\pi} d \theta \left( \rho_a G(x,\rho_a e^{i
\theta}) \p_\rho h_P^{(a)} - \rho_a (\frac{\p}{\p \rho} G(x,\rho
e^{i \theta}))\vert_{\rho=\rho_a} h_P^{(a)} \right)
\end{equation}
where we use   coordinates $z = z_a + \rho e^{i \theta}$. In
general, if $x$ is distance $\Delta$ away from $\p D_a$ in the
metric $\vert p(z)dz^2 \vert$ then the difference $\vert \delta h -
h_{appxt}\vert $  is -- very roughly speaking -- of order $e^{-2R
\Delta} e^{- \frac{8 R}{3}\rho_a^{3/2}} $, and so is exponentially
smaller than $h_P^{(a)}$ in the regions $D_a$.

The one place where this argument fails is when  $x\in \p D_a$. At
short distances we may replace $G(x,y) \sim - const. \log \vert
x-y\vert$. One finds that the   corrections for $x\in \p D_a$ are
therefore of order $h_{P}^{(a)}$, as is quite reasonable since the
true solution will be $\CC^{\infty}$ and our initial approximation
$h_{\rm ax}$ is discontinuous. So we expect the corrections to
smooth out the discontinuity. Everywhere else the corrections are
exponentially smaller than $h_P^{(a)}$.

Thus, we conclude that we have set up a good approximation scheme,
and we have shown that we can consistently take $h= h_{\rm ax}$.

Finally, let us consider the modifications to the above
$\theta_\gamma \neq 0$. One can further partially fix the gauge
freedom in the choice of flat gauge field $\tilde a$ by choosing it
to vanish near the turning points. Then, one  can check that there
exists a solution $f_c$ to (\ref{eq:DefFlat}) which is smooth and
consistent with our boundary condition on $h$. Thus, we can continue
to take the same boundary conditions $h \sim - \half \log \vert
z-z_a \vert$ in the neighborhood of the turning points. Away from
turning points we construct solutions to the pair of equations
(\ref{eq:DeformedpSG}), (\ref{eq:DefFlat}) as a perturbation series
in both $h$ and $f_c$. If $h=0$ we know that $f_c$ is nonzero. If
$a=0$ we also know that we can take $f_c = 0$.  If
$a$ cannot be gauged to zero, because $\theta_\gamma \neq 0$,
then we cannot set $f_c = 0$, but from the differential
equation we learn that away from turning points the derivatives of
$f_c$ are order $h\times a$ or smaller. Therefore, we expect that
$f_c$ will be exponentially small for $R\to \infty$ in the fatgraph
region. Note that the  presence of nonzero $\tilde a + \pb f_c - \p
\bar f_c$ simply increases the mass parameter in the Green's
function and therefore increases the exponential suppression which
makes $h$ and therefore $f_c$ small. It would be nice to do better
here and really solve the connection problem for $f_c$ connecting
exponentially small solutions in the fatgraph region to appropriate
solutions near the turning points.

\subsection{The large $R$ limit of $\CX_\gamma$}

The upshot of the last section is that, along the edges of the WKB
triangulation, the ``trivial'' solution described in Section \ref{sec:trivial} is equal to the
exact solution of Hitchin's equations, up to corrections that are
exponentially suppressed in $R$.

Now by using the $SU(2)$ gauge transformation\footnote{This gauge transformation is actually multi-valued even when lifted to $\Sigma$;
note however that it would be single-valued on $\Sigma$ when considered as an element of $PSU(2)$.  As we noted at
the beginning of this section, we are avoiding some subtleties by considering only gauge group $PSU(2)$.}
\begin{equation}\label{eq:unitarygtii}
g =  \frac{1}{\sqrt{2}} \begin{pmatrix} \bar \eta & - \bar \eta \\
\eta & \eta \\ \end{pmatrix},
\end{equation}
we can bring this trivial solution to a diagonal gauge in which
\begin{align}
\varphi &= p^{1/2} \sigma^3, \\
A &= (\tilde a + (\pb
f_c - \p \bar f_c) )\sigma^3.
\end{align}
Then, by a computation very similar to that used in Section
\ref{sec:asymptotics} above to compute the $\zeta \to 0$
asymptotics of $\CX_{\gamma}$, we find that
\begin{equation}
\CX^{\vartheta}_{\gamma} = \CX_{\gamma}^{\sf} \left(1 + \CO(e^{- c
R})\right),
\end{equation}
where
\begin{equation}
\CX_{\gamma}^{\sf} = \exp[\pi R\zeta^{-1} Z_\gamma + i \theta_\gamma
+ \pi R\zeta \overline{Z_\gamma} ].
\end{equation}
We conjecture that the optimal value for the above constant $c$ is
given by  the norm of the minimal period $\pi {\rm min}_{\gamma}
\vert Z_\gamma \vert$. Indeed, it would be extremely interesting to
compute these corrections! Comparison with equation (5.14) of
\cite{Gaiotto:2008cd} would allow one to extract the BPS
degeneracies $\Omega$.

\subsection{The real section}

In the above discussion, the case where all $\theta_{\gamma}=0$,
so that we can take $\tilde a = 0$ and $f_c=0$, was particularly simple
to analyze.  There is also something else interesting about this locus.

Note that by a rigid gauge transformation of \eqref{eq:RealSect}
taking $(\sigma^1, \sigma^2, \sigma^3) \to (\sigma^1, \sigma^3,
-\sigma^2)$ we can make $\varphi$ a symmetric matrix, and $A$ in
\eqref{eq:SpecGaugeField} with $a=0$ becomes an antisymmetric real
matrix. Therefore, if we choose $\zeta$ to be a phase, then $\CA$ is
traceless and real, that is, it is valued in $sl(2,\IR)$. From
\eqref{eq:monodromy-formula} it follows in particular that the
monodromy eigenvalues $\mu_i$ around each $\CP_i$ are real. Hence
the monodromy matrix is hyperbolic, and we can choose our
decorations (monodromy eigensections) to be real. It follows that
the Fock-Goncharov coordinates $\CX^T_E$ are real on this locus. It
defines a special ``real section'' of the Hitchin fibration. This
real section has been discussed in \cite{MR89a:32021,MR1174252} and
was also a very important ingredient in the considerations of
\cite{MR2233852}.

\subsection{Relation to Hitchin flows}

We conclude this section by checking that the Hitchin flows in the
holomorphic symplectic structure at $\zeta=0$ act by linear shifts of
the $\theta_\gamma$.

From equation (\ref{eq:symplectic}) and
\begin{equation}
\varpi_\zeta = - \frac{i}{2\zeta} \omega_+ + \omega_3 - \frac{i}{2}
\zeta \omega_-
\end{equation}
we learn that the holomorphic symplectic form in complex structure
$\zeta=0$ is:
\begin{equation}
\omega_+ = 2i R\int {\Tr} (\delta \varphi \wedge \delta  A^{0,1} ).
\end{equation}
Now, using the Higgs bundle point of
view, consider a linear flow holding $\varphi$ fixed while
\begin{equation}
a^{0,1} \to a^{0,1} -  i t  \omega_{\gamma}^{0,1}
\end{equation}
Here $\omega_{\gamma}$ is a real 1-form, Poincar\'e dual on $\Sigma$
to $\gamma$, and odd under the deck transformation.  This leads to
a linear flow $\theta_{\gamma'} \to \theta_{\gamma'} + t\langle
\gamma',\gamma\rangle $. Contracting this vector field with the
$(2,0)$ form gives
\begin{equation}
\iota(\frac{\p}{\p t} )\omega_+ =\delta \left(4 R\int_{C} \delta
\lambda  \omega_{\gamma} \right)= \delta \left( 2\pi R Z_{\gamma}
\right)
\end{equation}
From this result we conclude that the symplectic form can be nicely
written as
\begin{equation}\label{eq:simplesymplectic}
\omega_+ = -2\pi R\langle dZ,d \theta\rangle + \omega_+^{\rm
horizontal}
\end{equation}
where $ \omega_+^{\rm horizontal}$ has zero contraction with
the vertical vectors of the Hitchin fibration.

Equation (\ref{eq:simplesymplectic}) is a nice check on our
assertion about the $R\to \infty$ asymptotics of the coordinates
$\CX_{\gamma}$. Indeed, if we take the large $R$ limit of the
symplectic form $\varpi_\zeta := \half \langle d \log \CX,d \log \CX
\rangle$ \footnote{The normalization of $\varpi(\zeta)$ used in this
paper differs from the choice made in \cite{Gaiotto:2008cd}. Using
equation 5.16 of \cite{Gaiotto:2008cd} we find
\begin{equation}
\{ \log \CX_{\gamma}, \log \CX_{\gamma'} \}_\zeta = 4 \pi^2 R
\langle \gamma, \gamma' \rangle
\end{equation}
and therefore $\varpi_\zeta^{here} = 4 \pi^2 R
\varpi_\zeta^{there}$.} we get:
\begin{equation}\label{eq:RtoInfty3}
\varpi_\zeta \to \half   \langle \pi R\zeta^{-1} dZ  + \pi R \zeta
 d \bar Z + i d  \theta, \pi R\zeta^{-1} dZ
+ \pi R \zeta  d \bar Z + i d  \theta \rangle
\end{equation}
We can extract the various powers of $\zeta$. The coefficient
$\langle dZ, dZ \rangle$ of the double pole vanishes because the
$dZ_\gamma$ are periods of holomorphic differentials.\footnote{Note
that this would \emph{not} happen if we allowed $Z$ to vary
arbitrarily in $\Hom(\hat \Gamma,\IC)$, as one does in the theory of
Bridgeland stability conditions \cite{bridstab}.  The Coulomb branch
$\CB$ is locally a Lagrangian subspace of an appropriate space of
Bridgeland stability conditions.} The residue of the simple pole in
$\varpi_\zeta$ at $\zeta=0$ agrees with (\ref{eq:simplesymplectic}).

The Hamiltonians $Z_\gamma$ are closely related to the standard ``Hitchin Hamiltonians''
$h_a$. The latter are functions on $\CB$ defined by expanding the quadratic differential ${\Tr} \varphi^2$ as
\begin{equation}
{\Tr} \varphi^2 = \sum h_a \beta_a,
\end{equation}
where $\beta_a$ are a basis of quadratic differentials with second
order poles at $\CP_i$.  For the case of $\nsing$ regular singularities on $\IC\IP^1$,
$a = 1, \dots, 2\nsing - 3$.  On the other hand, we can expand $\lambda =
\sum_{k=1}^{3\nsing-6} Z_{\gamma_k} \alpha_k$, where $\alpha_k$ are
meromorphic one-forms on $\Sigma$ dual to a basis $\gamma_k$ for
$\hat\Gamma$, anti-invariant under the exchange of sheets,
$\sigma^*(\alpha_k) = - \alpha_k$.  Now, $\lambda^2 = \half {\Tr}
\varphi^2 = \sum_{k,j} Z_{\gamma_k} Z_{\gamma_j} \alpha_k \alpha_j$.
But $\alpha_k \alpha_j$ are invariant under $\sigma$
and hence can be expanded in the basis $\beta_a$.
In this way we can write the standard Hamiltonians
$h_a$ as quadratic polynomials in the periods $Z_{\gamma_k}$.

\section{Comparison with \cite{Gaiotto:2008cd}: differential equations
and the Riemann-Hilbert problem} \label{sec:comparison}

In this section we would like to compare the properties of the
Darboux coordinates $\CX_\gamma^\vartheta$ constructed in this paper
with the properties of the coordinates $\CX^\RH_\gamma$ whose existence was
established in \cite{Gaiotto:2008cd}.

One important property of the $\CX_\gamma^\RH$ of \cite{Gaiotto:2008cd} is that
(letting $\CX$ stand for any of the $\CX_\gamma^\RH$ or any
function of them) they satisfy a set of differential equations of the form
\begin{align}\label{eq:conn}
\partial_{u^j} \CX &= \left( \frac{1}{\zeta} {\CA^{(-1)}_{u^j}} + \CA^{(0)}_{u^j} \right) \CX,  \\
\partial_{\bar{u}^{\bar j}} \CX &= \left( {\CA}^{(0)}_{\bar{u}^{\bar j}} + \zeta {\CA}^{(1)}_{\bar{u}^{\bar j}} \right) \CX,\\
\Lambda \partial_\Lambda \CX &= \left( \frac{1}{\zeta} {\CA_\Lambda^{(-1)}} + \CA_\Lambda^{(0)} \right) \CX,\\
\bar\Lambda \partial_{\bar\Lambda} \CX &= \left( \CA_{\bar \Lambda}^{(0)} + \zeta \CA_{\bar \Lambda}^{(1)} \right) \CX, \\
R \partial_R \CX &= \left( \frac{1}{\zeta} {\CA_R^{(-1)}} + \CA_R^{(0)} + \zeta \CA_R^{(1)} \right) \CX, \\
\zeta \partial_\zeta \CX &= \left( \frac{1}{\zeta} {\CA_\zeta^{(-1)}} + \CA_\zeta^{(0)} + \zeta \CA_\zeta^{(1)} \right) \CX. \label{eq:zetaeq}
\end{align}
On the right hand side of \eqref{eq:conn} we have introduced a
connection $\CA$ on the space $\CB \times \IC^\times \times \IR_+$. It is
not to be confused with  the $\CA$ on $C$, defined in
\eqref{eq:def-CA}! The various pieces of the connection $\CA$   are
complex vector fields on the torus fibers of $\CM$, i.e.
differential operators in some basis of angular coordinates
$\theta_a$, evaluated at constant $u, \bar u$. On the right side we
have also  explicitly exhibited the $\zeta$ dependence of the
connection.

In writing these equations it is most natural to view the $\CX$ as
functions not just on $\CM$, but on an extended version of $\CM$,
where the parameters $m_i, m_i^{(3)}$ determining the residues at
the singularities $\CP_i$ are allowed to vary. The base $\CB$ is
then extended to include $m_i$ while the torus fiber is extended to
include $m_i^{(3)}$.

Notice that if any set of coordinates satisfies these equations,
then every other coordinate system related to it by a $(R,
\zeta)$-independent coordinate transformation also satisfies them.
So in particular, if any set $\CX^\vartheta_\gamma$ satisfy them, or
any $\CX^T_E$ for some triangulation does, then so do more
conventional coordinates such as traces of monodromy matrices (as
these are written as certain rational functions of the $\CX^T_E$
--- see Appendix \ref{app:MonodromyX}).

One can obtain these differential equations simply from our
asymptotic analysis. Here let us just discuss the most important example, the equation
\eqref{eq:zetaeq} controlling the $\zeta$ dependence.\footnote{If we view $\CM$ as the moduli space of Higgs bundles, extended
by the parameters $m_i$ and $m_i^{(3)}$, then this equation just represents the infinitesimal generator of the $\IC^\times$ action
infinitesimally rescaling the Higgs field $\varphi$.}
Consider a basis of coordinate functions $\CX_i$ and angles $\theta_a$, and define a vector field
on the torus fiber by
\begin{equation} \label{eq:def-azeta}
\CA_\zeta = \zeta \frac{\partial \CX_i}{\partial \zeta} \left[ \frac{\partial \CX_i}{\partial \theta_a} \right]^{-1} \frac{\partial}{\partial \theta_a}.
\end{equation}
By the chain rule, this definition is clearly independent both of the specific
parametrization of the torus fiber, and of the choice of coordinate
system $\CX_i$. The Jacobian $\frac{\partial \CX}{\partial \theta}$
is indeed invertible for a
good coordinate system $\CX$, as $\CM$ in any complex structure
$J^\pz$ away from $\zeta=0, \infty$ is locally the complexification
of the torus fiber.  So we have more or less tautologically
\begin{equation}
\zeta \partial_\zeta \CX = \CA_\zeta \CX. \label{eq:zetaeqtaut}
\end{equation}
Now depending on which properties of $\CA_\zeta$ have to be
determined, different choices of coordinate system $\CX$ are
appropriate. To show that $\CA_\zeta$ is holomorphic away from
$\zeta=0, \infty$, it is useful to use around each value of $\zeta$
and point in $\CM$ some coordinate system which is good around that
point.  As long as the sections $s_i$, $s_j$ given by the decoration
do not coincide along some edge $E_{ij}$ of $T$, the Fock-Goncharov
coordinate system $\CX^T$ is fine; moreover, the traces of monodromy
matrices around various cycles of $C$ provide a perfectly sensible
global choice of coordinates. To show that $\CA_\zeta$ has poles of
order at most one at $\zeta=0, \infty$, we consider the coordinate
system $\CX^\vartheta$, while letting $\zeta$ approach $0$ or
$\infty$ inside $\IH_\vartheta$. Plugging the known asymptotics of
$\CX^\vartheta$ into \eqref{eq:def-azeta} then gives the desired
information about $\CA_\zeta$. So we obtain the desired
\eqref{eq:zetaeq}.

The system of compatible differential equations \eqref{eq:conn}-\eqref{eq:zetaeq} is
quite powerful.  In a finite-dimensional context where the operators $\CA$ are
matrices instead of differential operators, such an equation would be
directly equivalent to a Riemann-Hilbert problem.
In our infinite-dimensional case, though, the differential
equation \eqref{eq:zetaeq} is not strong enough to guarantee that the
solutions are holomorphic in $\zeta$ away from $\zeta=0, \infty$.  (That is obvious
from the fact that any rational function of some $\CX$ satisfying
\eqref{eq:conn}-\eqref{eq:zetaeq} also satisfies \eqref{eq:conn}-\eqref{eq:zetaeq}.)

For sufficiently large $R$, the error of the WKB analysis can be
bounded with some work, so it should be possible to guarantee that the
small flat section at one end of a WKB curve will never coincide with the
small flat section at the other end, which means $\CX^\vartheta_\gamma$ will
not have a pole in the neighborhood of the ray $\zeta \in e^{i
\vartheta} \IR_+$.  For small $R$, though, we can see no clear way to rule out
this possibility.  Indeed, the WKB triangulation only carries information about the
Higgs field $\varphi$:  any constraint on the gauge connection $A$ comes only
very indirectly, from the solution of the Hitchin equations.  It is conceivable that
nevertheless there is never a flat section $s$ which is small at both ends
of an edge in $T_\WKB$, but we have not found any indications in favor of, or
against, such a conjecture.

As a result, we can only assert with certainty that for sufficiently
large $R$ the coordinates $\CX_\gamma$, and hence the metric on the
Hitchin system moduli space, can be determined from the general
Riemann-Hilbert problem formulated in \cite{Gaiotto:2008cd},
combined with the spectrum generator (Stokes matrix) computed here.
It would be interesting to find out whether the Riemann-Hilbert
problem simply fails to have a solution for small $R$, or if a
solution exists, but gives the wrong metric.  We hope to present
some numerical tests in a future publication.

\section*{Acknowledgements}

We thank Aaron Bergman, Tom Bridgeland, Ron Donagi, David Dumas, Alexander Goncharov, Sergei
Lukyanov, Sav Sethi, Ivan Smith, Yan Soibelman, J\"org Teschner, Valerio Toledano Laredo,
Edward Witten, Xi Yin, and Sasha Zamolodchikov for helpful discussions
and correspondence.

The work of GM is supported by the DOE under grant
DE-FG02-96ER40949. We thank the KITP at UCSB for hospitality during
the course of part of this work and therefore this research was
supported in part by DARPA under Grant No. HR0011-09-1-0015 and by
the National Science Foundation under Grant No. PHY05-51164. GM
would like to thank the Galileo Galilei Institute and the Aspen
Center for Physics for hospitality during the completion of this
work. The work of AN is supported by the NSF under grant numbers
PHY-0503584 and PHY-0804450.  D.G. is supported in part by the DOE grant DE-FG02-
90ER40542. D.G. is supported in part by the Roger Dashen membership in the Institute for Advanced
Study.

\appendix

\section{Expressing monodromy matrices in terms of Fock-Goncharov
coordinates}\label{app:MonodromyX}

In this appendix we give a simple argument that shows that the
Fock-Goncharov coordinates $\CX_{E}^T$ really do provide a system of
coordinates on a patch of $\CM$ defined by a decorated triangulation
$T$.
Of course, this result already appears
in the work of Fock and Goncharov \cite{MR2233852} (see Theorem 1.8 and Section
6.6 of that paper), as well as related literature on Teichm\"uller theory
(see e.g. \cite{qttcp} and references therein), but for completeness, we give a proof here.

We would like to compute the monodromy matrix for a closed path
$P$ in $C$. If we can show that the traces of the powers of the
monodromy matrix can be expressed in terms of $\CX_E^T$ then we are
done, since those functions provide a complete set of gauge
invariant functions on $\CM$. It suffices to consider a simple
closed curve with no self-intersections, and we restrict attention
to this case.

Choosing a basepoint, the path $P$ will begin in some triangle
$t_1$, and pass through a number of triangles $t_1, ..., t_m$ in
succession and then return to $t_1$. Associated to each triangle in
this sequence we can consider three bases of flat sections, defined
up to a common scale. To define these, focus on one particular
(oriented) triangle $t\in T$ and label the vertices $1,2,3$  in
counterclockwise order around the boundary of $t$. Then to each edge
$E_{ij}$ of the triangle (where $ij = 12, 23$ or $31$) we associate
an ordered basis of flat sections:
\begin{equation}
\CB(E_{ij},t):=\left( s_i (s_j\wedge s_k), s_j (s_k \wedge s_i)
\right)
\end{equation}
Here we have used the decoration to choose $s_i, s_j, s_k$
associated with each vertex, and $i,j,k$ is in counterclockwise
order. Of course, the decoration only defines each $s_i$ up to
scale, so we must make an \emph{ad hoc} choice of three scales.
Notice that different choices just change the basis $\CB(E_{ij},t)$
by an overall scalar.

\instwofigs{2.1in}{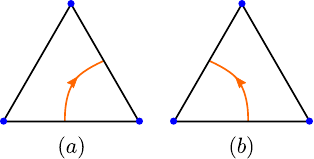}{The right and
left turns, given by simple matrices.}{1.6in}{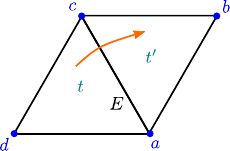}{Passing through an edge
$E$ between triangles $t$ and $t'$.}

Now, $s_i,s_j, s_k$ are initially defined in cut neighborhoods of
the vertices $i,j,k$. They can be continued to be single-valued in
the entire triangle $t$, and indeed can be continued into the next
triangle $t'$ met by the path $P$ to be well-defined and
single-valued in the larger region $t\cup t'$ and so on, as long as
the region remains simply connected. We will compute the monodromy
matrix by computing the changes of basis from one such basis to
another around the path $P$. There are two distinct kinds of
change of basis we will need. If our path enters a triangle $t$
through an edge $E$ and leaves triangle $t$ through edge $E'$ then
we will need the change of basis
$
M_{E}^{E'}(t)
$
computing the change of bases associated to the two edges within the
triangle $t$. This is illustrated in Figure
\ref{fig:mono-3}. On the other hand, if the path passes
through an edge $E$ from triangle $t$ to triangle $t'$ then we will
need the change of basis
$
M_{t}^{t'}(E)
$
between the two different bases, as shown in Figure
\ref{fig:mono-2}.

We first show how to compute $M_{E}^{E'}(t)$. Within a triangle $t$
we have the simple relation
\begin{equation}
s_1 (s_2 \wedge s_3) + s_2 (s_3 \wedge s_1)+ s_3 (s_1 \wedge s_2)=0
\end{equation}
Using this relation it is then trivial to compute the change of
basis
\begin{equation}
\CB(E,t) = \CB(E',t) M_{E}^{E'}(t)
\end{equation}
where
\begin{equation}\label{eq:left-right-turn}
M_{E}^{E'}(t) = \begin{cases} \begin{pmatrix} 0 & -1 \\ 1 & -1 \\
\end{pmatrix}& \langle E,E'\rangle = +1  \\
\begin{pmatrix} -1 &  1 \\ -1 & 0 \\
\end{pmatrix}& \langle E,E'\rangle = -1  \\
\end{cases}
\end{equation}
Note that $\det M_{E}^{E'}(t)=1$, and that the case $\langle E, E' \rangle = 1$
corresponds to a \ti{right} turn, labeled a) in Figure \ref{fig:mono-3}.  Note
also that the right-turn matrix is the inverse of the left-turn.

On the other hand, if $t\cup t'$ make a quadrilateral $Q_E$ with
vertices $a,b,c,d$ in counterclockwise order, (so $t$ has vertices
$acd$ and $t'$ has vertices $abc$, and  $E$ is the edge $ac$, as in
Figure \ref{fig:mono-2}) then
\begin{equation}
\CB(E,t') = \CB(E,t)\begin{pmatrix} 0 & \frac{s_b \wedge s_c}{s_c
\wedge s_d} \\ \frac{s_a \wedge s_b}{s_d \wedge s_a} & 0 \\
\end{pmatrix}
\end{equation}
and hence
\begin{equation}\label{eq:M-edge-matrix}
M_{t}^{t'}(E) =\frac{s_b \wedge s_c}{s_c \wedge s_d} \begin{pmatrix}
0 & 1 \\ - \CX_E^T & 0 \\ \end{pmatrix}
\end{equation}
In the argument below it will be useful to denote
\begin{equation}
\hat M_{t}^{t'}(E) :=
\begin{pmatrix} 0 & 1 \\ - \CX_E^T & 0 \\ \end{pmatrix}
\end{equation}
Note that $\det \hat M_{t}^{t'}(E) = \CX_E^T$.

\insfig{3.8in}{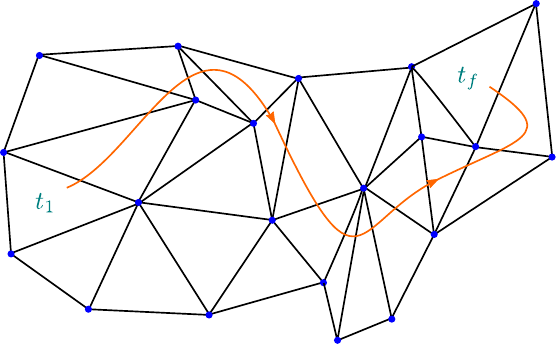}{We compute the monodromy along the orange
path by a series of basis changes within and between the successive
triangles.}

Now consider a path through several triangles as in Figure
\ref{fig:mono-1}. After parallel transport of the basis
$\CB(E_1,t_1)$ along an open path to a basis $\tilde\CB(E_1,t_1)$
which is single-valued throughout a simply connected region
assembled by the successive triangles we have the relation in the
final triangle $t_f$ of the open path:
\begin{equation}
\CB(E_f, t_f) =\tilde\CB(E_1,t_1) M_{E_1}^{E_2}(t_1)
M_{t_1}^{t_2}(E_2) M_{E_2}^{E_3}(t_2) \cdots M_{t_{f-1}}^{t_f}(E_f)
\end{equation}
Now suppose we close the path, taking $t_f=t_1$ and $E_f = E_1$.
Then $\tilde\CB(E_1,t_1) = \CB(E_1,t_1) \cdot \CU(P)$ where
$\CU(P)$ is the monodromy matrix for the closed path $P$
expressed in the basis $\CB(E_1,t_1)$. Since this basis is
well-defined up to scale and $\CU(P)$ changes by conjugation under
any change of basis, it follows that $\CU(P)$ is a well-defined
$SL(2,\IC)$ matrix, independent of the choices of scale of
$s_i,s_j,s_k$ made for $\CB(E_1,t_1)$. Therefore $\CU^{-1}$ is the
above product of $M$-matrices. Now, becuase of the scalar prefactor
$\frac{s_b \wedge s_c}{s_c \wedge s_d}$ in equation
(\ref{eq:M-edge-matrix}) it is not obvious that the monodromy matrix
$\CU(P)$ can be expressed in terms of the edge coordinates.
However, we can use the following trick. The matrix $\CU(P)^{-1}
$, is up to a scalar factor, the same as the product where we
replace the factors $M_{t}^{t'}(E)$ by $\hat M_{t}^{t'}(E)$. That
is, we have
\begin{equation}
\CU(P)^{-1} = \kappa M_{E_1}^{E_2}(t_1) \hat M_{t_1}^{t_2}(E_2)
M_{E_2}^{E_3}(t_2) \cdots \hat M_{t_{f-1}}^{t_f}(E_f)
\end{equation}
for some scalar factor $\kappa$. We can determine $\kappa^2$ by
taking a determinant:
\begin{equation}
\kappa^{-2}  =   \CX_{E_2}^T \cdots \CX_{E_f}^T
\end{equation}
thus, $\kappa$ involves a square-root of the product of $\CX_E^T$
along the edges met by the path $P$. In order to define this
square root we may invoke a kind of nonabelian version of Stokes
theorem: if $C$ is a punctured $\IC \IP^1$ then the closed curve $P$
bounds a region $\CR$, and by a generalization of Section
\ref{sec:monodromies}, we have
\begin{equation}
  \CX_{E_2}^T \cdots \CX_{E_f}^T \prod_{E\in {\rm
Int}(\CR)} (\CX_E^T)^2 = \prod_{v\in {\rm Int}(\CR)} (\mu_v)^2.
\end{equation}
This formula allows us to choose the square root in a canonical way, namely
\begin{equation}
\kappa = (  \CX_{E_2}^T \cdots \CX_{E_f}^T)^{-1/2} =
\frac{\prod_{E\in {\rm Int}(\CR)} \CX_E^T}{\prod_{v\in {\rm
Int}(\CR)} \mu_v}.
\end{equation}

We conclude with several remarks:

\begin{enumerate}

\item In the above discussion we have assumed that all the triangles
are nondegenerate. If there is a degenerate configuration such as
shown in Figure \ref{fig:degenerate-face}  then we think of the edge
$E$ as being doubled. Thus, we choose decorations $s_1$ at the
central vertex and $s_2$ at the outer vertex and use the triplet of
sections $s_1,M s_2, s_2$ in the above construction, where $M$ is
the clockwise monodromy around the central vertex. In particular we
regard $E$ as having two ``sides'' with basis $\CB_{\rm left} = \{
s_1 (M s_2 \wedge s_2), M s_2 ( s_2 \wedge s_1) \}$ on the left and
$\CB_{\rm right} = \{ s_2 (s_1 \wedge M s_2),s_1 (M s_2 \wedge s_2)
\}$ on the right. For a path that winds around avoiding the edge $E$
we use an edge-to-edge transformation to relate these two bases. For
a path that goes from left to right through the edge $E$ we
analytically continue $\CB_{\rm left}$ through the edge with
$M^{-1}$ and obtain the change of basis
\begin{equation}
\CB_{\rm right} = \tilde \CB_{\rm left} \begin{pmatrix} 0 & \mu \\ -
\mu^{-1} & 0 \\ \end{pmatrix}.
\end{equation}
A similar remark applies to the outer degenerate triangle in Figure
\ref{fig:degenerate-quadrilateral}.

\item In generalizing this result to the case where $C$ is a
punctured surface of higher genus, we need to choose a system
of square-roots $(\CX_{E_2}^T \cdots \CX_{E_f}^T)^{-1/2}$. It
suffices to provide them for a basis of nonbounding cycles, and hence
the set of such choices is a torsor for $H_1(\bar C, \IZ_2)$.

\item The monodromy matrix elements are evidently
\emph{polynomials}  in the $\CX_{E}^T$ (divided by factors of
$\mu_v$), and in particular, the traces of the monodromy matrices
for closed paths $P$  can be written as
\begin{equation}\label{eq:UV-IR-monodromy}
{\Tr}\ \CU(P) = \sum_\gamma a_\gamma(P) \CX_\gamma.
\end{equation}
The geometrical coefficients $a_\gamma(P)$ depend on the
triangulation and in particular depend on $\vartheta$. Notice that
the left-hand-side of \eqref{eq:UV-IR-monodromy} is independent of
$\vartheta$, so that $a_{\gamma}(P)$ and $\CX_\gamma^\vartheta$
must have compensating discontinuities.  In the 5d Yang-Mills
theory, ${\Tr}\ \CU(P)$ is the vev of a supersymmetric Wilson line
operator for $\CA$ along a path $P$ in $C$. Clearly, it originates
from a surface operator in the nonabelian six-dimensional $(2,0)$
theory, wrapping a one-cycle in $C$ times $S^1_R$. This represents a
line operator in the ultraviolet four dimensional gauge theory.
Depending on the cycle $\gamma$, this line operator can be a Wilson
loop operator for the four dimensional gauge fields, an 't Hooft
operator or a more general Wilson-'t Hooft operator. With this fact
in mind, we interpret equation (\ref{eq:UV-IR-monodromy}) as a
relation between the line operators defined in the UV theory and
their decomposition into a different basis of line operators which
carry quantum numbers for the IR abelian theory. Indeed, the
$\CX_{\gamma}$ are expectation values of line operators wrapped
around $S^1_R$, defined in the IR four-dimensional $\CN=2$ theory,
(or equivalently, surface operators in the IR abelian $(2,0)$ theory
on $\IR^{1,2} \times S^1_R\times \Sigma$.)

There is a $\zeta$-dependent nilpotent supersymmetry operator
$Q_\zeta$ which annihilates these operators, and hence they are
analogues of the supersymmetric 't Hooft-Wilson line operators
similar to those discussed in
\cite{Maldacena:1998im,Rey:1998ik,geom-lang-n4}. With this
interpretation, equation (\ref{eq:UV-IR-monodromy}) encodes some
interesting physics, which we hope to write about on another
occasion.

\item There is a subtle point regarding the distinction
    between the moduli space of $SL(2,\IC)$ and
    $PSL(2,\IC)$ connections. In this paper we are
    considering connections on a trivialized $SL(2,\IC)$
    principal bundle over $C$, or on its quotient by the
    center, which is a trivialized $PSL(2,\IC)$ bundle.
    There is a one-one correspondence between
    connections on these two bundles. However, there are
    $PSL(2,\IC)$ gauge transformations which do not lift to
    $SL(2,\IC)$ gauge transformations, so the moduli space
    of $SL(2,\IC)$ connections is a discrete cover of the
    moduli space of $PSL(2,\IC)$ connections. The
    $\CX_{\gamma}$ are defined on the quotient moduli space
    of $PSL(2,\IC)$ connections, and can be pulled back to
    the moduli space of $SL(2,\IC)$ connections.
    If we think of the moduli space as determined by specifying
    the monodromy eigenvalues $\mu_i$ and not
    just $\mu_i^2$, then, when $C = \IC \IP^1$,
    the $\CX_E^T$ uniquely determine a flat
    $SL(2,\IC)$ connection modulo $SL(2,\IC)$ gauge
    invariance. For $C$ of higher genus, the $\CX_E^T$ do not quite separate
    points on the moduli space of $SL(2,\IC)$ connections, even if we
    have specified the $\mu_i$:
    to determine the connection from
    the $\CX_E^T$, we also need to specify the
    system of square roots $(\CX_{E_2}^T \cdots
    \CX_{E_f}^T)^{-1/2}$. The existence of these choices
    raises the question of what is the
    ``correct'' moduli space of the four
    dimensional gauge theory on $\IR^3 \times S^1$. Should it be the
    moduli space of $SL(2,\IC)$ connections, of
    $PSL(2,\IC)$ connections or of some intermediate cover of
    the moduli space of $PSL(2,\IC)$ connections?
    Interestingly, defining the appropriate four dimensional gauge
    theory also involves a discrete choice which might correspond
    to the above
    ambiguity. As described in \cite{Gaiotto:2009we},
    the appropriate four-dimensional gauge theories are
    generalized quivers of $SU(2)$ gauge groups. A subgroup
    ${\cal C}$ of the center of the gauge groups acts
    trivially on the matter fields, and there is a freedom
    to quotient the gauge group by any subroup of ${\cal
    C}$. A comparison with the well known case of the
    $\CN=2^*$ $SU(2)$ gauge theory, associated to the Hitchin
    equations on a once-punctured torus, should be sufficient to
    establish the full dictionary, but we will not pursue this
    matter further here.

\end{enumerate}

\section{Computing the Hamiltonian flows} \label{app:poorman}

We begin with an auxiliary computation.
Choose a path $\gamma$, parameterized as $z(t)$, $0\leq t \leq 1$.  Define $z_i = z(0)$ and $z_j = z(1)$.
For the moment we take them to be regular points.  Let $s_i$ and $s_j$ denote two flat sections along $\gamma$,
with fixed boundary conditions $s_i(t = 0) = d_i$ and
$s_j(t = 1) = d_j$.
Given these two sections we may consider the section $s_i \wedge s_j$ of $\wedge^2(V)$.  For convenience we also choose a
flat section $\eps$ of $\wedge^2(V^*)$ (volume form); this may always be done since $\CA$ is $sl(2)$-valued.
If we normalize by contracting with $\eps$, we can regard $s_i \wedge s_j$ as a number, and we do this in what follows;
in a local frame for $V$,
\begin{equation}
s_i \wedge s_j = \epsilon_{ab} s_i^a(t) s_j^b(t).
\end{equation}
Since $\epsilon$ and $s_{i,j}$ are all flat, $s_i \wedge s_j$ is independent of $t$.
In what follows it is also convenient to use
$\epsilon$ to relate the section $s_i^a$ of $V$ to a dual section
$s_{ia}$ of $V^*$.  In a local frame
\begin{equation}
s_{ia}(t) = \epsilon_{ab} s_i^b(t).
\end{equation}

Now define a function on the space $\CN$ of ``all gauge fields'' by
\begin{equation}
\CO_{\gamma, d_i, d_j}:= \log( s_i \wedge s_j ).
\end{equation}
We want to compute $\{\CO_{\gamma, d_i, d_j}, \CA^a_{\lambda b}(z)
\}$. We choose the sign of the Poisson bracket associated to
\eqref{eq:symplectic} to be
\begin{equation} \label{eq:spoisson}
\{ \CO ,  \CA^a_{\lambda b}(z) \} = \epsilon_{\lambda\mu}
\frac{\delta
 \CO}{\delta \CA^b_{\mu a}(z)}.
\end{equation}
Choose some $t_*$ with $z(t_*) \neq z$, and evaluate $\CO_{\gamma, d_i, d_j}$ at this point;
then \eqref{eq:spoisson} becomes
\begin{equation}
\{\CO_{\gamma, d_i, d_j}, \CA^a_{\lambda b}(z) \}=
\frac{\epsilon_{\lambda\mu} }{s_i\wedge s_j} \epsilon_{a'b'}\left(
\frac{\delta s_i^{a'}(t_*)}{\delta \CA^b_{\mu a}(z) } s_j^{b'}(t_*)
+ s_i^{a'}(t_*) \frac{\delta s_j^{b'}(t_*)}{\delta \CA^b_{\mu a}(z)
} \right).
\end{equation}
To evaluate this we first introduce a bit of notation for the parallel transport:
write $\CA(t) = \CA_\mu(t) \dot z^\mu(t)$, and let $U(t_1, t_2)$ be the parallel-transport matrix
from $t_2$ to $t_1$, obeying
\begin{align}
\frac{d}{dt_1} U(t_1,t_2) &= - \CA(t_1) U(t_1,t_2), \\
\frac{d}{dt_2} U(t_1,t_2) &= U(t_1,t_2) \CA(t_2),
\end{align}
and the boundary condition $U(t,t)=\id$.
Then we have
\begin{equation}
\frac{\delta}{\delta \CA^a_{\mu b}(z)} U(t_1,t_2)^c_{~d} =
\int_{t_1}^{t_2} dt \delta^{(2)}(z-z(t)) \dot z^\mu(t)
U(t_1,t)^c_{~a} U(t,t_2)^b_{~d}.
\end{equation}
It follows that the variations of $s_{i,j}$ with respect to $\CA$ are given by
\begin{align}
 \frac{\delta
s_i^{a'}(t_*)}{\delta \CA^b_{\mu a}(z) } &= - \int_0^{t_*} dt\,
\delta^{(2)}(z-z(t)) \dot z^\mu(t) \left[ U(t_*,t)^{a'}_{~b}
s_i^a(t)- \half \delta^{a}_{~b} s_i^{a'}(t_*) \right], \\
 \frac{\delta
s_j^{b'}(t_*)}{\delta \CA^b_{\mu a}(z) } &= \int_{t_*}^1 dt\,
\delta^{(2)}(z-z(t)) \dot z^\mu(t) \left[ U(t_*,t)^{b'}_{~b}
s_j^a(t)-\half \delta^{a}_{~b} s_j^{b'}(t_*) \right].
\end{align}
(The subtraction term arises because we are varying $\CA$
within the space of $sl(2)$ connections, not $gl(2)$
connections, so we should project out the trace.)
Now using
\begin{equation}
\label{eq:rrrng}
 s_i^a(t) s_{jb}(t) - \half \delta^{a}_{~b} s_i \wedge s_j = \half
 (s_{ib} s^a_{j} + s_{jb} s_i^a )(t)
\end{equation}
(proven by contracting the dual
index with a basis $s_i^a, s_j^a$), we arrive at
\begin{equation}\label{eq:poormandeltaA}
\{\CO_{\gamma, d_i, d_j}, \CA^a_{\lambda b}(z) \} =  \half
\epsilon_{\lambda\mu}
 \int_{0}^1 dt\,
\delta^{(2)}(z-z(t)) \dot z^\mu(t)  M(s_i,s_j)^a_{~b}(t),
\end{equation}
where we defined the operator
\begin{equation}
M(s_i,s_j)^a_{~b}(t) := \frac{s_{ib}(t) s^a_{j}(t) + s_{jb}(t)
s_i^a(t) }{s_i\wedge s_j}.
\end{equation}

Now let us consider two paths. The first $z_1(t)\in \gamma_1$ is a
path from $z_i$ to $z_j$ such as we have been considering thus far.
The second $z_2(t) \in \gamma_2$ intersects $\gamma_1$ transversally
(or not at all).  Now consider a vector $v^a(t)$
parallel-transported along $\gamma_2$, with a fixed initial boundary
condition.  We want to compute $\{\CO_{\gamma_1, d_i, d_j},
v^a(t_*) \}$, given by
\begin{equation}
\{ \CO_{\gamma_1, d_i, d_j}, v^a(t_*) \} = - \int_0^{t_*} dt \, \tilde U^{a}_{~b}(t_*, t)
\{\CO_{\gamma_1, d_i, d_j}, \CA(t)^{b}_{~c} \} v^c(t),
\end{equation}
where $\tilde U$ is the parallel transport along $\gamma_2$. Using
\eqref{eq:poormandeltaA} we find
\begin{multline}
\{ \CO_{\gamma_1, d_i, d_j}, v^a(t_*) \} =  \\
\half \int_0^1 dt_1 \int_0^{t_*} dt_2\,
\delta^{(2)}(z_2(t_2)-z_1(t_1))\epsilon_{\lambda\mu} \dot
z_2^\lambda(t_2) \dot z_1^\mu(t_1) \tilde{U}^a_{~b}(t_*,t_2)
M(s_i,s_j)^{b}_{~c}(t_1) v^c(t_2).
\end{multline}
Now note that
\begin{equation}
\delta^{(2)}(z_2(t_2)-z_1(t_1))\epsilon_{\lambda\mu} \dot
z_2^\lambda(t_2) \dot z_1^\mu(t_1)  = -(\gamma_1\cap
\gamma_2)\delta(t_1-t_1^{int}) \delta(t_2-t_2^{int})
\end{equation}
where $t_i^{int}$ is the value of $t$ at which the curves intersect,
and $\cap$ denotes the oriented intersection number.
Thus we get
\begin{equation} \label{eq:flow-single}
\{ \CO_{\gamma_1, d_i, d_j}, v(t_*) \} =  \half (\gamma_1 \cap
\gamma_2) \left( \frac{s_i \wedge v}{s_i \wedge s_j}(t_2^{int})
s_j(t_*) + \frac{s_j \wedge v}{s_i \wedge s_j}(t_2^{int}) s_i(t_*)
\right) \theta(t_* - t_2^{int}).
\end{equation}
where $s_i, s_j$ have been parallel transported along $\gamma_2$
from the intersection point to $t_*$.

Our goal in this Appendix was
to describe the flow on $\CM$ generated by a function $\log
\CX^T_E$. To lift this function up to $\CN$ involves combining four
functions $\CO_{\gamma, d_i, d_j}$ where we identify $\gamma$
successively with the four edges of the quadrilateral $Q_E$, and
take $d_i, d_j$ to agree with our choices of decoration at the
vertices. Combining \eqref{eq:flow-single} for the four edges gives
the flow we described in Section \ref{sec:ham-flows}.

Strictly speaking we actually have to regularize by deforming
the quadrilateral slightly,
pushing its vertices away from the singular points, and then take a limit where the vertices approach the singular points.
The regularized functions $\CO_{\gamma, d_i, d_j}$ are not gauge invariant (because $d_{i,j}$ are not),
but their combination does become gauge invariant in the limit:  recall that
the gauge group $\mathcal G$ only includes transformations which at the singularities are
restricted to the maximal torus $\IC^\times \subset SL(2,\IC)$; this group preserves $d_{i,j}$
up to overall rescaling, and that overall rescaling cancels out when we sum over the four edges.

\section{WKB error analysis} \label{app:wkb-error}

In this Appendix we consider the propagation of the exponentially growing flat sections
along a WKB curve.  Our goal is to see that the $\zeta \to 0$ asymptotics of
this propagation are just obtained by integrating the eigenvalue $\lambda$ of the Higgs
field $\varphi$.

More precisely:  choose a gauge in which $\varphi$ is diagonal,
\begin{equation}
\varphi = \begin{pmatrix} \lambda & 0 \\ 0 & -\lambda \end{pmatrix}.
\end{equation}
Then let $z(t)$ be a WKB curve with phase $\vartheta$, and let $\zeta \in \IH_\vartheta$,
so that $\Re \lambda_z z'(t) / \zeta < 0$.  Let $s$ be a flat section with
\begin{equation}
s(z(0)) = \begin{pmatrix} 1 \\ 0 \end{pmatrix}.
\end{equation}
The statement of the WKB approximation is that as $\zeta \to 0$ we have
\begin{equation} \label{eq:wkb}
s(z(t)) \sim c(t) \begin{pmatrix} e^{- \frac{R}{\zeta} \int_{z(0)}^{z(t)} \lambda} \\ 0 \end{pmatrix},
\end{equation}
for some function $c(t)$ independent of $\zeta$.

To prove \eqref{eq:wkb}, begin by defining the WKB remainder $\psi$ by
\begin{equation} \label{eq:wkb-remainder}
\psi(z) = s(z) \exp \left( \frac{R}{\zeta} \int_{z(0)}^z \lambda \right).
\end{equation}
The flatness equations $(d + \CA) s = 0$ become
\begin{align}
(\partial_{\bar z} + A_{\bar z} + R \zeta \bar\varphi_{\bar z}) \psi &= 0, \\
(\partial_z + A_z + \frac{R}{\zeta} (\varphi_z - \lambda_z \id)) \psi &= 0.
\end{align}
So along the curve $z(t)$ the evolution of $\psi$ is
\begin{equation} \label{eq:reduced-ode}
\left( \frac{d}{dt} + B(t) \right) \psi(z(t)) = 0
\end{equation}
where
\begin{equation}
B(t) = \frac{R}{\zeta} z' \begin{pmatrix} 0 & 0 \\ 0 & - 2 \lambda_z \end{pmatrix} + z' A_z + \bar{z}' (A_{\bar z} + R \zeta \bar\varphi_{\bar z}).
\end{equation}

The desired \eqref{eq:wkb} is equivalent to saying that $\lim_{\zeta \to 0} \psi$ exists
and moreover its second component vanishes.
Why should this be so?
The intuition is that the only term in $B(t)$ that is not finite as $\zeta \to 0$ is the
term $-2 R z' \lambda_z / \zeta$ in the bottom right corner, and the only effect of this term
will be to introduce a factor like $e^{- \# / \zeta}$ (with $\Re \# > 0$) in
the second component of $\psi$, which thus vanishes as $\zeta \to 0$.

To justify this intuition, let $B_0(t) = g(t) B(t) g^{-1}(t)$ be the diagonalization of $B(t)$.
We would like to show that
\begin{equation} \label{eq:wkb-interm}
\lim_{\zeta \to 0} \bigg\lVert{ \Pexp \int_0^t B(t) - \Pexp \int_0^t B_0(t) \bigg\rVert} = 0.
\end{equation}
Having established \eqref{eq:wkb-interm} we may use $B_0$ instead of $B$ to evaluate the
parallel transport of $\psi$ in the $\zeta \to 0$ limit; that would prove the desired \eqref{eq:wkb}.

A direct computation, using
only the fact that $B(t)$ is a $2 \times 2$ matrix for which the real part of the bottom right entry approaches $- \infty$ while all others
remain finite, shows that $\exp B_0(t)$ is bounded
as $\zeta \to 0$ and that we can choose $g(t)$ such that $g(t) \to \id$ as $\zeta \to 0$.
From these two facts \eqref{eq:wkb-interm} follows.
To see this, we first break the interval into small pieces over which $B$ is slowly varying, to
reduce to the case where $B$ is $t$-independent.  In that case \eqref{eq:wkb-interm} reduces to
\begin{equation}
\lim_{\zeta \to 0} \lVert e^{t B} - e^{t B_0} \rVert = 0.
\end{equation}
But this follows from
$B = g  B_0 g^{-1}$, $\lim_{\zeta \to 0} g = \id$, and
the existence of $\lim_{\zeta \to 0} e^{t B_0}$.\footnote{This
is the moment where we use the fact that, as $\zeta \to 0$, the real part of the bottom right entry of $B(t)$
approaches $- \infty$ rather than $+ \infty$.}
This finishes the proof of \eqref{eq:wkb-interm} and hence of \eqref{eq:wkb}.

More precisely, we have shown \eqref{eq:wkb} in the case where the WKB curve begins at
a regular point.
For our application we need to consider the case where the WKB curve begins at a singularity,
located at say $z(0)$.  In that case we clearly cannot hope for \eqref{eq:wkb} to hold on the nose since
$\int_{z(0)}^{z(t)} \lambda$ diverges.  Instead choose
some other function $I(z)$, defined on a neighborhood of the WKB curve (excluding the singularity),
with $d I = \lambda$.  Then an argument similar to the
above shows there is a flat section $s$ which as $\zeta \to 0$ behaves as
\begin{equation} \label{eq:wkb-sing}
s(z(t)) \sim \begin{pmatrix} c(t) e^{- \frac{R}{\zeta} I(z(t))} \\ 0 \end{pmatrix}.
\end{equation}
Moreover, one can take $s$ to be the small flat section associated to the WKB curve $z(t)$ and singularity $z(0)$.

\section{Holomorphic coordinates on multi-center Taub-NUT} \label{app:HoloTN}

We begin with the Gibbons-Hawking ansatz. The Taub-NUT space   $TN$
has a map $\pi: TN \to \IR^3$, with generic fiber a circle, and
metric
\begin{equation}
ds^2 = V^{-1} \Theta^2 + V (d\vec r)^2,
\end{equation}
with $d \Theta = \pi^*( *dV)$. The globally well-defined one-form $\Theta$ is
normalized by $\pi_* \Theta = 4\pi$.

The Taub-NUT centers are at $\vec r_a$ and
\begin{equation}
V = 1 + \sum_a \frac{1}{\vert \vec r - \vec r_a \vert}.
\end{equation}
Let $z=x^1 + i x^2$. Holomorphic functions in one complex structure
are annihilated by
\begin{equation}
\p_{\bar z} - \Theta_{\bar z} \p_\psi
\end{equation}
\begin{equation}
\p_3 + (i V - \Theta_3) \p_\psi
\end{equation}
where $\p_\psi$ is the globally well-defined vector field generating
rotations in the fiber, normalized to $\langle \Theta, \p_\psi
\rangle = 1$.

To write the holomorphic coordinates it is useful to cover the
manifold by patches. Introduce angular coordinates $(\theta_a,
\phi_a)$ associated with each center. We will define patches
$\CU_\epsilon$ where $\epsilon$ is a vector with one component for
each center and $\epsilon_a = \pm 1$. We let $\CU_\epsilon$ be the
set of $\vec r$, $\vec r\not=\vec r_a$ so that $\theta_a \not=
\half(1-\epsilon_a)\pi$. Over each patch we can define a fiber
coordinate $\psi_\epsilon$ of period $4\pi$ so that
\begin{equation}
\Theta = d \psi_\epsilon + A_\epsilon
\end{equation}
\begin{equation}
A_\epsilon = \sum_a \epsilon_a (1 + \epsilon_a\cos\theta_a) d \phi_a
\end{equation}
Since $\Theta$ is globally well-defined one reads off the transition
functions for $\psi_\epsilon$. Now, it is convenient to define:
\begin{equation}
R_{a,\pm}: = \vert \vec r - \vec r_a\vert \pm (x_3 - x_{3,a} )
\end{equation}
Note that $R_{a,+}R_{a,-} = \vert z-z_a\vert^2$. It is
straightforward to verify that
\begin{equation}\label{eq:holoTN1}
\tilde U_\epsilon =   \prod_{\epsilon_a=+1} (R_{a,-})^{-1/2}
\prod_{\epsilon_a = -1} (R_{a,+})^{1/2} e^{\half (i \psi_\epsilon +
x_3)}
\end{equation}
is annihilated by the antiholomorphic vector fields. Holomorphy is
preserved if we multiply (\ref{eq:holoTN1}) by
$\prod_{\epsilon_a=+1}(z-z_a)$, and we do this to  define
\begin{equation}
U_\epsilon = \prod_a (R_{a,+})^{1/2} \prod_{\epsilon_a=+1}
\frac{z-z_a}{\vert z-z_a\vert} e^{\half (i \psi_\epsilon + x_3)}
\end{equation}
One verifies that $U_\epsilon$ is in fact independent of $\epsilon$,
i.e., it is \emph{globally well-defined}.  As we have said, it is
holomorphic on $TN$.  Its divisor is a disjoint union of holomorphic
disks which project to the lines $\theta_a = \pi$. We henceforth
drop the subscript $\epsilon$ and simply write $U$.  Any global
holomorphic function on $TN$ with winding number 1 around the fibers
must be a polynomial in $z$ times $U$.

Now, one may also check that
\begin{equation}
W = \prod_a (z-z_a)/U
\end{equation}
is globally well-defined and holomorphic.  Its divisor is a union of
holomorphic disks projecting to the lines $\theta_a = 0$. One can
write explicitly
\begin{equation}
W = \prod_a (R_{a,-})^{1/2} \prod_{\epsilon_a=-1}\frac{ z-
z_a}{\vert z-z_a\vert} e^{-\half ( i \psi_\epsilon + x_3)}.
\end{equation}

In Section \ref{subsec:WittenConstruction} we consider the limit
$x_{3,a} \to \pm \infty$ at fixed $\vec r$.  Suppose we take $x_{3,a}
\to +\infty$ for $a \in \CR$ and $x_{3,a} \to -\infty$ for $a \in
\CL$. Then $\vec r$ is in the patch with $\epsilon_a = +1$ for $a
\in \CR $ and $\epsilon_a = -1$ for $a\in \CL$. One finds that
\begin{equation}
\tilde U_\epsilon \to \prod_{a\in \CR} (2 x_{3,a})^{-1/2}
\prod_{a\in \CL} (2 \vert x_{3,a}\vert)^{+1/2} e^{\half (i \psi
+x_3)}
\end{equation}
Therefore, to have a good limit, we normalize $U$ so
that
\begin{equation}
U = \prod_{a\in \CR} (z-z_a) \prod_{a\in \CR} \left(
\frac{2x_{3,a}}{R_{a,-}} \right)^{1/2} \prod_{a \in \CL}
\left(\frac{ R_{a,+}}{2\vert x_{3,a}\vert}\right)^{1/2}   e^{\half
(i \psi_\epsilon + x_3)}
\end{equation}
in this distinguished coordinate patch.

Now, we can apply this to the discussion in Section
\ref{subsubsec:LinConfQuivMatt}. We identify the cylindrical
coordinate $t = e^{-\half (i \psi_\epsilon + x_3)}$, and $z$ is
identified with $v$, while $x^3$ is identified with $x^6$.
Therefore, in the limit we have
\begin{align}
U &\to t^{-1} \prod_{a\in \CR} (v-v_a), \\
W &\to t \prod_{a\in \CL} (v-v_a),
\end{align}
in terms of the natural holomorphic coordinates $(t, v)$ on
$T^*\IC^\times$.

\section{Configurations of integers with nonpositive second discrete derivative} \label{app:ConvexKay}

We summarize here some simple observations about the collections of
integers $\{ k_\a \}$, $0 \leq \a \leq n+1$, which arise in the
D4/NS5-brane configurations of Section
\ref{subsec:WittenConstruction}.
According to \eqref{eq:NegBetas} these configurations obey
\begin{equation} \label{eq:NegBetas2}
- 2k_\a + k_{\a+1} + k_{\a-1} + d_\a \leq 0, \qquad\qquad 1\leq \a \leq n.
\end{equation}
Since $d_\a \geq 0$, \eqref{eq:NegBetas2} implies that if we think of $k_\a$ as a
function of $\a$, its second discrete derivative is nonpositive, i.e. its graph is convex.

\insthreefigs{1.15in}{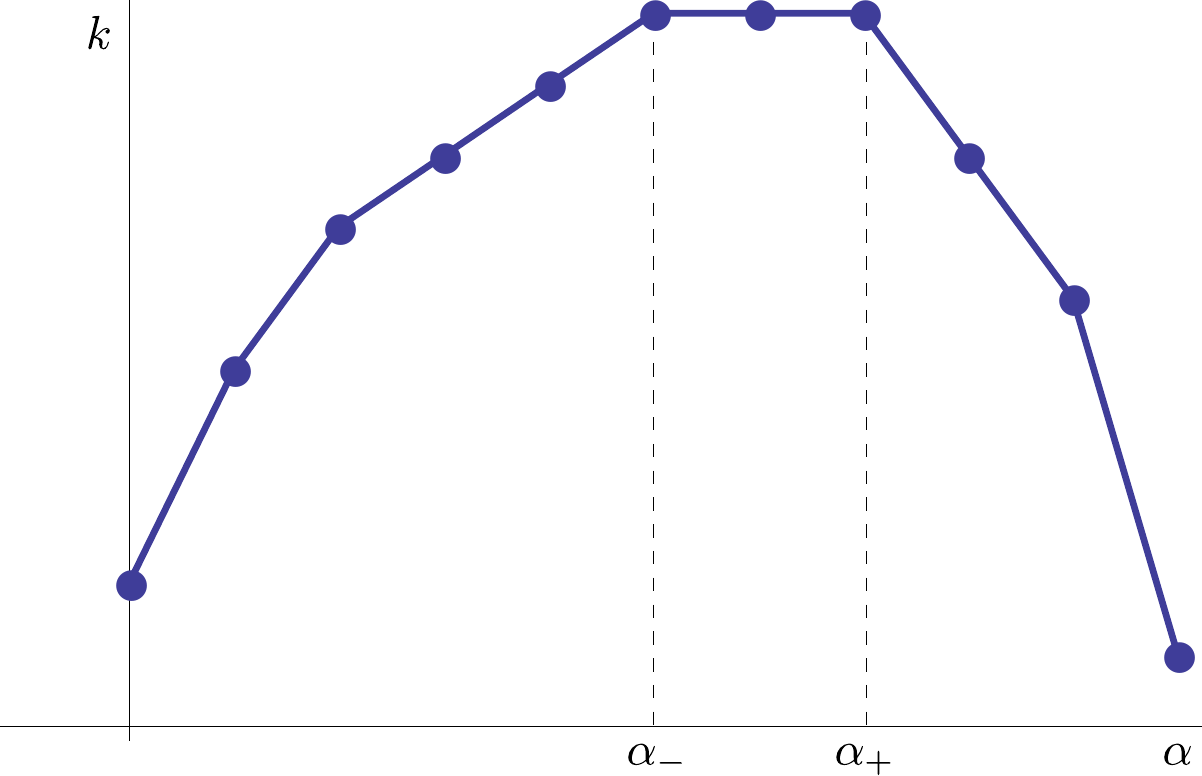}{The typical profile of a set of integers
$k_\a$ with $2k_\a-k_{\a-1} - k_{\a+1}\geq 0$ when $k_1-k_0>0$. }
{1.15in}{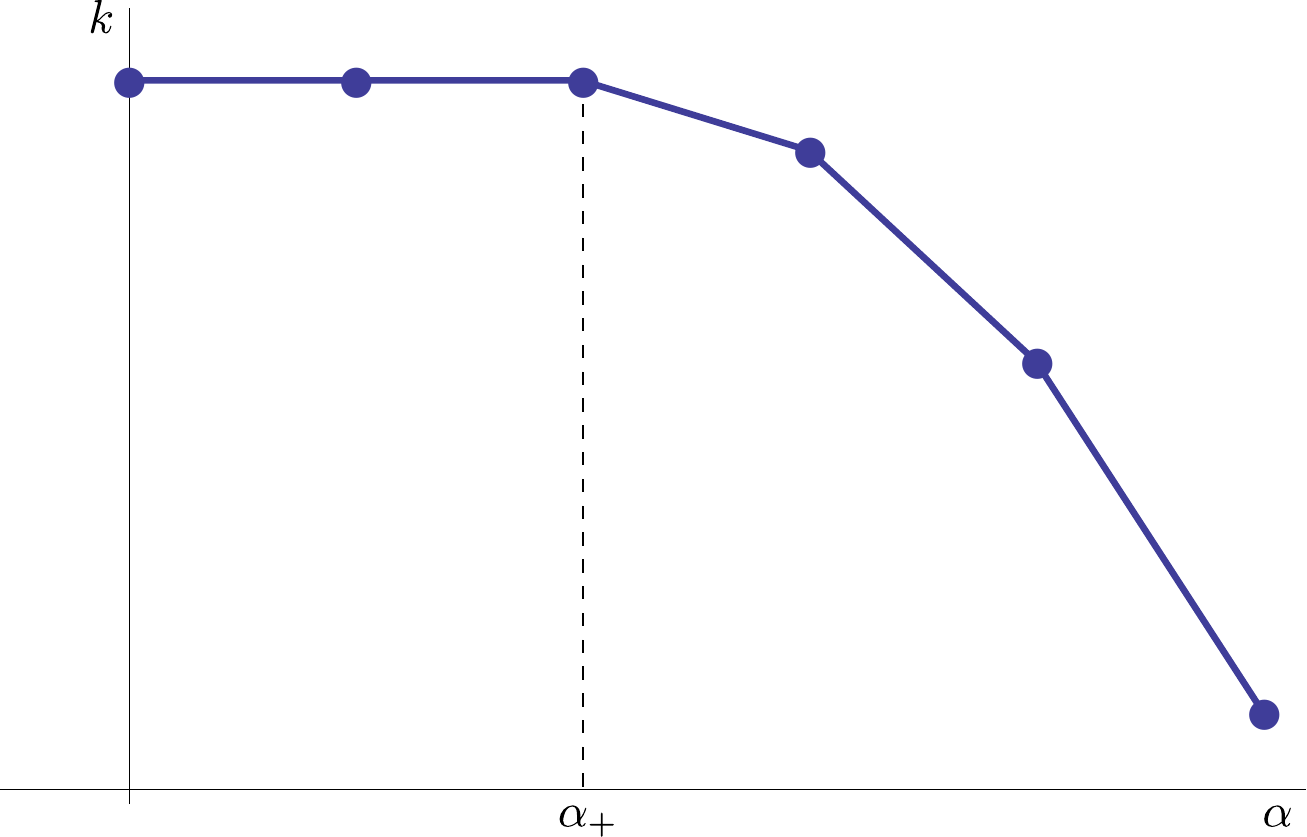}{The typical profile of a set of integers
$k_\a$ with $2k_\a-k_{\a-1} - k_{\a+1}\geq 0$ when $k_1-k_0=0$. }
{1.15in}{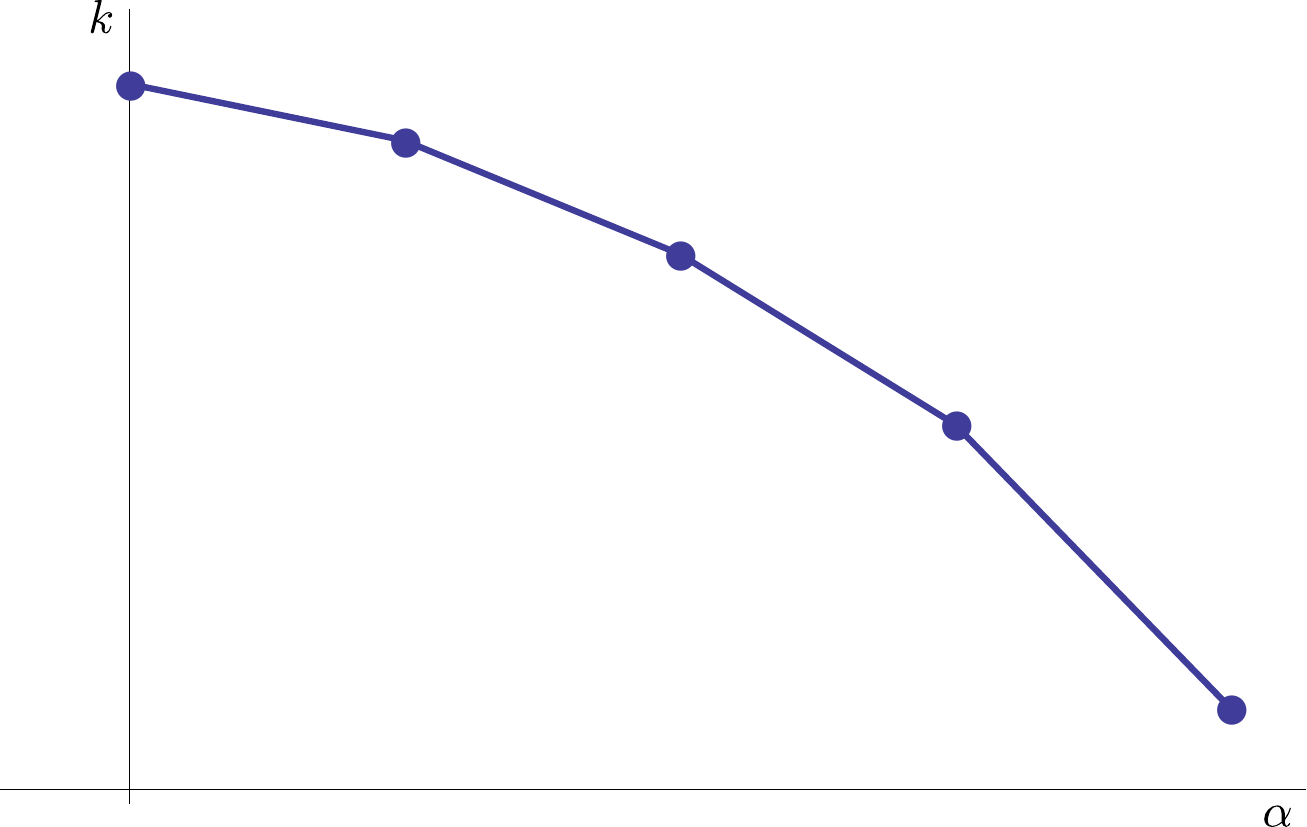}{The typical profile of a set of integers
$k_\a$ with $2k_\a-k_{\a-1} - k_{\a+1}\geq 0$ when $k_1-k_0<0$. }

There are three basic cases to be distinguished.
The first case is $k_1-k_0 > 0$. Then the $k_\a$ are strictly
increasing for $0\leq \a \leq \a_-$, attain a maximal value
$K$ for $\a_-\leq \a \leq \a_+$, and
are strictly decreasing for $\a \geq \a_+$. See Figure \ref{fig:convexk-1}.
If $k_1 - k_0 = 0$ then we have a similar behavior except that $\a_- = 0$, shown in Figure \ref{fig:convexk-2}.
Finally, if $k_1 - k_0 < 0$ then we have $\a_- = \a_+ = 0$, as in Figure \ref{fig:convexk-3}.

There is also an important special case where the second discrete derivative vanishes,
$-2k_\a + k_{\a+1} + k_{\a-1} = 0$ for $1 \leq \a \leq n$.  In this case the graph is just a line,
i.e.
\begin{equation}
k_\a = k_0 + \a r, \qquad\qquad 0 \leq \a \leq n+1.
\end{equation}

\bibliography{wkb-paper}

\end{document}